\documentclass[acmtog,nonacm]{acmart}

\acmSubmissionID{papers915\_s1}

\DeclareGraphicsExtensions{.pdf,.jpg,.jpeg,.png}
\usepackage[ruled,vlined]{algorithm2e} % For algorithms

\SetAlFnt{\small}
\SetAlCapFnt{\small}
\SetAlCapNameFnt{\small}
\SetAlCapHSkip{0pt}

\acmJournal{TOG}
\usepackage{subcaption} % Required for creating subfigures

\usepackage{mathtools}
\usepackage{siunitx}
\usepackage{microtype}
\usepackage{enumitem}
\usepackage{colortbl}

\hypersetup{%
  colorlinks=false,% hyperlinks will be black
  linkbordercolor=red,% hyperlink borders will be red
  pdfborderstyle={/S/U/W 1}% border style will be underline of width 1pt
}

\definecolor{myRed}{rgb}{.5,0..01,0.03}
\definecolor{myGray}{rgb}{.99,0.99,0.9}
\definecolor{myBlue}{rgb}{.3,0.4,0.8}

\newcommand{\Q} {\ensuremath{\mathbf{q}}}
\newcommand{\X} {\ensuremath{\mathbf{x}}}
\newcommand{\Xt} {\ensuremath{\mathbf{x}_t}}
\newcommand{\Y} {\ensuremath{\mathbf{y}}}
\newcommand{\Xtp} {\ensuremath{\mathbf{x}_{t+1}}}
\newcommand{\Xtm} {\ensuremath{\mathbf{x}_{t-1}}}
\newcommand{\Vt} {\ensuremath{\mathbf{v}_t}}
\newcommand{\Vtp} {\ensuremath{\mathbf{v}_{t+1}}}
\let\F\undefined % -- to avoid clash
\newcommand{\F} {\ensuremath{\mathbf{f}}}
\newcommand{\G} {\ensuremath{\mathbf{g}}}
\newcommand{\D} {\ensuremath{\mathbf{d}}}

\newcommand{\Qv} {\ensuremath{\mathbf{q}}}
\newcommand{\Eelast} {\ensuremath{{E}}}
\newcommand{\Eel} {\ensuremath{\mathbf{e}}}
\newcommand{\Eelnf} {\ensuremath{{\mathbf{e}}^{\mathrm{n}}}}
\newcommand{\Eelff} {\ensuremath{{\mathbf{e}}^{\mathrm{f}}}}
\newcommand{\Eela} {\ensuremath{\hat{\mathbf{e}}}}

\newcommand{\Eelaff} {\ensuremath{\hat{\mathbf{e}}^{\mathrm{f}}}}

\newcommand{\Del}[2]{\ensuremath{\frac{\partial #1}{\partial #2}}}
\newcommand{\Delf}[2]{\ensuremath{{\partial #1}/{\partial #2}}}
\newcommand{\Loss}{\ensuremath{\mathcal{L}}}
\newcommand{\Param}{\ensuremath{\theta}}

\newcommand{\Um}{\ensuremath{\mathbf{u}}}

\newcommand{\uC}{\ensuremath{\mu C}}

\newcommand{\mathcolorbox}[2]{\colorbox{#1}{$\displaystyle #2$}}

\SetCommentSty{mycommfont}

\newcommand{\IGNORE}[1] {} % suppressed 

\newcommand{\BZ}[1] {\IGNORE{\textcolor{blue}{Bruce: #1}}}

\newcommand{\KG}[1] {\IGNORE{\textcolor{purple}{Chris: #1}}}

\begin{document}
% Title portion
\title{Implicit-Explicit simulation of Mass-Spring-Charge Systems}

\author{Zhiyuan Zhang$^*$}
\affiliation{
  \institution{University of Edinburgh}
  \country{UK}
}

\author{Krzysztof Grykiel$^*$}
\affiliation{
  \institution{University of Edinburgh}
  \country{UK}
}

\author{Zhaocheng Liu}
\affiliation{
  \institution{University of Edinburgh}
  \country{UK}
}

\author{Stefanos Papanicolopulos}
\affiliation{
  \institution{University of Edinburgh}
  \country{UK}
}

\author{Kartic Subr}
\affiliation{
  \institution{University of Edinburgh}
  \country{UK}
}

\thanks{$^*$Zhiyuan Zhang and Krzysztof Grykiel contributed equally as co-first authors.}

% DO NOT ENTER AUTHOR INFORMATION FOR ANONYMOUS TECHNICAL PAPER SUBMISSIONS TO SIGGRAPH 2019!
%\author{Gang Zhou}
%\orcid{1234-5678-9012-3456}
%\affiliation{%
%  \institution{College of William and Mary}
%  \streetaddress{104 Jamestown Rd}
%  \city{Williamsburg}
%  \state{VA}
%  \postcode{23185}
%\email{gang_zhou@wm.edu}
%\affiliation{%
%  \institution{Inria Paris-Rocquencourt}
%\email{beranger@inria.fr}
%\affiliation{%
% \institution{Rajiv Gandhi University}
%\email{aprna_patel@rguhs.ac.in}
%\affiliation{%
%  \institution{Tsinghua University}
%  \streetaddress{30 Shuangqing Rd}
%  \country{China}
%}
%\email{chan0345@tsinghua.edu.cn}
%\author{Ting Yan}
%\affiliation{%
%  \institution{Eaton Innovation Center}
%  \city{Prague}
%  \country{Czech Republic}}
%\email{yanting02@gmail.com}
%\author{Tian He}
%\affiliation{%
%  \institution{University of Virginia}
%  \department{School of Engineering}
%  \city{Charlottesville}
%  \state{VA}
%  \postcode{22903}
%  \country{USA}
%}
%\affiliation{%
%  \institution{University of Minnesota}
%\email{tinghe@uva.edu}
%\author{Chengdu Huang}
%\author{John A. Stankovic}
%\author{Tarek F. Abdelzaher}
%\affiliation{%
%  \institution{University of Virginia}
%  \department{School of Engineering}
%  \city{Charlottesville}
%  \state{VA}
%  \postcode{22903}
%  \country{USA}
%}

%\renewcommand\shortauthors{Zhou, G. et al}

\begin{teaserfigure}
	\begin{center}		
     \begin{tabular}{ccccc}
             \includegraphics[width=0.18\columnwidth]{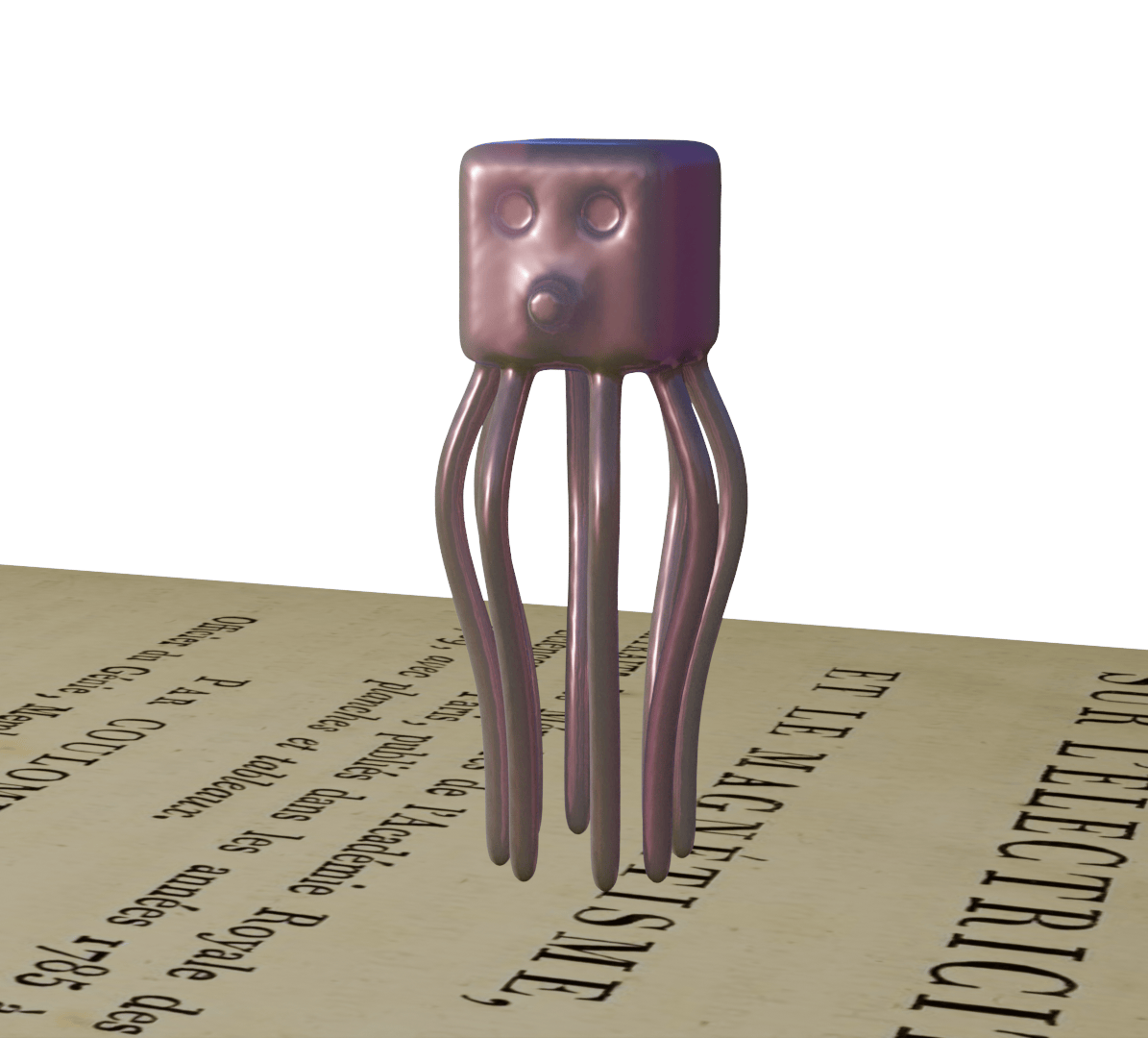} &
             \includegraphics[width=0.18\columnwidth]{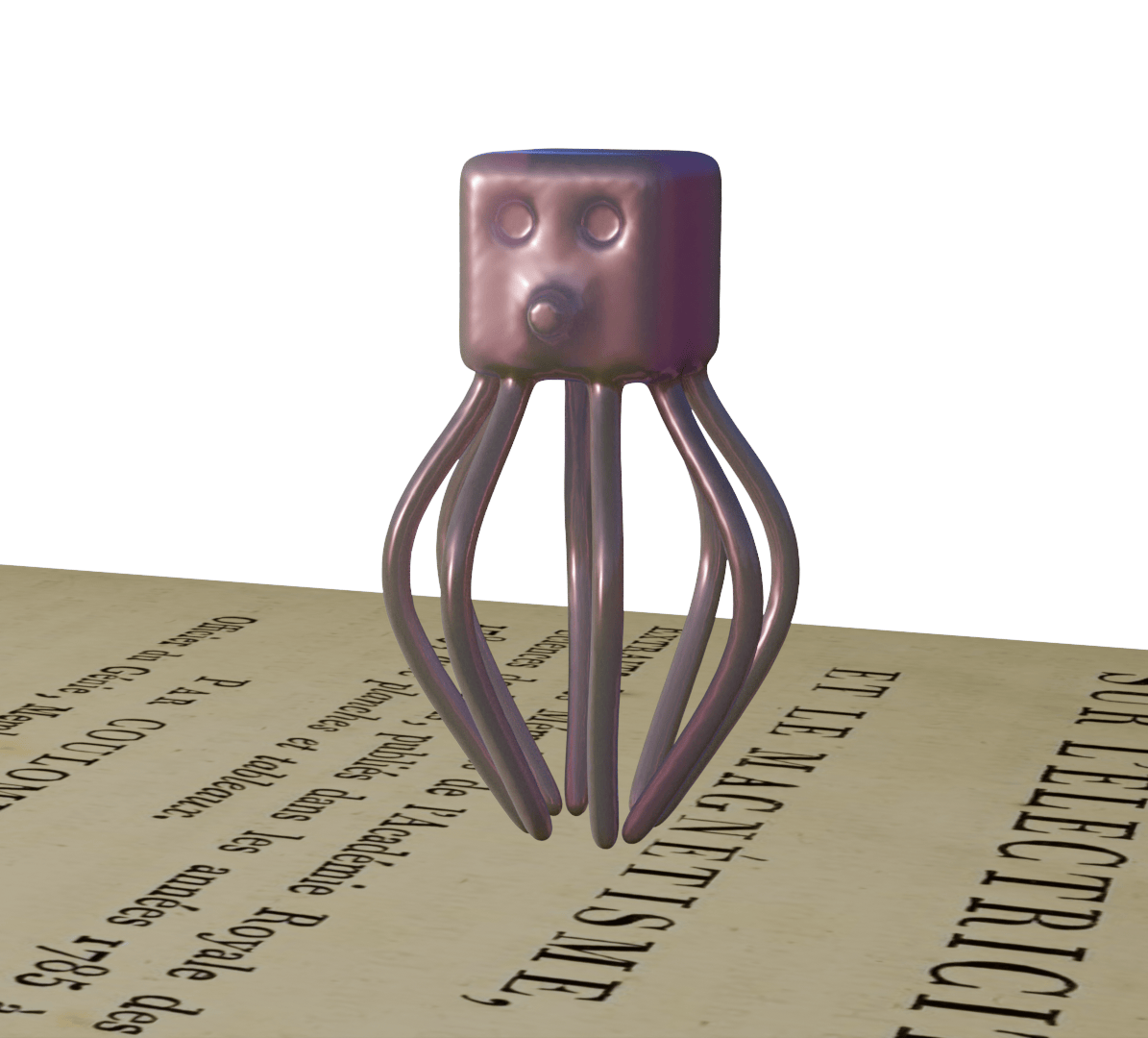} &
             \includegraphics[width=0.18\columnwidth]{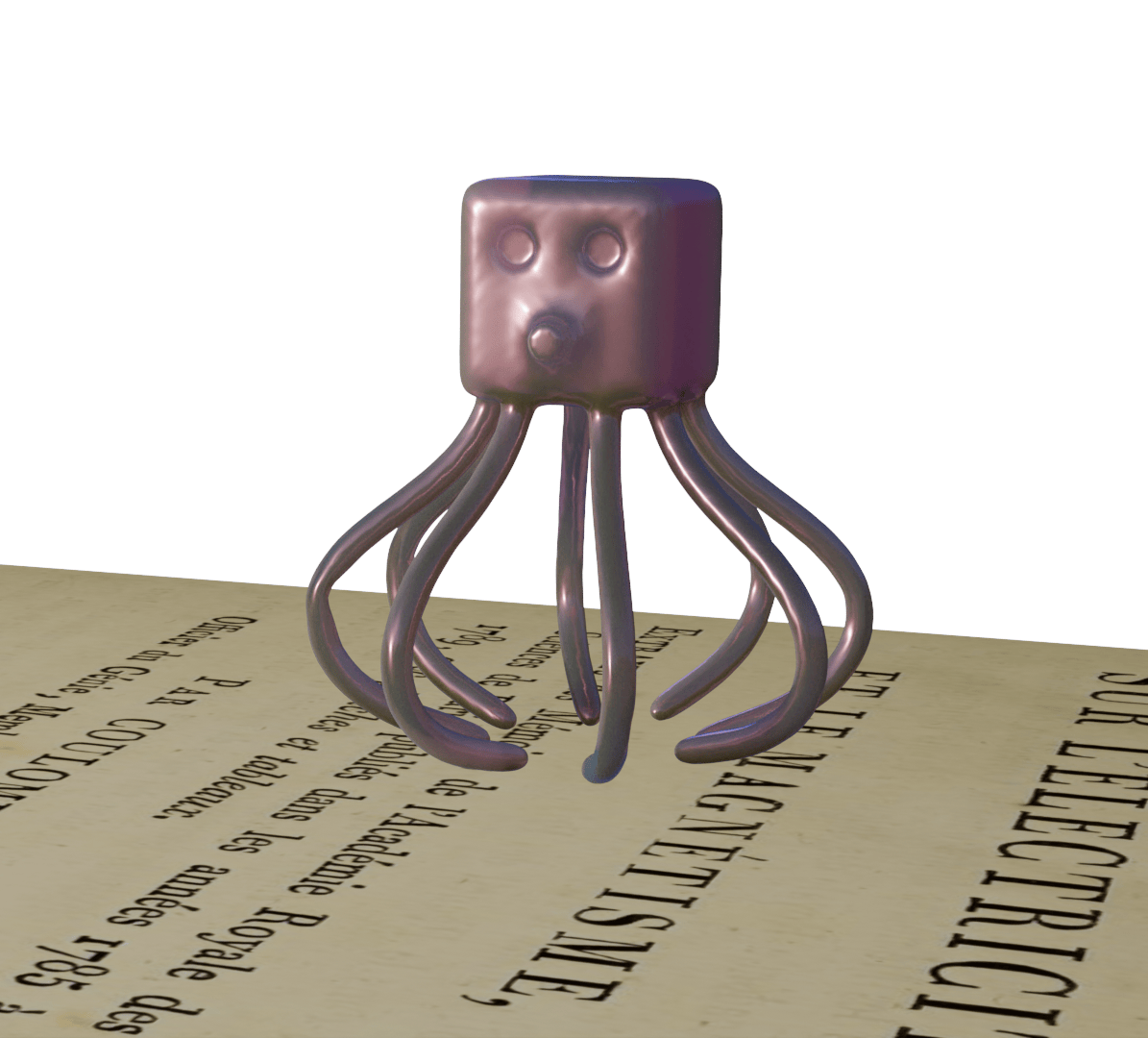} &
             \includegraphics[width=0.18\columnwidth]{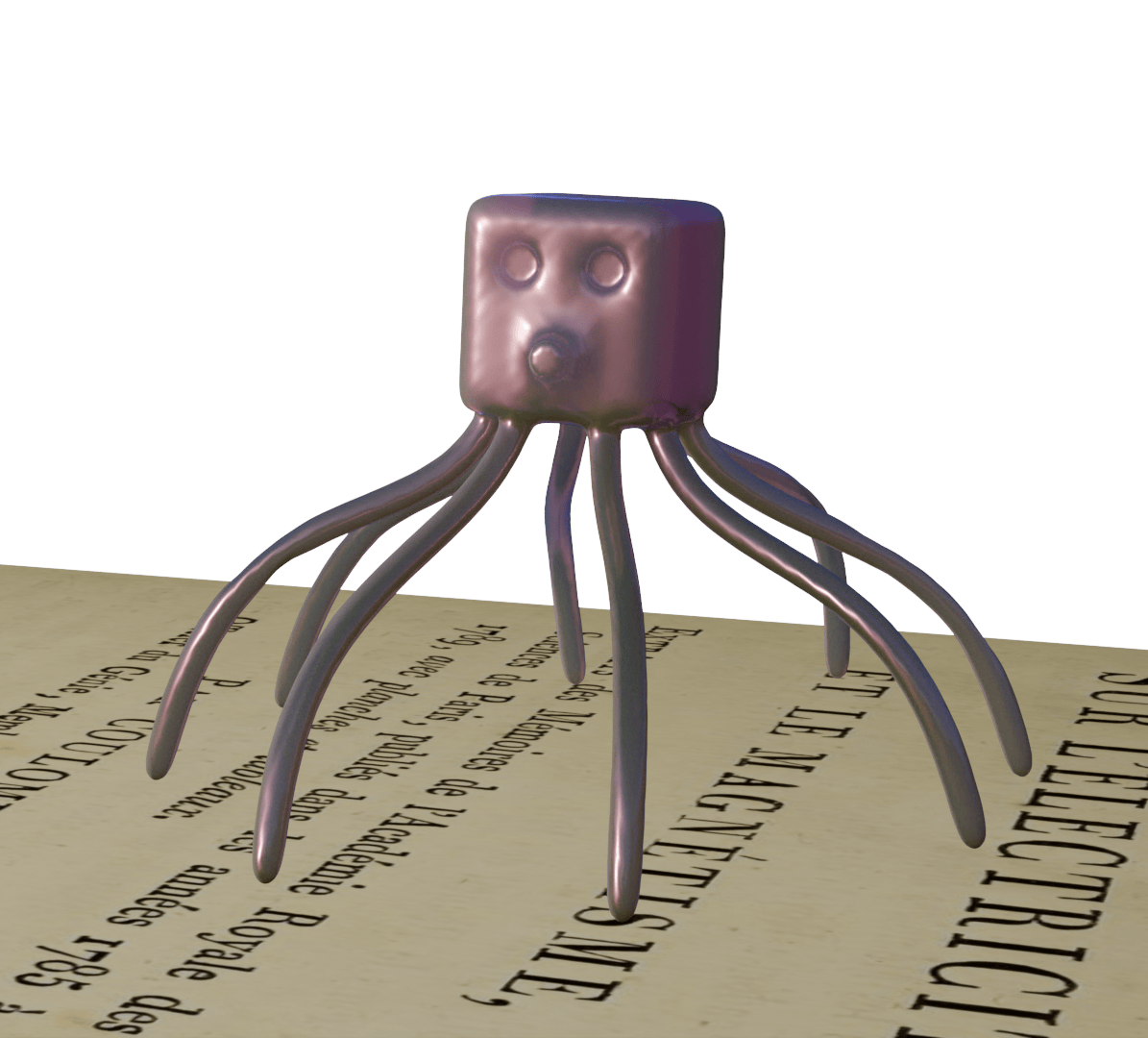} &
             \includegraphics[width=0.18\columnwidth]{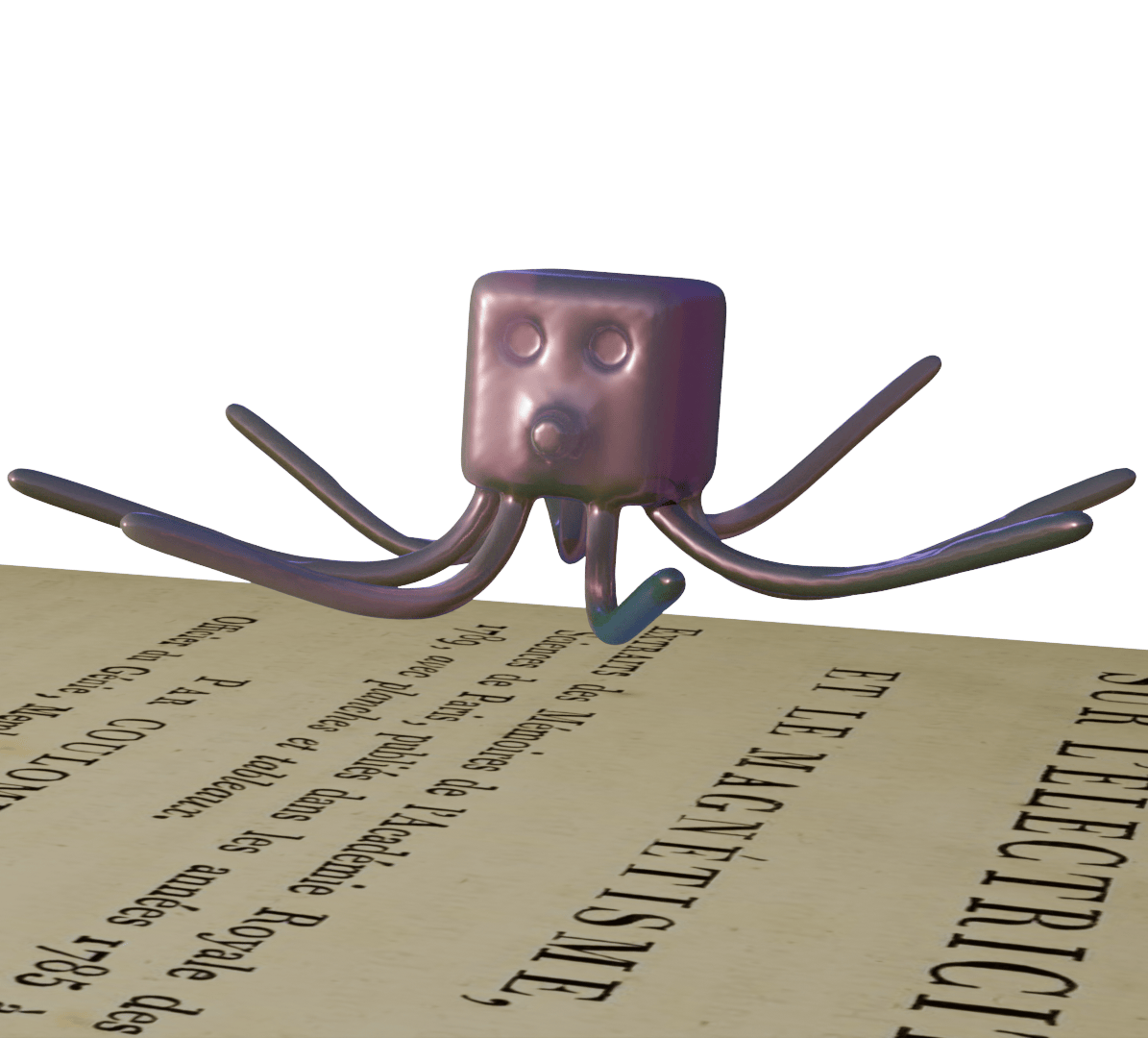} 
     \end{tabular}
	\caption{\label{fig:teaser} This paper presents charged point-masses connected via  springs as a model for computer animation These frames are shown from an example animation (please see accompanying video) created easily and intuitively by key framing charge values of groups of vertices on the octopus' tentacles.}
	\end{center}
\end{teaserfigure}

\begin{abstract}
The simulation of dynamically deforming objects plays a vital role in computer animation. The toolkit for artistic animation of 3D objects is rich with representations such as mass-spring systems, magnetic particles and volumetric finite-elements. We propose to augment this array of tools with electrostatic forces due to charged particles. Specifically, we imbue the masses in mass-spring systems with electrostatic charge and call these \emph{charged mass-spring systems}. We design a simple simulation method that combines an efficient algorithm for calculating pairwise interactions between charges and an implicit-explicit integrator that combines elastic and electrostatic potentials. We perform quantitative evaluations to validate stability and low error with our method for coarse simulation time-steps. We demonstrate that despite its simplicity, charged mass spring systems intuitively enable animation effects that could otherwise be cumbersome to author.
\end{abstract}

    




\maketitle
% temperarely use single column for drafting

\section{Introduction}
Modelling the deformation of objects is an established problem that has received considerable attention. While much of the relevant physics was developed centuries ago, a number of open questions remain unanswered with respect to efficient and scalable computational models for discretization and simulation. Deformations can be approximated at various levels of abstraction ranging from finite element methods to simulation at the atomic level~\citep{PhysRevB.60.11971}, depending on the application. In this paper, we propose charged mass-spring systems as a promising model for computer graphics applications. 

The representation of dynamically deforming objects using vertices connected by springs is ubiquitous in computer animation. A key property for such applications is stability and fidelity when the simulation time step is increased. The associated simulation methods have been honed over almost four decades-- from the early days of applying Euler's method, to implicit integration \citep{terzopoulos1987elastically, BaraffAndWitkin98}, position-based dynamics \citep{PBD2007} and projective dynamics \cite{bouaziz2023projective}. We adopt an implicit integrator for the elastic term in our model.

While springs offer a natural means of modelling forces between discrete points representing a continuum, the topology induced by the connectivity of the springs is limiting. Two vertices that are not connected together by a spring cannot influence (apply forces on) each other without relying on other springs (hence vertices). This is a well-studied limitation for applications that model biological or chemical systems where molecules may interact directly without being `connected'. The introduction of electrostatic charges and therefore Coulomb forces between vertices is common in those domains. We posit that the introduction of charges at vertices can provide a novel mechanism for artistic expression-- for example by key-framing charge values,  designing guiding electric fields, etc. 

We propose a novel model for simulating charged mass-spring systems. In this paper, we adopt a straightforward approach that decouples the elastic and Coulomb potentials, by integrating the former in classical fashion while using an explicit integrator for the latter.  We demonstrate that despite its simplicity, such an implicit-explicit integrator is surprisingly stable. We also show that the system is differentiable, allowing gradient-based optimization for inverse (parameter) estimation. In this paper, we:
% Expected contributions:
\begin{enumerate}
    [leftmargin=2em]
    \item introduce charged mass-springs for computer animation;
    \item develop a parallelisable approximation algorithm for estimating the Coulomb potential by using a meshed domain to partition near- and far-fields;
    \item derive an implicit-explicit integrator for charged mass-springs and show that it is differentiable; and
    \item demonstrate experimentally that the above can lead to stable simulation at large time steps.  
\end{enumerate}

\section{Background}\label{sec:relatedwork}
% We first summarise relevant work from the vast literature on simulation techniques. Then, in Section~\ref{sec:msreview} we review the specific implicit integrator for mass-spring systems which we use in this paper. 

\paragraph{Long-range potentials} Common examples of long-range potentials include the Lennard-Jones potential, which models atomic interactions (e.g., van der Waals forces); gravitational potential, which governs astrophysical dynamics; Coulomb potential, which describes interactions between charges; and magnetic potential, which characterizes the behavior of magnets. These potentials pose significant computational challenges by introducing an $O(n^2)$ complexity.  Each model has unique characteristics: gravitational potential is exclusively attractive, while the Lennard-Jones potential accounts for both repulsion and attraction depending on distance. Magnetic fields, similar to Coulomb fields, can produce both attraction and repulsion. However, modeling a magnetic field requires defining the magnetization on a volumetric mesh, whereas Coulomb potentials can be defined point-wise. In computer graphics, magnetic behavior has been used for animation \cite{thomaszewski2008magnets} using the Barnes-Hut algorithm \cite{barnes1986hierarchical} to subdivide a 3D mesh. Meanwhile, \cite{huang2019accurate} leverages the Fast Multipole Method (FMM) \cite{greengard1988rapid} to model the magnetic field of ferrofluids. However, these methods have limitations: Barnes-Hut suffers from reduced accuracy and FMM performs poorly in non-uniformly distributed systems\cite{blelloch1997practical}. A combined method, the Parallel Multipole Tree Algorithm (PMTA) \cite{board1994scalable}, mitigates these disadvantages by employing an adaptive tree structure along with multipole expansions to model grouped fields more efficiently.

\paragraph{Systems of charged particles} 
The simulation of charges is an important and long-standing problem in many scientific disciplines such as crystal structure~\cite{madelung1919,park2022topological}, solutions~\cite{ermak1975computer}  and molecular dynamics~\cite{LibreTexts1,wang2021fixed}. A recurring problem across these applications is that the na\"ive calculation of the potential (or forces) of a system with $n$ point charges involves all pairs of considerations ($n^2$). 
%
% Typical strategies to overcome this difficulty is to exploit specific geometries~\cite{birdsall1991particle}, separate short-range from long-range periodic behaviour~\cite{Ewald} or exploit symmetry or homogeneity considerations~\cite{WolfSum}. 
This is a well-studied problem~\cite{CFMD12,fennell2006ewald,lekner1991summation} with many solutions. Two standard approaches are Ewald summation~\cite{Ewald}, which separates short-range from long-range periodic behavior, and the Wolf sum~\cite{Wolf} which exploits symmetry or homogeneity considerations. Other methods that address this problem are the particle mesh and particle-particle particle-mesh methods~\cite{hockney1988}. Truncation-based methods such as the cutoff method and the reaction field method suffer from high bias~\cite{van1990computer}, particularly when the number of point charges is large. 
Inspired by acceleration data structures in computer graphics, \citet{Max2009} proposed an efficient method to identify critical points in the electric field due to a collection of charges.

\paragraph{Time integration} At each simulation step, the position of a dynamical system is calculated from its velocity and acceleration by integrating them over time. The equations characterizing a dynamical system are either derived starting from Newton's second law or via the principle of least action. The equations are then discretized to obtain \emph{update rules} for the state (position, velocity, momentum, etc.) at each time step. In this paper, we focus only on single-step updates, where a state is calculated using only its preceding state. The most straightforward method to simulate the evolution of a dynamical system numerically is to discretize time in intervals of $h$ units and to calculate the state at $t+h$ as a consequence of the forces on the system at time $t$. This method, invented by Euler~\shortcite{euler1769institutionum,hairer2006geometric}, performs \emph{explicit} updates on the state as the simulation is unrolled and is therefore straightforward to implement. An alternative approach is to use the forces at time $t+h$ instead, which yields an \emph{implicit} equation in the state variable at $t+h$. This formulation often requires the solution of a system of non-linear equations to calculate the updated position. An alternative approach is to discretize a quantity called the \emph{Lagrangian of a system}---the difference between the system's kinetic and potential energies---rather than Newton's equations. The principle of least action, which states that the evolution of a system from time $t$ to $t+h$ necessarily minimizes the integral of the Lagrangian over that time, may be used in conjunction with the discretized Lagrangian to derive a system of equations which yield update rules for the state of the system~\cite{hairer1997variable,stern2006discrete}. Such \emph{variational or symplectic integrators} exhibit desirable properties such as energy conservation.

% Need to incorporate this 
% For example, Cao et al. \cite{msc2010} has simulation the cloth with electrostatics with explicit method. Implicit method is much more stable than explicit method. \cite{terzopoulos1987elastically, largestepinCloth} has showed the implicit method can would maintain stable in stiff system and with large time steps. However, the implicit would introduce numerical damping. 

% Energy conservation is desirable in many applications for biological and chemical simulations. Symplectic integrators \cite{stern2006discrete,stern2009implicit} conserved energy and easy to implement. 

\paragraph{Mass-spring systems}
Masses connected by springs have long been used in scientific computing and physics-based animation~\cite{PBACourse19}. They are used as a proxy for volumetric solids \cite{teschner2004versatile}, surfaces such as cloth~\cite{bridson2002robust,clothsimms} as well as linear elements such as hair~\cite{MSHair91,MSHair08}. In computer graphics these systems are simulated via explicit or fast implicit~\cite{liu2013fast} approaches. Alternatively, positions at each time step can be calculated directly as the solution to a quasi-static problem for \emph{position-based-dynamics} methods~\cite{PBD2007,PBDSurvey17} which sacrifice accuracy for efficiency and controllability. 
%  \cite{liu2017quasi, overby2017admm} further extend this method to hyperelastic materials.

\paragraph{Differentiable simulation}
Many downstream applications necessitate reasoning about physical systems.  \emph{System identification}, or the estimation of parameters (coefficient of friction, viscosity, restitution, etc.) of a physical system, plays an important role in boosting the generalization of learned policies~\cite{beck1977parameter,yu2017preparing}. Differentiable rendering is another example of parameter estimation via inversion of a physical process. The general approach is to design the simulation procedure enabling the calculation of the derivative of state information with respect to the parameters of interest. Several useful differentiable simulation frameworks have been developed over the past few years~\cite{jatavallabhula2021gradsim,diffpd,li2022diffcloth,todorov2012mujoco}.

%\BZ{\paragraph{Neural Physics} this paragraph is removed. Currently we dont do any neural task and see if we have enough space}

\subsection{Review: Implicit integrator for Mass-spring systems}\label{sec:msreview}
%\BZ{should this be a seperate section?}
 Let $\Xt \in \mathbb{R}^{3n}$ and $\Vt \in \mathbb{R}^{3n}$ represent the position and velocity of the system at time $t$, where $n$ represents the  number of vertices in the system. Let $\Q\in\mathbb{R}^n$ denote the point charges located at each of the vertices and $\F:\mathbb{R}^{3n}\rightarrow\mathbb{R}^{3n}$ represent the net (external and internal) forces at vertices.

A number of fast approximations have been proposed in the computer graphics literature~\cite{hauser2003interactive,muller2002stable,BaraffAndWitkin98,liu2017quasi}, for simulating elastic potentials. We use a fast implicit-integrator~\cite{liu2013fast} which we summarize here. The method first uses Newton's second law of motion to formulate implicit update equations for positions and velocities as
\begin{align}
    \Xtp = \Xt + h\Vtp \quad \mathrm{and} \quad
    \Vtp = \Vt + h M^{-1} \F(\Xtp),  \label{eq:v}
\end{align}
where $h$ is the time step and $M$ is a $3n\times 3n$ mass matrix. Substituting \Vtp\ into the equation for \Xtp, and using $h\Vt = \Xt - \Xtm$ yields an implicit equation~\cite{BaraffAndWitkin98} (\Xt\ and \Xtm\ are known):
\begin{equation}
    M\left(\Xtp - (2\Xt - \Xtm)\right) = h^2  \F(\Xtp). \label{eq:opt}
\end{equation}

We adopt the variational implicit Euler formulation~\cite{martin2011example} (renamed as optimization implicit Euler~\cite{liu2013fast} to avoid confusion with variational integrators) to solve this equation. This approach calculates \Xtp\ as the critical point of a function $\G(\Xt)$ constructed so that $\nabla \G =0$ yields Equation~\eqref{eq:opt}:
\begin{equation}
    \G(\Xtp) \;\;= \;\;\frac{1}{2} \left(\Xtp - \Y \right)^T M \left(\Xtp - \Y \right) \;\;+ \;\;h^2 \; \Eelast(\Xtp), \label{eq:G}  
\end{equation}
$\Y = 2\Xt - \Xtm$ is known (at $t$) and \Eelast\ is the elastic potential energy . 
% \SP{Attention, \Eel\ also depends on the external forces, e.g.\ in Equation~\eqref{eq:Eelx}, since its gradient is \F\ that also includes the external forces.}

For a spring between vertices $i$ and $j$, located at $\X^i$ and $\X^j$ respectively, its elastic potential (Hooke's law) is  $k_{ij} (\|\X^i - \X^j\| - \ell_{ij})^2/2$ where $k_{ij}$ and $\ell_{ij}$ are the spring constant and rest length. The non-linearity in $\|\X^i - \X^j\|$ was factored by \citet{liu2013fast} into a step that solves a non-linear optimization to find a vector $\D_{ij}\in\mathbb{R}^3$ per spring, which when assembled into $\D\in\mathbb{R}^{3s}$ ($s$ is the number of springs) enables expressing the potential minimization of a quadratic:
% \begin{equation}
%     \D_{ij} = \min_{\|\Pd\|=\ell_{ij}}  \frac {k_{ij}} 2 (\X^i - \X^j - \Pd)^2. \label{eq:subopt}
% \end{equation}
% 
\begin{eqnarray}\label{eq:Eelx}
    \Eelast(\X) &=&  \min_{\D\in U} \; \frac 1 2  \X ^T L \X  \;-\;  \X^T J \D \;-\; \X^T\F_{\mathrm{ext}} \;+\; c 
\end{eqnarray}
where $U$ is the set of rest-length spring directions, $L$ is a stiffness-weighted Laplacian of the graph representing spring connectivities, $J$ is a matrix that assembles the mixed terms in $k_{ij}(\X^i-\X^j-\ell_{ij})^2$ across all springs and $\F_{\mathrm{ext}}$ is the net external force. $c$ is the constant terms that would disappear after differentiation. 
% \KS{Provide explicit definitions for $L$ and $J$}
We include definitions~\cite{liu2013fast} of $L \in \mathbb{R}^{3n\times3n}$ and $J \in \mathbb{R}^{3n\times3s}$ :
\begin{eqnarray}
    L=\left(\sum^s_{i=1}k_iA_iA_i^T\right) \otimes I_3 \quad \mathrm{and} \quad
    J=\left(\sum^s_{i=1}k_iA_iS_i^T\right) \otimes I_3
\end{eqnarray}
where \(k_i\) is the spring constant, \( A_i \in \mathbb{R}^n \) is the incidence vector of \( i \)-th spring, \( S_i \in \mathbb{R}^s \) is the \( i \)-th spring indicator,  \( I_3 \in \mathbb{R}^{3 \times 3} \) is the identity matrix and \( \otimes \) denotes the Kronecker product. 

Once \D\ is optimized, $\Eelast$ can be substituted into Equation~\eqref{eq:G} to obtain a quadratic $\G(\Xtp)$ whose critical point is obtained via solving $\nabla\G(\Xtp) = 0$, which results in a linear system in \Xtp
\begin{eqnarray}
    % \nabla\G(\Xtp) \; = \; 
    \label{eq:critpoint} 
    M (\Xtp-\Y) \; +\;  h^2\;  ( L \Xtp -  J\D - \F_{\mathrm{ext}}) \; =\; 0 \\
    \label{eq:xfwdstep} 
    \mathrm{solved \; as\; } \Xtp = (M+h^2L)^{-1} \bigl(M\Y + h^2 (J\D+\F_{\mathrm{ext}})\bigr). 
\end{eqnarray}
% is a linear system that can be solved to calculate the updated position 

The overall optimization is solved via block-coordinate descent~\cite{sorkine2007rigid,liu2013fast} that alternates between a local optimization step to solve for \D\ and a global linear step to solve for \Xtp\ given \D. The local optimization step is solved via Newton's method and the global step is solved via a Cholesky decomposition of the matrix $M + h^2 L$.

\subsection{Review: Forces due to point-charges}
The Coulomb force $\F_c^i(\Xt)$ at vertex $i$ is given as the superposition 
\begin{equation}
    \F_c^i(\Xt) =  \sum_{j\neq i} \;\; \frac{k_c \; q_i \; q_j}{\|\Xt^i - \Xt^j\|^3} \; (\Xt^i - \Xt^j), \label{eq:coulomb}
\end{equation}
of forces due to all other point charges, where $k_c$ is the Coulomb constant and $q_i$ is the charge at vertex $i$. Thus, calculation of the Coulomb force (or the corresponding potential) at $n$ vertices requires $O(n^2)$ evaluations of this potential. The presence of an external electric field $\Eel: \mathbb{R}^3\rightarrow\mathbb{R}$  introduces an additional force of $q_i \Eel(\Xt^i)$ at vertex $i$. Similarly, the presence of an external charge $q_e$ at position $\mathbf{x}_e$ introduces a force of $k_c q_i q_e (\Xt^i - \mathbf{x}_e)/\|\Xt^i - \mathbf{x}_e\|^3$ at vertex $i$. 
% \SP{We use \Eel\ both here and previously as the potential that generates the total forces in the mass-spring system.}

\section{Methods: Charged Mass-Spring Systems}

This chapter presents our simulation method for charged mass-spring systems. We first describe our novel method for approximating the electric field at any point in the domain due to a system of point charges. The key novelty is to define neighborhoods within which exact computation is performed using tetrahedra obtained by meshing the domain. This leads to a simple algorithm that lends itself to parallelization. Then we describe the integrator that we use to simulate the dynamics of the system.     

\subsection{Domain-Discretized Electric Field (DDEF) algorithm}\label{sec:DDEF}
The direct computation of the electric field at each vertex due to all other charges is computationally expensive. The electric field $\Eel(\X^i)$ calculated at each vertex $i$ is used to obtain the force $\F_c^i(\X)=q_i \Eel(\X^i)$. Inspired by previous work, we partition the field calculation into near-field and far-field $\Eel=\Eelnf+\Eelff$ and use an approximation for the latter. We use the exact calculation for nearby charges (to limit error) leading to the approximation $\Eela=\Eelnf + \Eelaff\approx\Eel$. The key novelty is in the definition of `nearby' charges and the method of approximating the far-field. We sample the domain at $m$ grid sites, calculate the contribution of each charge to the sites \emph{that are not `nearby'} and obtain $\Eelff$ via interpolation. We now detail each of these three steps and provide pseudocode (Algorithm~\ref{alg:DDEF}). 

% This results in a worst case performance of $O(mn)$ instead of $O(n^2)$.

\subsubsection{Domain discretization} 
We use a low-discrepancy sequence (Halton) to generate $m$ samples $\Um_j\in\mathbb{R}^3,\;j=1,2,\cdots,m$ within the bounding box of the point charges, followed by a Delaunay tetrahedralization $R$ to partition the domain. Each element tetrahedron $\tau\in\mathbb{N}^4$ is a four-tuple containing the indices of its vertices (grid points). Although the complexity of this step is theoretically $O(m^2)$, it has been shown that ``For all practical purposes, three-dimensional Delaunay triangulations appear to have linear complexity''~\cite{ED2001}, which is $O(m)$. For clarity, we will call \Um\ grid points and \X\ charge locations.

\begin{figure}
	\begin{center}
		\includegraphics[width=\linewidth]{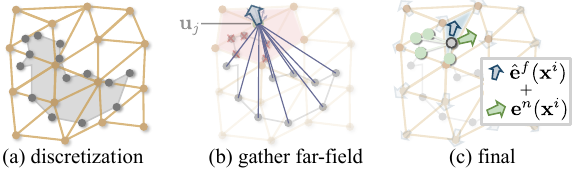}
		\caption{\label{fig:algo} A 2D illustration of our Domain-Discretization Electric Field (DDEF) algorithm. (a) The domain is discretized into $m$ grid points (orange) and triangulated. (b) The far-field due to the $i^{th}$ charge is calculated at all grid points except those that are nearby (red polygon). (c) The electric field at each charge location is then calculated by interpolating the grid-based far-fields and explicit near-field calculation. }
	\end{center}
\end{figure}

\subsubsection{Approximate far-field on the grid} 
For the $i^{th}$ point-charge, we consider a grid point $\Um_j\in\mathcal{N}_i$ to be `\textbf{nearby}' if it is a vertex of the tetrahedron ($\tau_i$) containing $\X^i$ or a vertex of an adjacent tetrahedron. Then we define the far-field to be the electric field at each grid point $\Um_j$ due to all charges except those that are nearby. That is, over the set $\{\Um_j\}\setminus\mathcal{N}_i$ where $\{\Um_j\}$ is the set of vertices of the tetrahedralization $T$. The approximate far field at each grid point $\Um_j$ is 
\begin{equation}
    \Eelaff(\Um_j) = \sum\limits_{\Um_j\in \{\Um_j\} \setminus \mathcal{N}_i} k_c q_i \frac{\Um_j-\X^i} {||\Um_j-\X^i||^3}. 
\end{equation}

% For the $i^{th}$ point-charge, we consider it $\X^i\in\mathcal{T}_j$ to be `\textbf{nearby}' if it is a vertex of the tetrahedron containing $\X^i$ or of an adjacent tetrahedron. The far-field is the electric field of tetrahedron $\tau_j$ due to all charges except those that are nearby. And for all the nodes within the tetrahedron, are sharing the same far-field, we estimate the far-field on point-charge using barycentric coordinates. The approximate far field of tetrahedron $\tau_j$ at node $\Um_\alpha, \alpha = 0,1,2,3$ is 
% \begin{equation}
%     \Eelaff_j(\Um_\alpha) = \sum\limits_{\X^i\notin \mathcal{T}_j } k_c q_i \frac{\Um_\alpha-\X^i} {||\Um_\alpha-\X^i||^3}. 
% \end{equation}

This definition avoids spikes since far-field computations at grid point $\Um_j$ are no longer arbitrarily close to a charge (for finite $m$). The far-field node is at least one ring away of $\Um_j$. The computational complexity of obtaining the sum is $O(mn)$ in practice. In the implementation, we first identify neighborhoods associated with each charge location $\X^i$ and then limit the gather to only vertices outside the neighborhood. Avoiding the if-statement enables an efficent GPU implementation. 

\subsubsection{Domain-discretized Electric Field} 
Finally, we calculate the approximate electric field at each charge location $\Eela(\X^i)$ in two steps. First, we approximate the far-field as $\Eelaff(\X^i)$ via interpolation of \Eelaff\ over the tetrahedron $\tau_i$ containing $\X^i$\BZ{ do we need to mention the interpolation needs to be careful due to near- far- field overlapping in tet?}. Then, we calculate $\Eelnf(\X^i)$ directly for nearby charges.  Here, nearby charges are defined as for vertex $\X^s$ whose $\mathcal{N}_s$ contain all vertices of  $\tau_i$ . The theoretical complexity of this step is $O(n^2)$ since it is possible that all charges are nearby and the algorithm reduces to exact computation. We found this unlikely in practice especially as $m$ is increased (see Figure~\ref{fig:DDEF_time}).

\subsection{Implicit Explicit Integrator} \label{sec:fwdsim} 
We propose a simple and direct implicit-explicit integrator for mass-spring-charge systems which is stable and suitable for interactive computer graphics applications at large simulation time-steps. The net force in our system is   
\begin{equation}
    \F(\Xt) = \F_{s}(\Xtp) + \F_c(\Xt) +\F_{\mathrm{ext}}. 
\end{equation}
% \BZ{hmmm I think $t$ here is a bit misleading, since it should be$\F_{s}(x_{t+1}) + \F_c(\Xt)$}
where $\F_s$ is the force on vertices due to springs, $\F_c$ is the total Coulomb force on vertices and $\F_{\mathrm{ext}}$ is the net external force. 

We demonstrate the benefits of a simple integration method where the elastic potential is integrated implicitly and the Coulomb potential is integrated explicitly. That is, Equation~\eqref{eq:opt} becomes
\begin{equation}
    M\left(\Xtp - \Y \right) \;\;=\;\; h^2  (\F_{s}(\Xtp) + \F_{c}(\Xt) + \F_{\mathrm{ext}}). \label{eq:optcoul}
\end{equation}

Further, since Equation~\eqref{eq:critpoint} obtains the critical point of the system by setting $\nabla\G(\Xtp) = \partial \G(\Xtp) / \partial \Xtp$ to zero, and since Equation~\eqref{eq:optcoul} features $\F_c(\Xt)$ rather than $\F_c(\Xtp)$, the only change is to incorporate the Coulomb force in the calculation of \Xtp, which now becomes
\begin{equation}
    \label{eq:fwdsolwcoulomb}
    \Xtp= (M+h^2L)^{-1} \bigl(M\Y + h^2 (J\D+\Q_t+\F_{\mathrm{ext}})\bigr)
\end{equation}
where $\Qv_t\in\mathbb{R}^{3n}$ is a vector that stacks $\F_c^i(\Xt), i=1,\cdots,n$ from Equation~\eqref{eq:coulomb}.

The solver for this integrator therefore proceeds exactly as for the mass-spring system in Section~\ref{sec:msreview} including the use of the same system matrix. The only change is that the final calculation of \Xtp\ involves the vector $\Q_t$ containing the Coulomb force on each mass due to all other charges in the system. For large $n$, this is the most expensive part of the simulation. We use our DDEF algorithm to calculate this.

%%%%%%%%%%%%%%%%%%%%%%%%%%%%%%%%%%%%%%%%%%%%%%%%%%%%%%%%%%%%%%%%%%%%%%%%%%%%%%%%%%%%%%%%%%%%%%%%%%%%%%%%%%%%%%%%
%%%%%%%%%%%%%%%%%%%%%%%%%%%%%%%%%%%%%%%%%%%%%%%%%%%%%%%%%%%%%%%%%%%%%%%%%%%%%%%%%%%%%%%%%%%%%%%%%%%%%%%%%%%%%%%%
%%%%%%%%%%%%%%%%%%%%%%%%%%%%%%%%%%%%%%%%%%%%%%%%%%%%%%%%%%%%%%%%%%%%%%%%%%%%%%%%%%%%%%%%%%%%%%%%%%%%%%%%%%%%%%%%
%%%%%%%%%%%%%%%%%%%%%%%%%%%%%%%%%%%%%%%%%%%%%%%%%%%%%%%%%%%%%%%%%%%%%%%%%%%%%%%%%%%%%%%%%%%%%%%%%%%%%%%%%%%%%%%%

We also show that our integrator can be differentiable and thus can be used for parameter estimation tasks. We derive the analytical gradient using our integrator and the analytical method of the charge potential.  If $\Xtp = \phi(\Xt, \Vt, \theta)$ represents the forward simulation described in Section~\ref{sec:fwdsim} where $\Param$ is a parameter of the system (e.g. spring constant or charge), then we can use the chain rule to compute the gradient of some loss function \Loss\ with respect to $\theta$ as 
\begin{eqnarray}
    \label{eq:diff1}
    \nonumber
    \Del{\Loss}{\Param} &=& \Del{\Loss}{\Xtp} \Del{\Xtp}{\theta} \;\;\;\mathrm{where}  \\
    \Del{\Xtp}{\Param}  &=& \mathcolorbox{myGray}{\Del{\phi}{\Param}} \;\; + \;\; 
                            \mathcolorbox{myGray}{\Del{\phi}{\Xt}} \Del{\Xt}{\Param} \;\; + \;\;
                            \mathcolorbox{myGray}{\Del{\phi}{\Vt}} \Del{\Vt}{\Param}\\
    \Del{\Vtp}{\Param} &=& \left({\Del{\Xtp}{\Param}} \;\; - {\Del{\Xt}{\Param}}\right)/h.
\end{eqnarray}
% \BZ{I know \Delf{\Vt}{\Param} is trivial, but I think we still should have it somewhere mention how it is computed}

Often  \Loss\ is chosen as a function of \Xtp\ so that \Delf{\Loss}{\Xtp} can be calculated from an analytical formula. Of the terms on the RHS of Equation~\eqref{eq:diff1}, \Delf{\Xt}{\Param} and \Delf{\Vt}{\Param} are known from the previous step. The gradients of the forward-step (shown highlighted) 
% $\Delf{\phi}{\Param}$,  $\Delf{\phi}{\Xt}$ and $\Delf{\phi}{\Vt}$  
are straightforward starting from Equation~\eqref{eq:fwdsolwcoulomb}:
\begin{eqnarray}
    \nonumber\Delf{\phi}{\Param} &=& \begin{cases}
                                H^{-1}\;h^2\;\Delf{\Qv}{\theta},  &\mathrm{for\;charges}\\
                                H^{-1}\;(h^2 \; \Delf{L}{\Param} \Xtp - M \; \Delf{J}{\Param} \; \D), & \mathrm{for \;springs}
                            \end{cases} \\
    \Delf{\phi}{\Xt} &=& H^{-1}\;(M + h^2\Delf{\Qv}{\Xtp})\\
    \Delf{\phi}{\Vt} &=& H^{-1}hM \quad \mathrm{where}  \\
    H &=& M+h^2(L\;-\;J\Delf{\D}{\Xtp}) \label{eq:Hinv}
\end{eqnarray} 
is the Hessian of potential energy. % and $I_{3n}$ is an $3n\times3n$ identity matrix.  
% H &=& [I+h^2M^{-1}(L\;-\;J\Delf{\D}{\Xtp})]\\
% \Delf{\phi}{\Param} &=& H^{-1}\;h^2M^{-1}\;\Delf{\Qv}{\theta}\\
% \Delf{\phi}{\Xt} &=& H^{-1}\;(I + h^2M^{-1}\Delf{\Qv}{\Xtp})\\
% \Delf{\phi}{\Vt} &=& H^{-1}\;hI
With these gradients defined, the na\"ive procedure involves  calculating all the gradients as highlighted above.  Inverting $H$ is the most expensive part of the gradient computation, for which \citet{diffpd} proposed to use adjoint method to replace the direct inversion with iterative linear solving.
% \begin{algorithm}
%     \caption{Backward simulation one step}
%     \label{alg:backward_singlestep}
%     \DontPrintSemicolon
%     $\Xtp = \phi(X_t,V_t,\Param)$ \;
%     $\Vtp = (\Xtp-\Xt)/h$ \;
%     Calculate $H^{-1}$ according to Equation~\eqref{eq:Hinv} \;
%     $partial\_grad = EvalPartialGradients(H\_inv, X_{t+1}, X_t,k_s,q)$ \;
%     $step\_grad = EvalStepGradients(partial\_grad)$\;
% \end{algorithm}

At each frame of the forward simulation, we perform three operations. First, we optimise \D\ exactly as in previous work\cite{liu2013fast}. Then we calculate \Q, the electrostatic forces at each point mass. Finally, we calculate \Xtp\ using Equation~\eqref{eq:fwdsolwcoulomb}. The straightforward (brute force) method to calculate \Q\ involves  calculating each of its elements by iterating through all other charges, via Equation~\eqref{eq:coulomb}. While this works in practice, its computational complexity of $\Theta(n^2)$ makes it impractical for large $n$. 

\BZ{do we mention that DDEF differentiability is not yet been verified? it is hard, because when we need to differentiate wrt to the position to backpropagte. }

\let\oldnl\nl% Store \nl in \oldnl
\newcommand{\nonl}{\renewcommand{\nl}{\let\nl\oldnl}}% Remove line number for one line
\begin{algorithm}[h]
	 \SetKwInOut{Input}{inputs}
	\SetKwInOut{Output}{output}
	
	\Input
	{		
		\par 
		\Indp
        $m$ \tcp*[r]{number of samples for domain discretization}
        $n$ \tcp*[r]{number of charge locations}
        $\{\X^i\}$ \tcp*[r]{locations of charges}
		\par
	} 

	\Output
	{
		\par 
		\Indp
        $\{\Eela(\X^i)\}$ \tcp*[r]{approximate electric field at mass locations} 
	}
	\vspace{.25em}
	{\hrule}

	\vspace{.45em}
    \textcolor{myBlue}{// 1. Discretize domain}\;
	$\{\Um_j\}\gets$ set of $m$ Halton samples within bounding box of $\{\Xt^i\}$\;
	$T\gets$ Delaunay triangulation of $\{\Um_j\}$ \tcp*[r]{tetrahedral grid}
	% $\tau\gets$ \{\}\;
	% $\psi\gets$ \{\}\;
	$\mathcal{N}_i\gets \{\},\; i=1,2,\cdots,n$\tcp*[r]{nearby grid points}
	$\mathcal{M}_j\gets \{\},\; j=1,2,\cdots,m$\tcp*[r]{inverse of nearby lookup}
	$\Eelaff(\Um_j)\gets 0,\; j=1,2,\cdots,m$ \tcp*[r]{far-field approx.}
	
	\vspace{.45em}

	\textcolor{myBlue}{// 2. Gather field due to charges onto non-nearby grid points}\;

        \For {$j\gets 1$ to $m$} {
            \For {$i\gets 1$ to $n$} {
                $\tau_i\gets$ tetrahedron in $T$ containing $\Xt^i$\;
    		$\psi_{i}\gets$ barycentric coordinates of $\Xtp^i$ in $T$\;
    		$\mathcal{N}_i \gets$ \{vert. indices of $\tau_i$\} $\cup$ \{vert. indices of neighbors of $\tau_i$\}\;
                \If{$\Um_j\in \mathcal{N}_i$}{
				$\mathcal{M}_j \gets \mathcal{M}_j \cup \{i\}$\;
			}\Else{
				$\Eelaff(\Um_j) \gets \Eelaff(\Um_j) \; + \; k_c q_i {(\Um_j-\X^i)} /{||\Um_j-\X^i||^3}$\;
			}                
            }
        }
	
	\vspace{.45em}

	\textcolor{myBlue}{// 3. Final: Interpolate far-field from grid and add exact near-field }\;
	$\Eelnf(\X^i)\gets 0,\; i=1,2,\cdots,n$ \tcp*[r]{near-field}
	\For {$i\gets 1$ to $n$} {
		$\Eelaff(\X^i) \gets$ Interpolate ($\tau_i$, $\psi_i$)\tcp*[r]{far-field approximation}
		$\mathcal{M}_u\gets\{\}$\tcp*[r]{initialize union of nearby points}	
		\ForEach{$l\in \mathcal{N}_i$}{
			$\mathcal{M}_u \gets \mathcal{M}_u \cup \mathcal{M}_l$\;
		}

		\ForEach{$s \in \mathcal{M}_u$}{
			$\Eelnf(\X^i) \gets \Eelnf(\X^i) \; + \; k_c q_s {(\X^i-\X^s)} /{||\X^i-\X^s||^3}$\;
		}

		$\Eela(\X^i) \gets \Eelnf(\X^i) + \Eelaff(\X^i)$\tcp*[r]{final approximation}
	}
	\caption{DDEF (Domain-discretized electric field)}
	\label{alg:DDEF}
\end{algorithm}

\section{Experiments}

\subsection{Validation}
\subsubsection{DDEF error}
\begin{figure}[htbp!]
  \centering
\includegraphics[width=0.95\columnwidth]{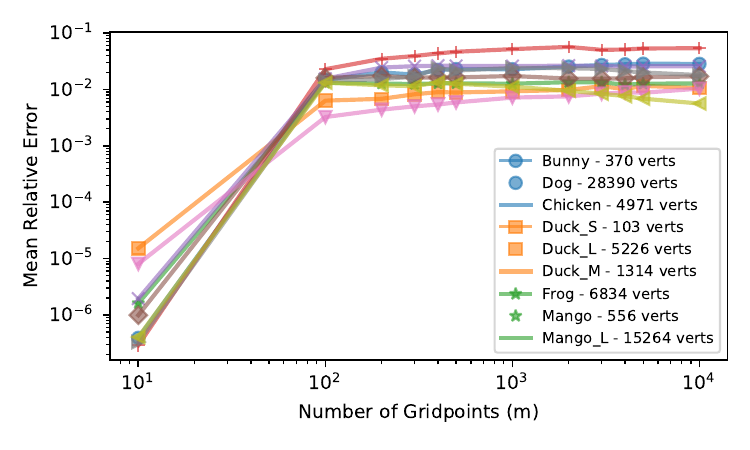}
\caption{Error validation of DDEF with increasing grid points ($m$). Simulations were conducted on different meshes with $m$ ranging from 10 to 10,000. For small $m$, a significant portion of the mesh lies in the near-field, leading to high accuracy due to brute force computation. The error converges around $m=100$, remaining below 0.1 for larger $m$. This allows for the selection of a larger $m$ to improve far-field parallelization efficiency. \KG{This is just restating the paragraph in 4.1.1 almost word for word}
}
\label{fig:DDEF_m}
\end{figure}

We validate that the error of DDEF would converge with larger grid points($m$). Figure \ref{fig:DDEF_m} plots the relative error vs number of grid points. We are running on different meshes for value of $m$ between $10$ and $10,000$. For small m, most of the mesh exists within the near-field, and therefore it is mainly performing brute force computation, resulting in high accuracy. The error converges at around 100 grid points with all examples below $0.1$ relative error. This enables us to choose a large m for better far-field parallelization without detriment to accuracy.  

\subsubsection{Error across time steps}\label{4_1} 
\begin{figure}
\centering
\setlength{\tabcolsep}{2pt} % Reduce space between columns
\renewcommand{\arraystretch}{0.8} % Reduce space between rows
\begin{tabular}{ccc}
     % Consider removing [6pt] to reduce vertical space    
     \multicolumn{3}{c}{\includegraphics[width=0.96\columnwidth]{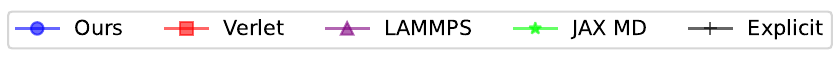}}\\
     \includegraphics[width=0.32\columnwidth]{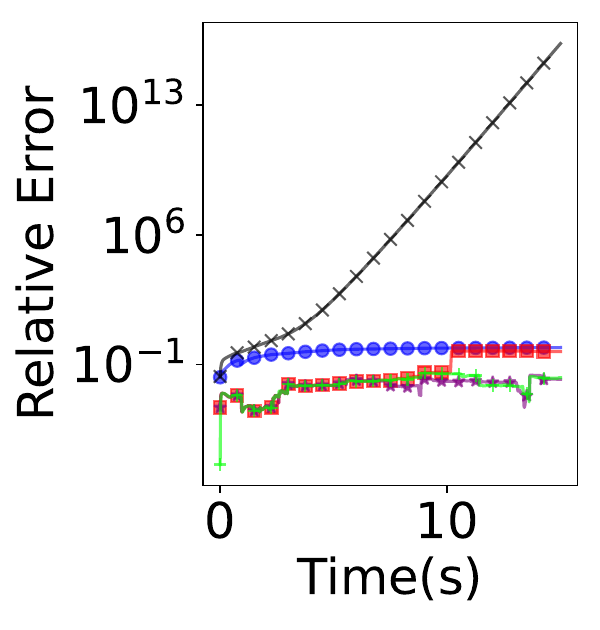} & 
  \includegraphics[width=0.32\columnwidth]{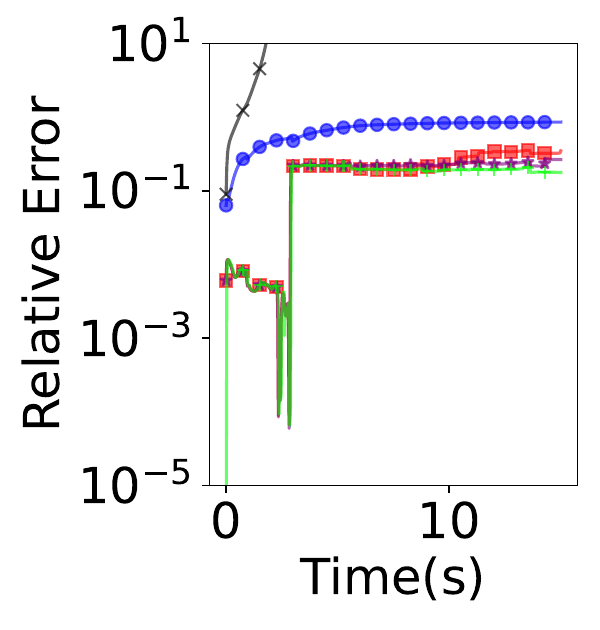} & \includegraphics[width=0.32\columnwidth]{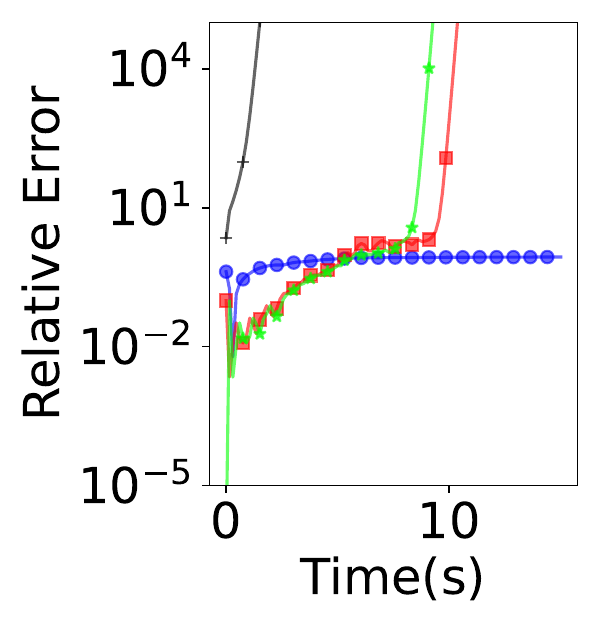}   \\
   (a) $h=\qty{0.015}{s}$ & (b) $h=\qty{0.03}{s}$ & (c) $h=\qty{0.15}{s}$ 
\end{tabular}
\caption{Our method is stable across resolutions (time-steps). Relative error during simulation of a charged torus mesh moving freely is plotted using different time-steps (h=$\qty{0.015}{s}$, $h=\qty{0.03}{s}$ and $h=\qty{0.15}{s}$). Although Verlet, LAMMPS and JAX MD are more accurate than ours in (a) and (b),  they become unstable for large time steps.\label{fig:validation}}
\end{figure}

 We validated the accuracy of our simulation across parameter choices for several meshes and found the relative error to be less than $15\%$ in all cases. Although our method results in error higher than Verlet for small time steps, it outperforms all variants for large time steps. Figure~\ref{fig:validation} plots the evolution of relative error for an example simulation of \qty{15}{s} across different choices of $h$. The mesh used is a sampled torus with 145 vertices, spring constants $k=\qty{10}{N/m}$ and point charges $q=\qty{6}{\uC}$. The plots show that our simulation is less accurate but more stable, especially as $h$ is increased. The black curve shows an explicit-explicit integrator~\cite{msc2010}, which is quickly unusable. LAMMPS\cite{LAMMPS} is not shown for $h=0.15s$ because its instability causes nodes to exceed a cutoff parameter, resulting in a runtime error.

\begin{figure}
    \centering
    \begin{subfigure}[b]{0.48\columnwidth}
        \includegraphics[width=\textwidth]{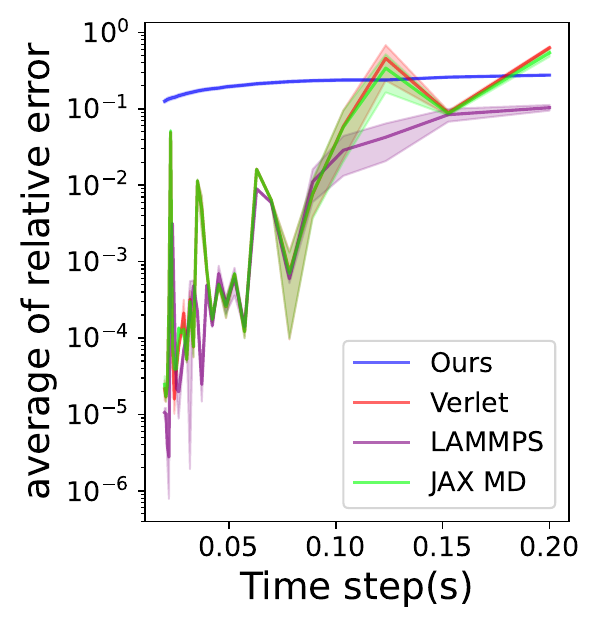}
        \caption{$q = \qty{6}{\uC}$}
        \label{lowq1}
    \end{subfigure}
    \begin{subfigure}[b]{0.48\columnwidth}
        \includegraphics[width=\textwidth]{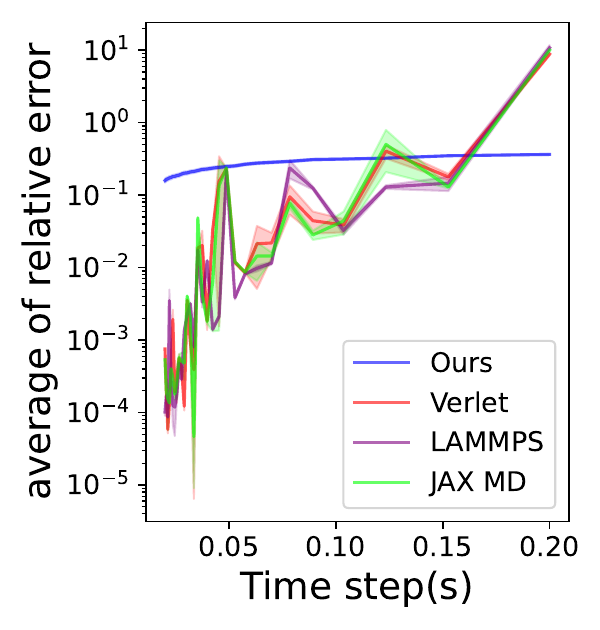}
        \caption{$q = \qty{8}{\uC}$}
        \label{lowq2}
    \end{subfigure}
    \caption{Relative error vs time-step resolution for a meshed bunny with two different charge parameters. Our method maintain a much more controlled error level across different $h$.}
    \label{4_1_2}
\end{figure}

We further investigated this behavior by plotting the statistics (average and standard deviation as a shaded area) of relative error versus time step for a larger mesh (bunny) with two choices of parameters ($k=\qty{0.5}{N/m}$, $q=\qty{6}{\uC}$ and $k=\qty{0.5}{N/m}$, $q=\qty{8}{\uC}$). We limit ourselves to the last \qty{5}{s} (total \qty{30}{s}) so that the variance is not too large for existing techniques. Figure~\ref{4_1_2} demonstrates that our simulation is stable in terms of error compared to other methods as the time step is increased. Variance in relative error manifests as `jiggle' during animation. 

\subsubsection{Effect of parameters}\label{sec413} 
\begin{figure}[htbp]
    \centering
    \includegraphics[width=0.9\columnwidth]{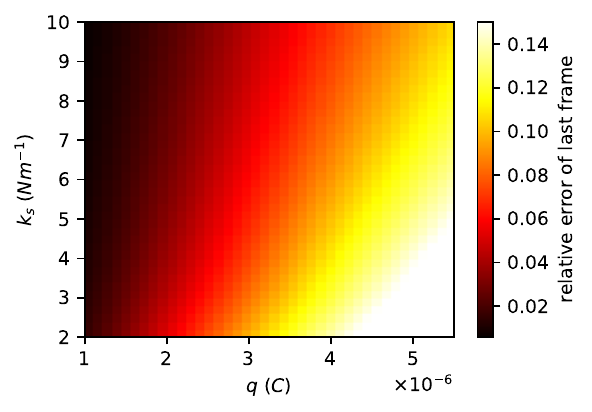}
    \caption{Relative error across different parameter settings for $k_s$ and $q$ for a meshed bunny. The error is generally larger when the system is dominated by electrostatics.
    }
    \label{fig:paramsheat}
\end{figure}

\begin{figure}[htbp]
    \centering
    \setlength{\tabcolsep}{2pt} % Reduce space between columns
    \renewcommand{\arraystretch}{0.8} % Reduce space between rows
    \begin{tabular}{cccccc}
        % Consider removing [6pt] to reduce vertical space
        
    \rotatebox{90}{Reference}&  %\hspace{-4mm}&
    \includegraphics[width=0.175\columnwidth]{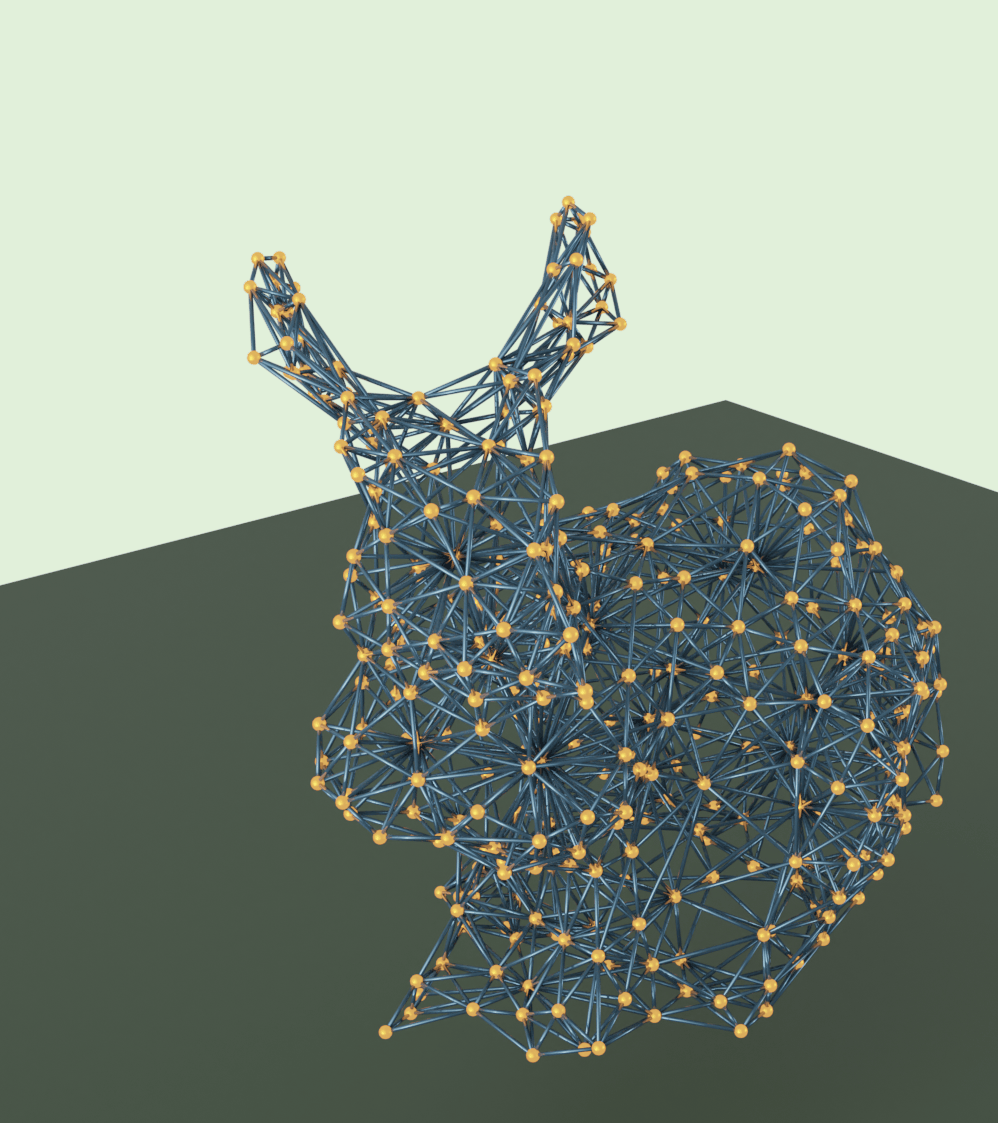} & 
    \includegraphics[width=0.175\columnwidth]{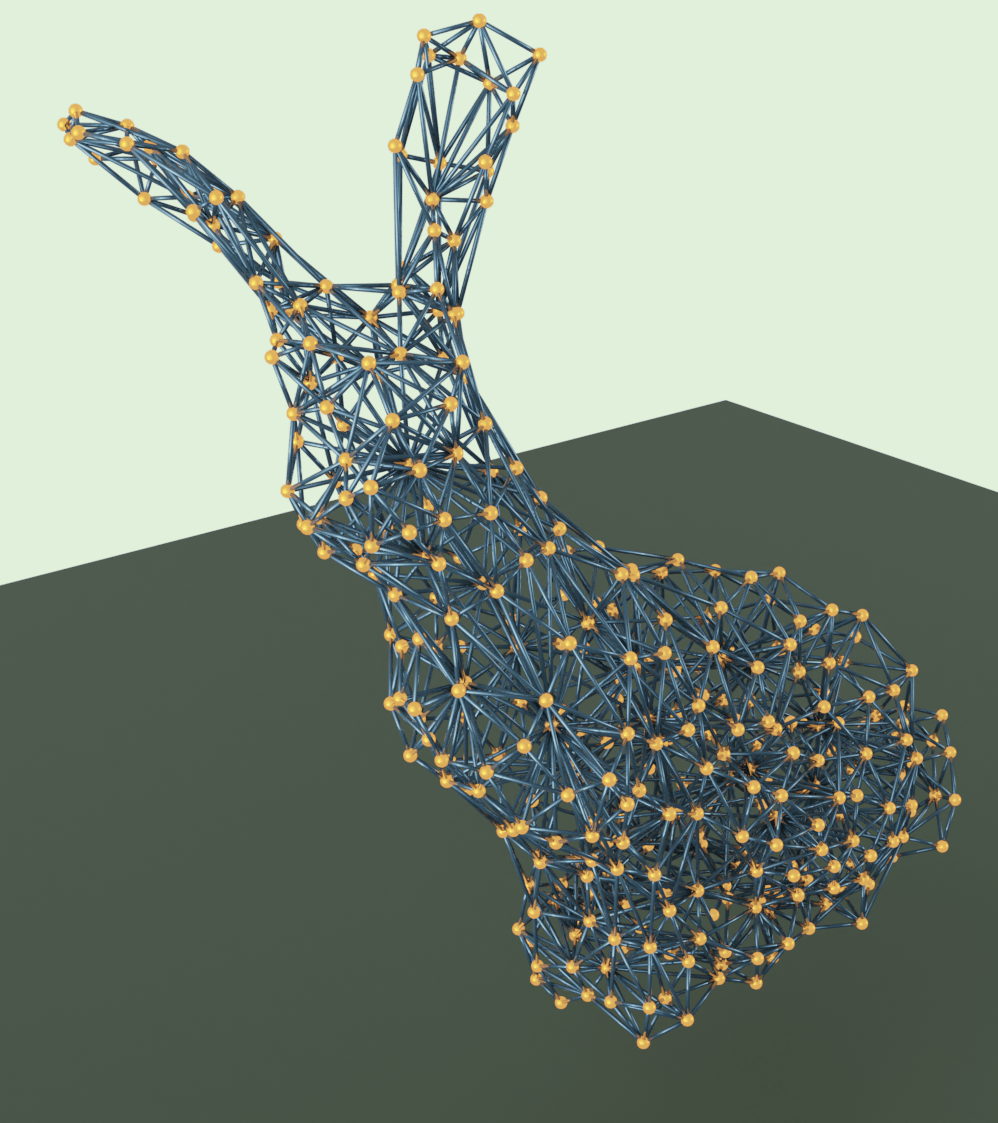} & 
    \includegraphics[width=0.175\columnwidth]{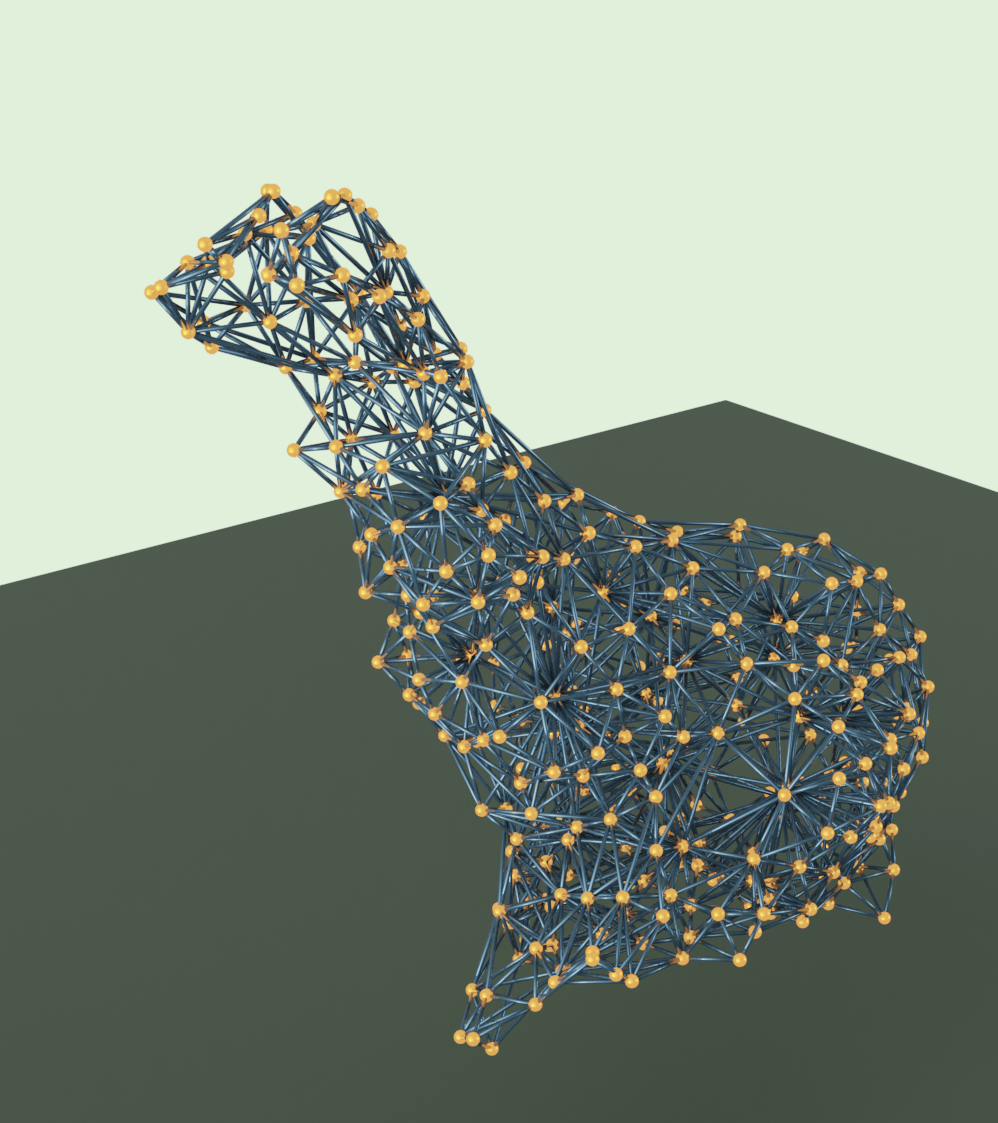} & 
    \includegraphics[width=0.175\columnwidth]{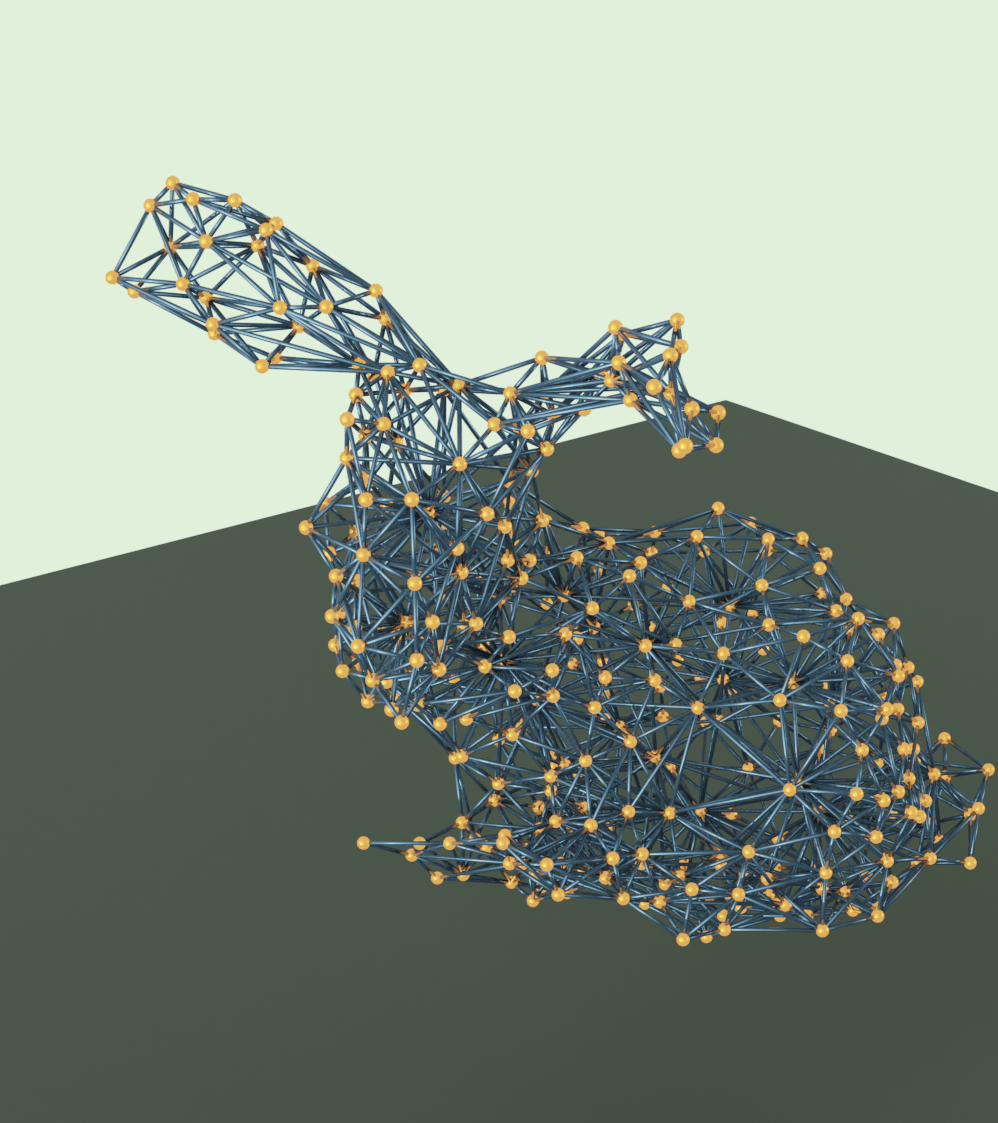} & 
    \includegraphics[width=0.175\columnwidth]{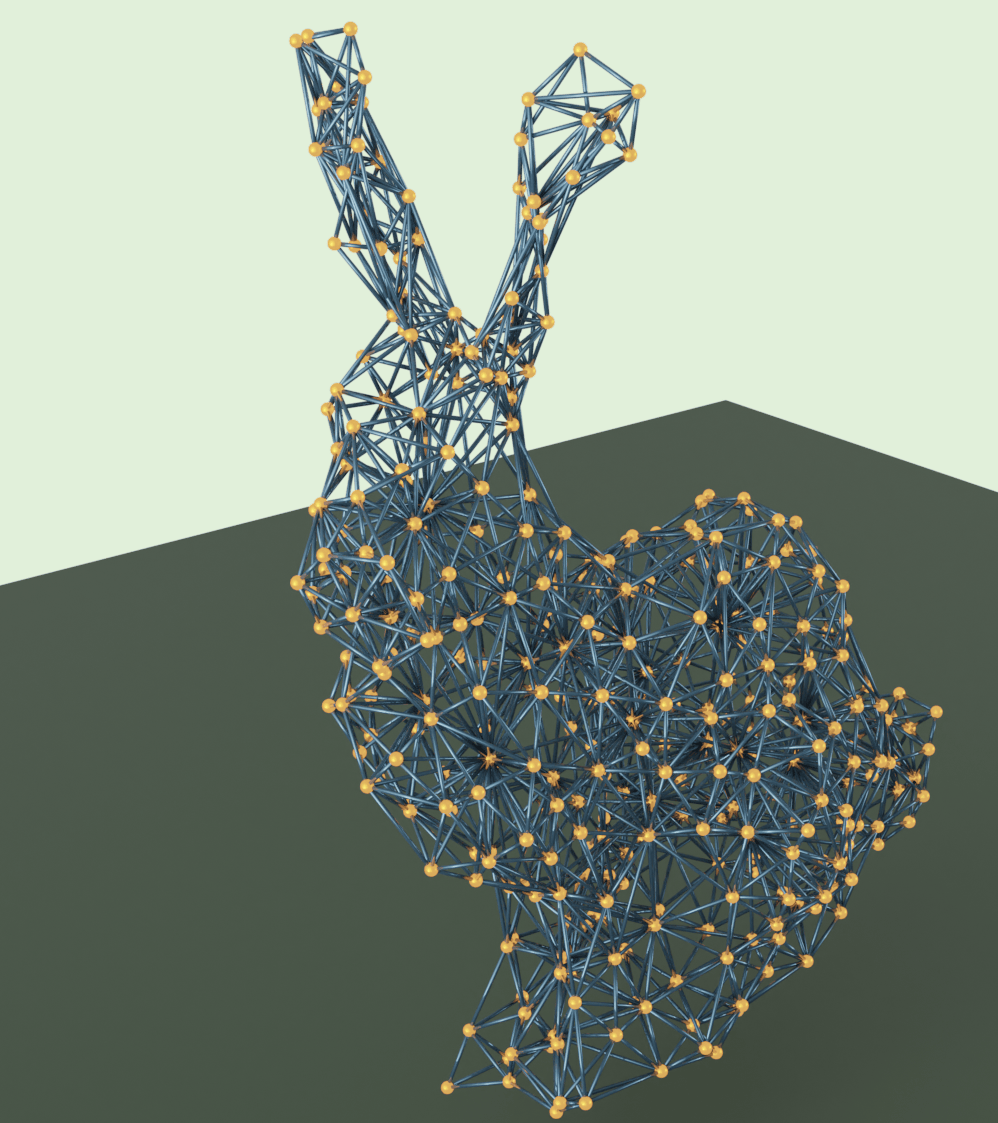}   \\
    \rotatebox{90}{Ours}& 
    \includegraphics[width=0.175\columnwidth]{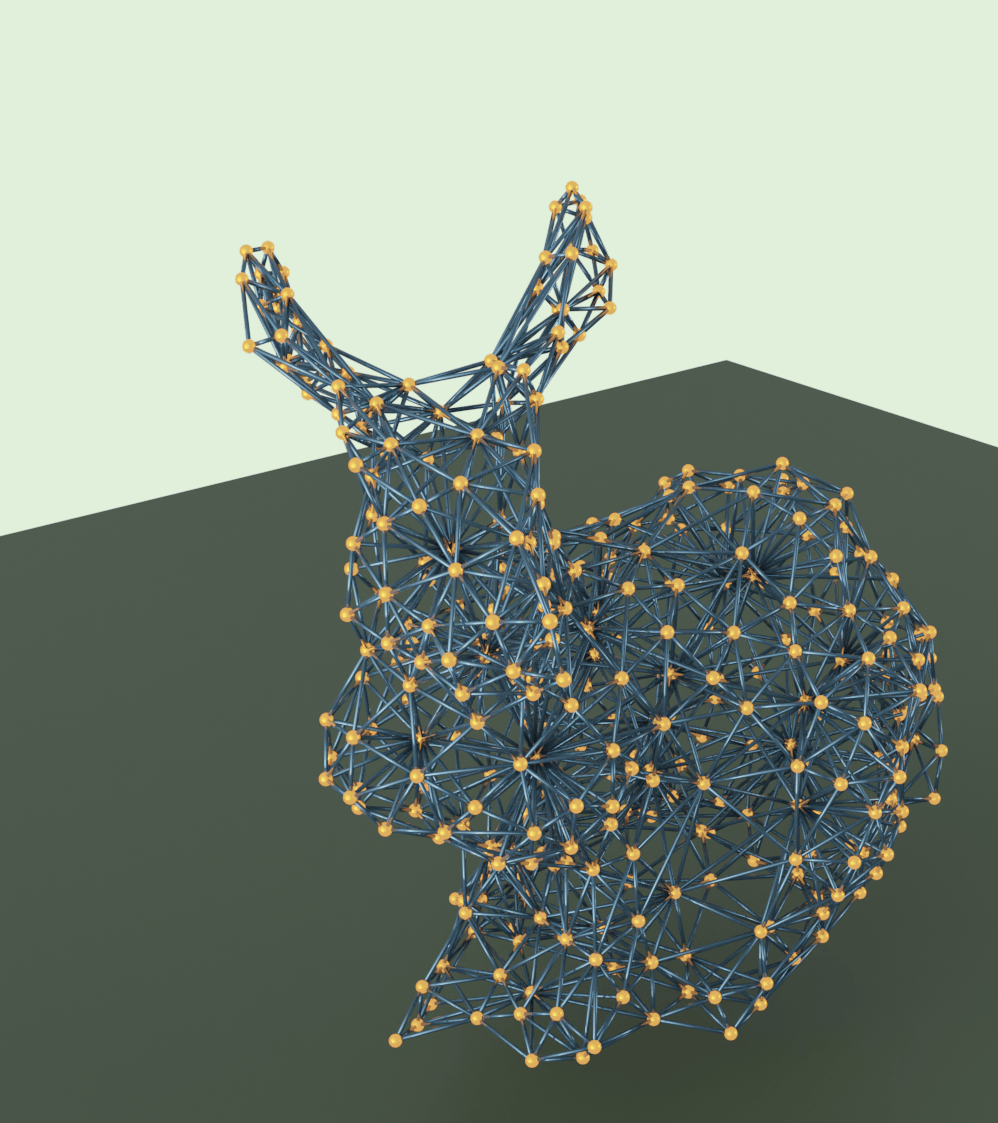} & 
    \includegraphics[width=0.175\columnwidth]{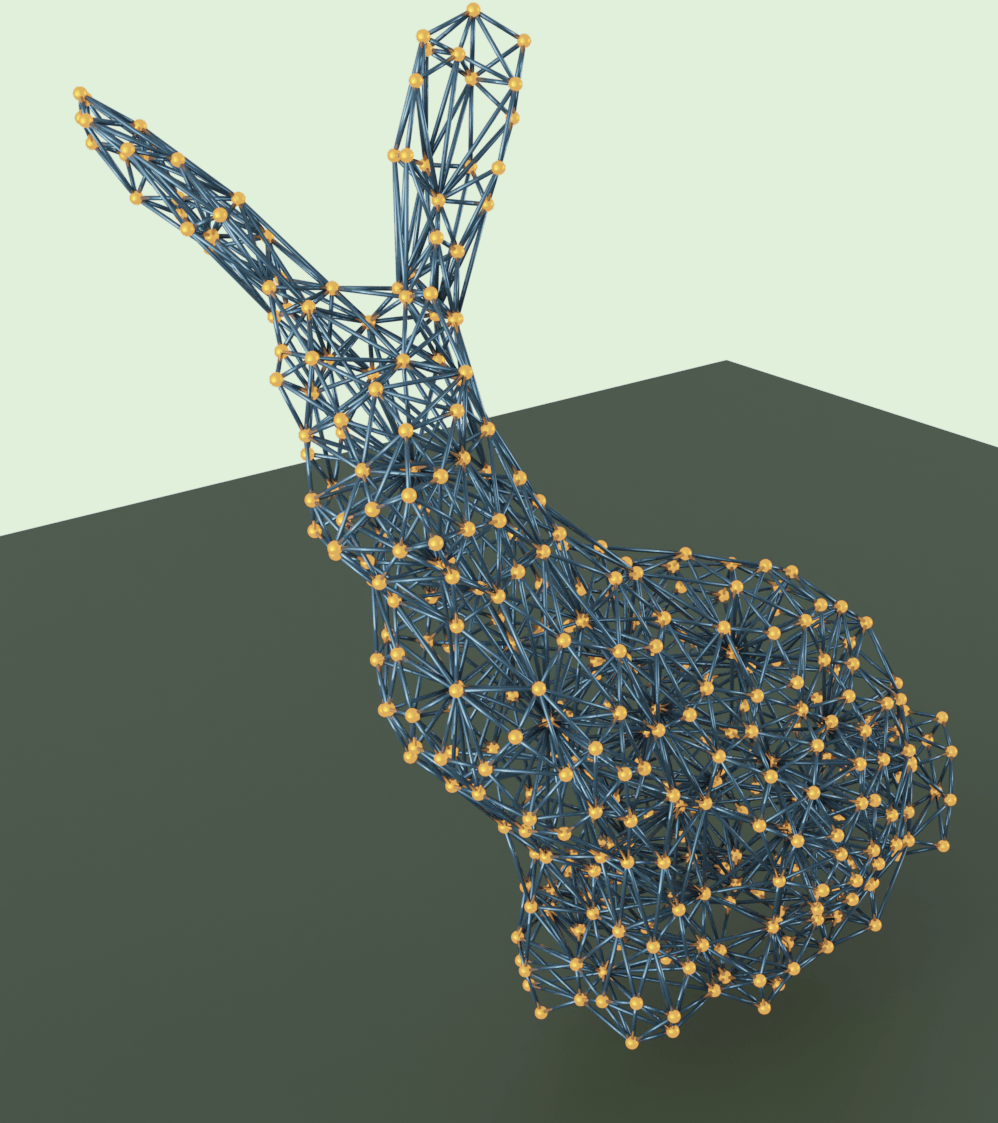} & 
    \includegraphics[width=0.175\columnwidth]{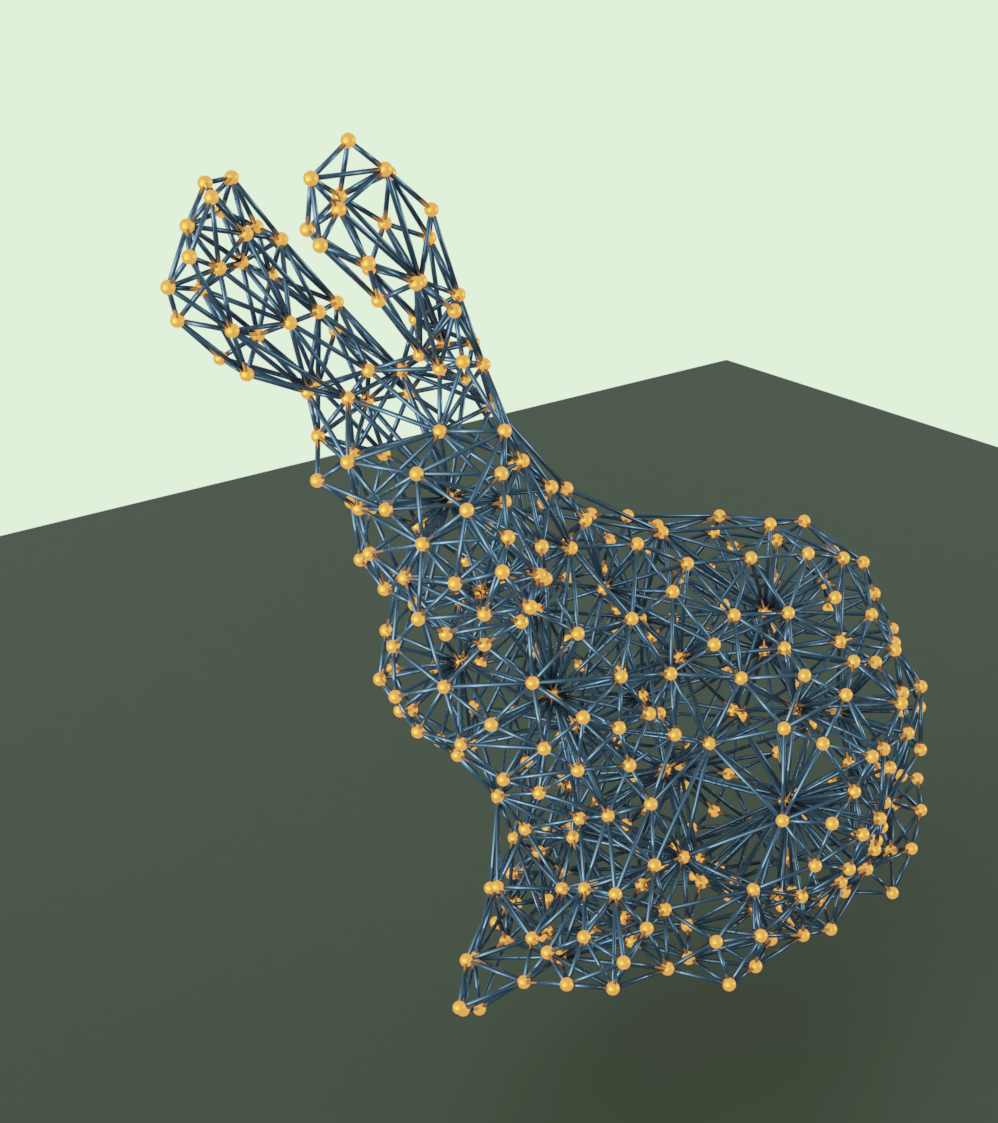} & 
    \includegraphics[width=0.175\columnwidth]{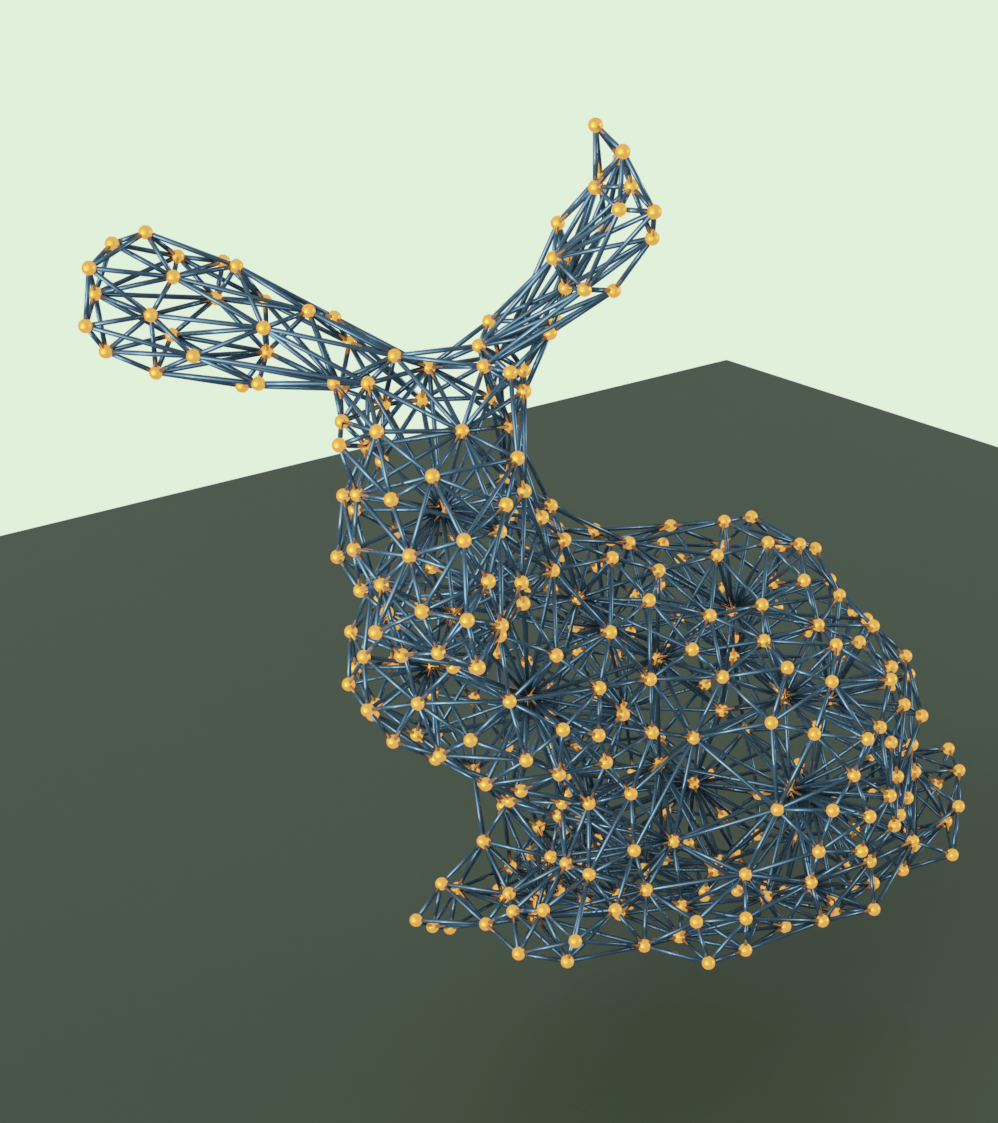} & 
    \includegraphics[width=0.175\columnwidth]{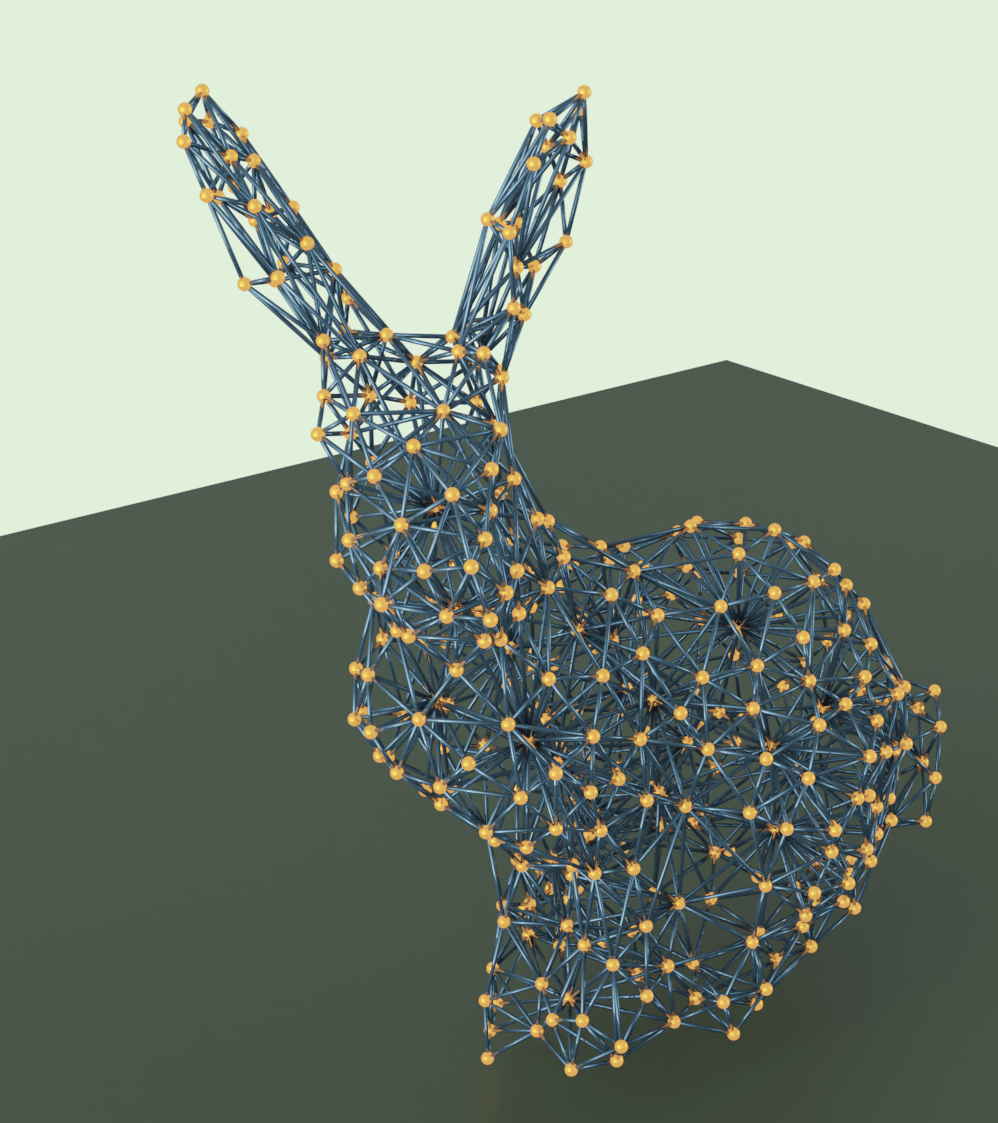}  \\
    & 1.5s &  3.75s &  7.5s & 11.25s &15s\\

    \end{tabular}
    \caption{We show the setup results in 14\% error that picked from Fig. \ref{fig:paramsheat}. The spring constant is $4.3N/m$ and charge is $5\mu C$. The experiment uses time step of $0.15s$ and the reference use time step of $0.05$. We can see that the reference has stronger forces with more dramatic movements. That is because the reference conserves the energy. Our method even uses the large method, due to the artificial damping results in smooth shapes.\label{4_1_3_2}}
\end{figure}

We studied the  error resulting from our implicit-explicit solver over the space of parameters (spring constant $k_s$ and charge $q$). For the bunny model, we measured relative error of the $100^{th}$ frame of simulation, using $h=\qty{0.05}{s}$ for a range of spring constants (\qty{2}{N/m} to \qty{10}{N/m}) and charge (\qty{1}{\uC} to \qty{6}{\uC}). We evaluated the last frame since  trajectories over time vary with choice of parameters. Figure~\ref{fig:paramsheat} plots the errors (color map) over $k_s$ vs $q$. The maximum error is around 20\% in this plot. The  reference was generated with Verlet at $h=\qty{0.01}{s}$. We found this behaviour to be representative of observations across models. Figure~\ref{4_1_3_2} shows that simulation results remain reasonable at an error of 15\%.

\subsection{Comparison}

\subsubsection{Forward simulation}
\begin{figure}[htbp]
    \centering
    \setlength{\tabcolsep}{2pt} % Reduce space between columns
    \renewcommand{\arraystretch}{0.8} % Reduce space between rows
    \begin{tabular}{ccccccc}
         % Consider removing [6pt] to reduce vertical space
        \multicolumn{7}{c}{Large time step}\\
       \rotatebox{90}{LAMMPS} & 
      \includegraphics[width=0.14\columnwidth]{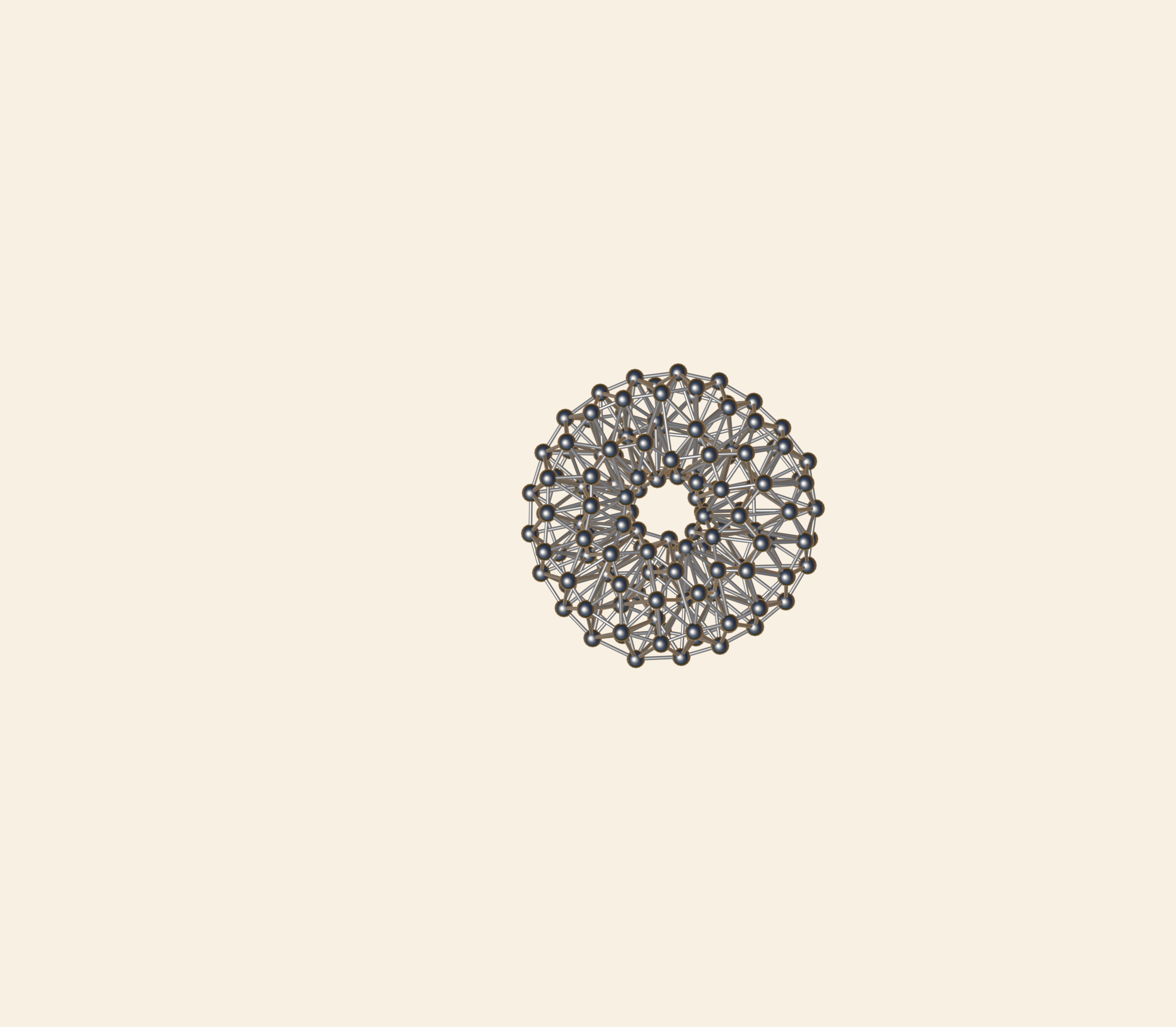} & 
      \includegraphics[width=0.14\columnwidth]{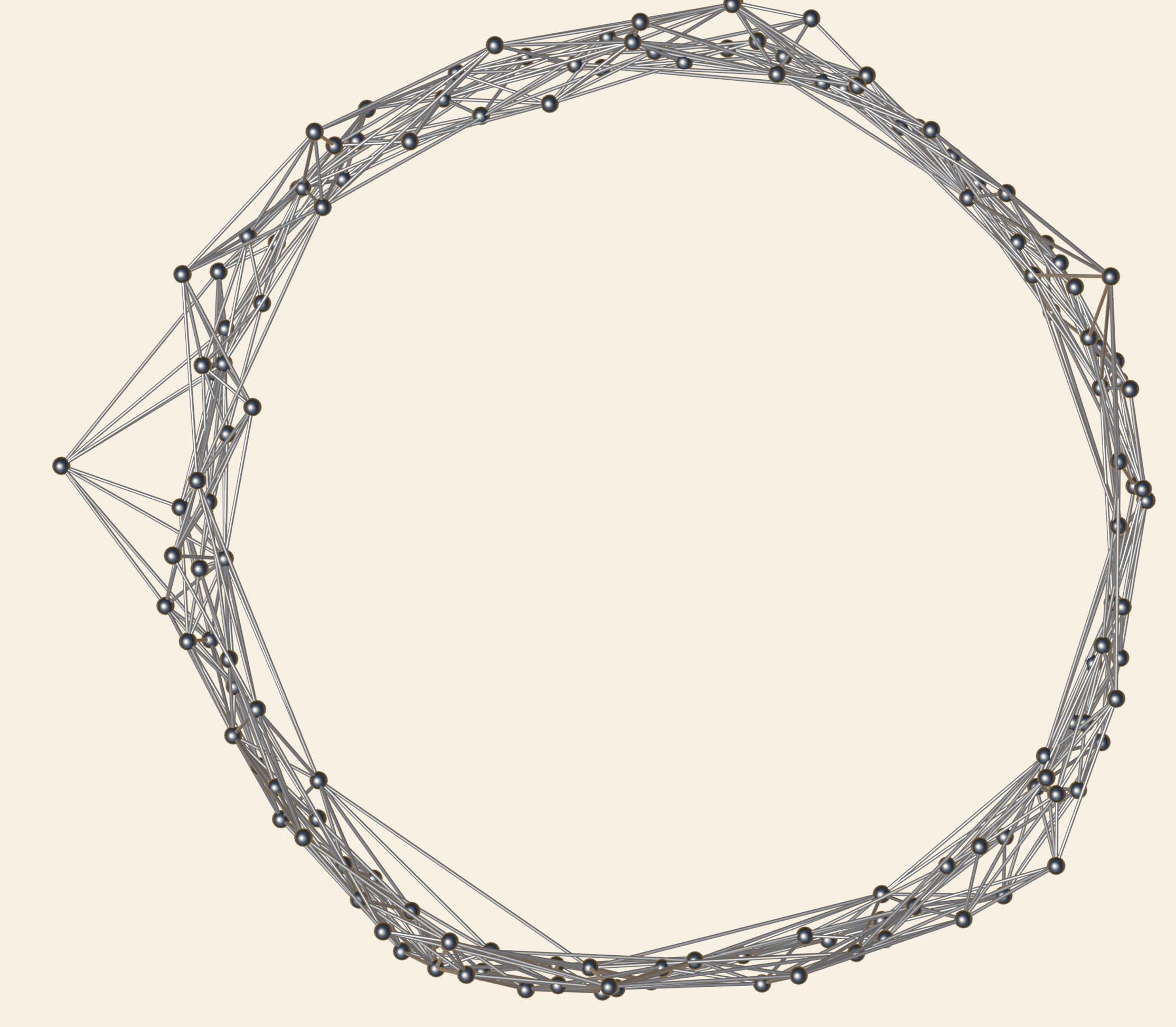} & 
      \includegraphics[width=0.14\columnwidth]{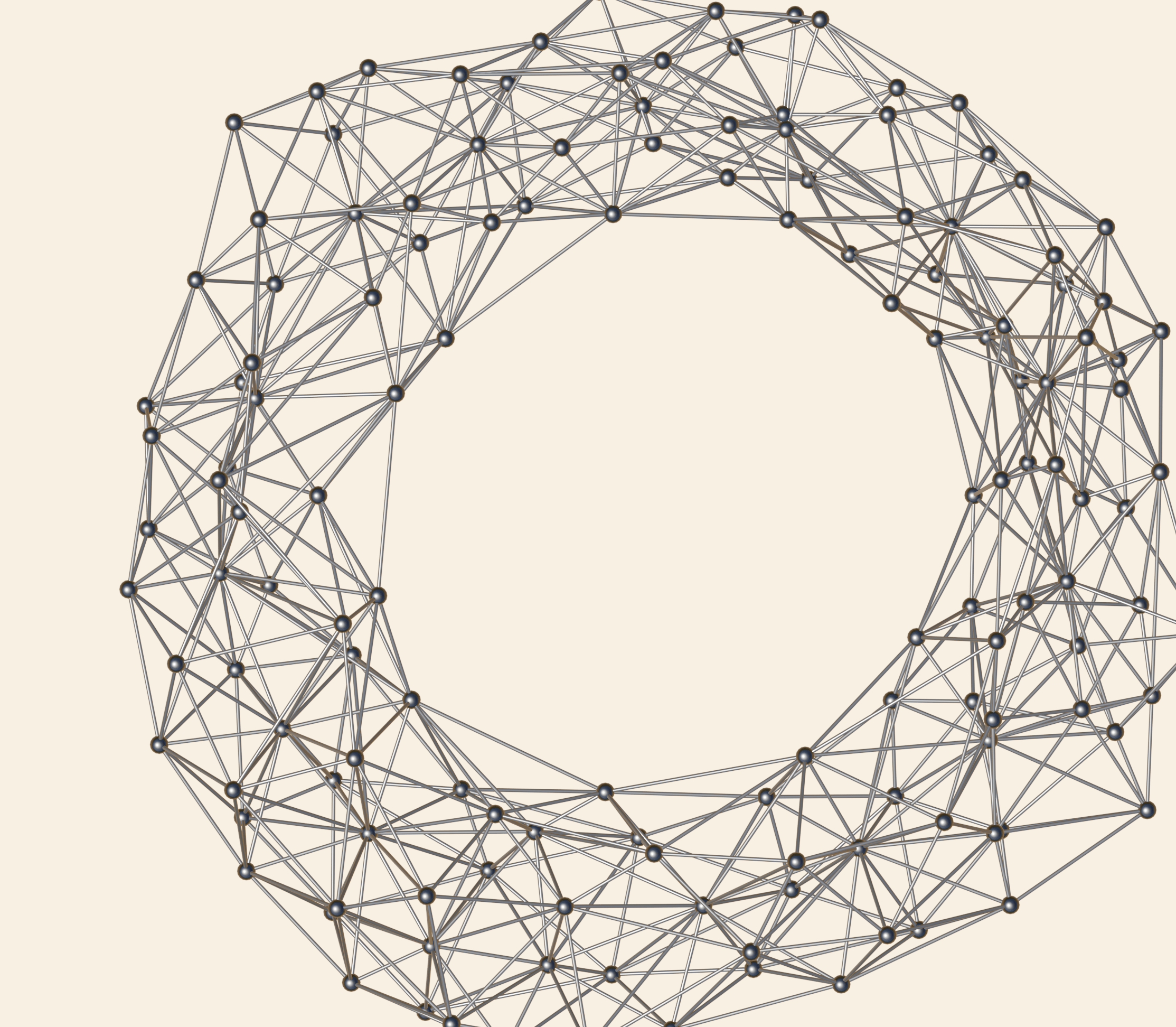} & 
      \includegraphics[width=0.14\columnwidth]{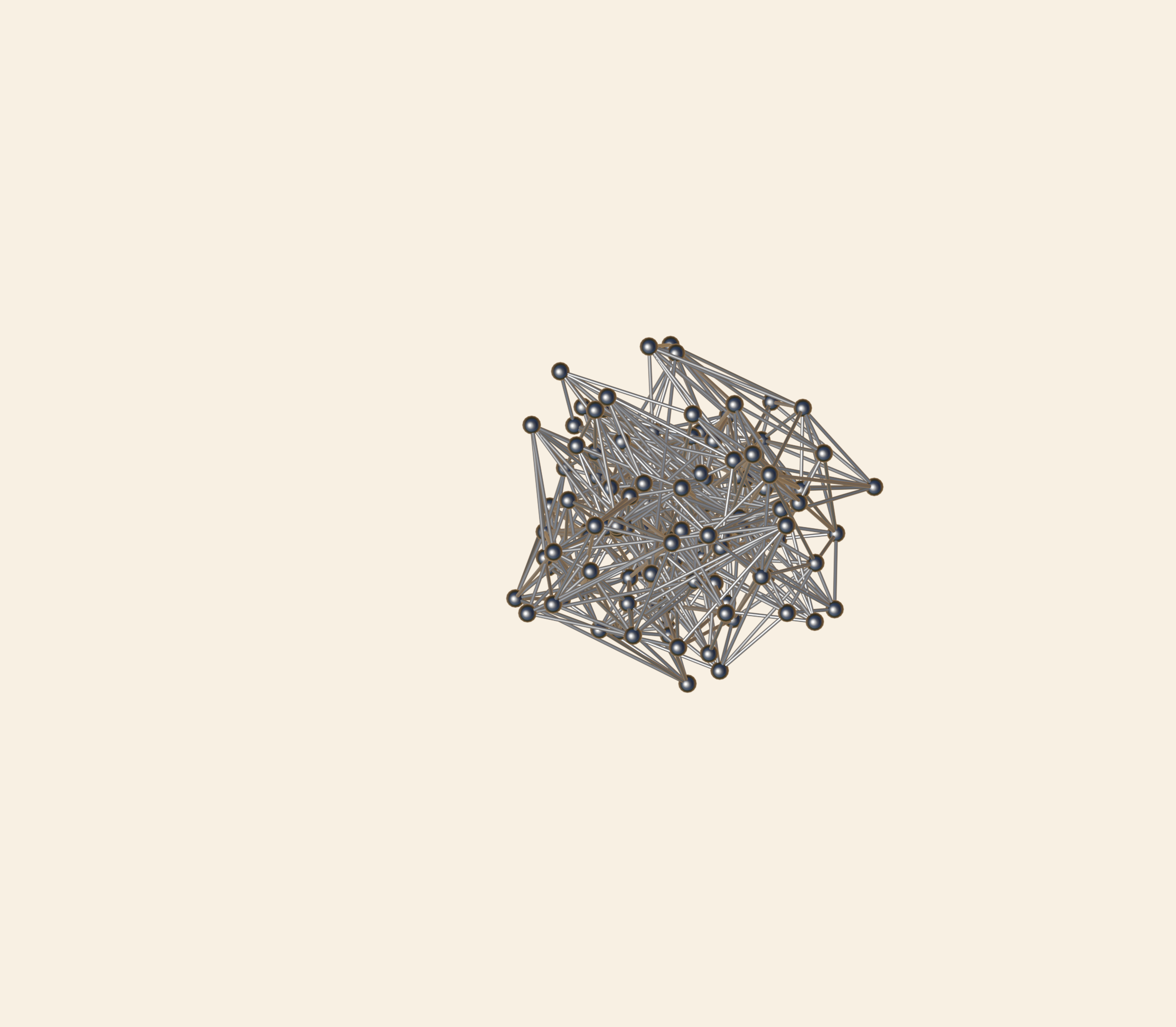} & 
      \includegraphics[width=0.14\columnwidth]{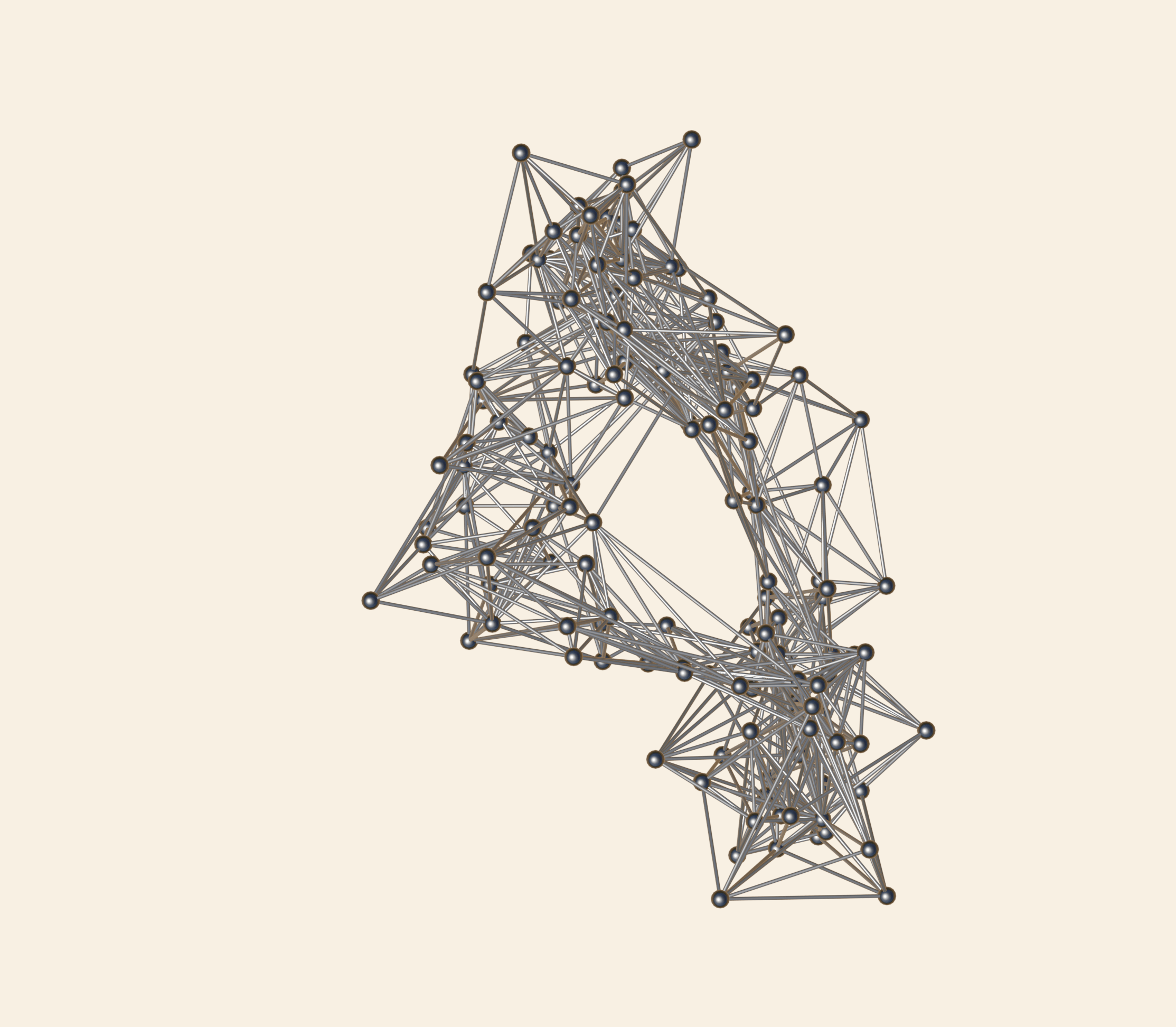} & \includegraphics[width=0.14\columnwidth]{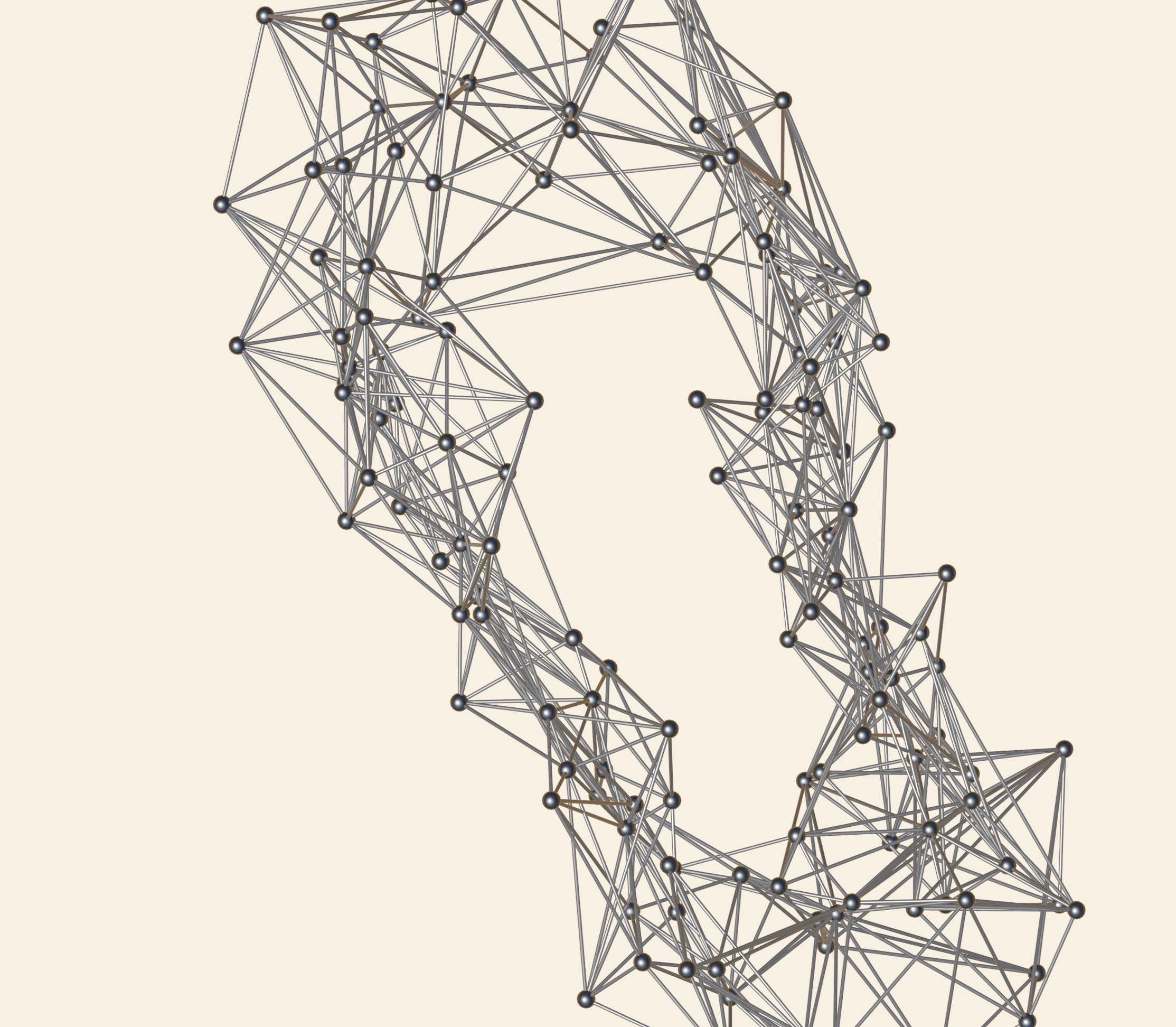}   \\
       \rotatebox{90}{Ours} & 
      \includegraphics[width=0.14\columnwidth]{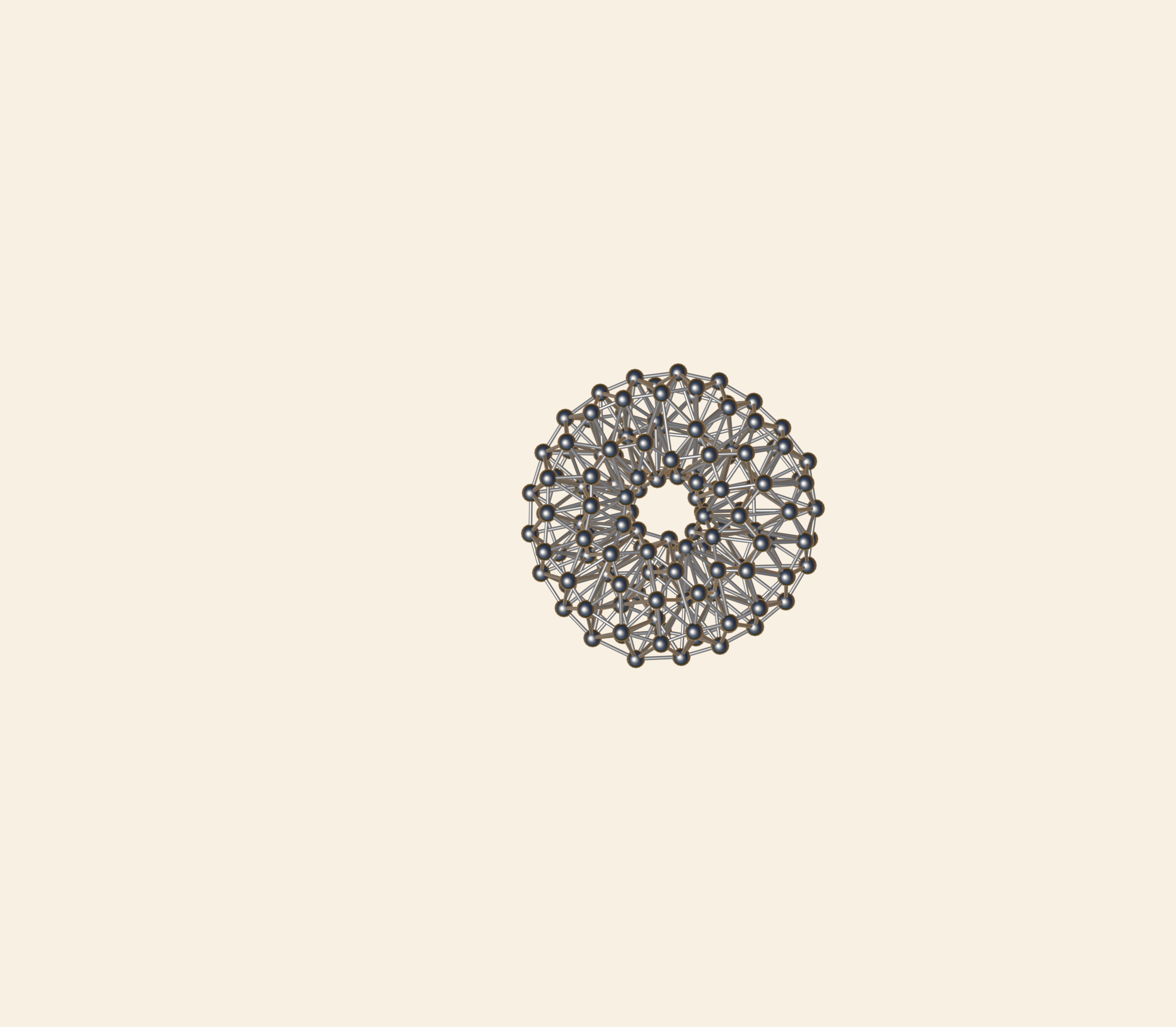} & 
      \includegraphics[width=0.14\columnwidth]{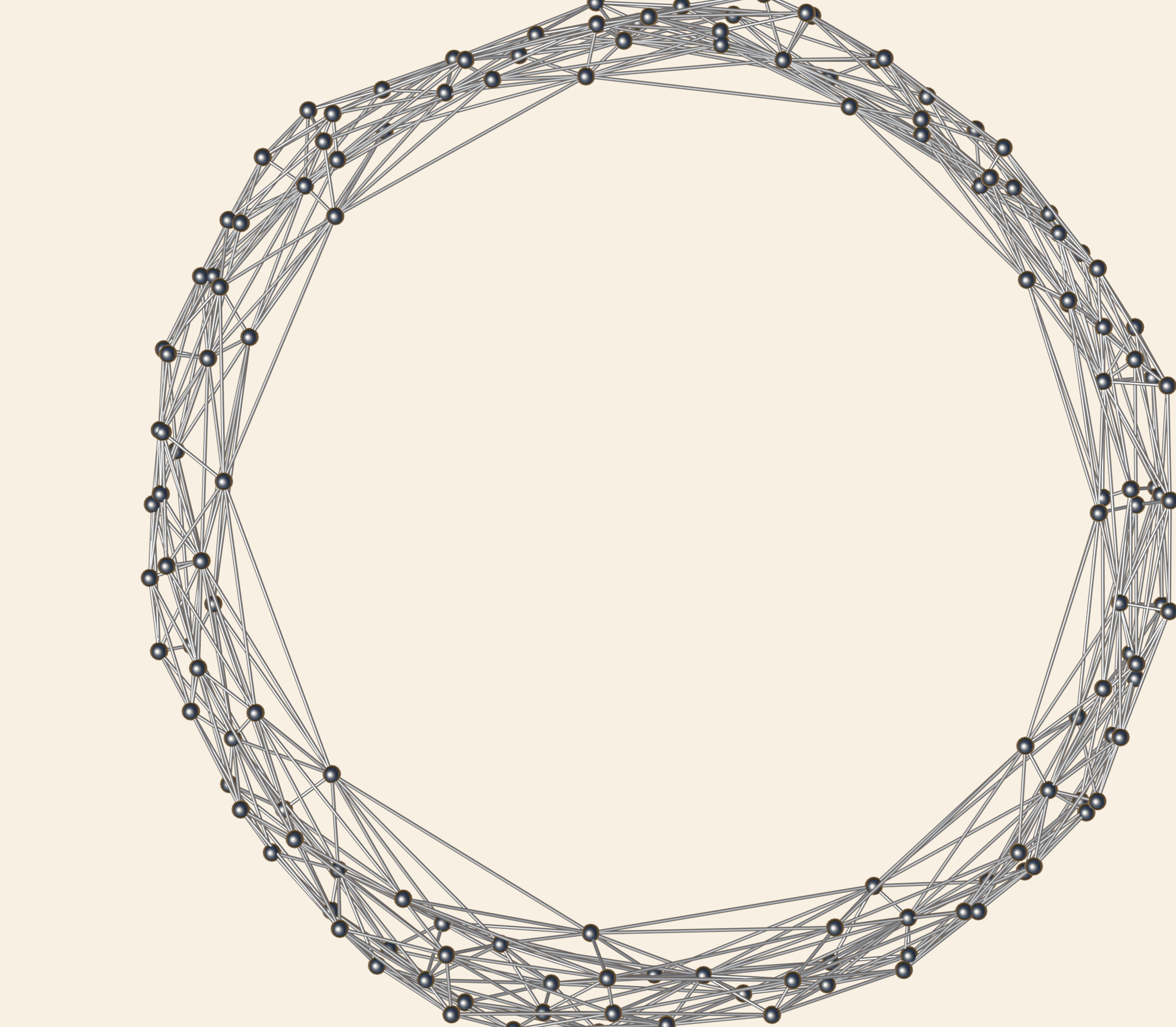} & 
      \includegraphics[width=0.14\columnwidth]{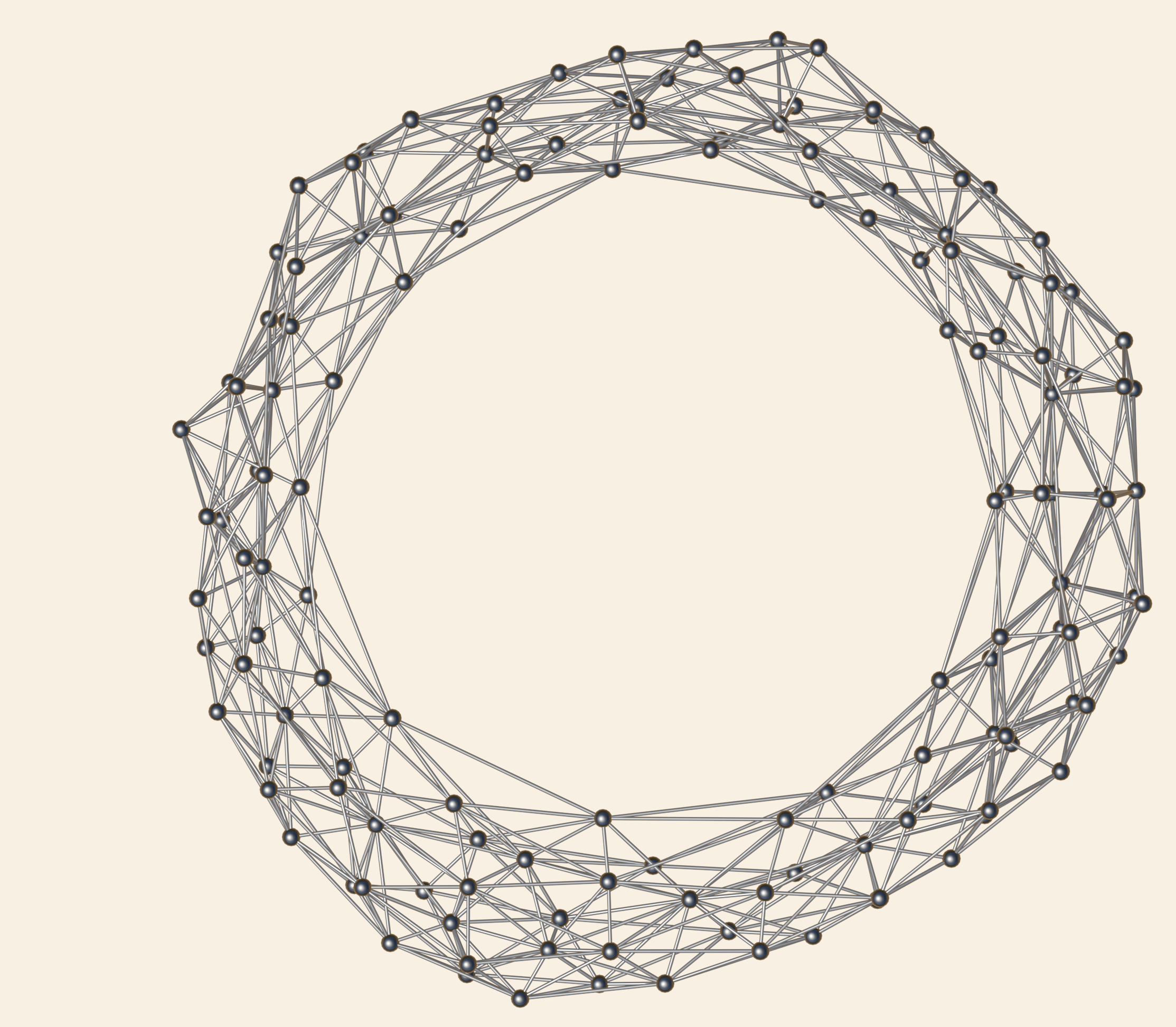} & 
      \includegraphics[width=0.14\columnwidth]{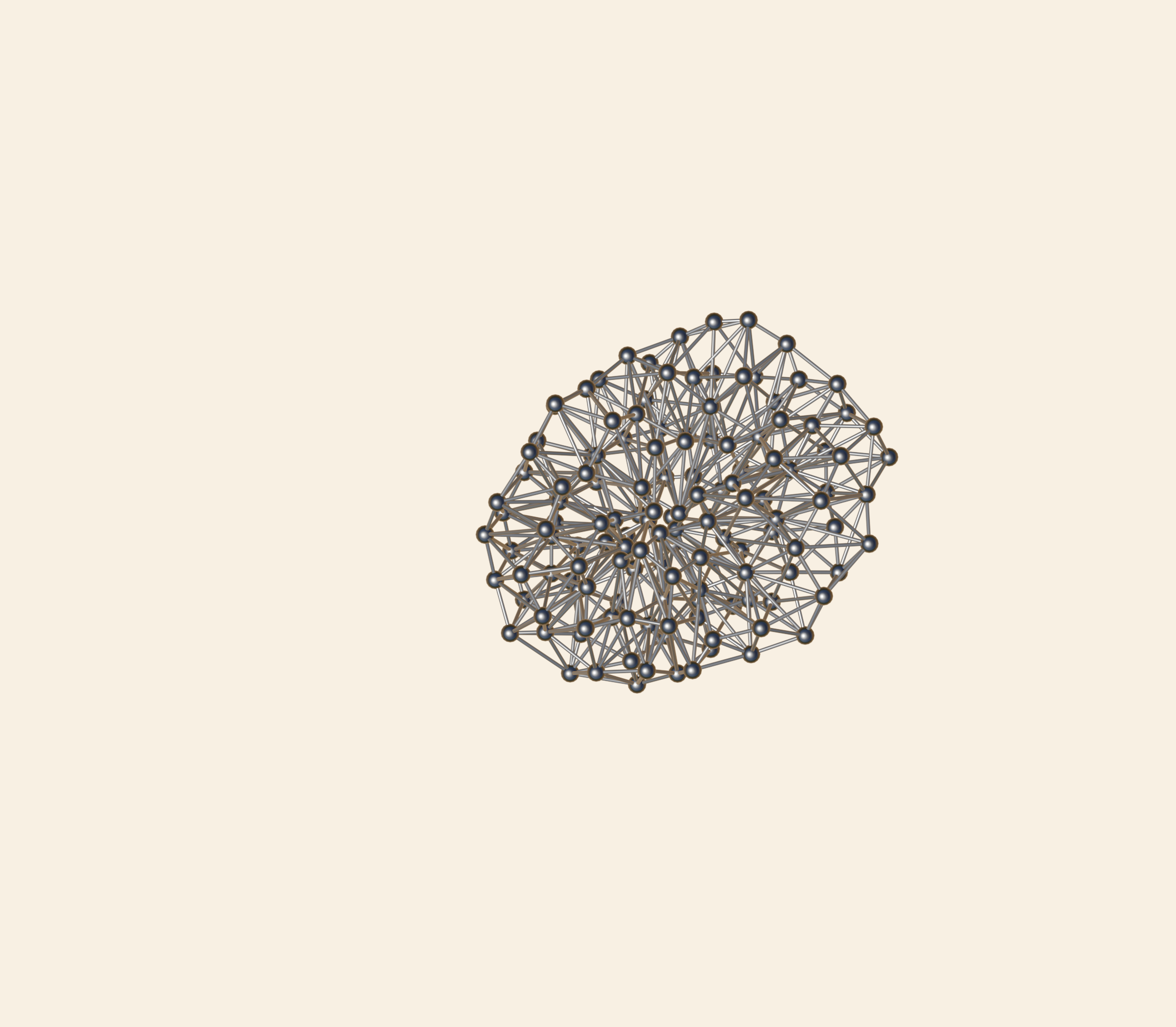} & 
      \includegraphics[width=0.14\columnwidth]{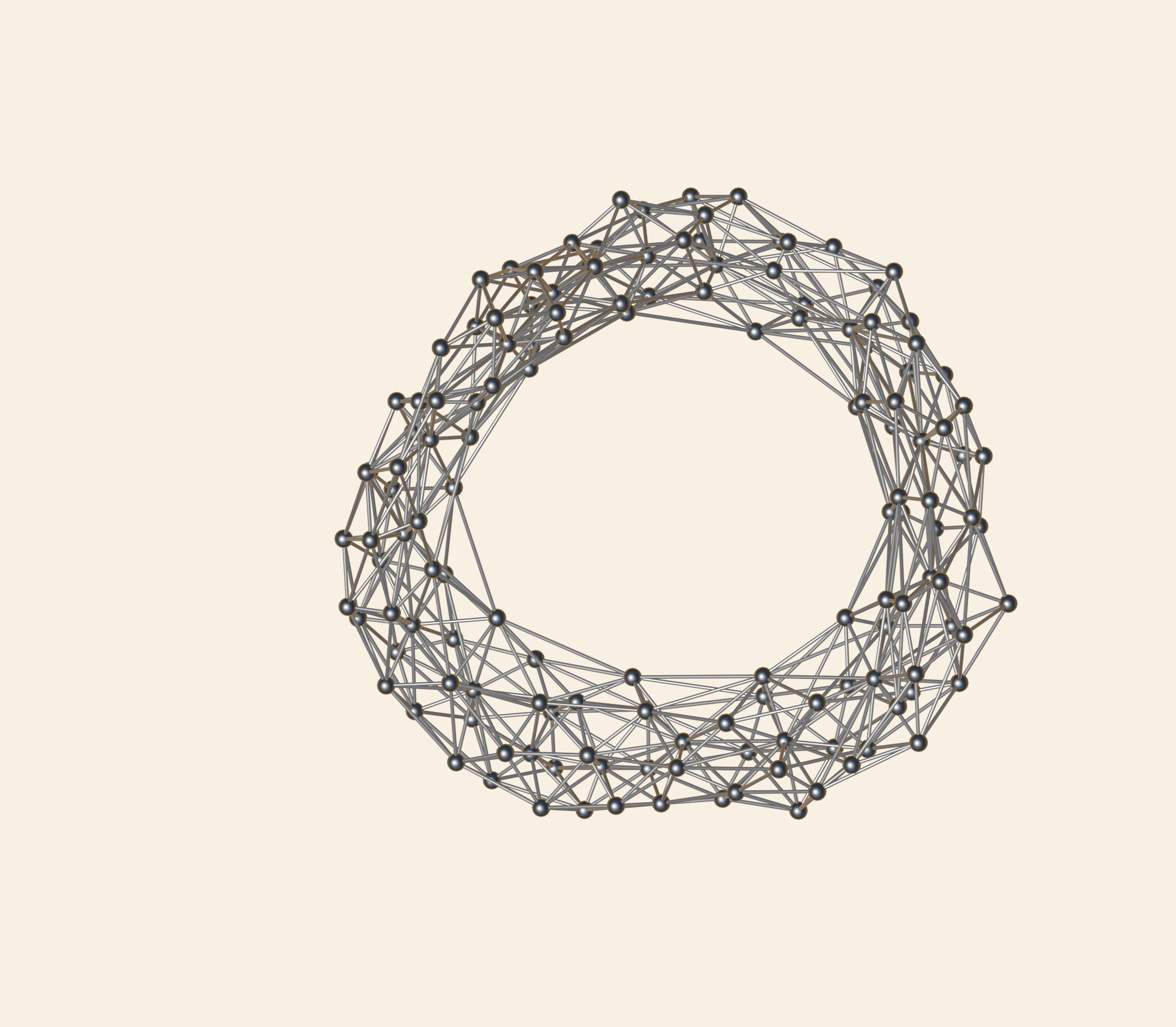} & \includegraphics[width=0.14\columnwidth]{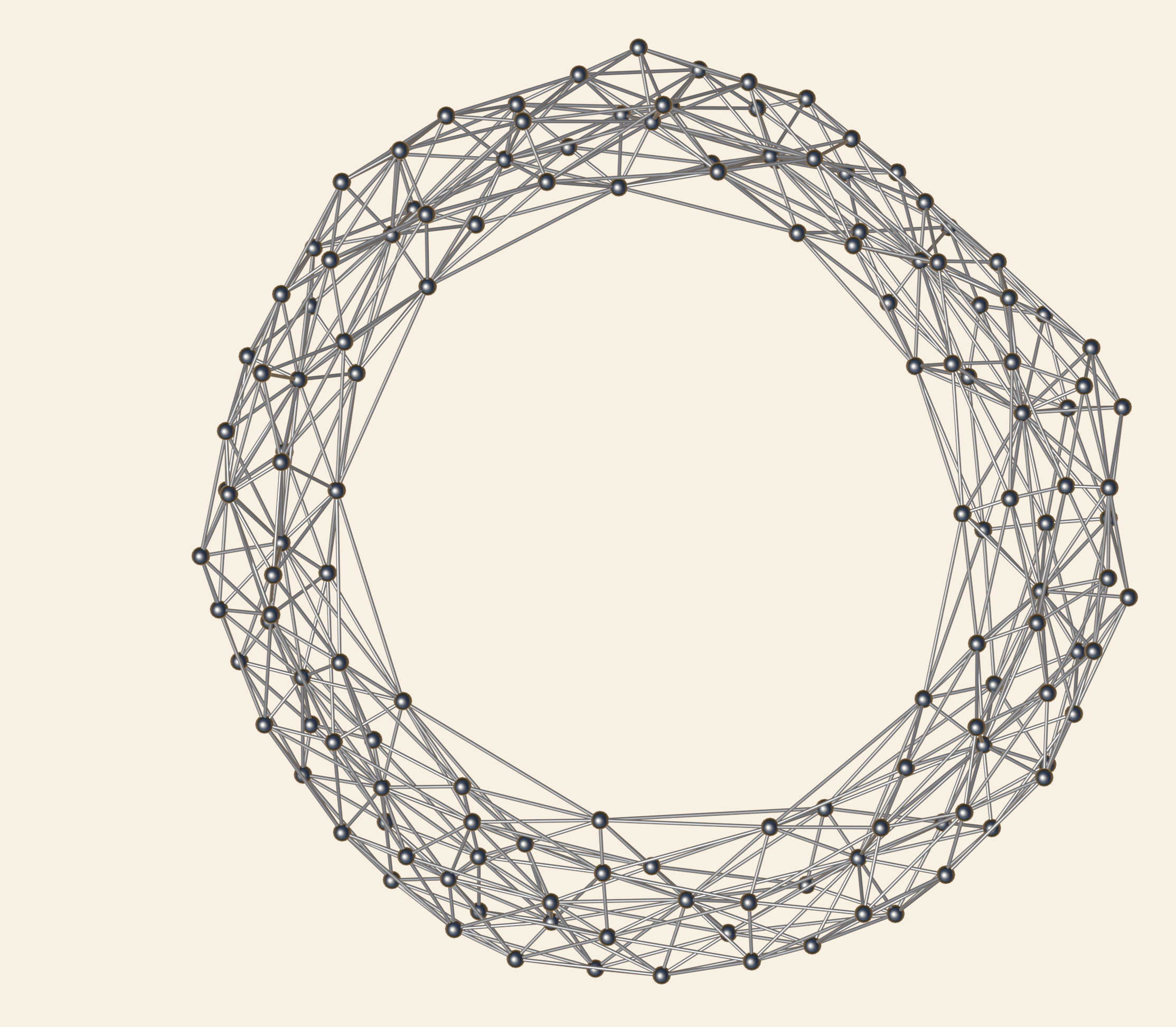}   \\
      \multicolumn{7}{c}{Small time step}\\
         \rotatebox{90}{LAMMPS} & 
      \includegraphics[width=0.14\columnwidth]{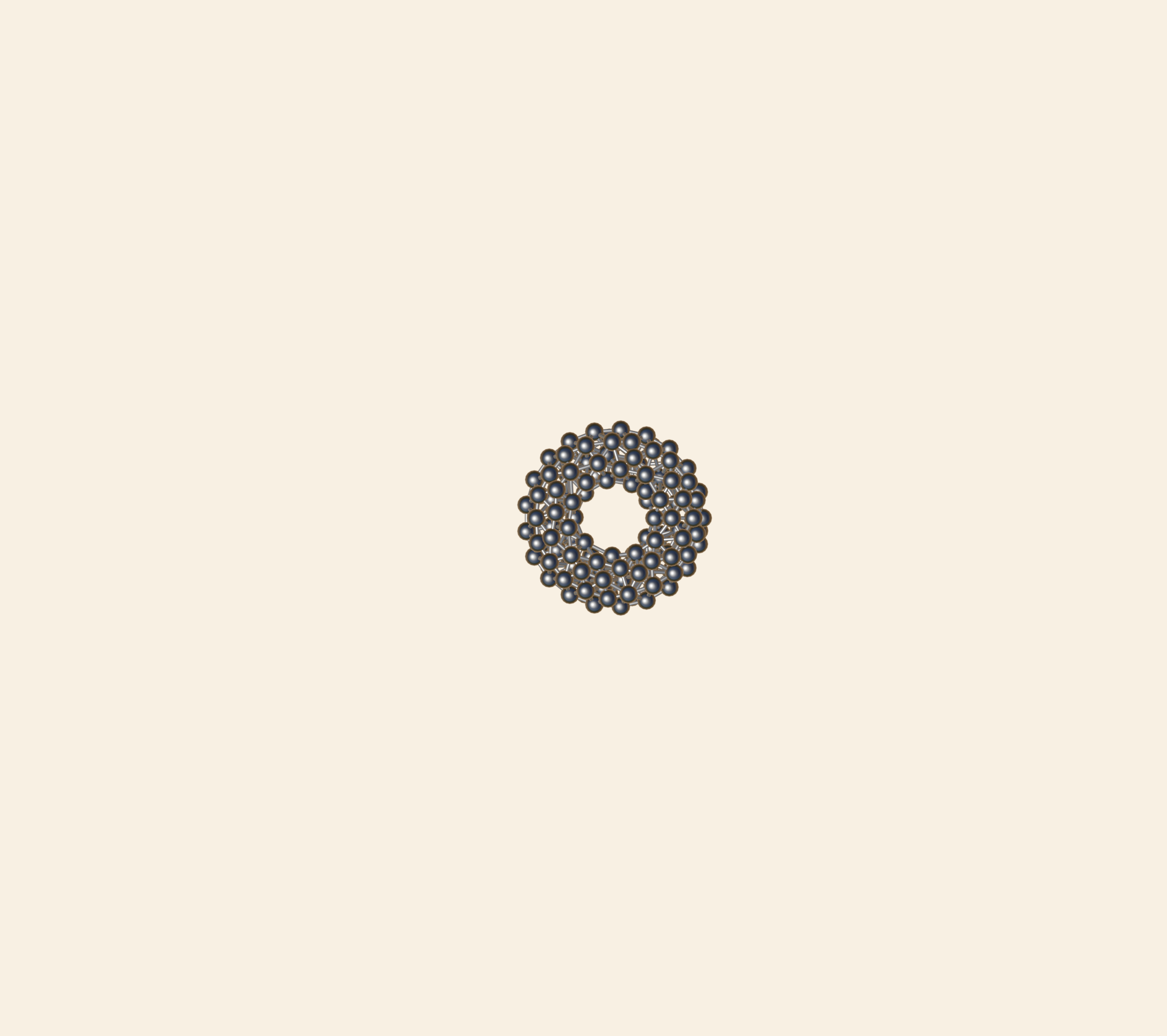} & 
      \includegraphics[width=0.14\columnwidth]{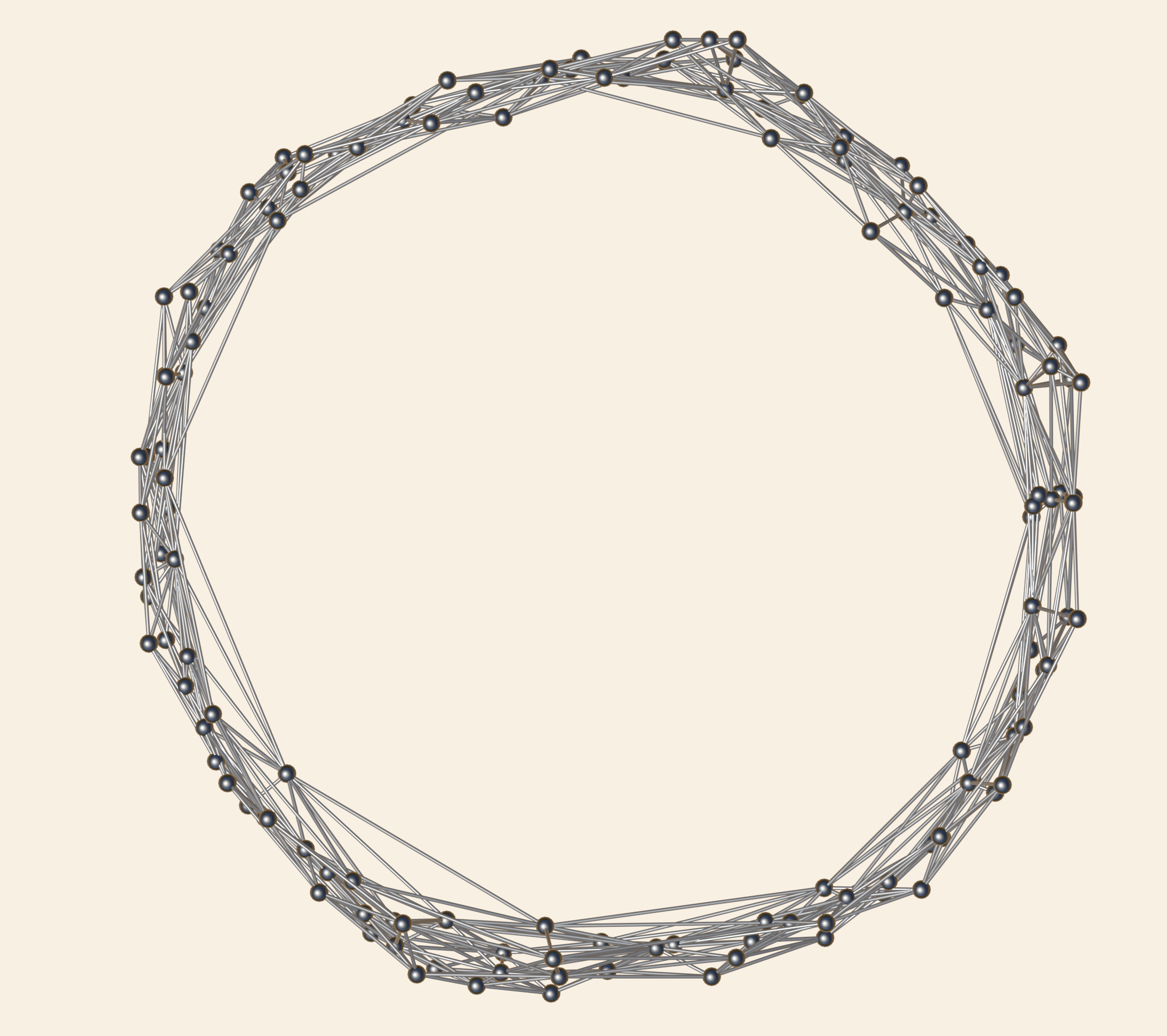} & 
      \includegraphics[width=0.14\columnwidth]{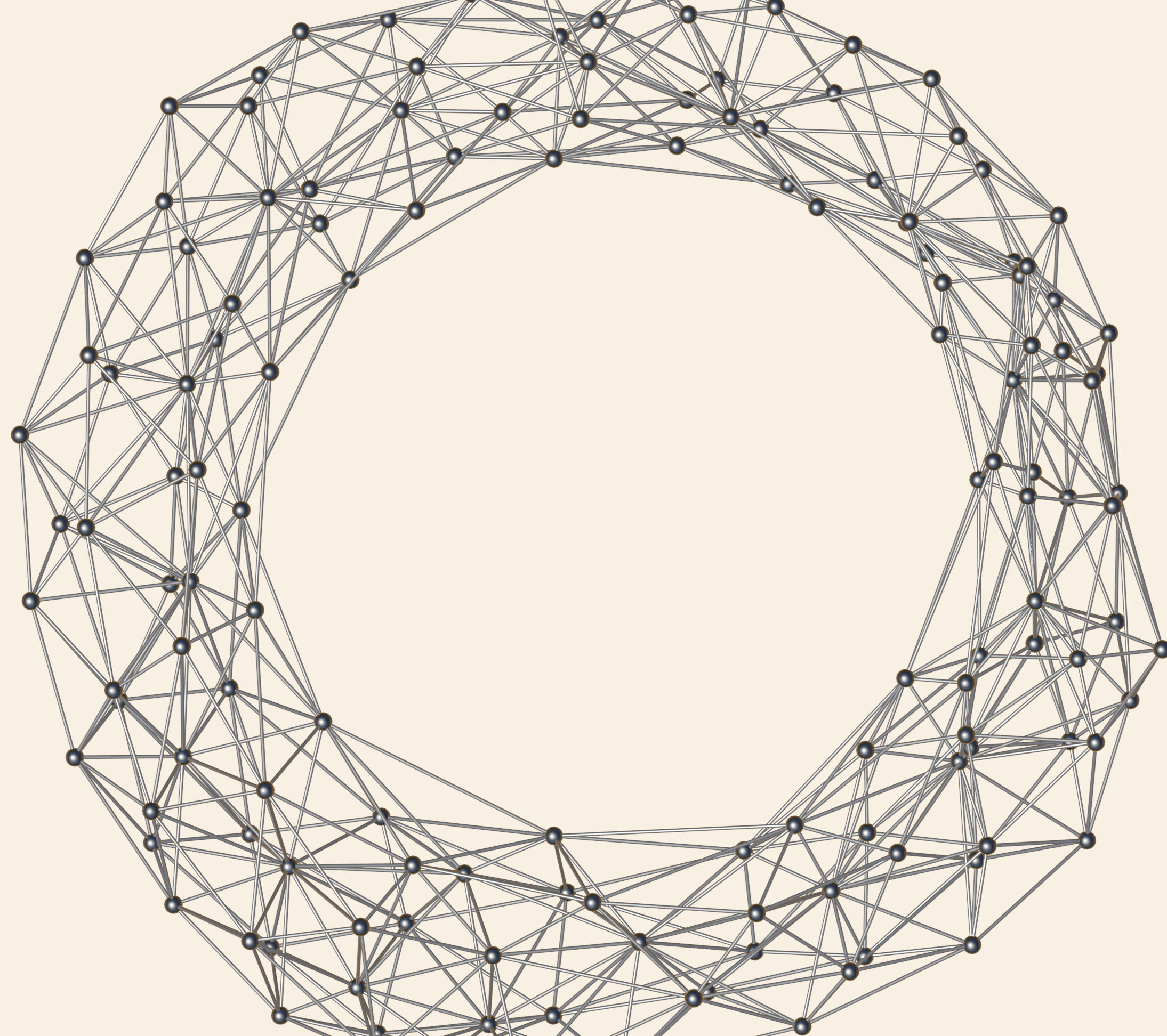} & 
      \includegraphics[width=0.14\columnwidth]{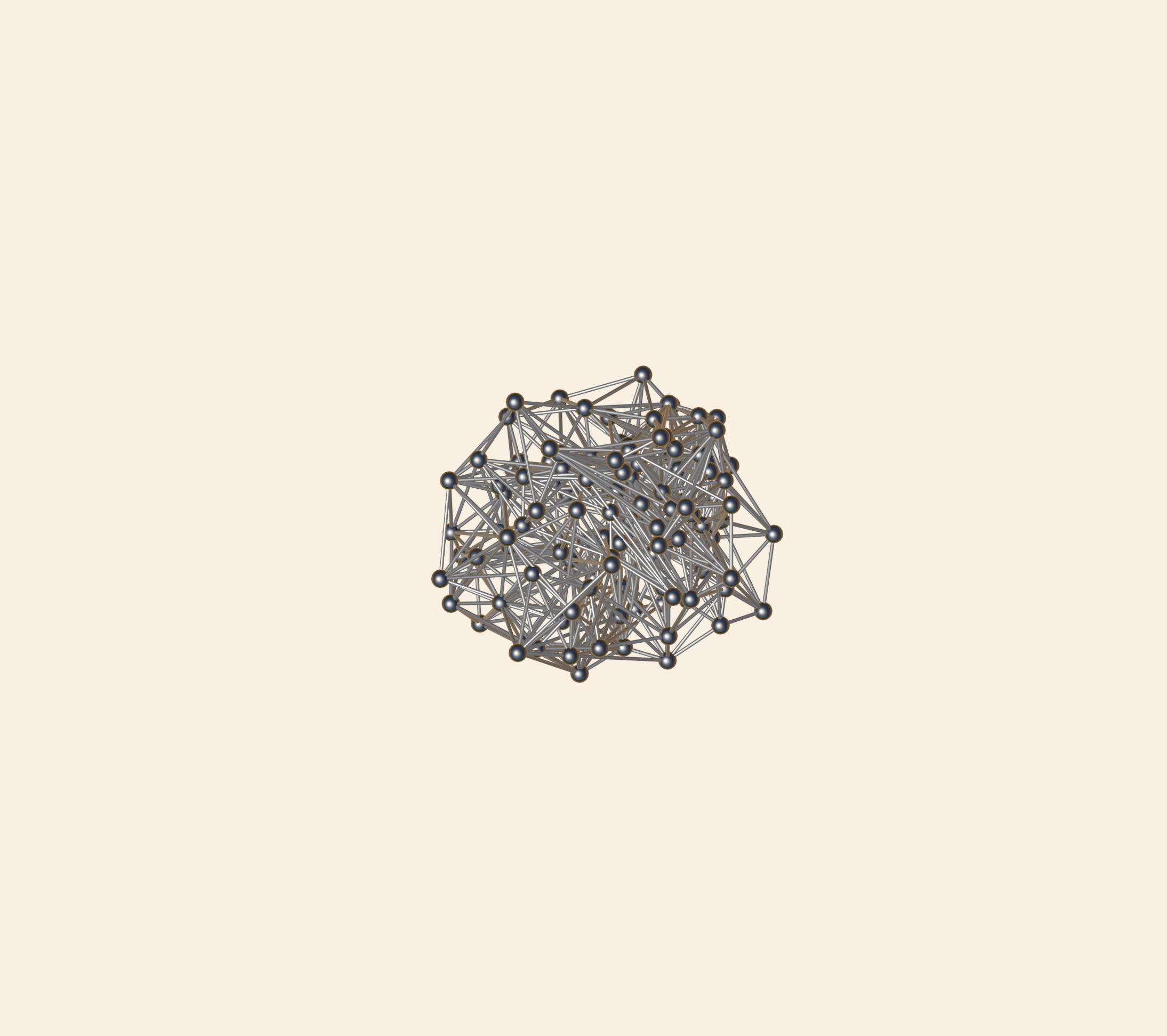} & 
      \includegraphics[width=0.14\columnwidth]{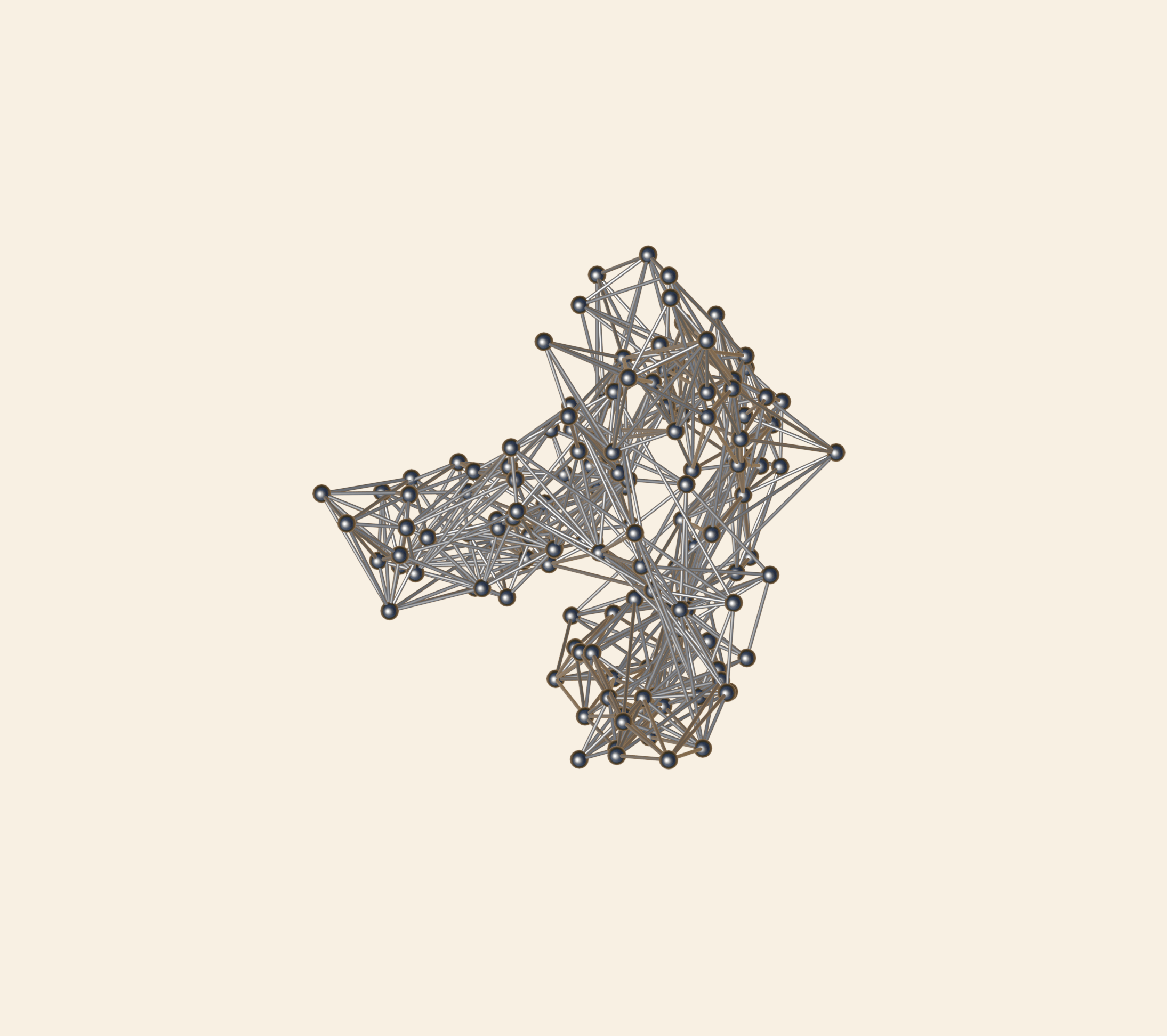} & \includegraphics[width=0.14\columnwidth]{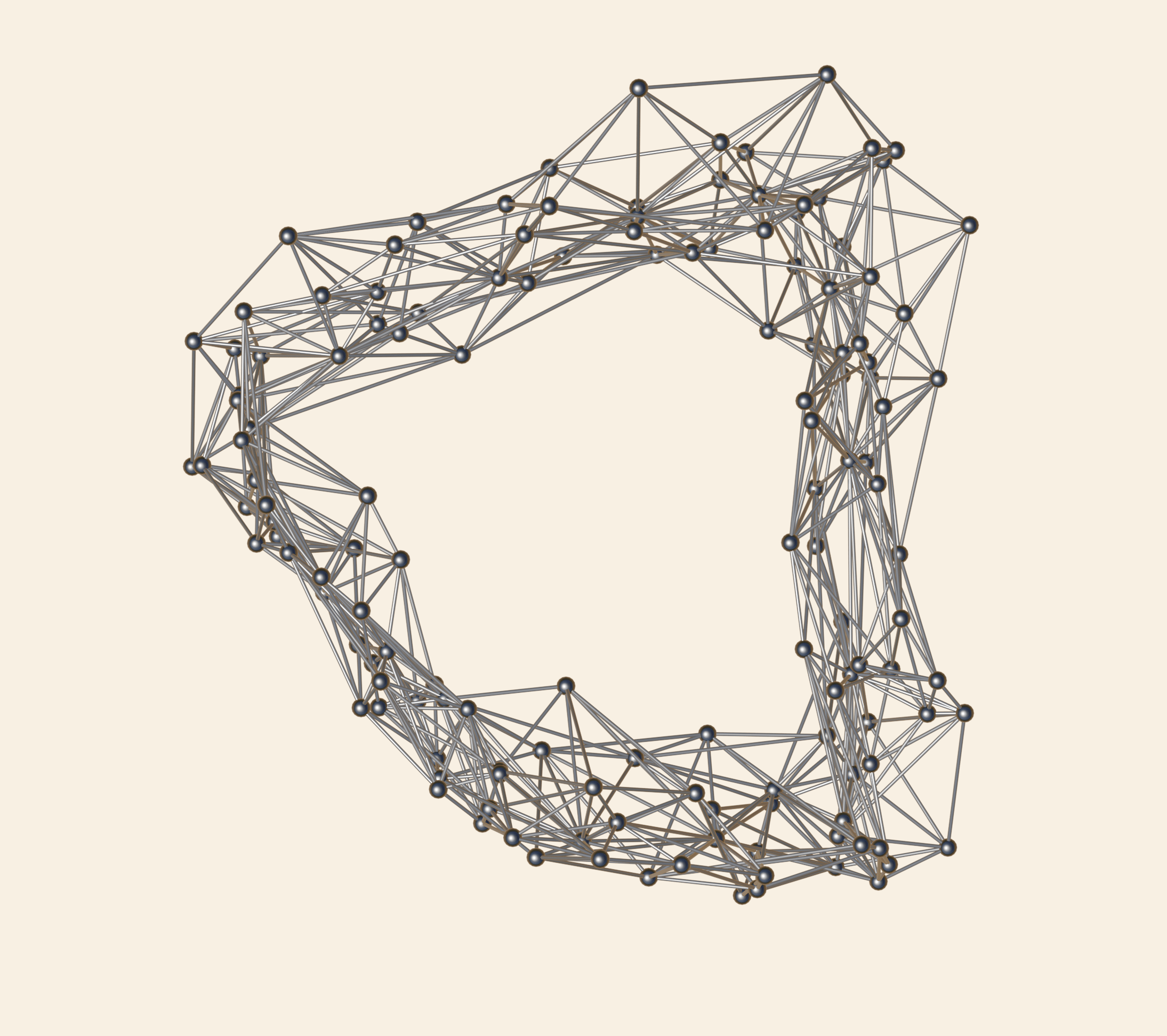}   \\
       \rotatebox{90}{Ours} & 
      \includegraphics[width=0.14\columnwidth]{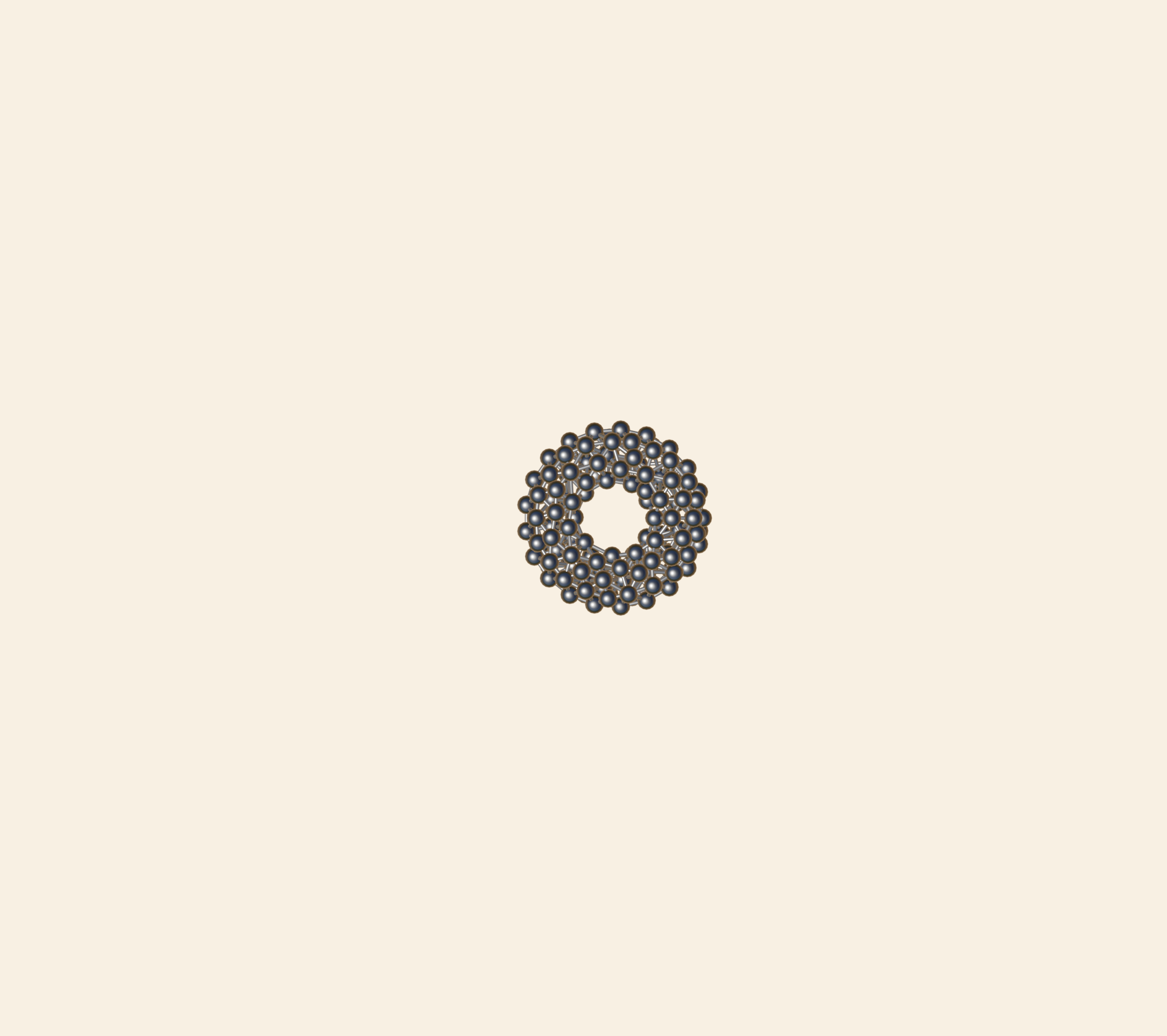} & 
      \includegraphics[width=0.14\columnwidth]{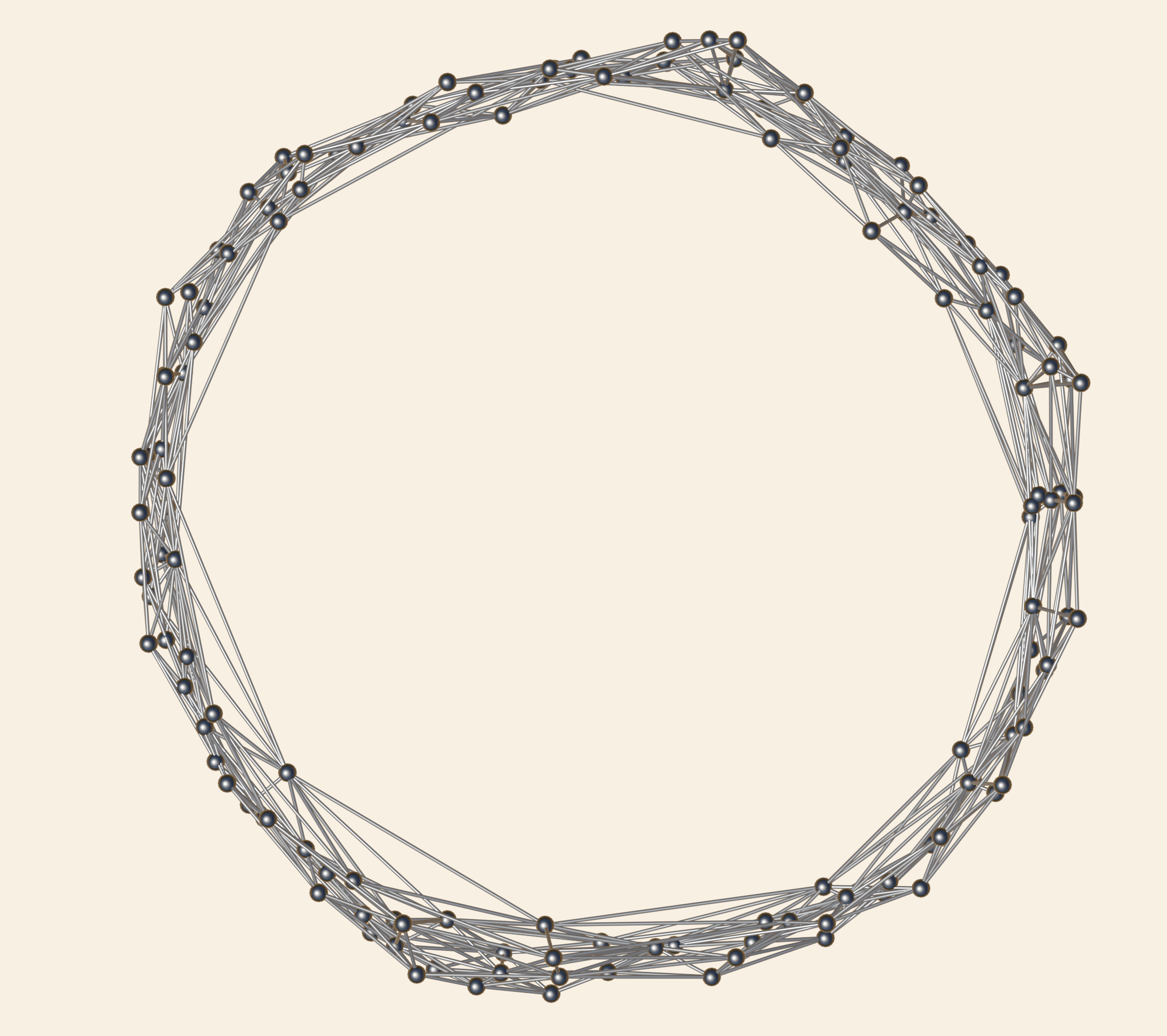} & 
      \includegraphics[width=0.14\columnwidth]{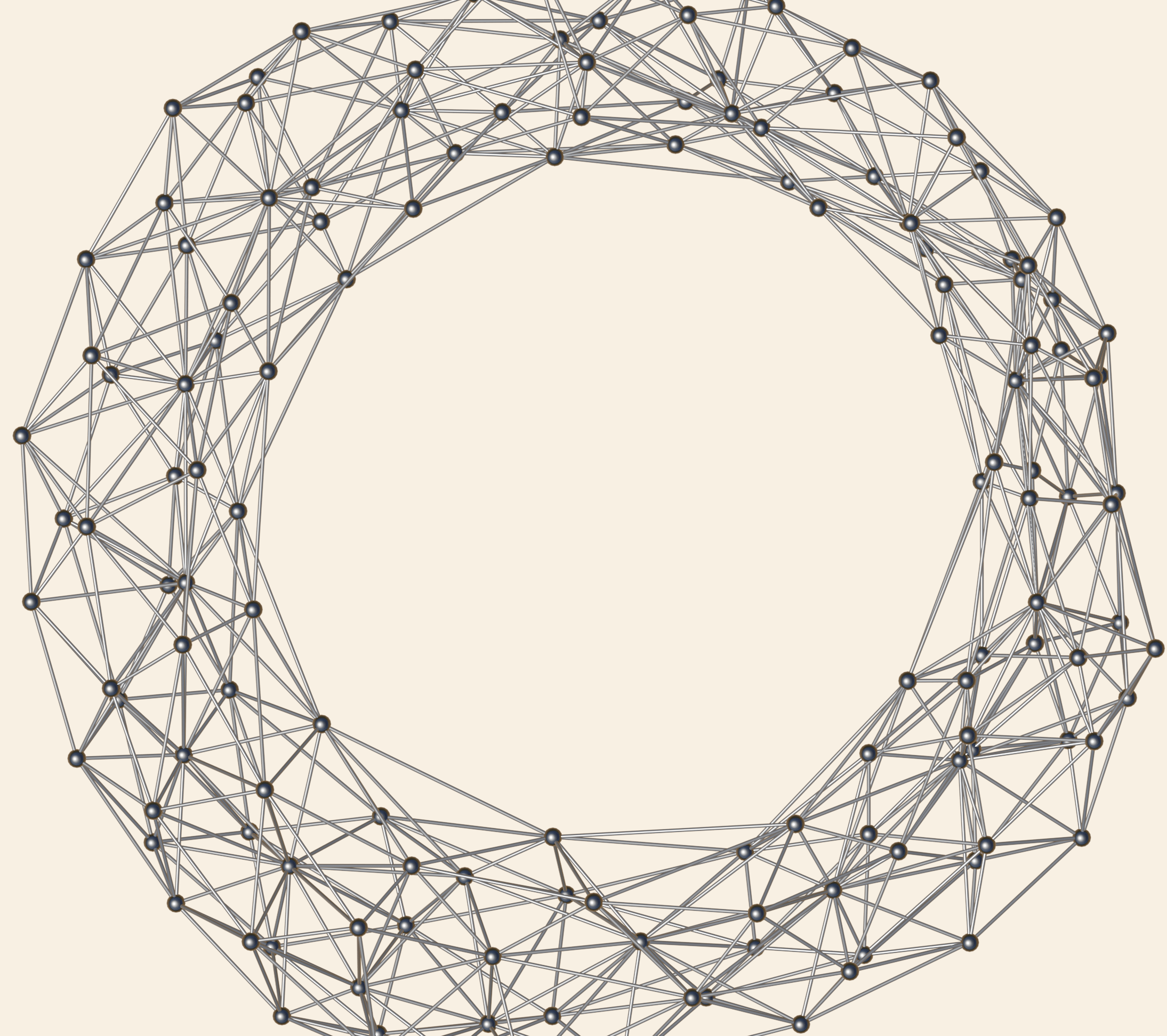} & 
      \includegraphics[width=0.14\columnwidth]{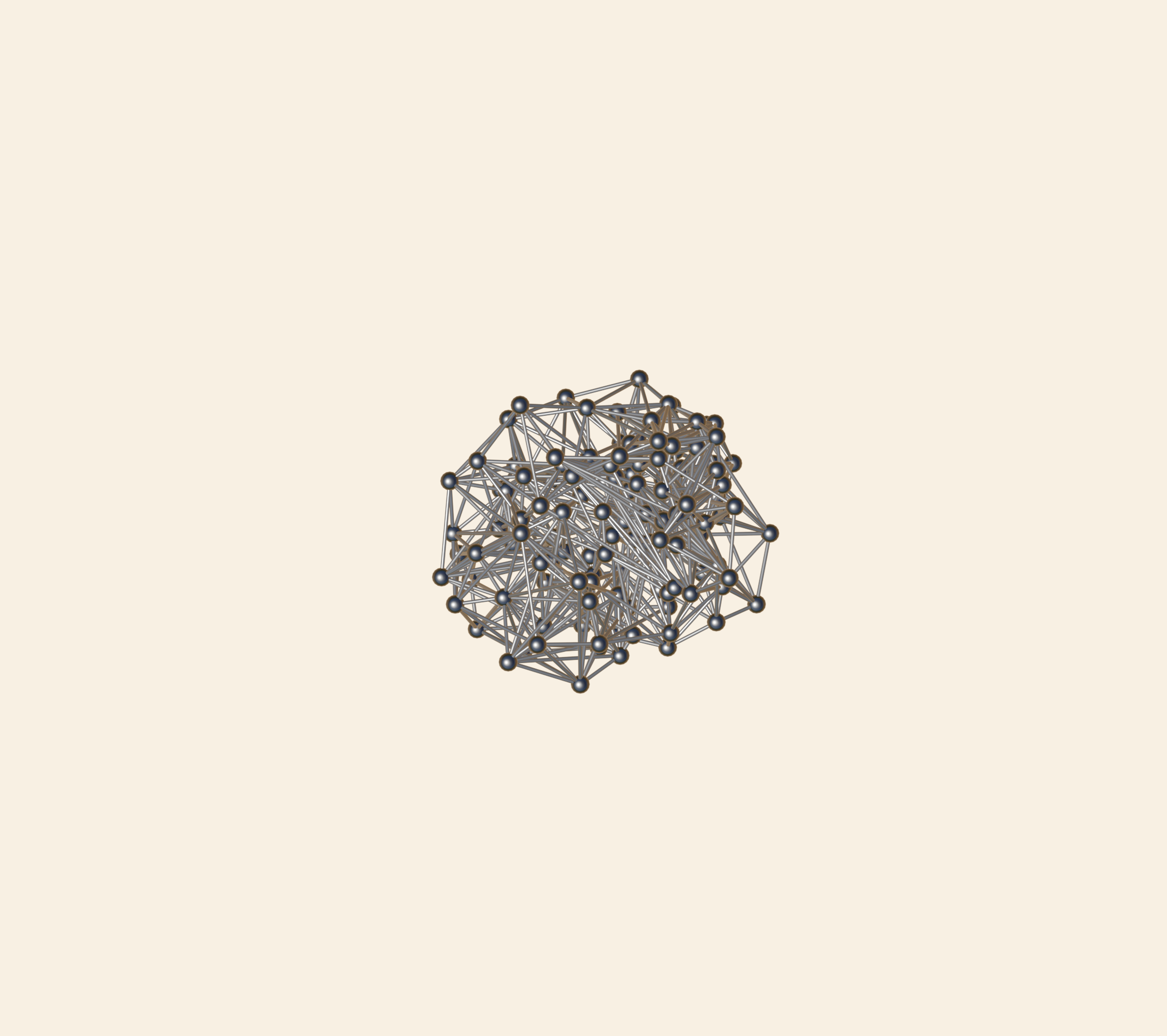} & 
      \includegraphics[width=0.14\columnwidth]{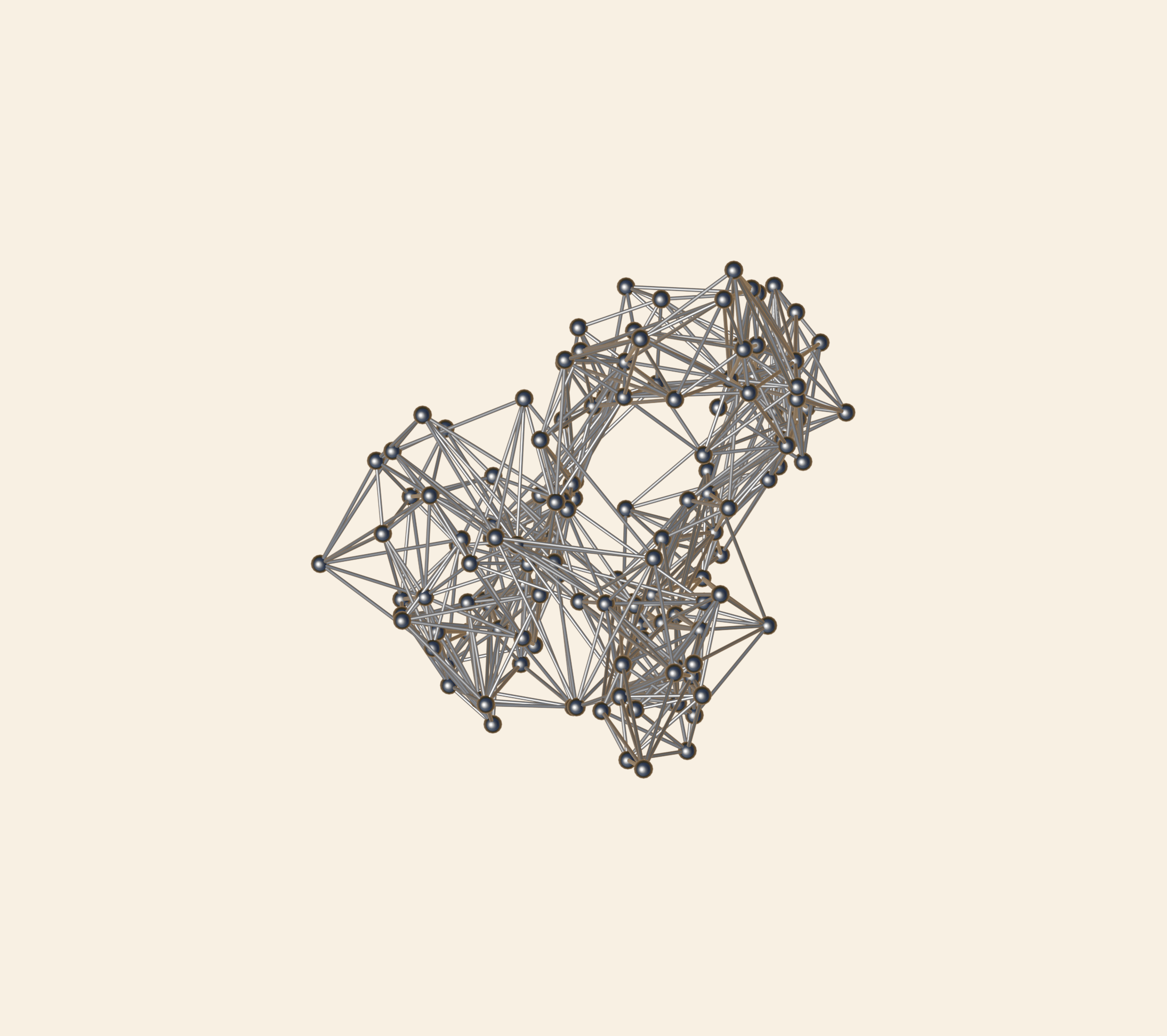} & \includegraphics[width=0.14\columnwidth]{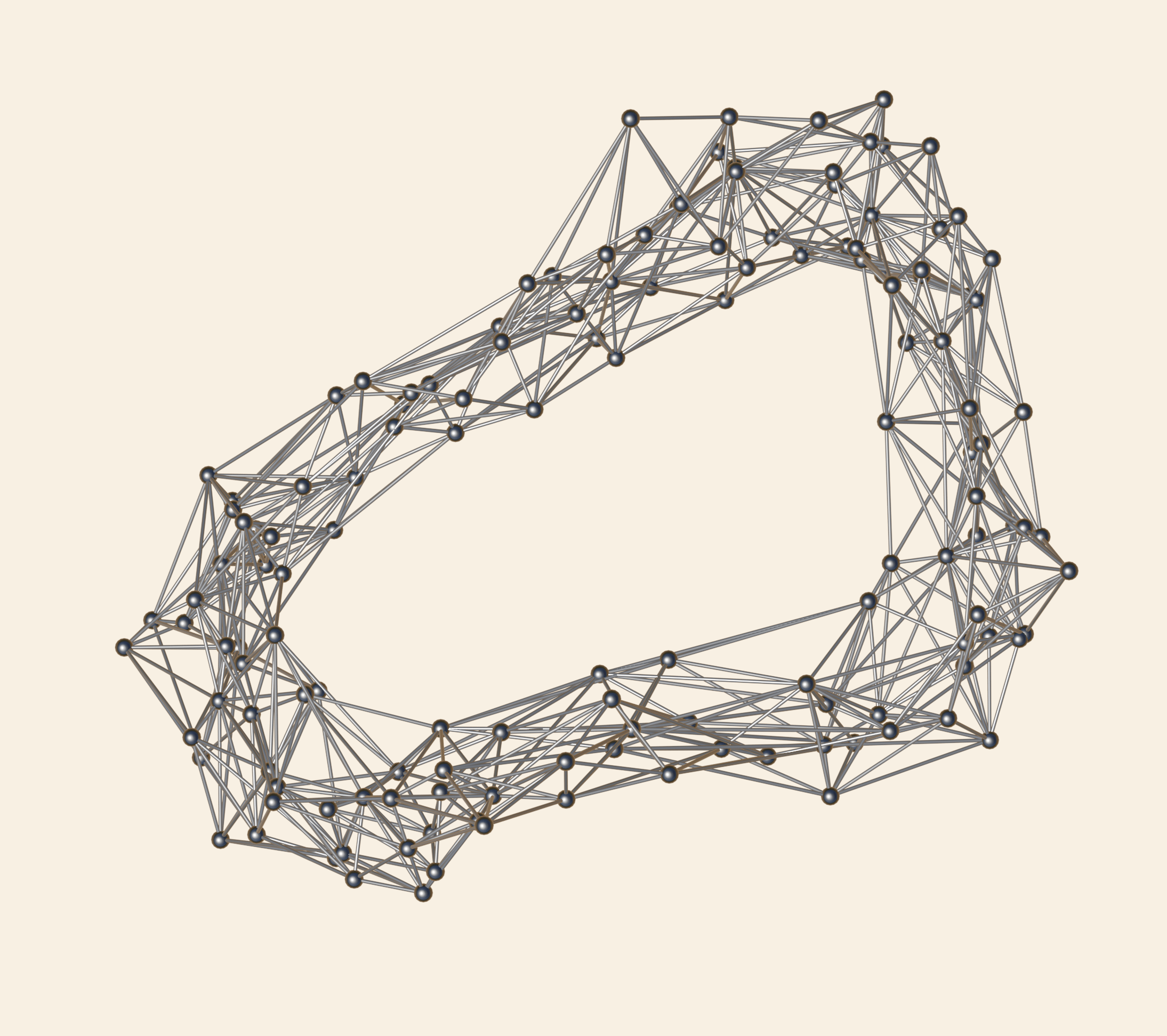}   \\
      \multicolumn{7}{c}{Reference}\\
       \rotatebox{90}{Reference} & 
      \includegraphics[width=0.14\columnwidth]{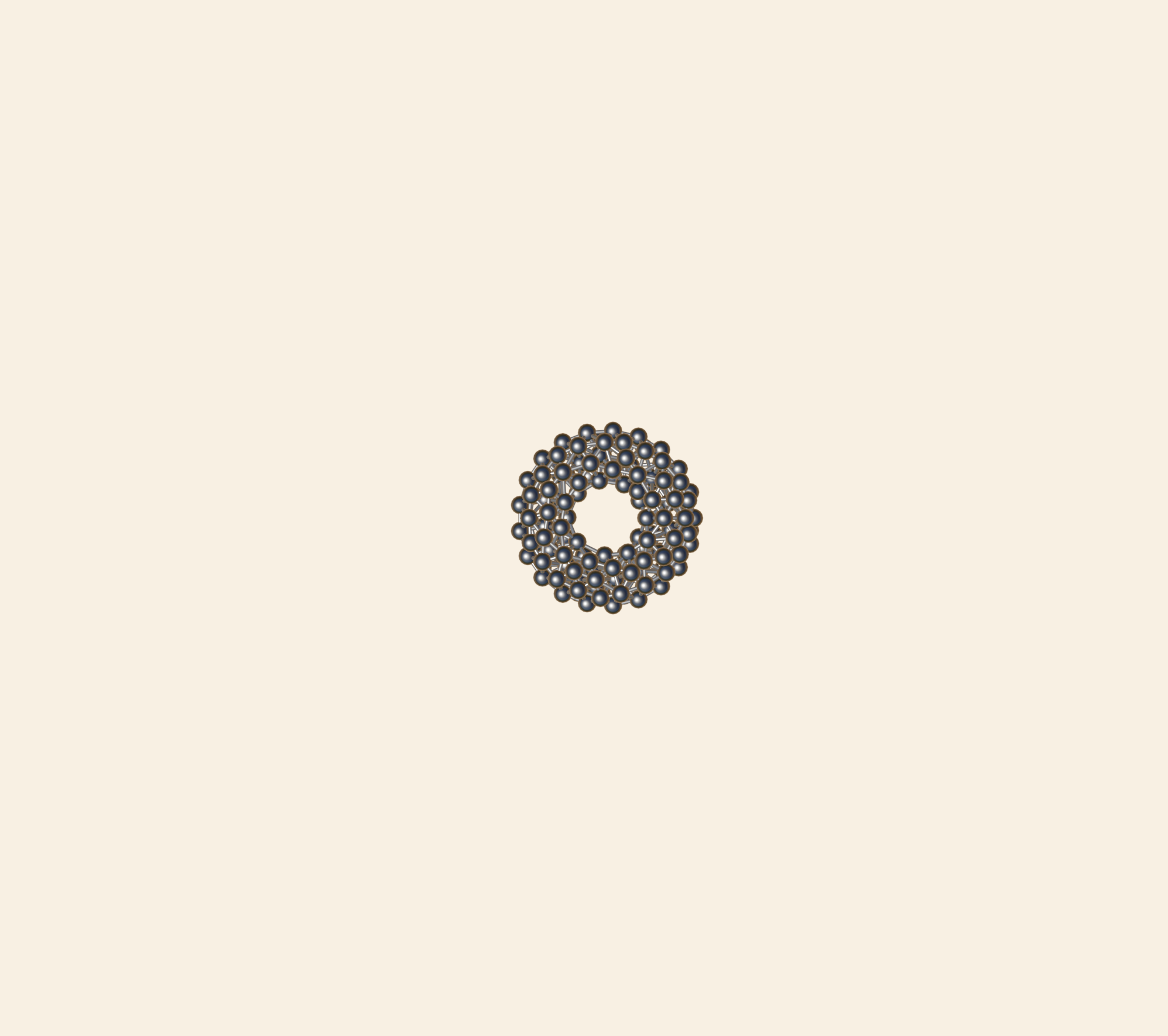} & 
      \includegraphics[width=0.14\columnwidth]{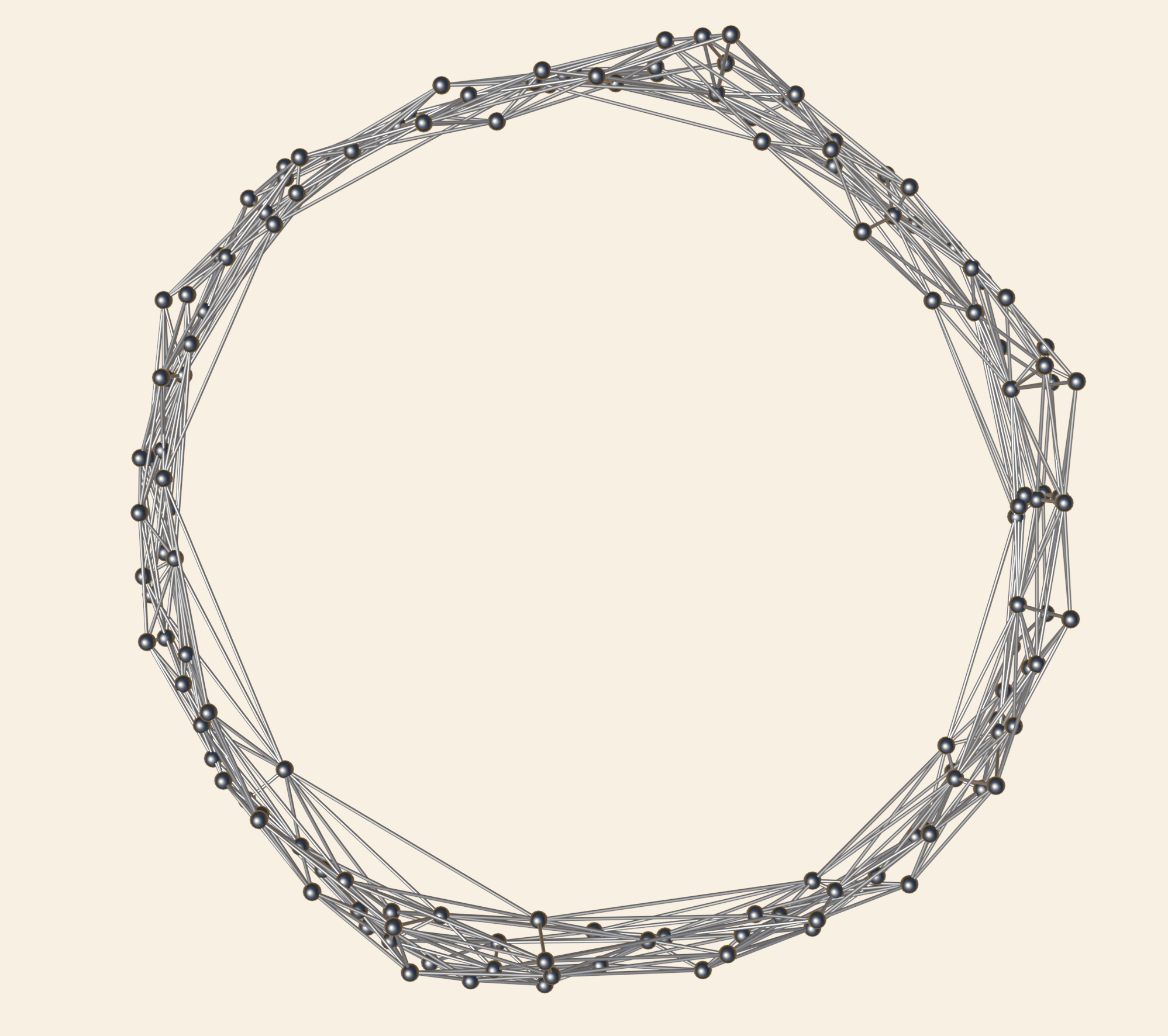} & 
      \includegraphics[width=0.14\columnwidth]{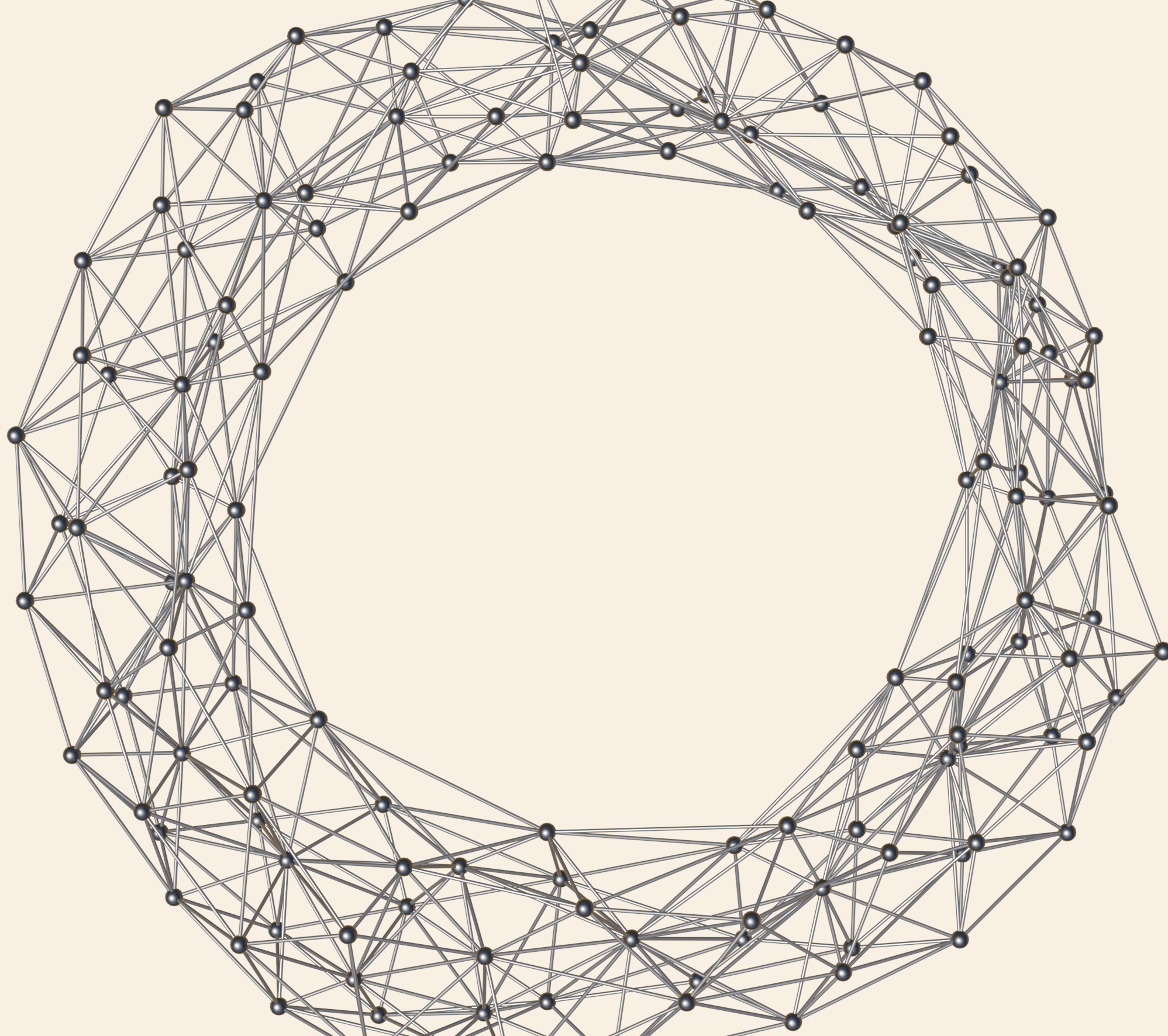} & 
      \includegraphics[width=0.14\columnwidth]{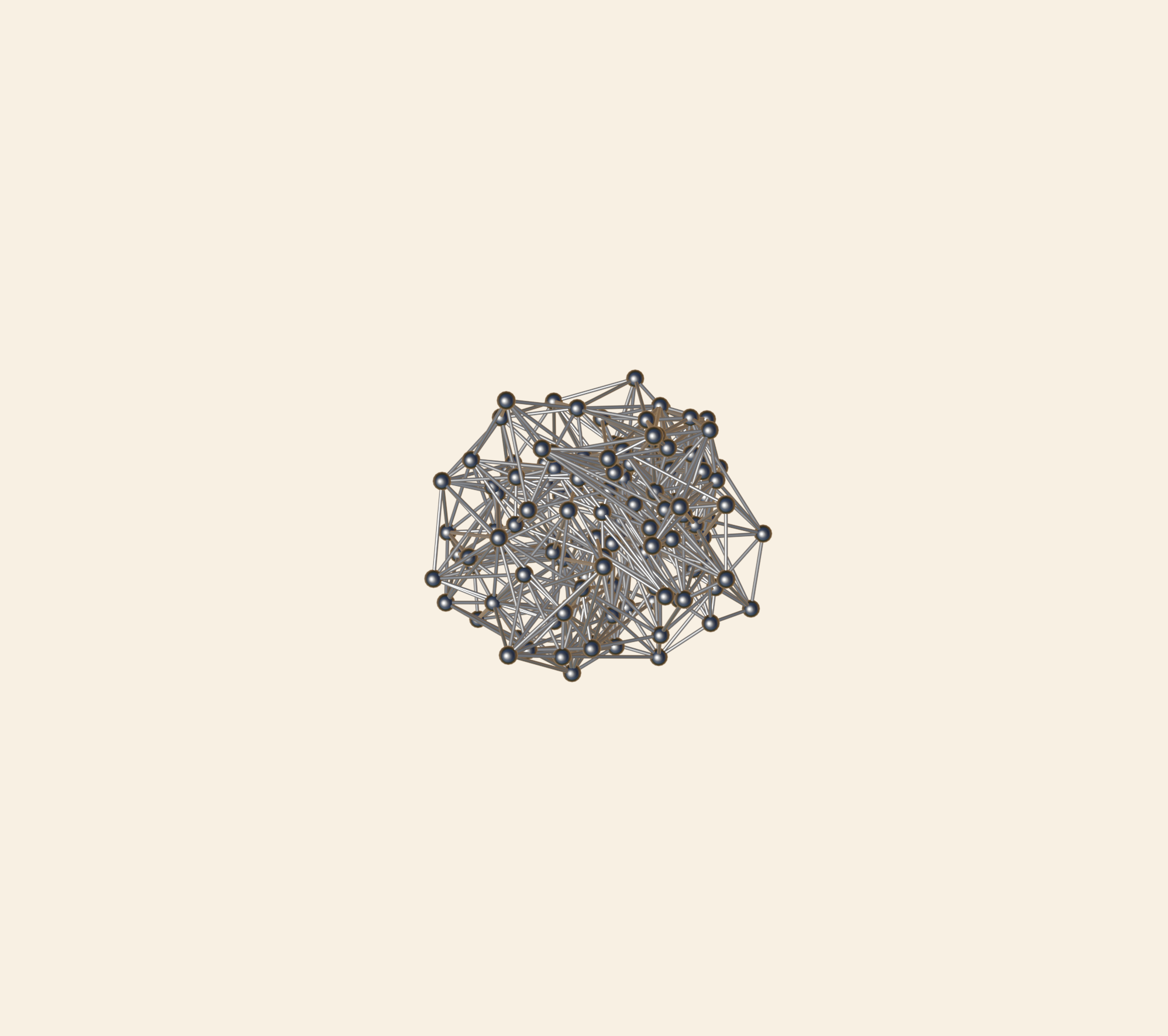} & 
      \includegraphics[width=0.14\columnwidth]{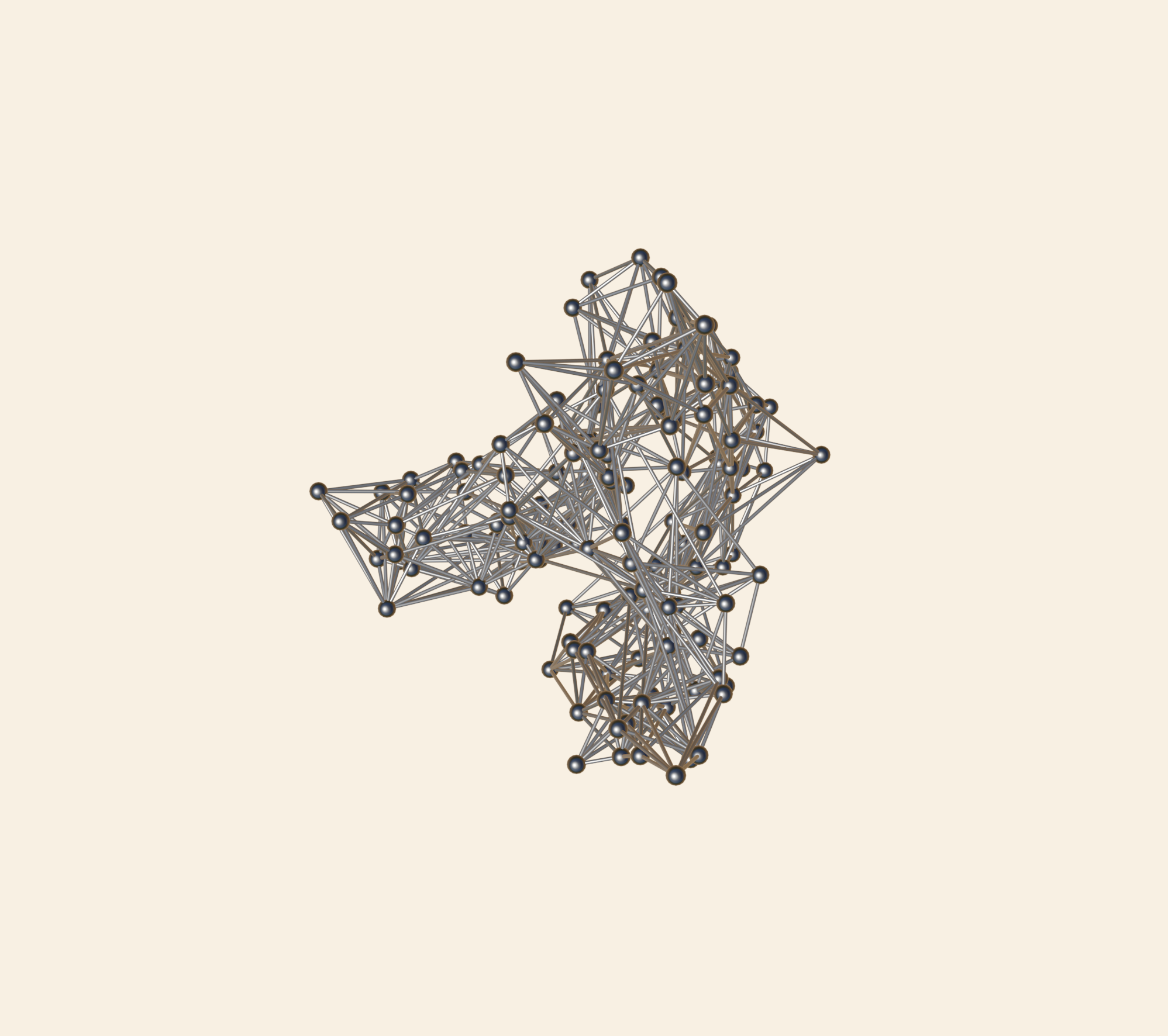} & \includegraphics[width=0.14\columnwidth]{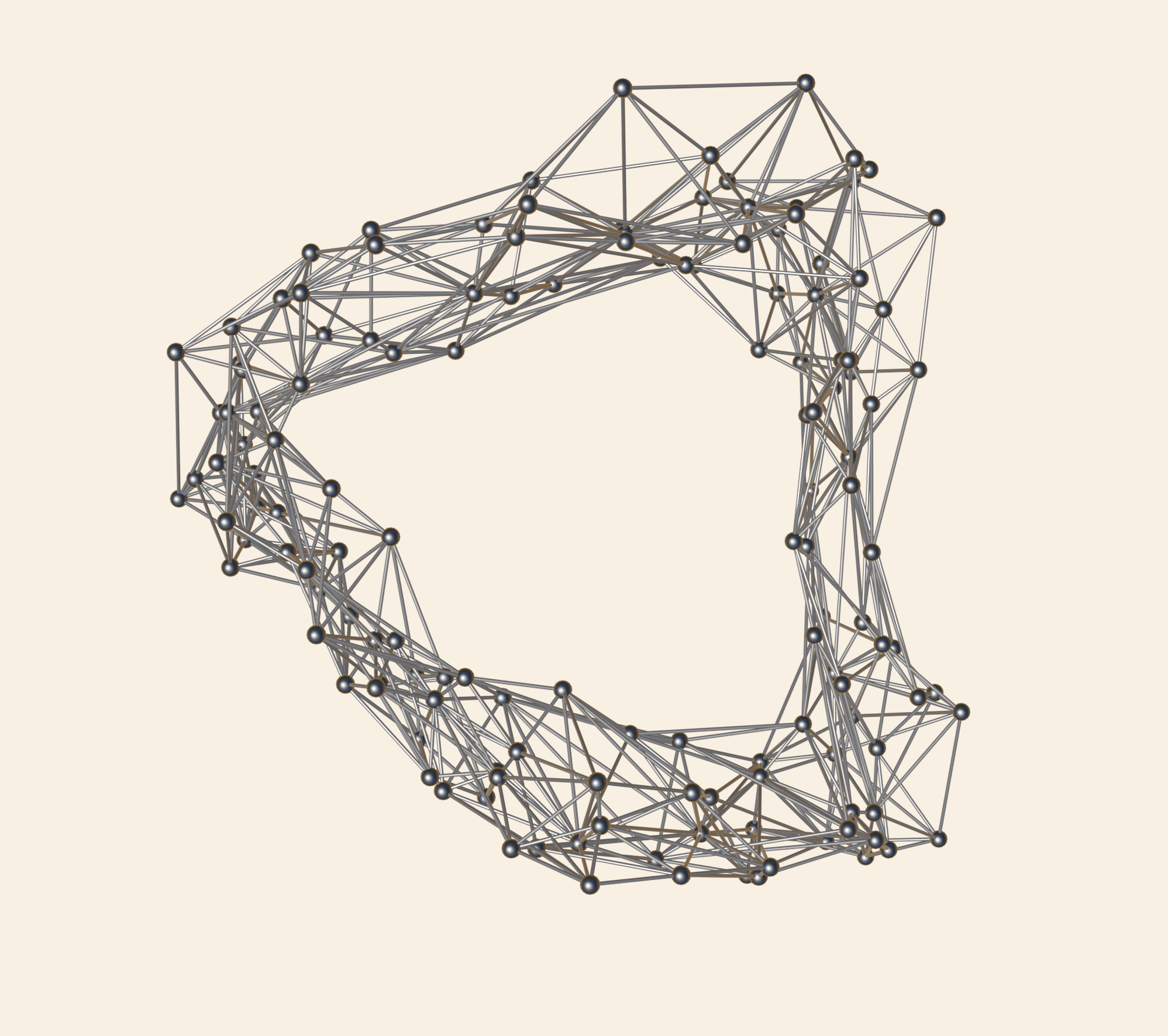}   \\
      &  0 & \qty{2}{s} & \qty{4}{s} &  \qty{6}{s} &  \qty{8}{s} &  \qty{10}{s}\\
    \end{tabular}
    \caption{We compare the evolution (columns) of our implementation with a molecular dynamics package called LAMMPS in simulating a charged mesh of a torus for large ($0.1s$) and small ($0.005$) time-steps. 
    % For large time-steps, we simulate the dynamics for \qty{5}{s} with 50 steps. The time step is therefore $\qty{0.1}{s}$. We can see our method can still generate reasonable results while LAMMPS starts to process in irregular shapes, and in the small time step (\qty{0.005}{s}), we can see LAMMPS can maintain accurate for longer time than ours.
    % \SP{The horizontal axis seems to be time in seconds, but for large time step we say we run only for \qty{5}{s}; not sure what happens for small time step. Please check!}
    \label{fig:qualLAMMPS}}
\end{figure}

We compare our mass-spring-charge simulation against LAMMPS\cite{LAMMPS} on a torus that is released freely. The results of the simulations are shown in Fig~\ref{fig:qualLAMMPS} for two different choices of $h$. While these qualitative results demonstrate that our system results in plausible behavior, it underscores the difficulty in using an error metric based on vertex positions to compare results. Hence our choice to compare relative error of system energies for quantitative experiments.  

\begin{figure}[htbp]
    \centering
    \setlength{\tabcolsep}{2pt} % Reduce space between columns
    \renewcommand{\arraystretch}{0.8} % Reduce space between rows
    \begin{tabular}{ccccccc}
            % Consider removing [6pt] to reduce vertical space
            
        \rotatebox{90}{DiffCloth} & 
        \includegraphics[width=0.14\columnwidth]{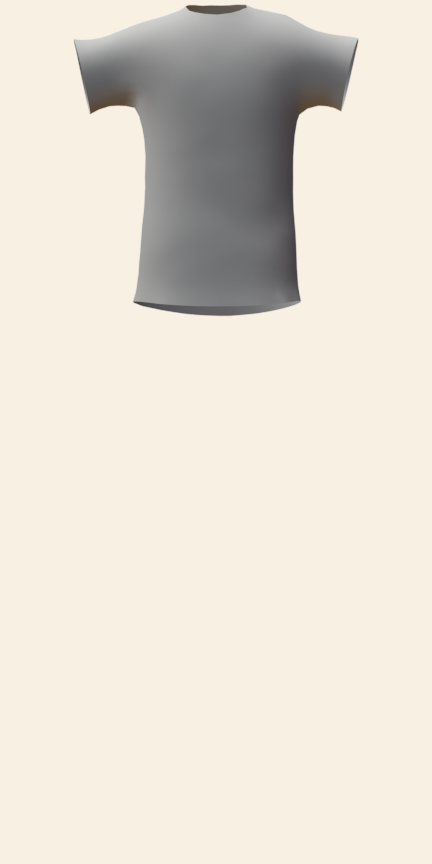} & 
        \includegraphics[width=0.14\columnwidth]{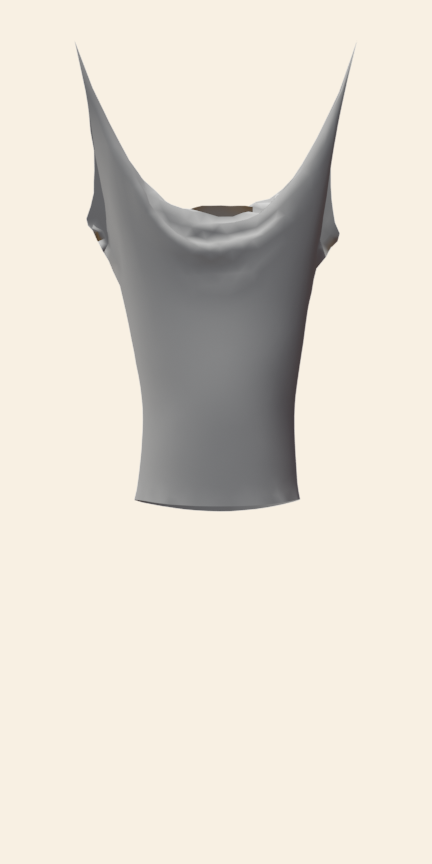} & 
        \includegraphics[width=0.14\columnwidth]{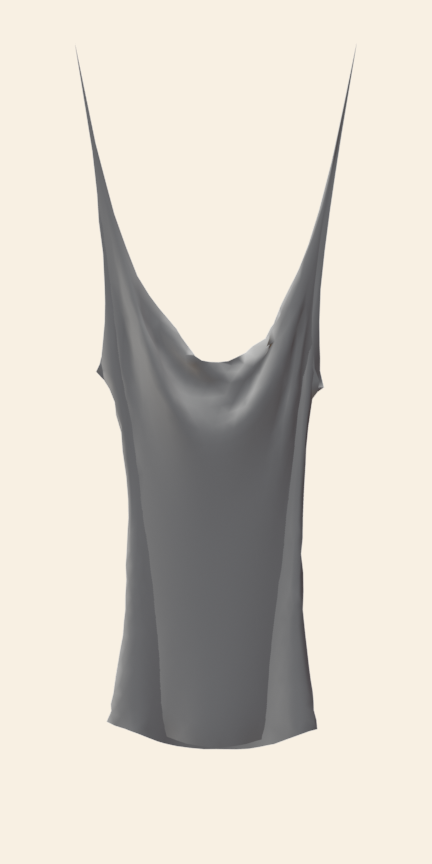} & 
        \includegraphics[width=0.14\columnwidth]{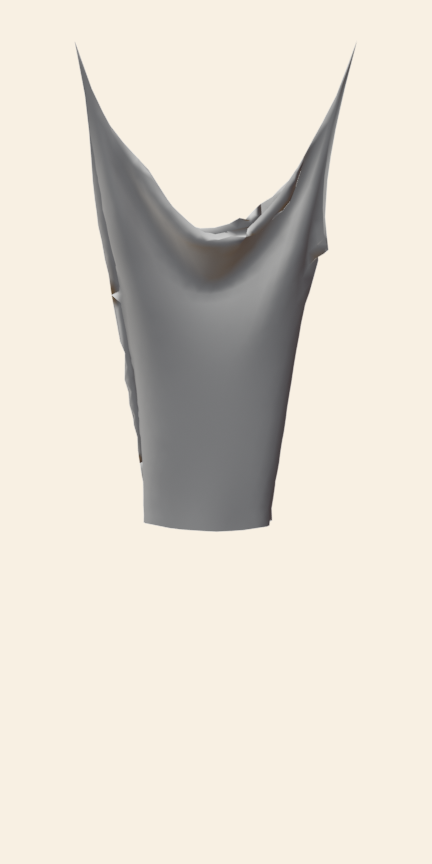} & 
        \includegraphics[width=0.14\columnwidth]{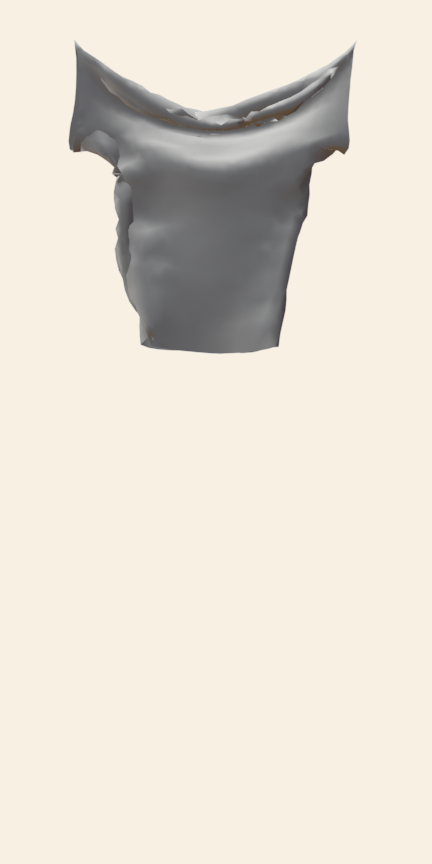} & \includegraphics[width=0.14\columnwidth]{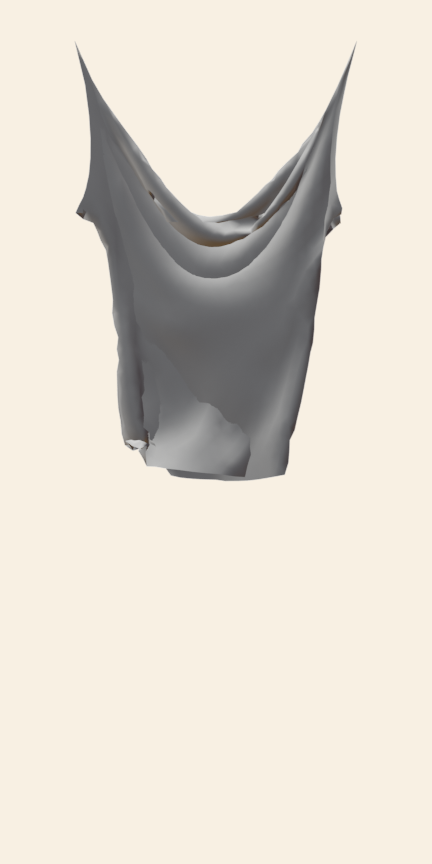}   \\
        \rotatebox{90}{Ours}& 
        \includegraphics[width=0.14\columnwidth]{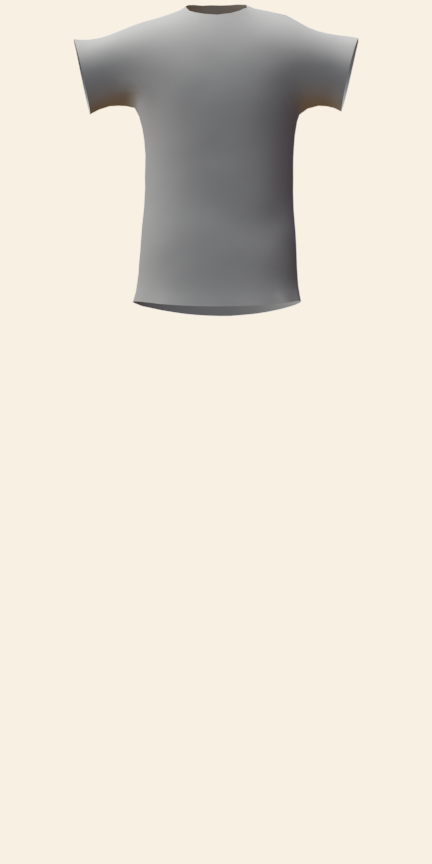} & 
        \includegraphics[width=0.14\columnwidth]{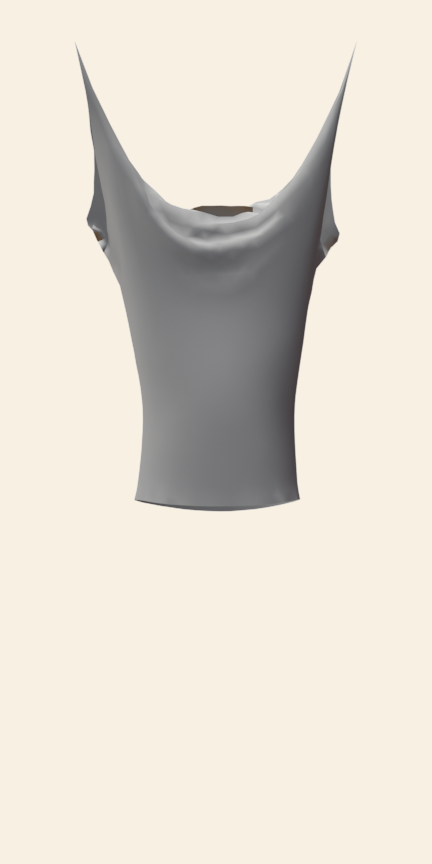} & 
        \includegraphics[width=0.14\columnwidth]{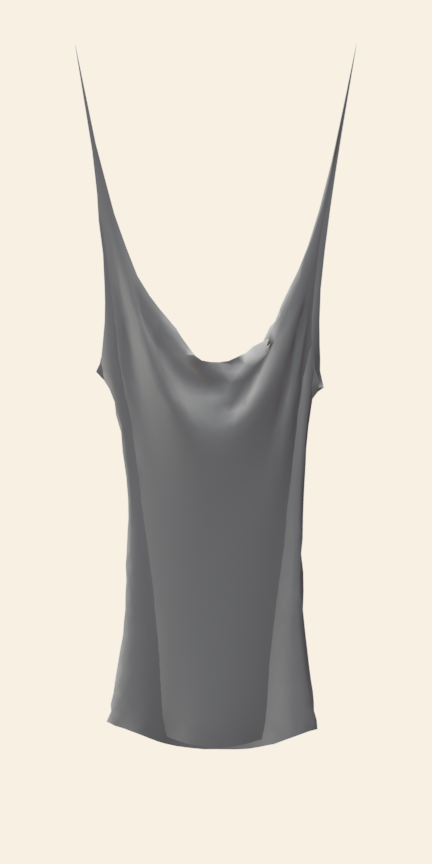} & 
        \includegraphics[width=0.14\columnwidth]{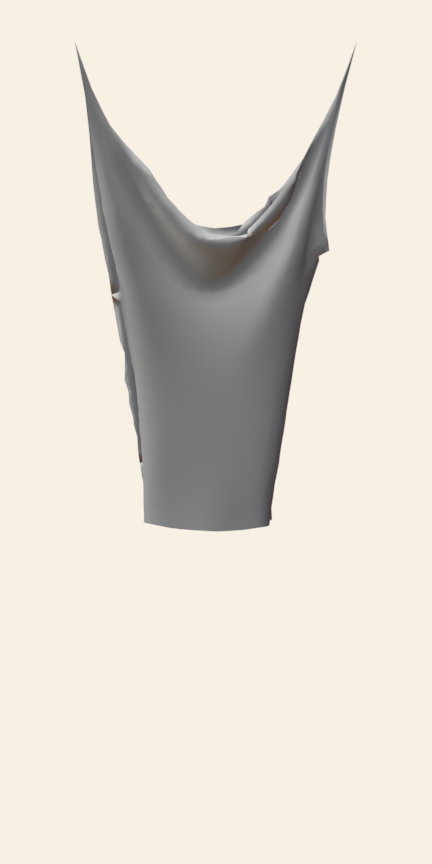} & 
        \includegraphics[width=0.14\columnwidth]{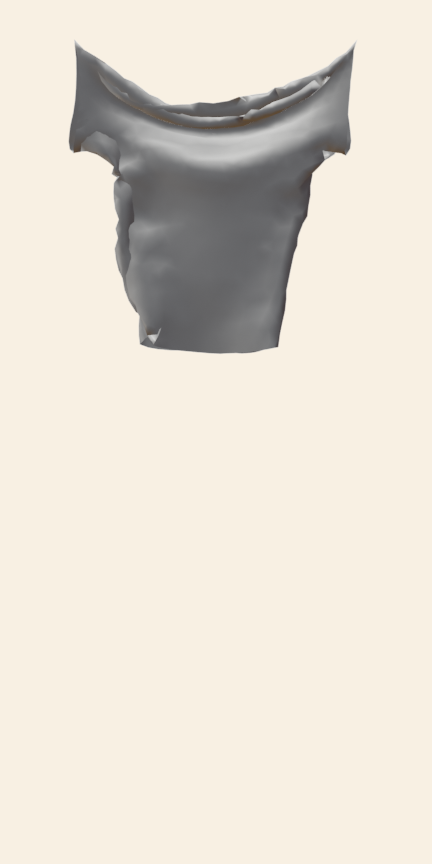} & \includegraphics[width=0.14\columnwidth]{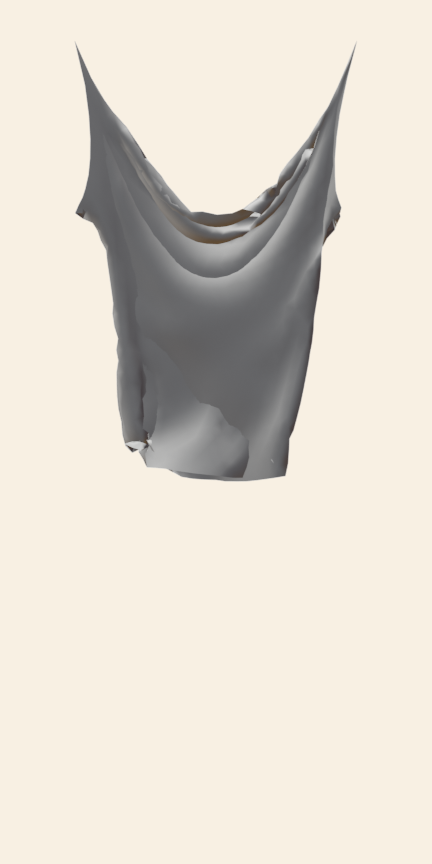}  \\
        &  1 & 25 & 50 &  75 &  100 &  120\\
    \end{tabular}
    \caption{Qualitative comparison of multiple frames (columns) of our simulation (bottom) against DiffCloth (top) showing no visual difference. The simulation is of a meshed t-shirt  stretching under $10\times$ gravity with spring constant $k_s=\qty{54}{N/m}$. \label{fig:qualdiffcloth}}
\end{figure}

\BZ{mark as potential removal if run out of space}
We compared our forward simulation of springs (charges set to zero) with DiffCloth\cite{li2022diffcloth}, on a t-shirt mesh suspended from two points and stretching due to gravity ($k_s = \qty{54}{N/m}$). Our forward simulation exhibits no visual difference in Figure~\ref{fig:qualdiffcloth}. This is not surprising since the forward simulation for DiffCloth uses an adaptation of projective dynamics that was shown to be equivalent to the fast implicit solver we use. Since parameter estimation has been depracated in their project, we limit ourselves to forward simulation only.  

% We also trying to compare the optimisation of the spring constant with DiffCloth. However, the released code of DiffCloth shows the spring constant is depretiated. We have modified the code where spring constant is supported and gradient is computed similar to ours. The DiffCloth is still failed to optimise, the BFGS solver stops at the first iteration. We think there should be bugs in their codes. So showing the comparison in such is not fair. We only show the optimisation results of our method.

\subsubsection{Inverse problem qualitative comparison spring only} 
\begin{figure}[htbp]
    \centering
    \setlength{\tabcolsep}{2pt} % Reduce space between columns
    \renewcommand{\arraystretch}{0.8} % Reduce space between rows
    \begin{tabular}{ccccccc}
            % Consider removing [6pt] to reduce vertical space
            
        \rotatebox{90}{Target} & 
        \includegraphics[width=0.14\columnwidth]{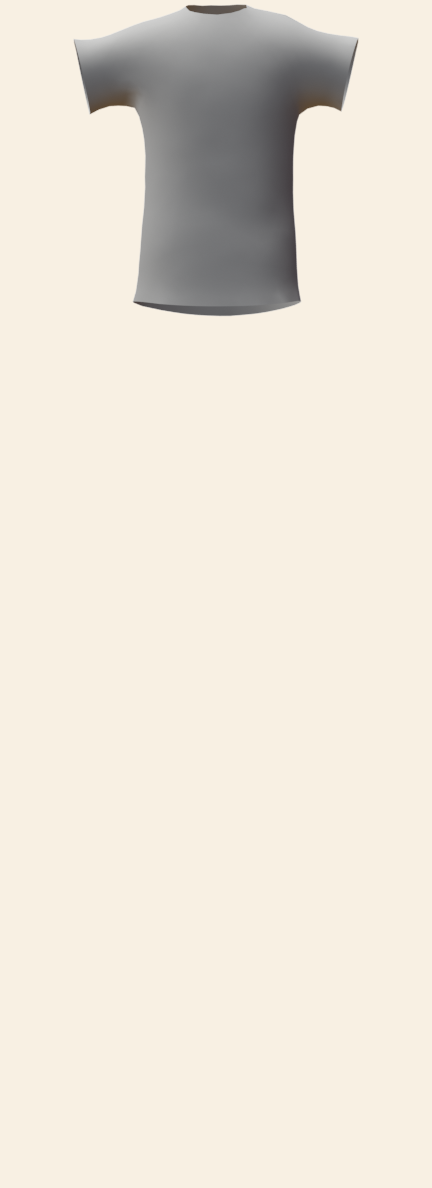} & 
        \includegraphics[width=0.14\columnwidth]{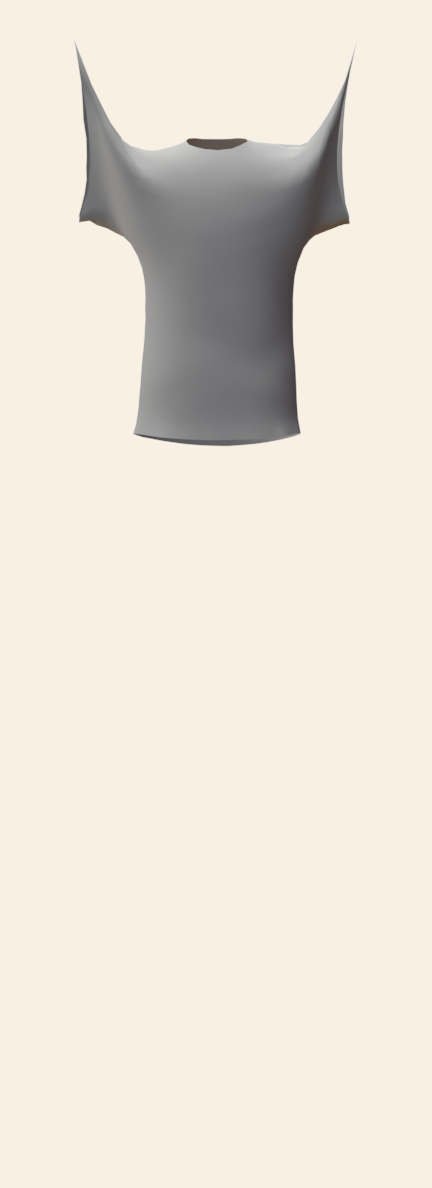} & 
        \includegraphics[width=0.14\columnwidth]{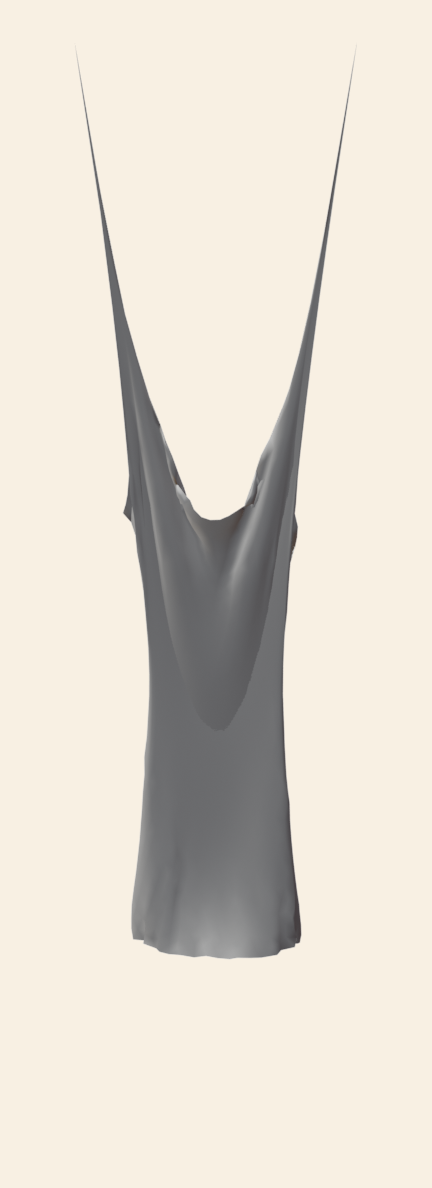} & 
        \includegraphics[width=0.14\columnwidth]{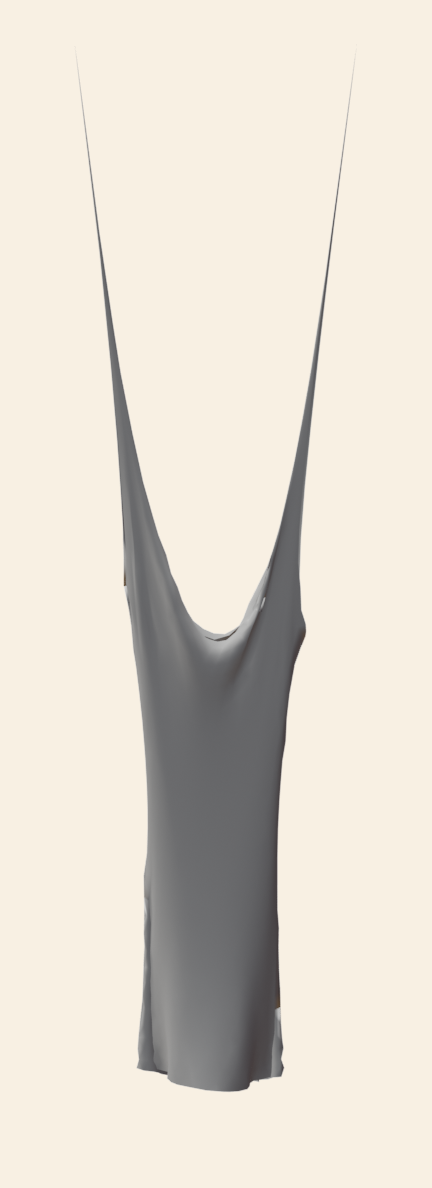} & 
        \includegraphics[width=0.14\columnwidth]{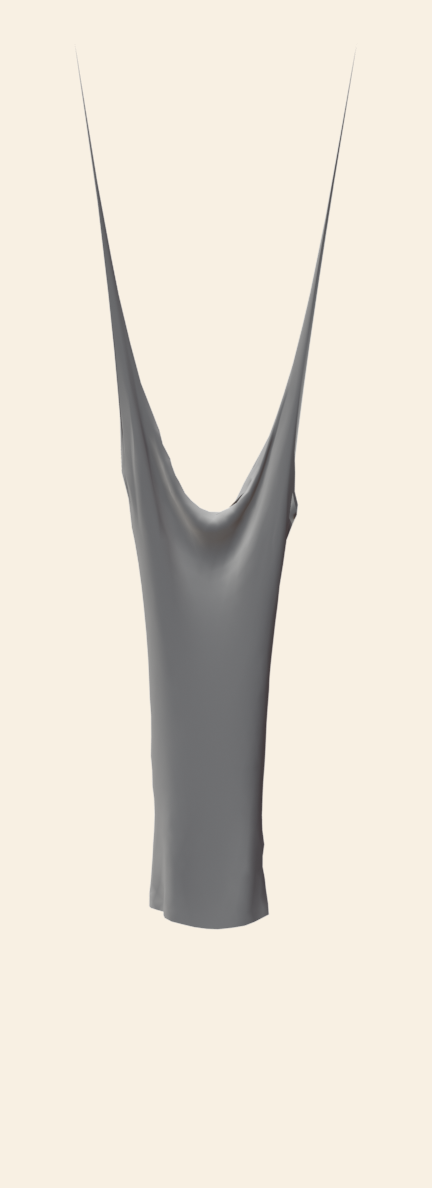} & \includegraphics[width=0.14\columnwidth]{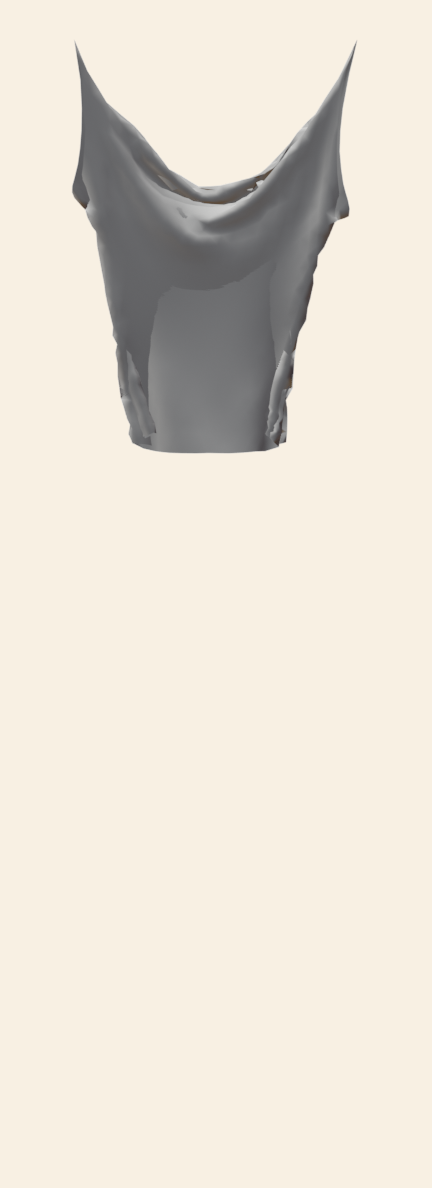}   \\
        \rotatebox{90}{Initial guess}& 
        \includegraphics[width=0.14\columnwidth]{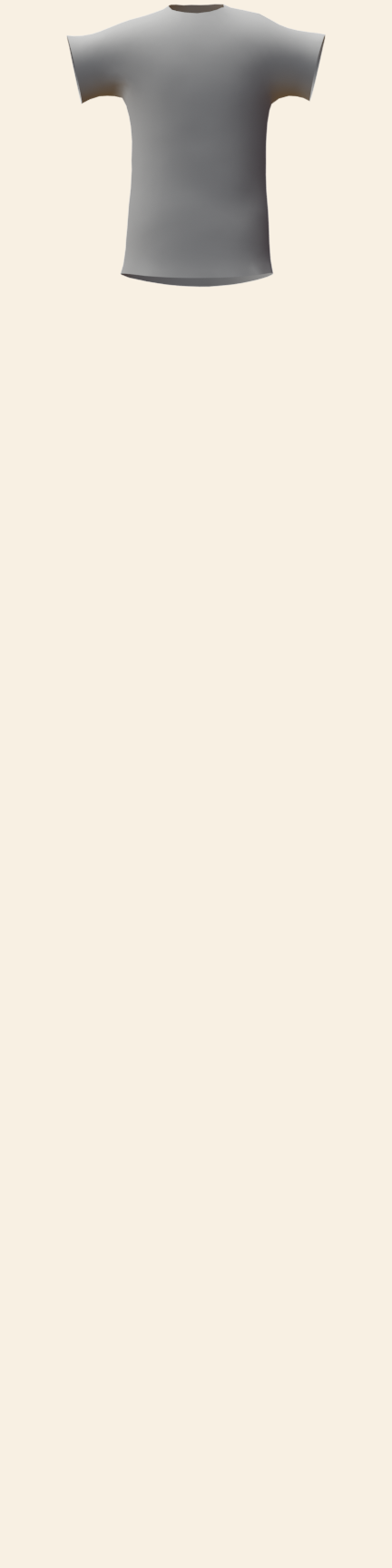} & 
        \includegraphics[width=0.14\columnwidth]{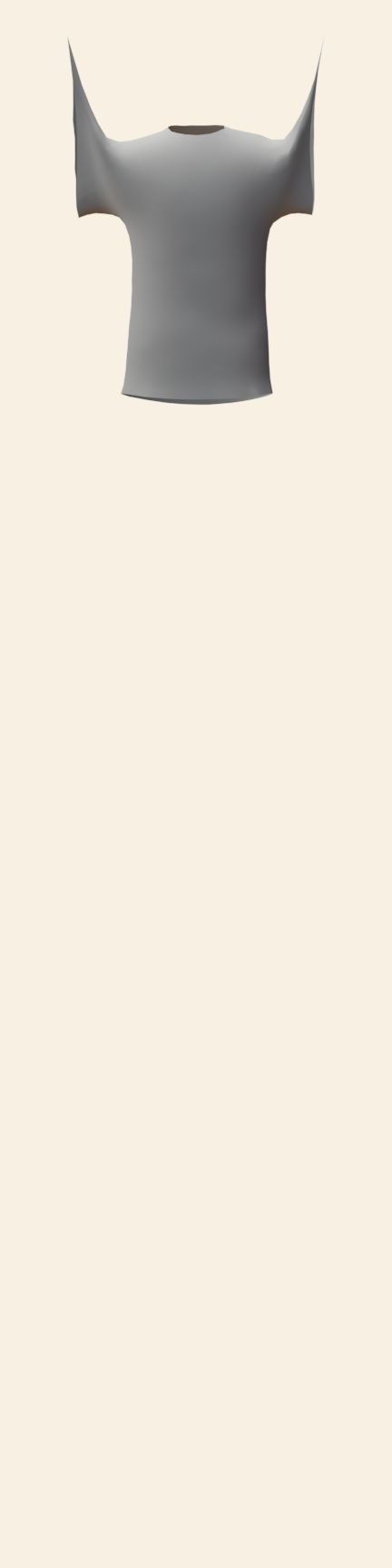} & 
        \includegraphics[width=0.14\columnwidth]{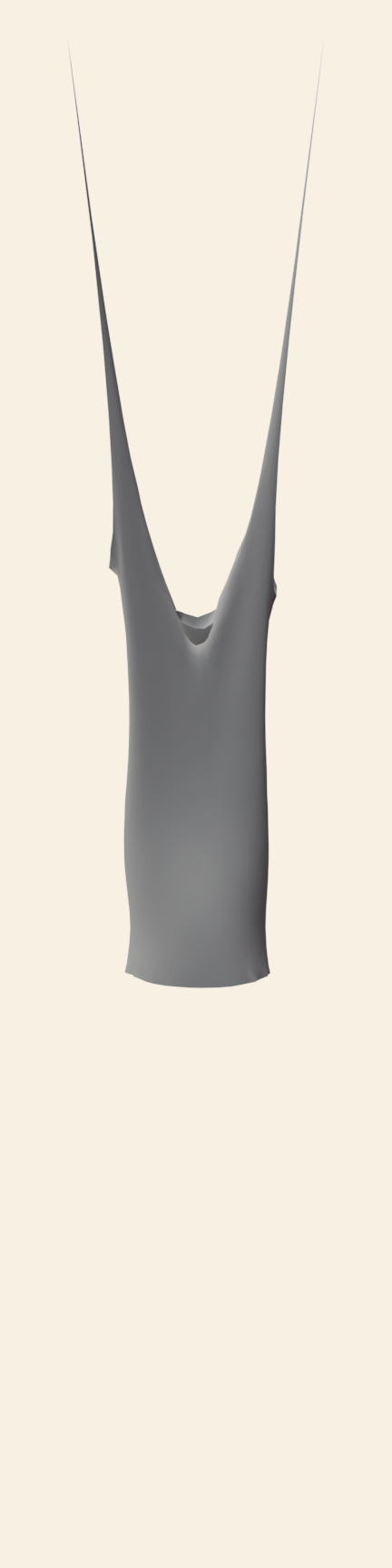} & 
        \includegraphics[width=0.14\columnwidth]{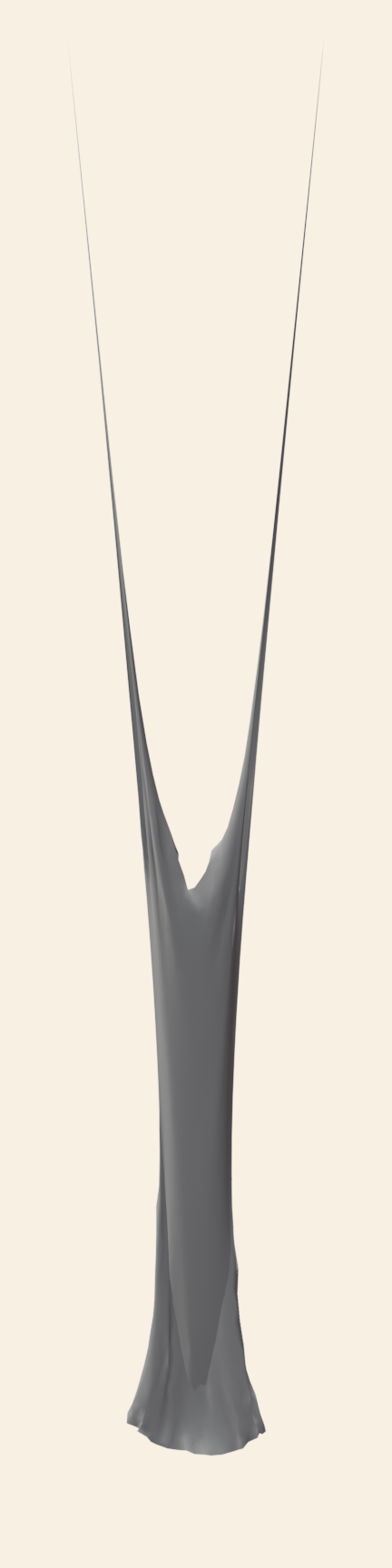} & 
        \includegraphics[width=0.14\columnwidth]{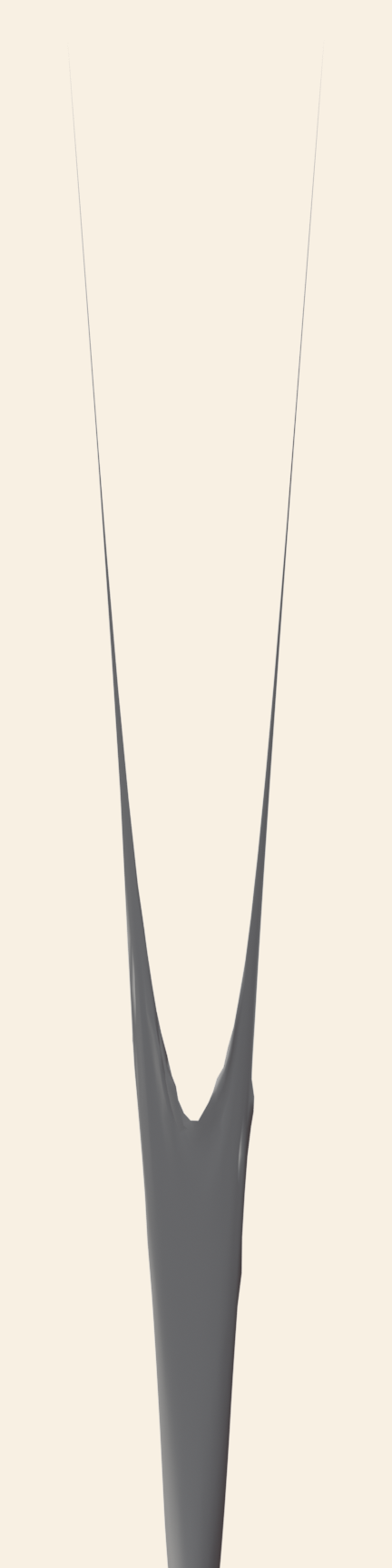} & \includegraphics[width=0.14\columnwidth]{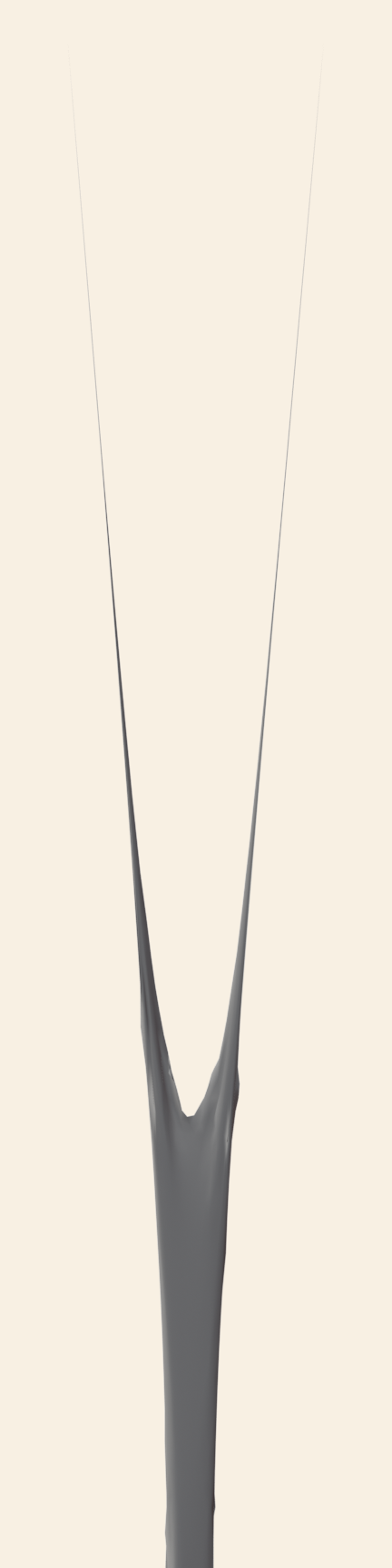}   \\
            \rotatebox{90}{After parameter estimation} & 
        \includegraphics[width=0.14\columnwidth]{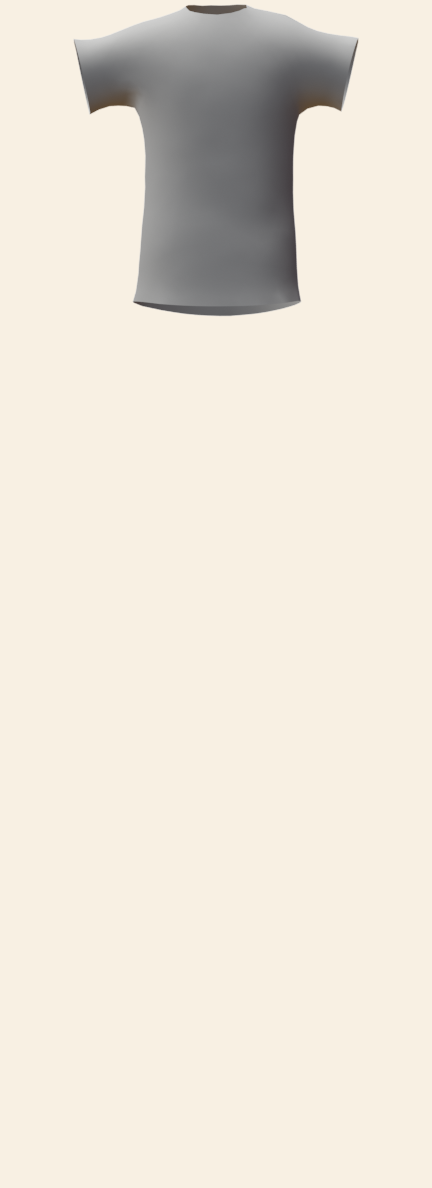} & 
        \includegraphics[width=0.14\columnwidth]{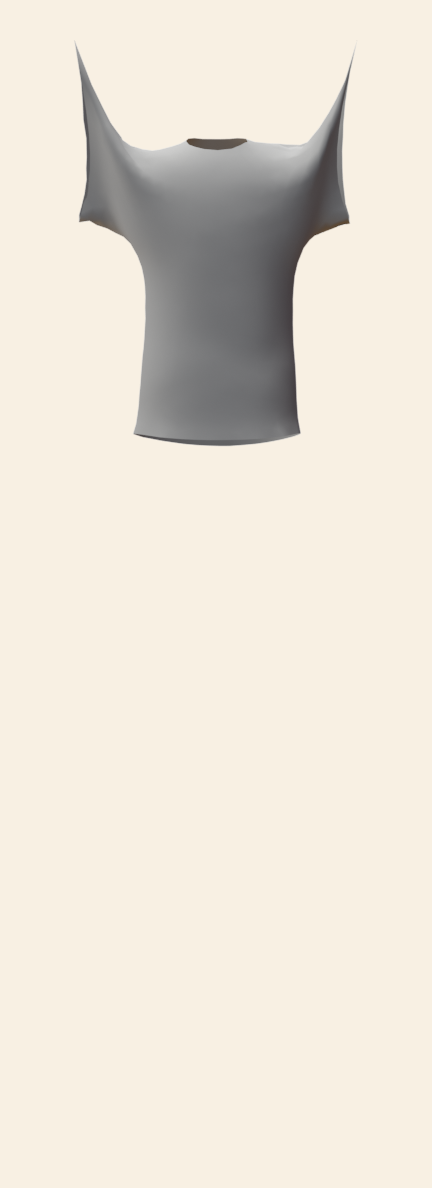} & 
        \includegraphics[width=0.14\columnwidth]{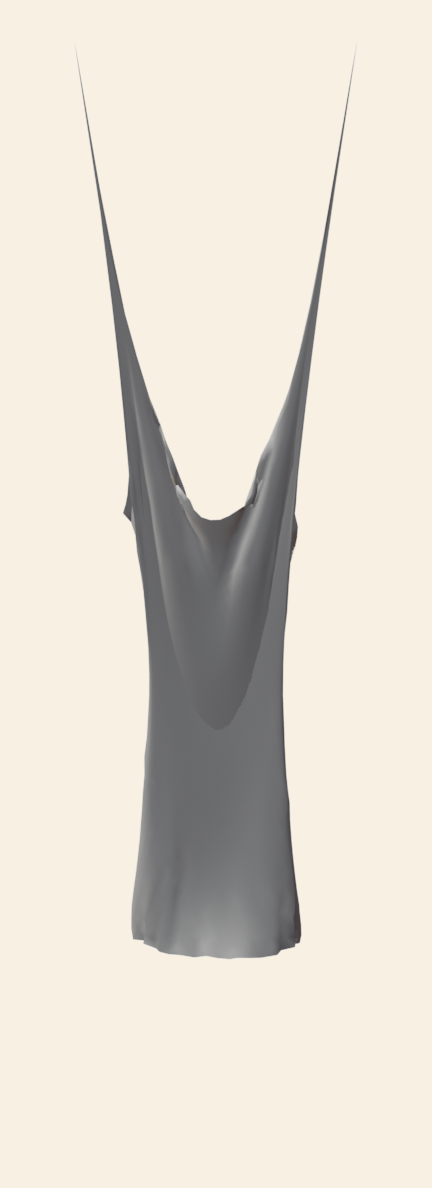} & 
        \includegraphics[width=0.14\columnwidth]{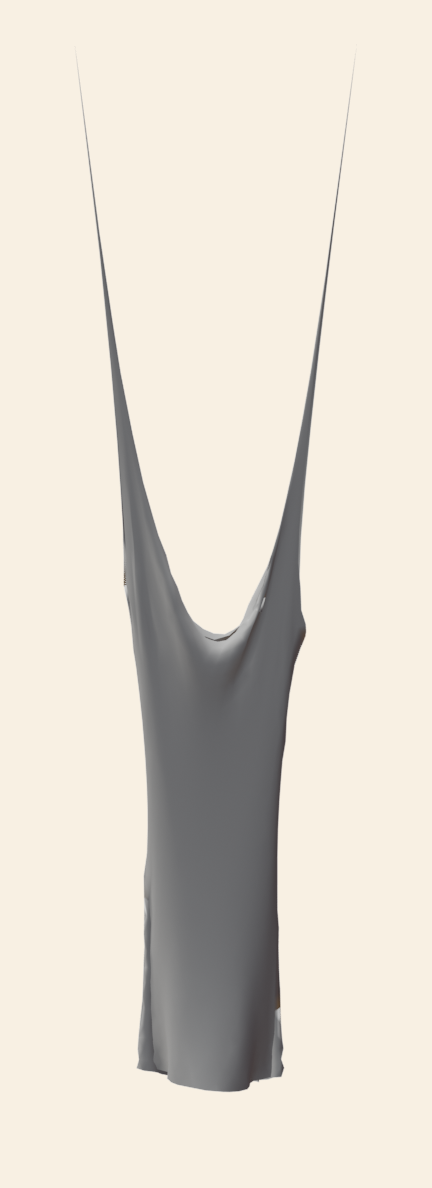} & 
        \includegraphics[width=0.14\columnwidth]{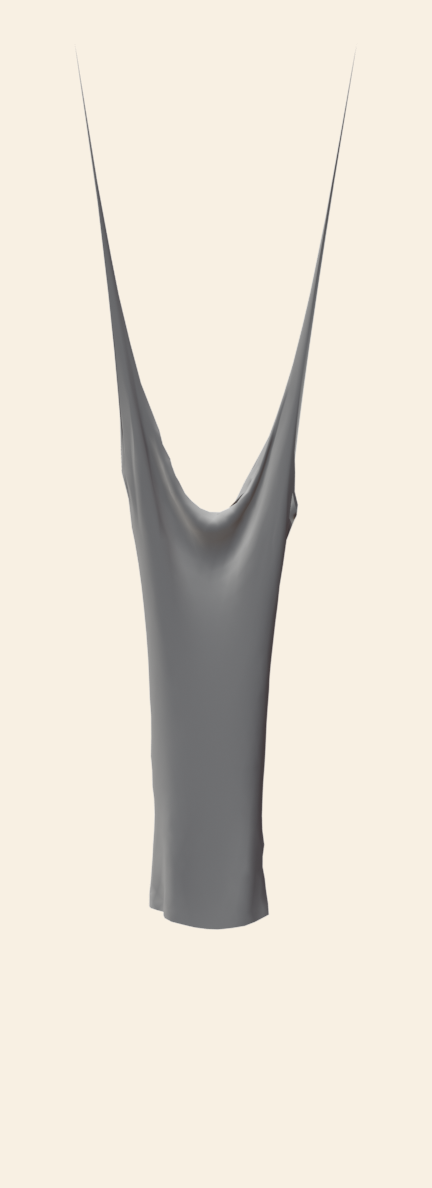} & \includegraphics[width=0.14\columnwidth]{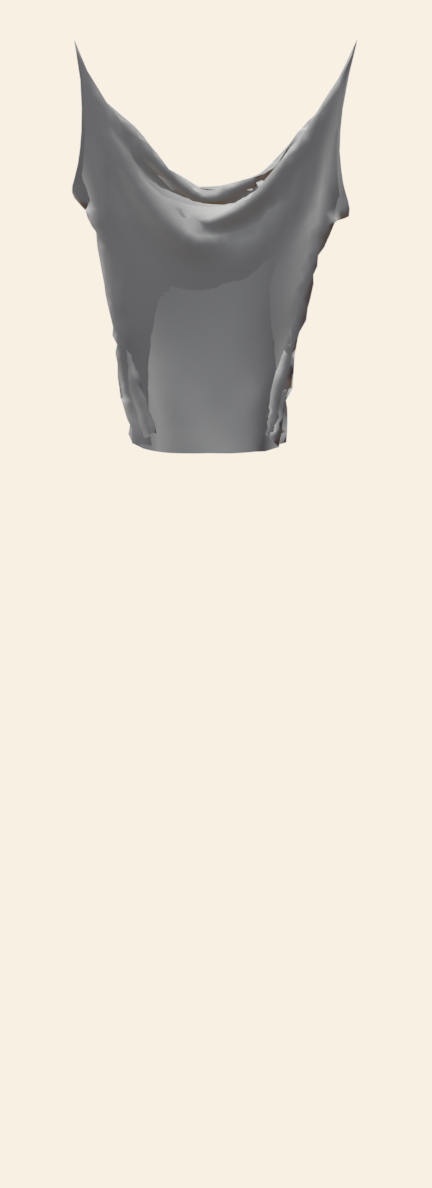}   \\
        &  1 & 20 & 50 &  70 &  90 &  120
    \end{tabular}
    \caption{Estimating the spring constant which leads to given target dynamics (top) generated with a spring constant of \qty{50}{N/m}. With an initial guess of \qty{20}{N/m}, the t-shirt drops to a lower position, while our estimated parameter leads to the required dynamics.\label{fig:qualsysid}}
\end{figure}

Figure~\ref{fig:qualsysid} compares the results of the estimation of spring constants given a target behavior (top row). The reference spring constant was \qty{50}{N/m} and the initial guess provided is \qty{20}{N/m}. The optimization results in finding the same value as the reference in this instance. Even when the parameter is not recovered accurately, our estimated parameter results in trajectories that qualitatively match the reference.

\subsubsection{Inverse problem quantitative comparison spring and charge}

\begin{figure}
    \centering
    \begin{subfigure}[b]{0.48\columnwidth}
        \includegraphics[width=\textwidth]{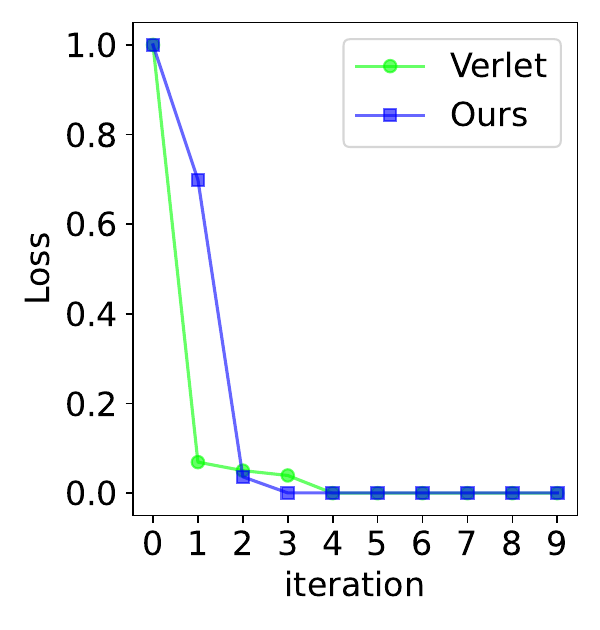}
        \caption{time step = \qty{0.009}{s}}
        \label{lowh}
    \end{subfigure}
    \begin{subfigure}[b]{0.48\columnwidth}
        \includegraphics[width=\textwidth]{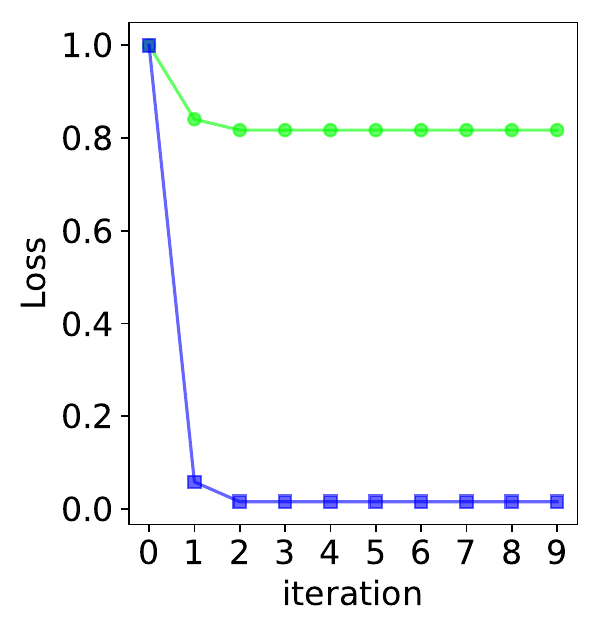}
        \caption{time step = \qty{0.18}{s}}
        \label{highh}
    \end{subfigure}
    \caption{Quantitative comparison of parameter estimation. The plots show the error in parameter values estimated using our method (blue) vs Verlet (green). We use a charged duck mesh deformed due to repulsive force between charges. When the time-step is small (\qty{0.009}{s}) both methods converged to the target value, while for the large time step (\qty{0.18}{s}) only our method converges to the correct value. 
     \label{fig:quantsysidJaxx}
    }
\end{figure}

\begin{figure}[htbp]
    \centering
    \setlength{\tabcolsep}{2pt} % Reduce space between columns
    \renewcommand{\arraystretch}{0.8} % Reduce space between rows
    \begin{tabular}{ccccccc}
        % Consider removing [6pt] to reduce vertical space
        \multicolumn{7}{c}{Optimised Results with small time step}\\
        \rotatebox{90}{Verlet} & 
    \includegraphics[width=0.14\columnwidth]{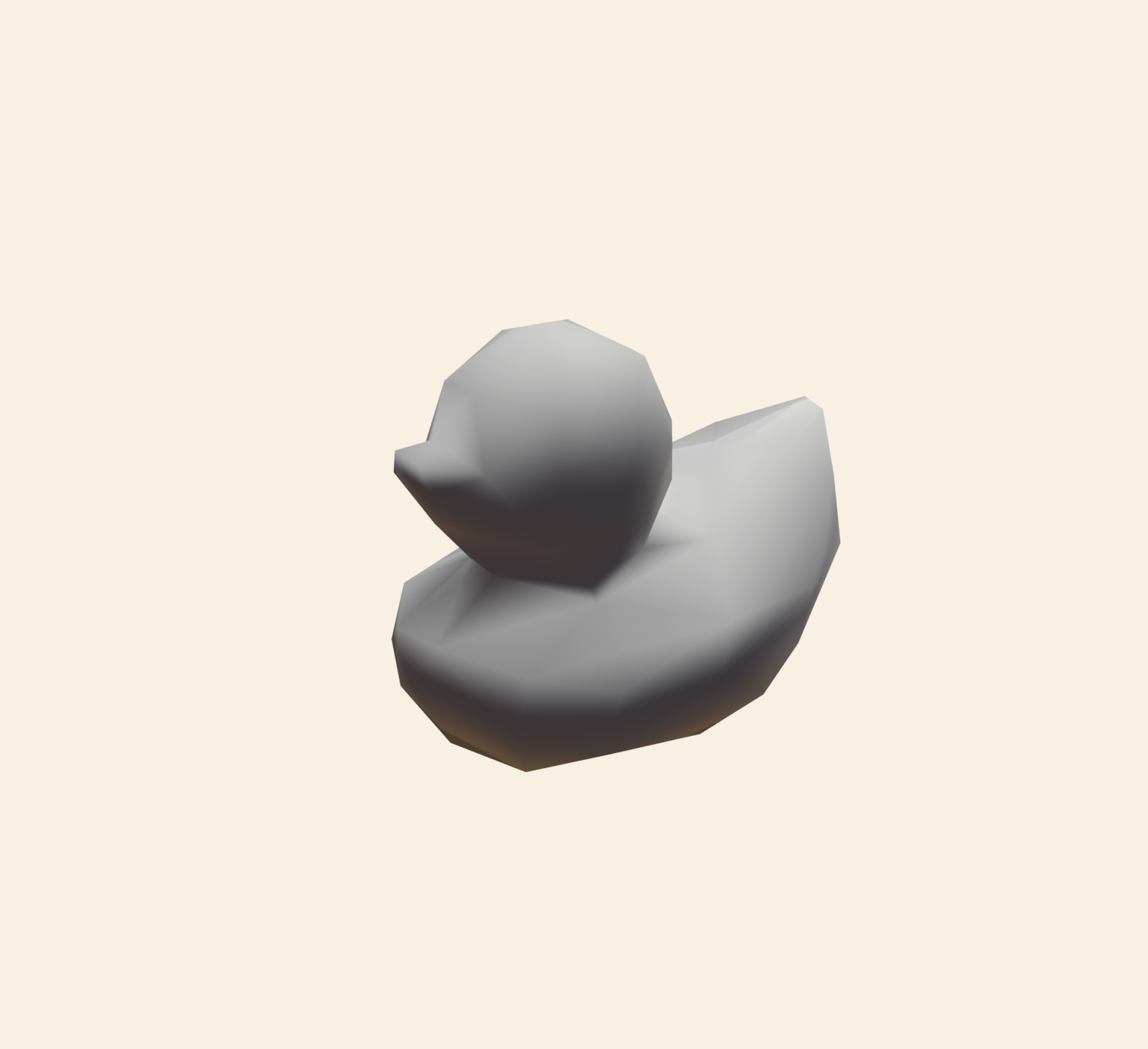} & 
    \includegraphics[width=0.14\columnwidth]{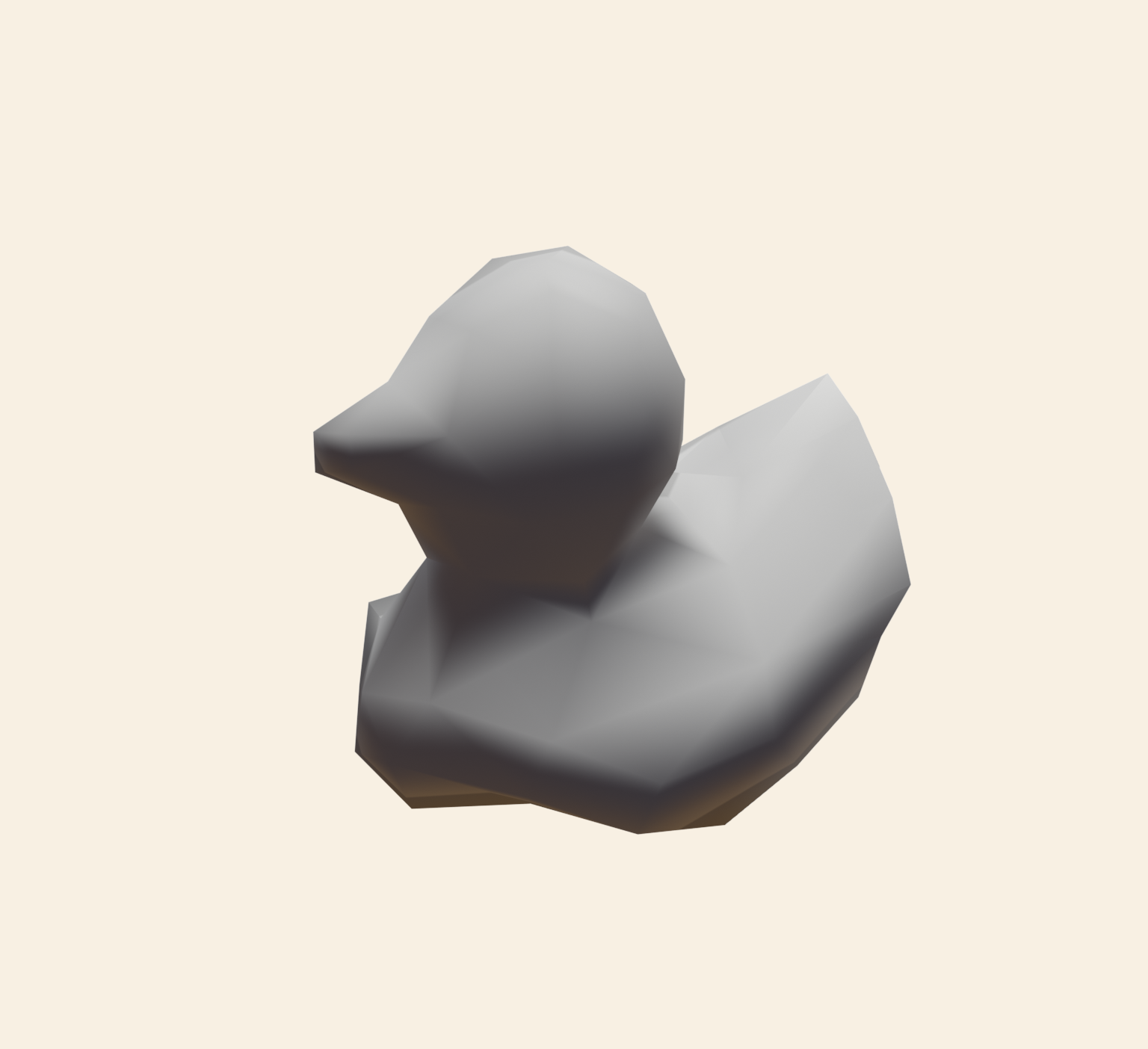} & 
    \includegraphics[width=0.14\columnwidth]{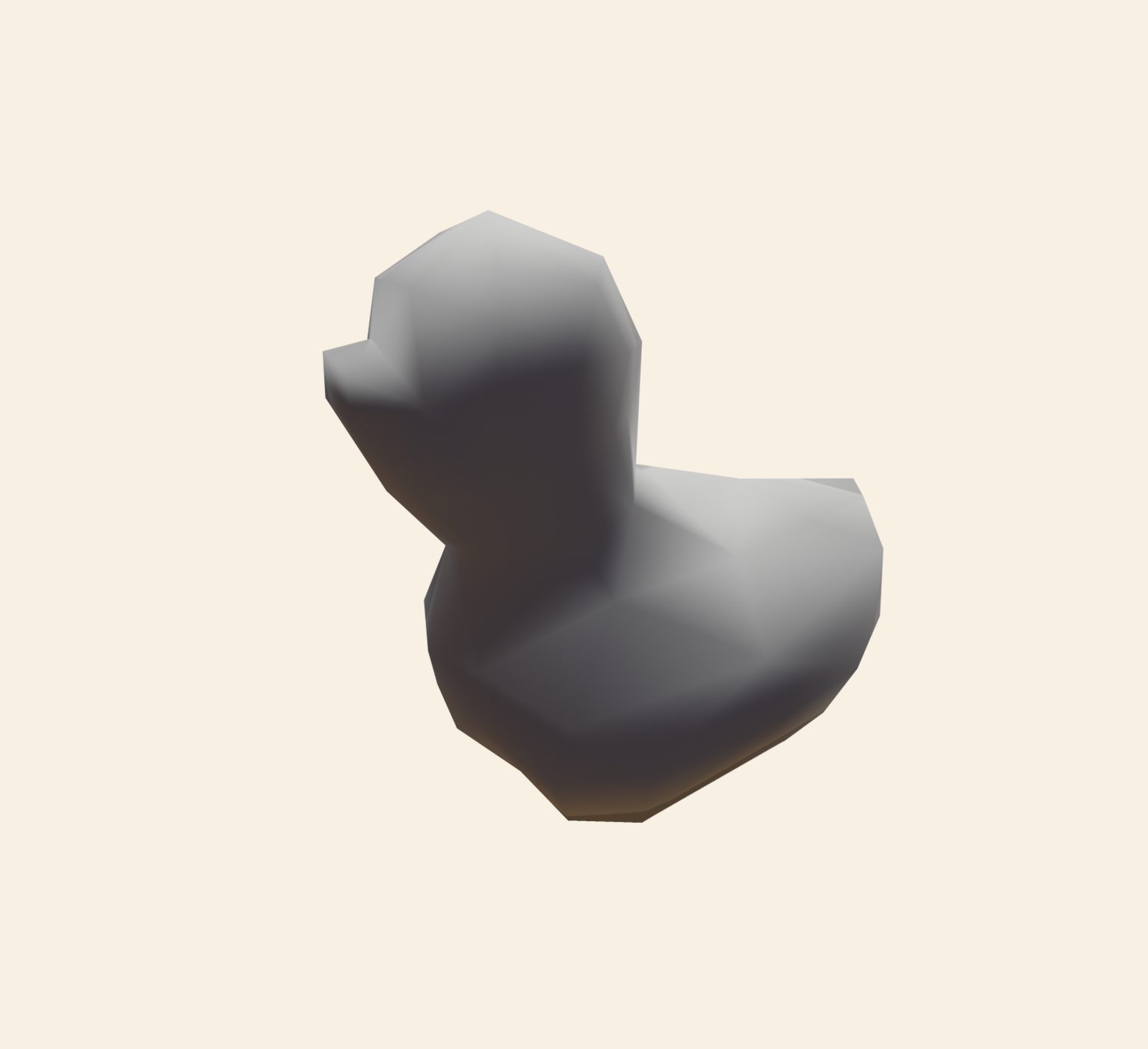} & 
    \includegraphics[width=0.14\columnwidth]{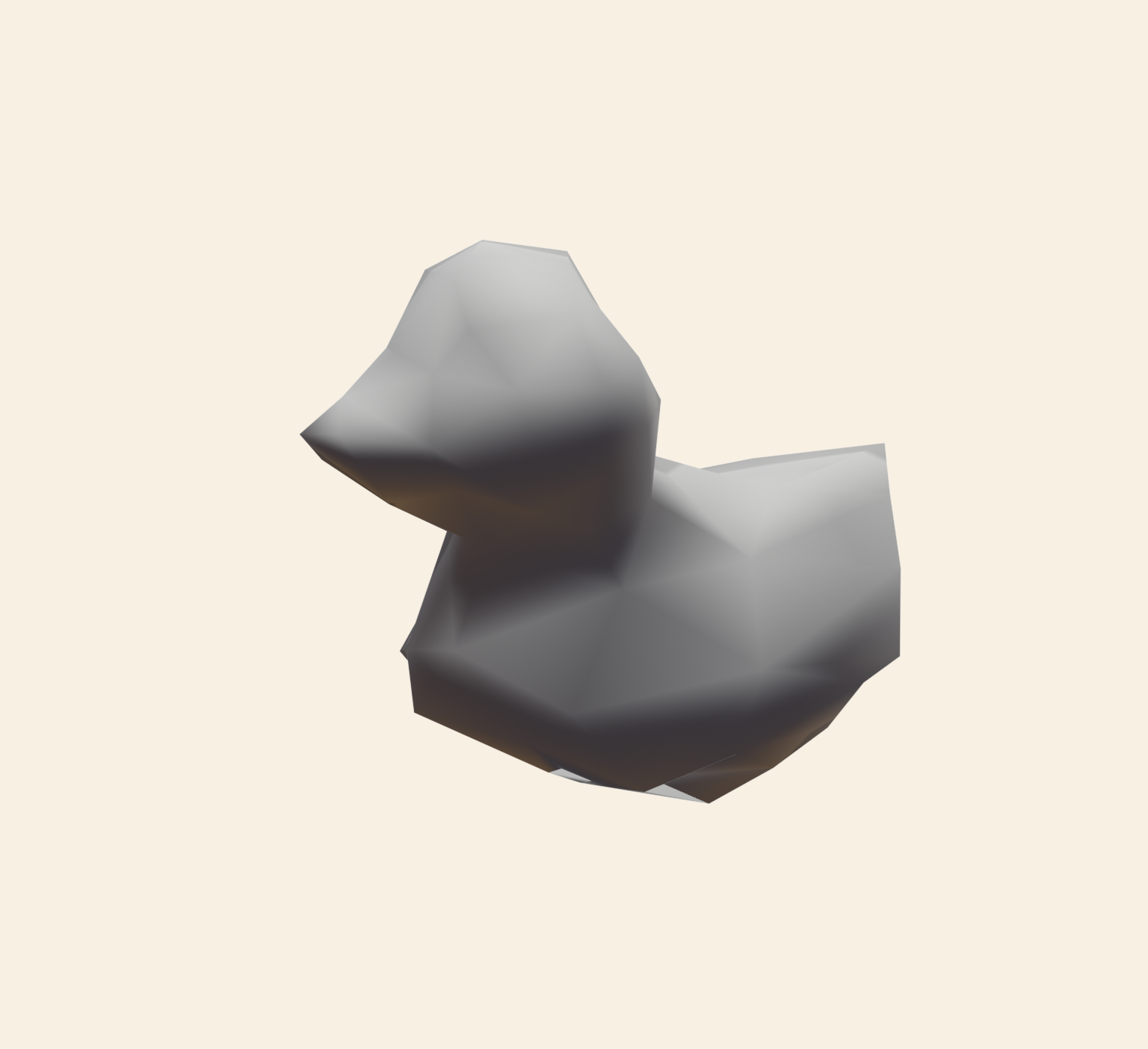} & 
    \includegraphics[width=0.14\columnwidth]{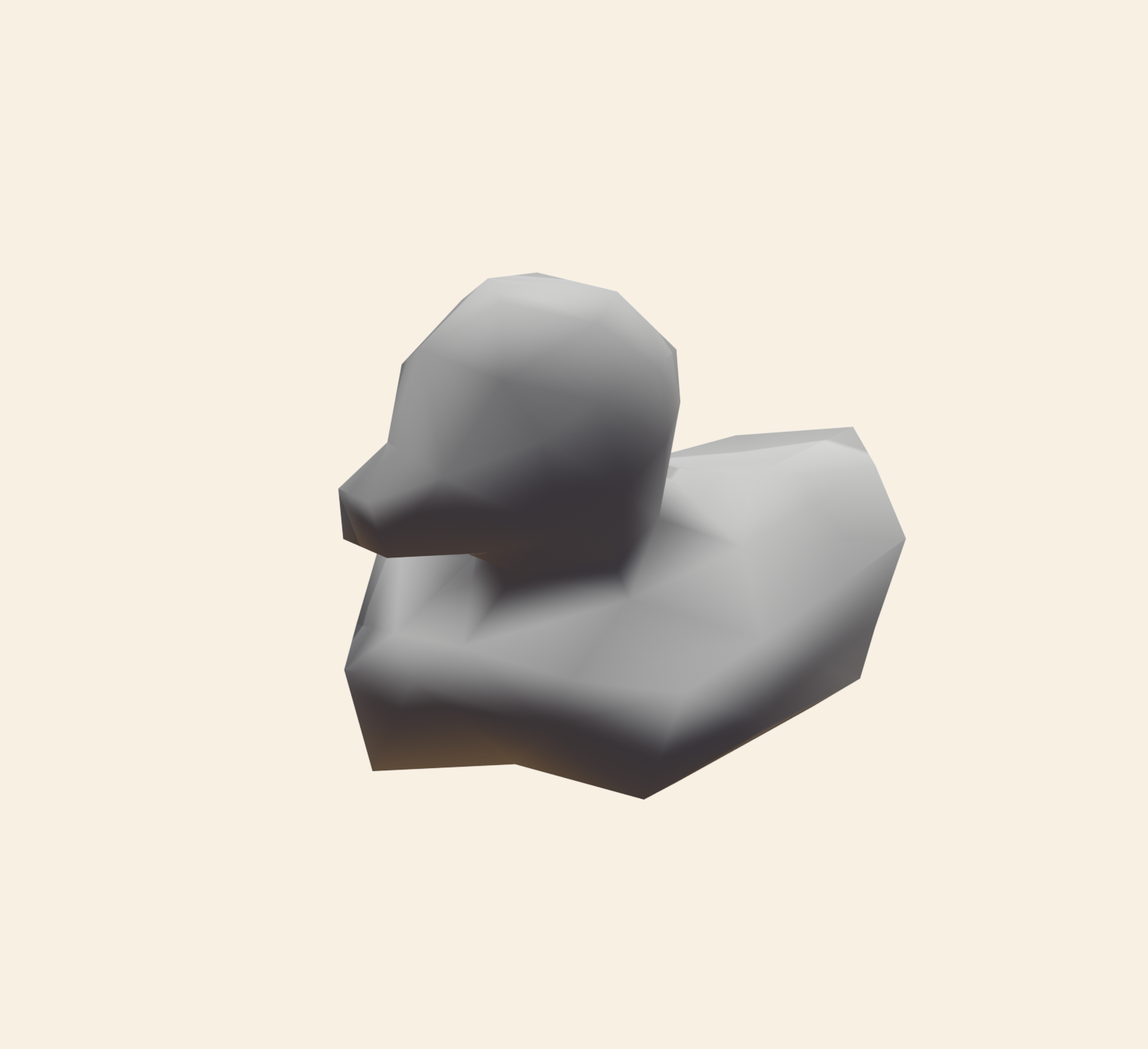} & \includegraphics[width=0.14\columnwidth]{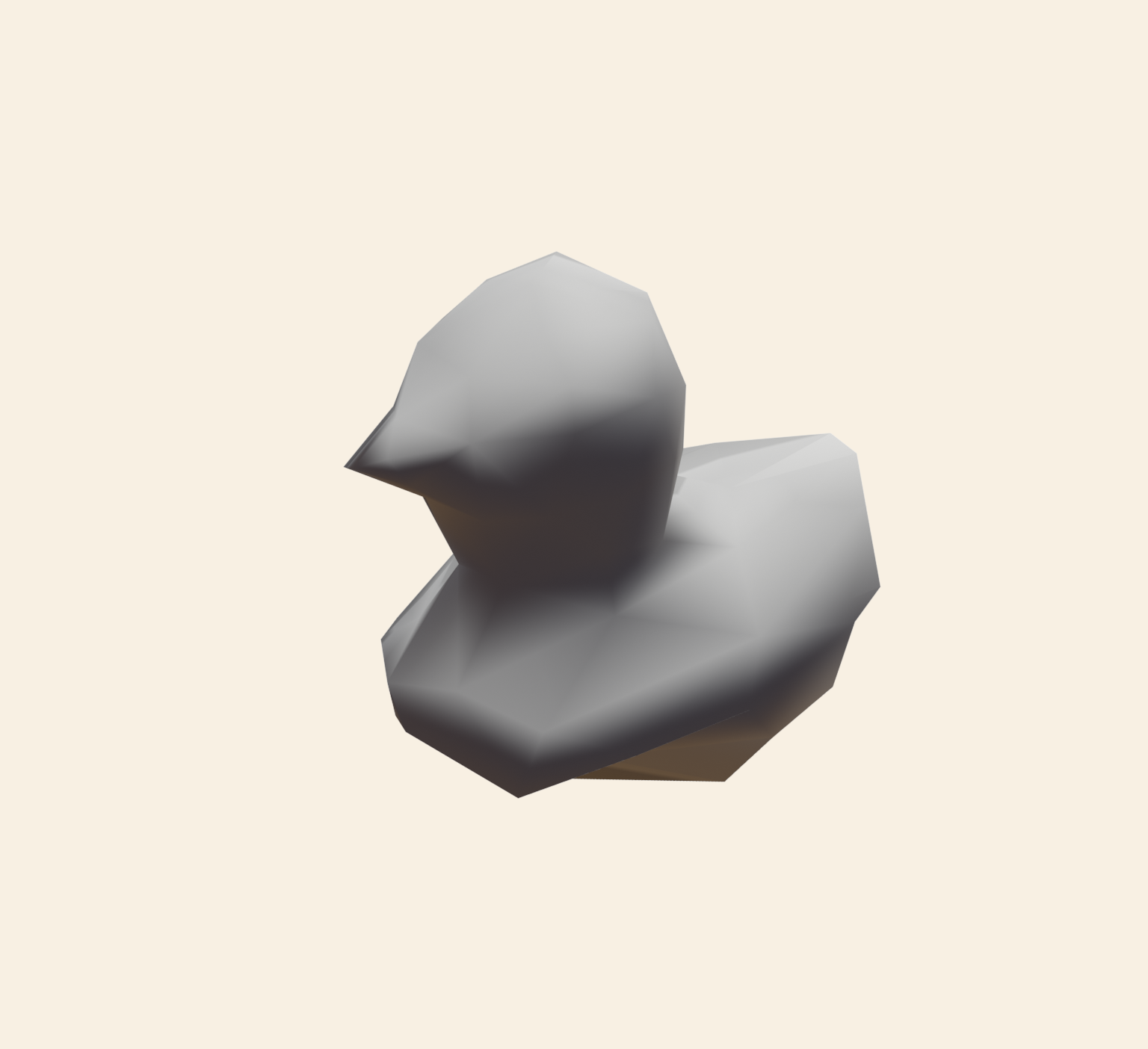}   \\
        \rotatebox{90}{Ours} & 
    \includegraphics[width=0.14\columnwidth]{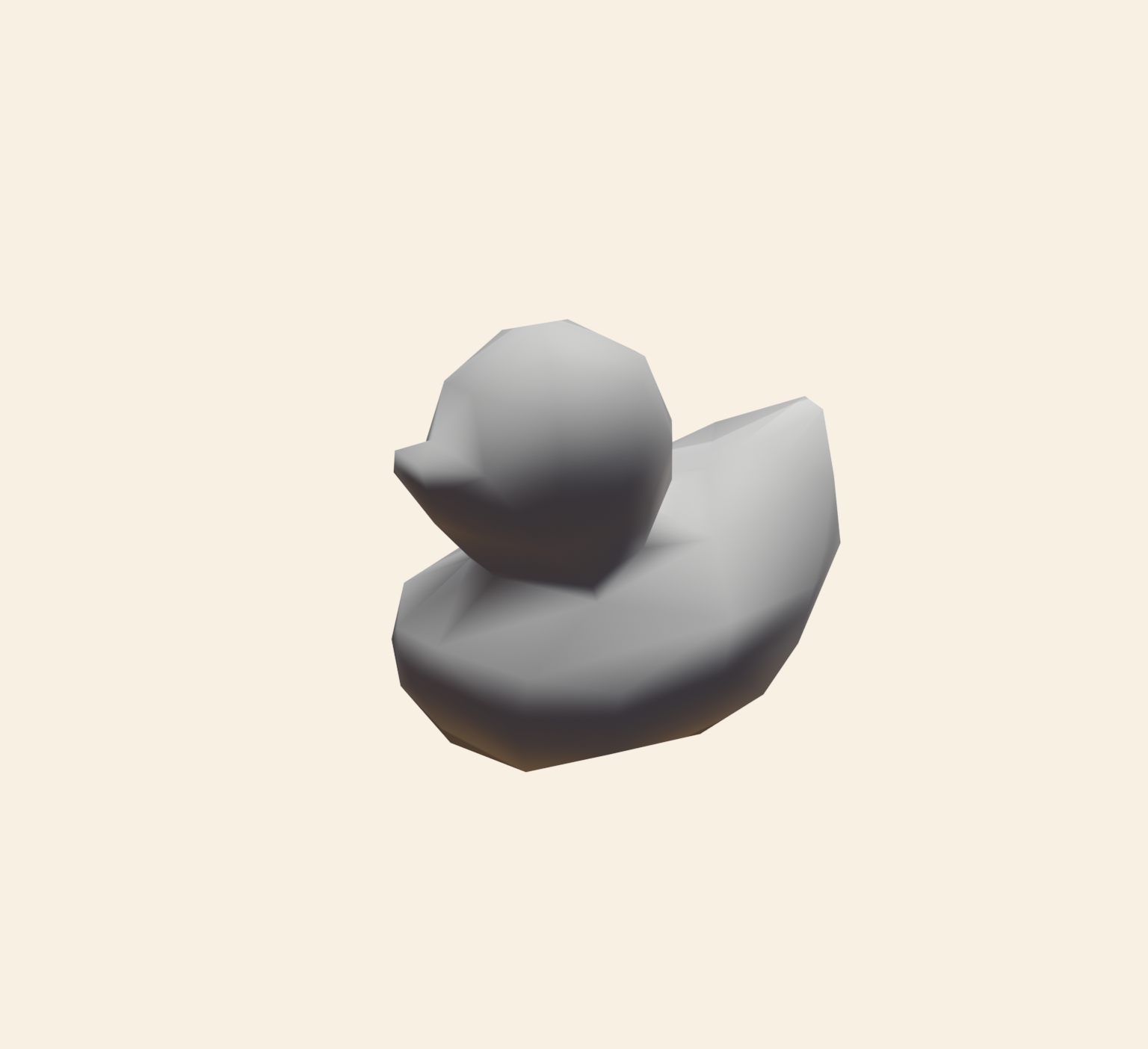} & 
    \includegraphics[width=0.14\columnwidth]{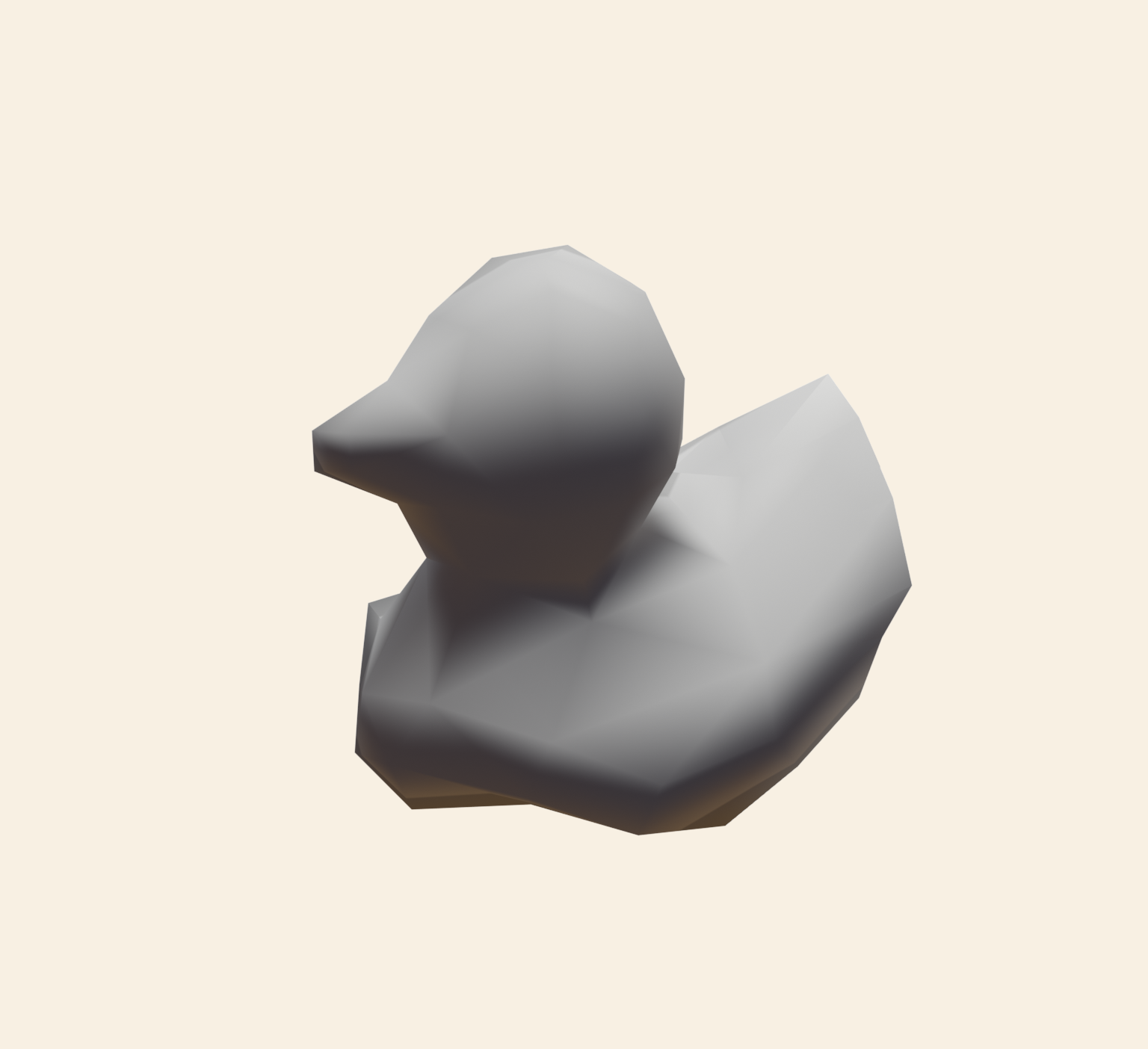} & 
    \includegraphics[width=0.14\columnwidth]{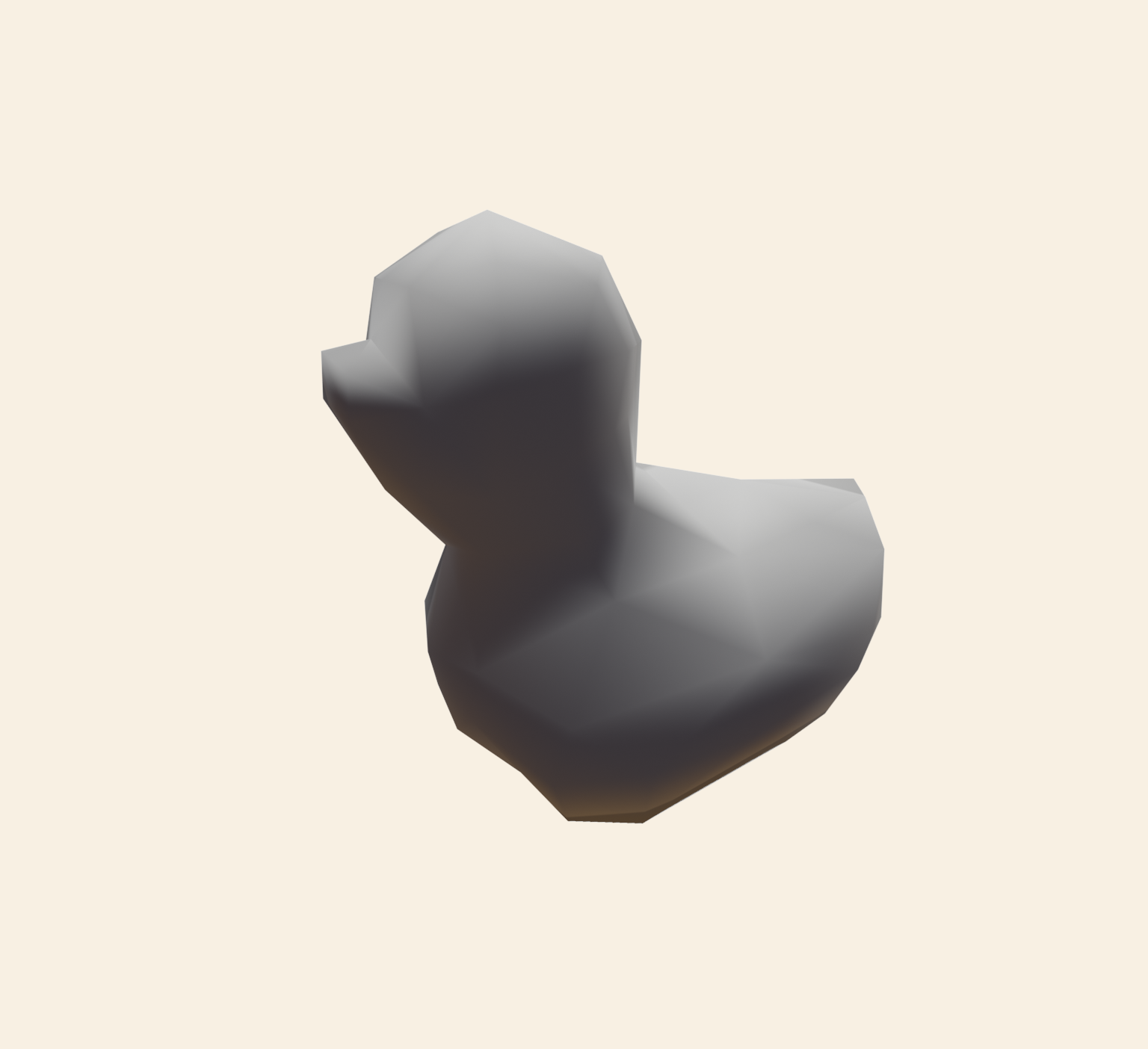} & 
    \includegraphics[width=0.14\columnwidth]{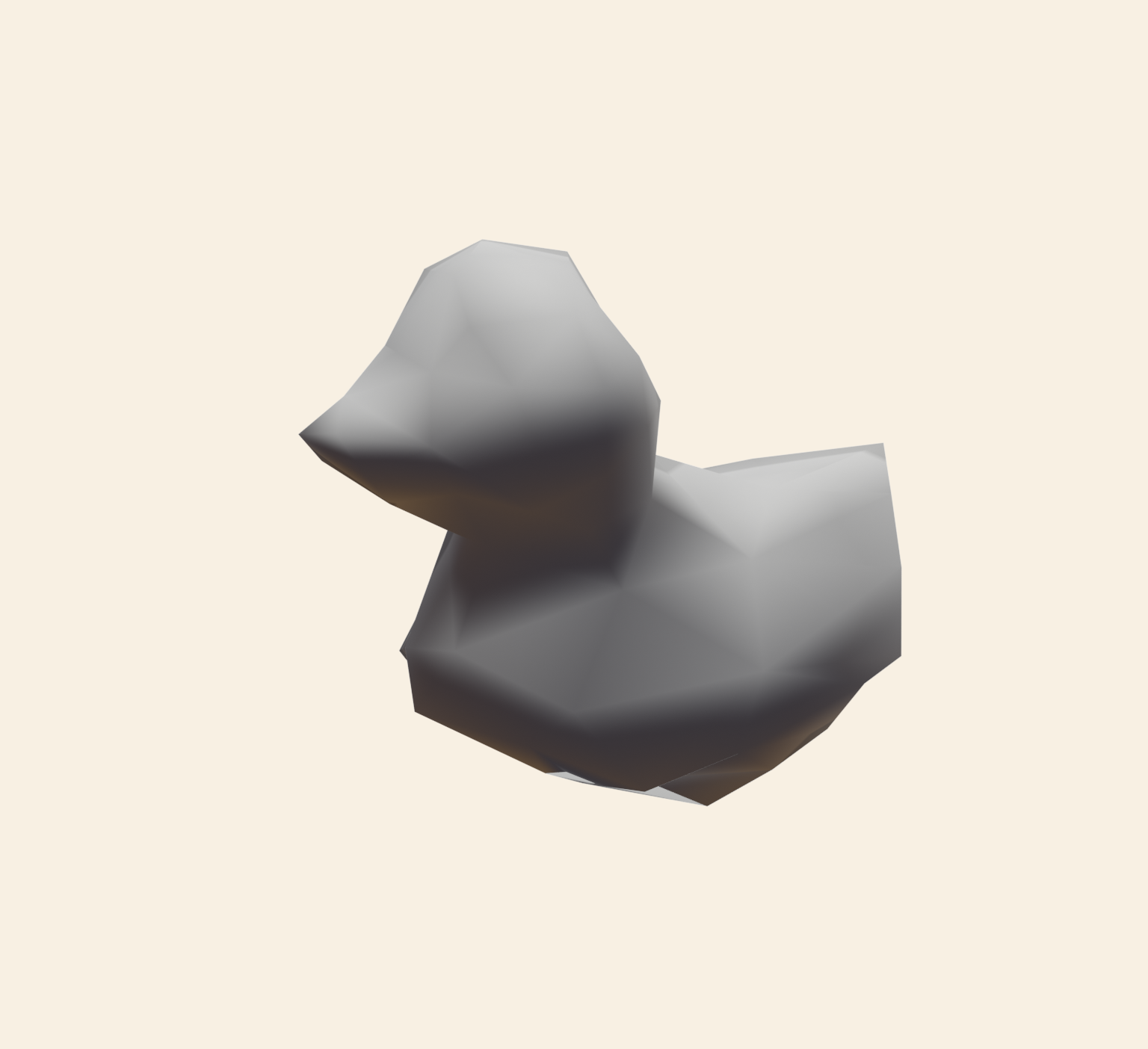} & 
    \includegraphics[width=0.14\columnwidth]{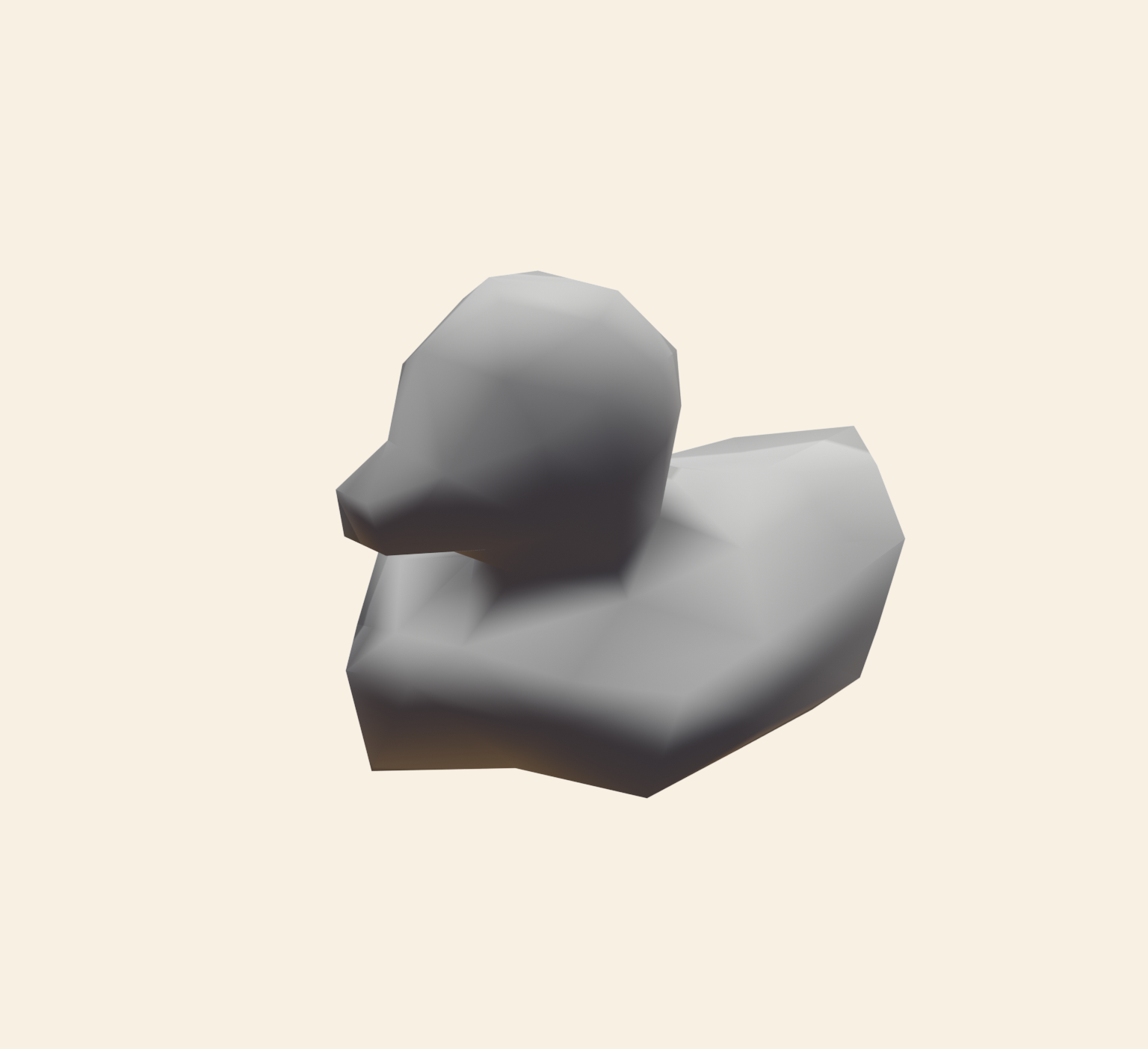} & \includegraphics[width=0.14\columnwidth]{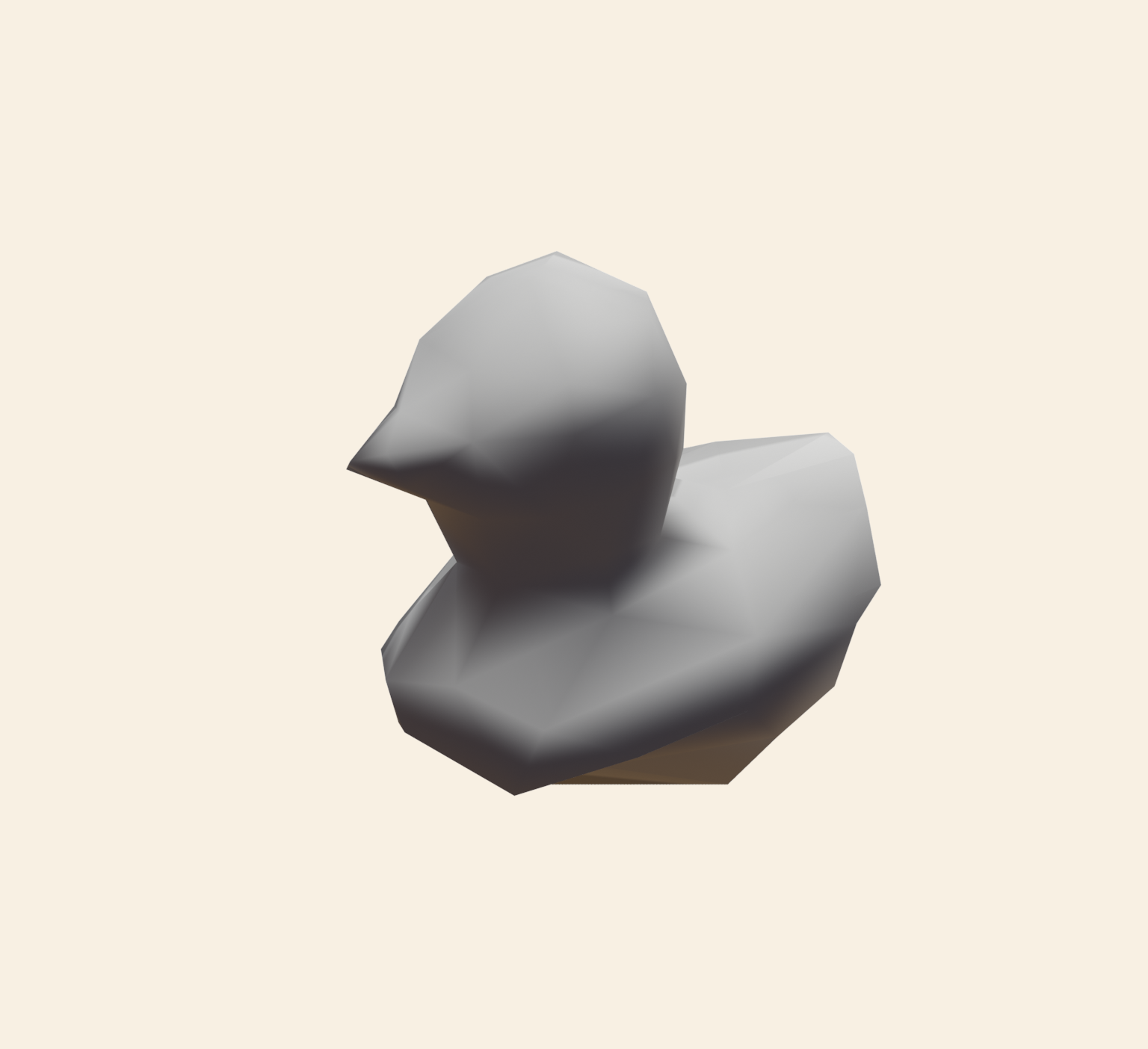}   \\
    \multicolumn{7}{c}{Optimised Results with large time step}\\

        \rotatebox{90}{Verlet} & 
    \includegraphics[width=0.14\columnwidth]{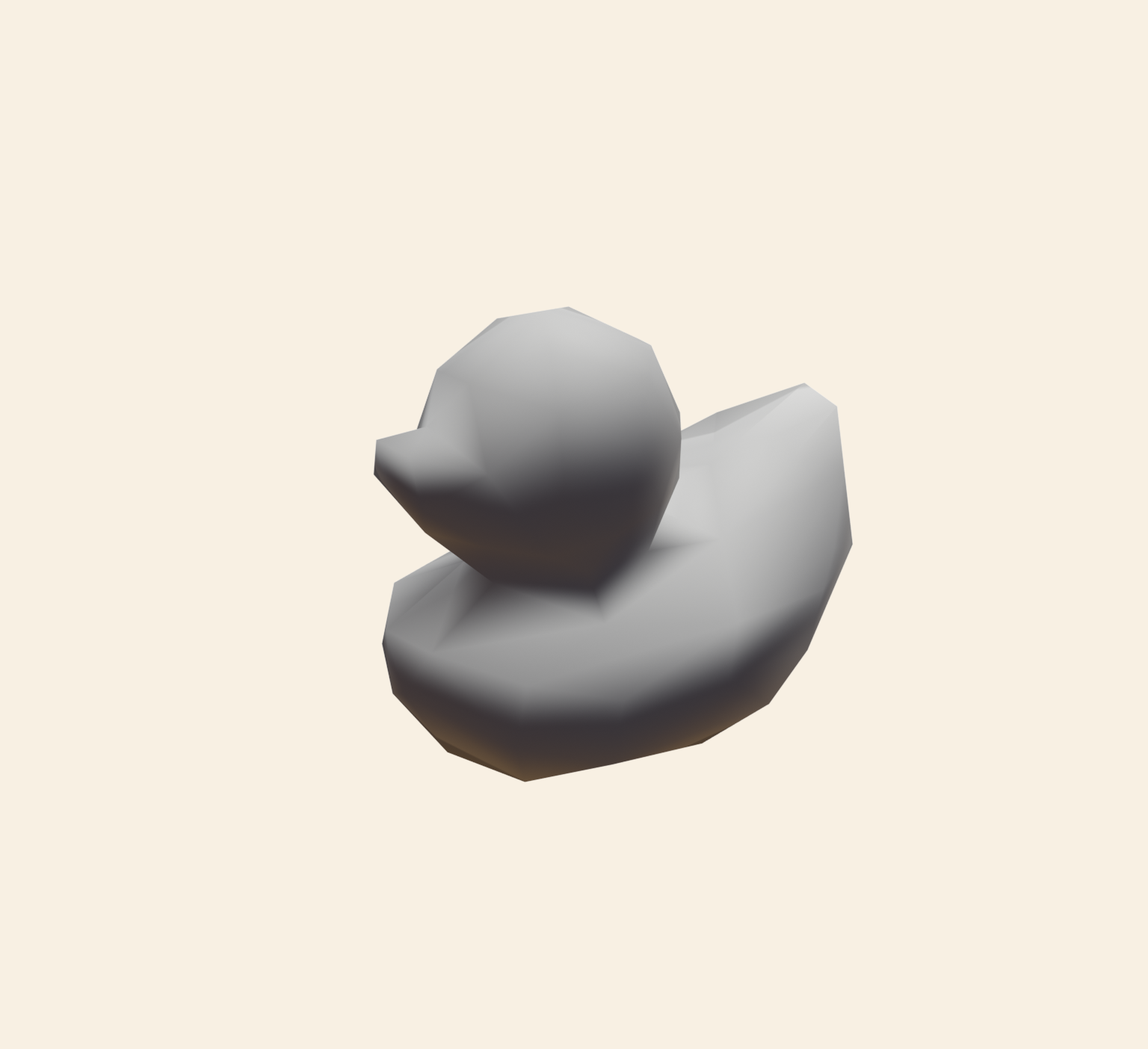} & 
    \includegraphics[width=0.14\columnwidth]{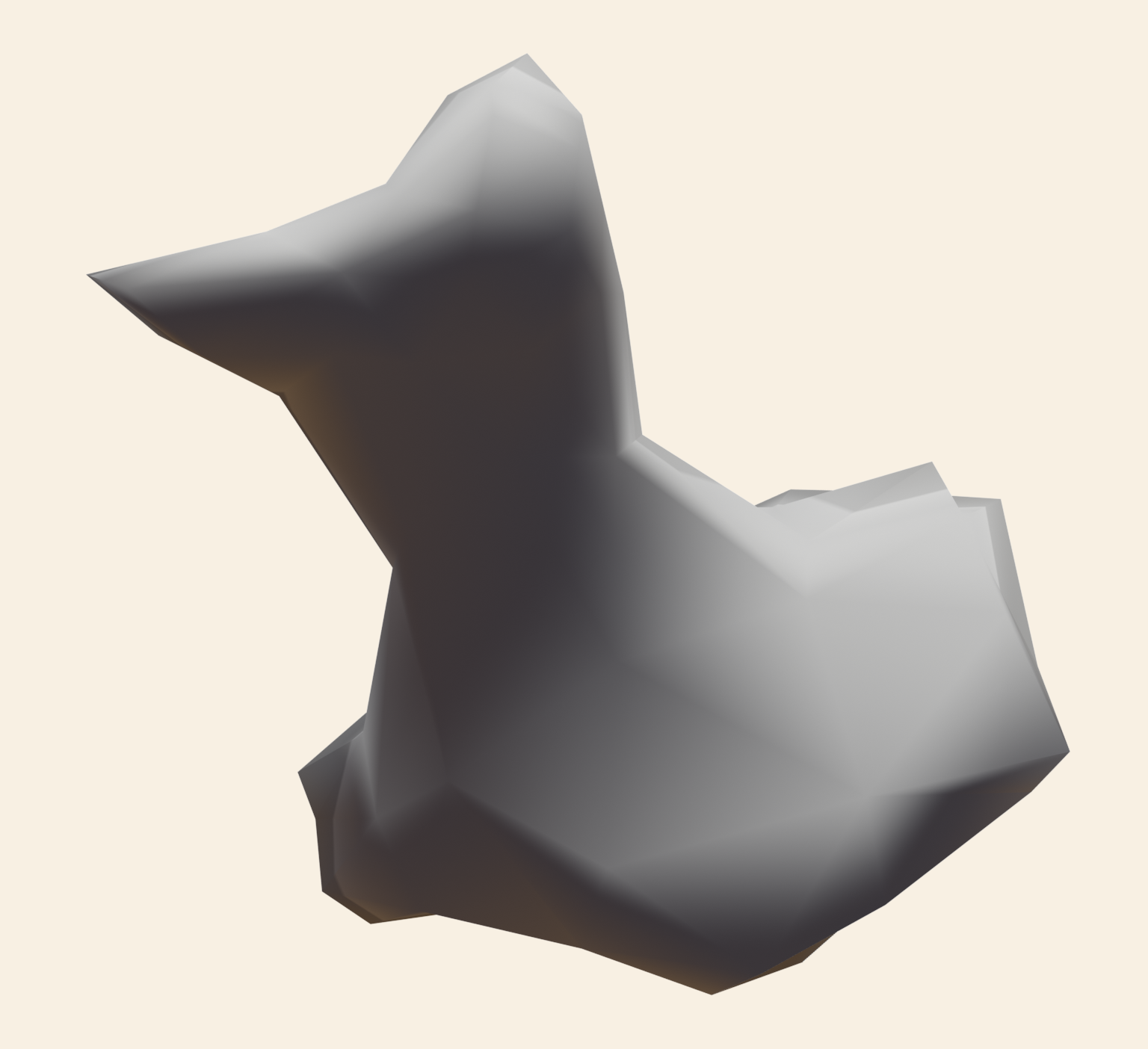} & 
    \includegraphics[width=0.14\columnwidth]{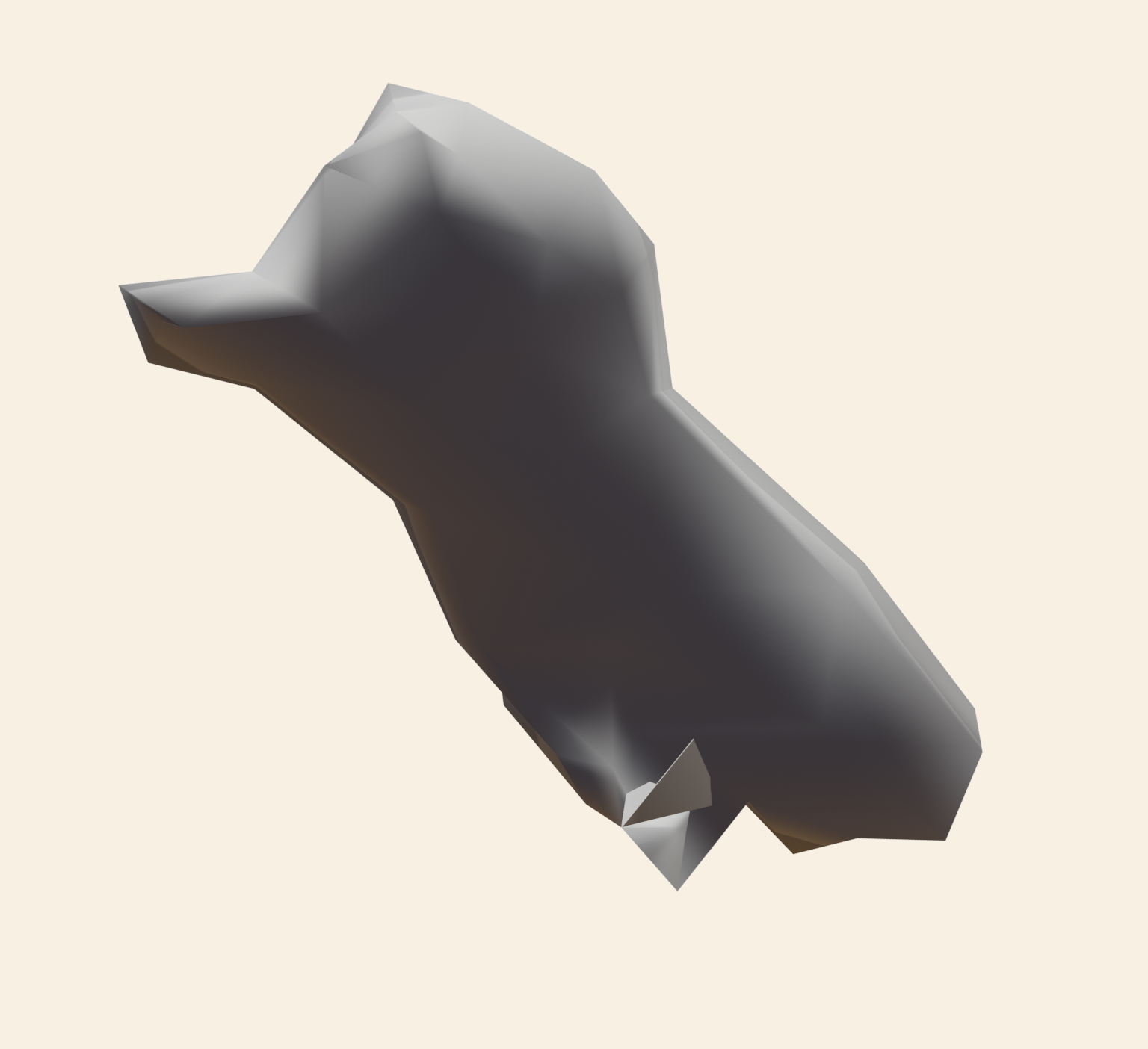} & 
    \includegraphics[width=0.14\columnwidth]{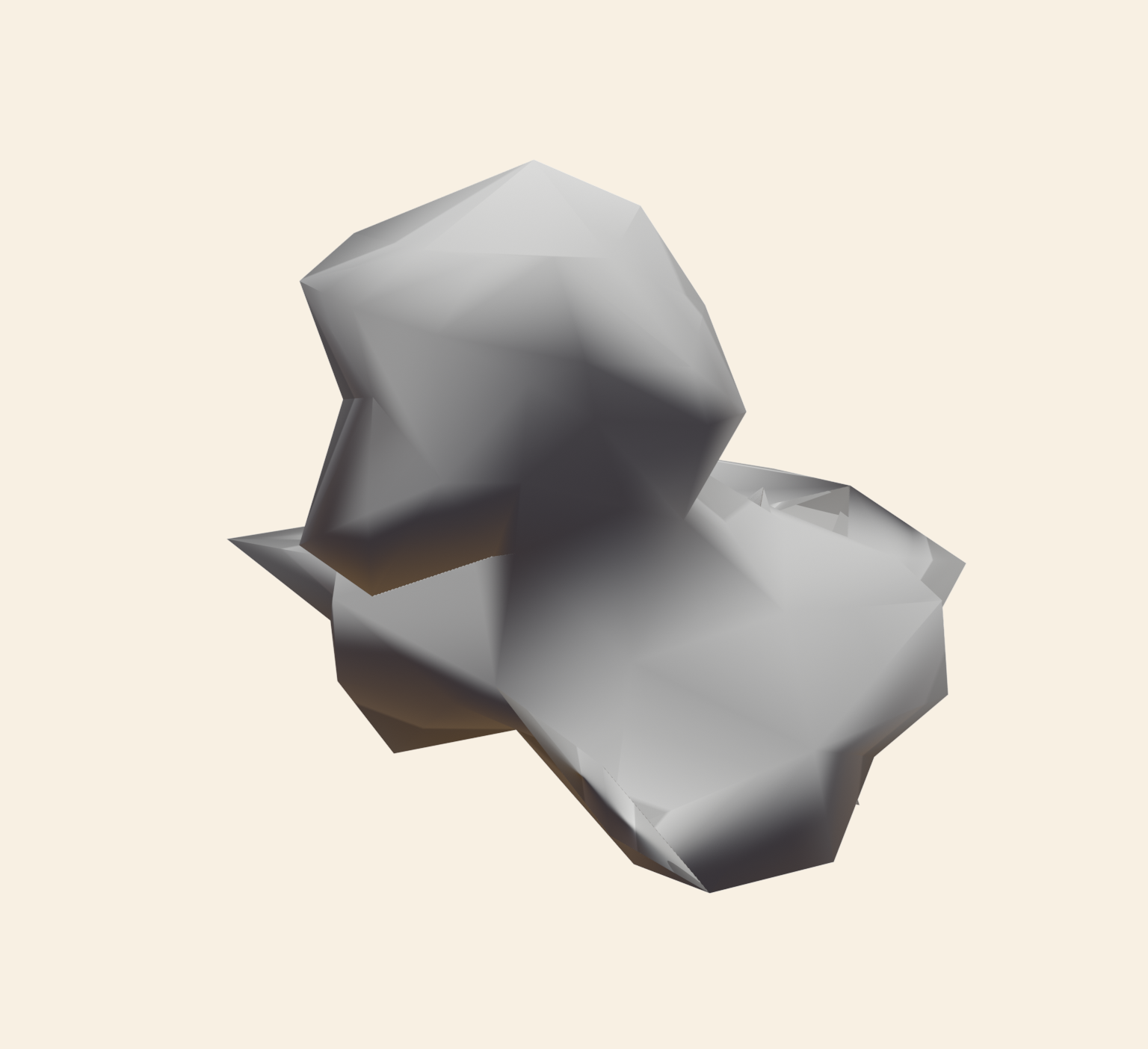} & 
    \includegraphics[width=0.14\columnwidth]{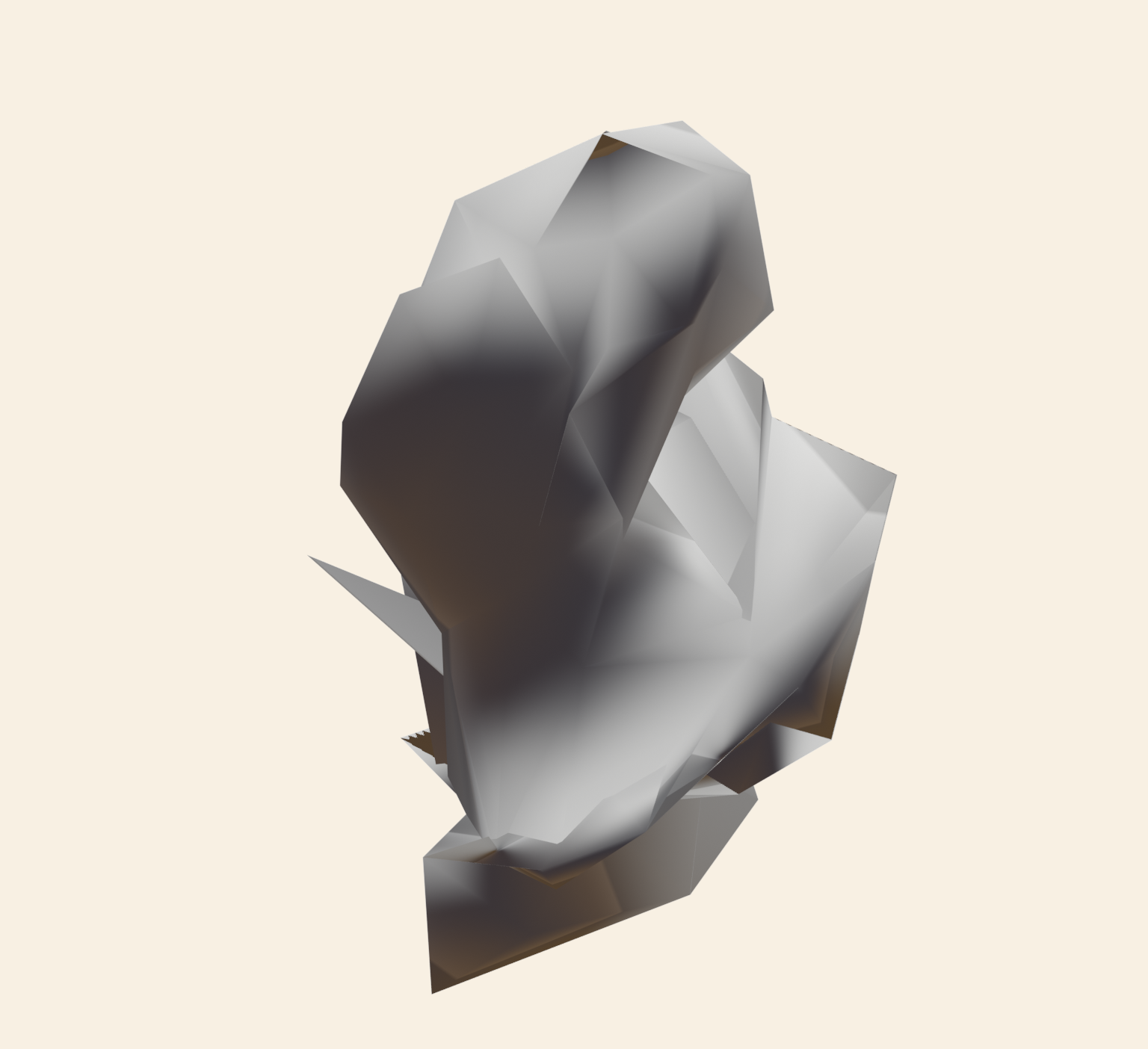} & \includegraphics[width=0.14\columnwidth]{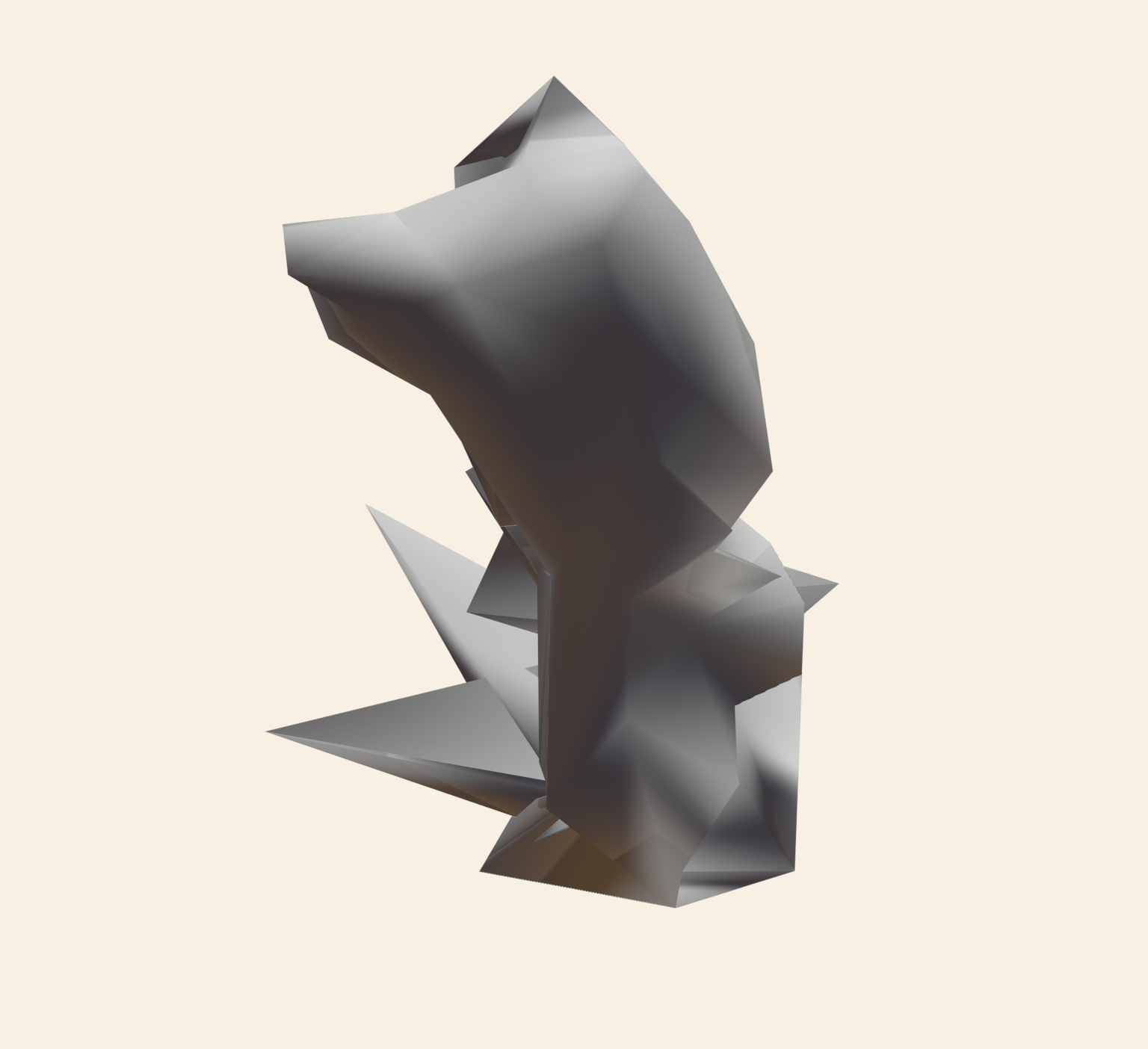}   \\
        \rotatebox{90}{Ours} & 
    \includegraphics[width=0.14\columnwidth]{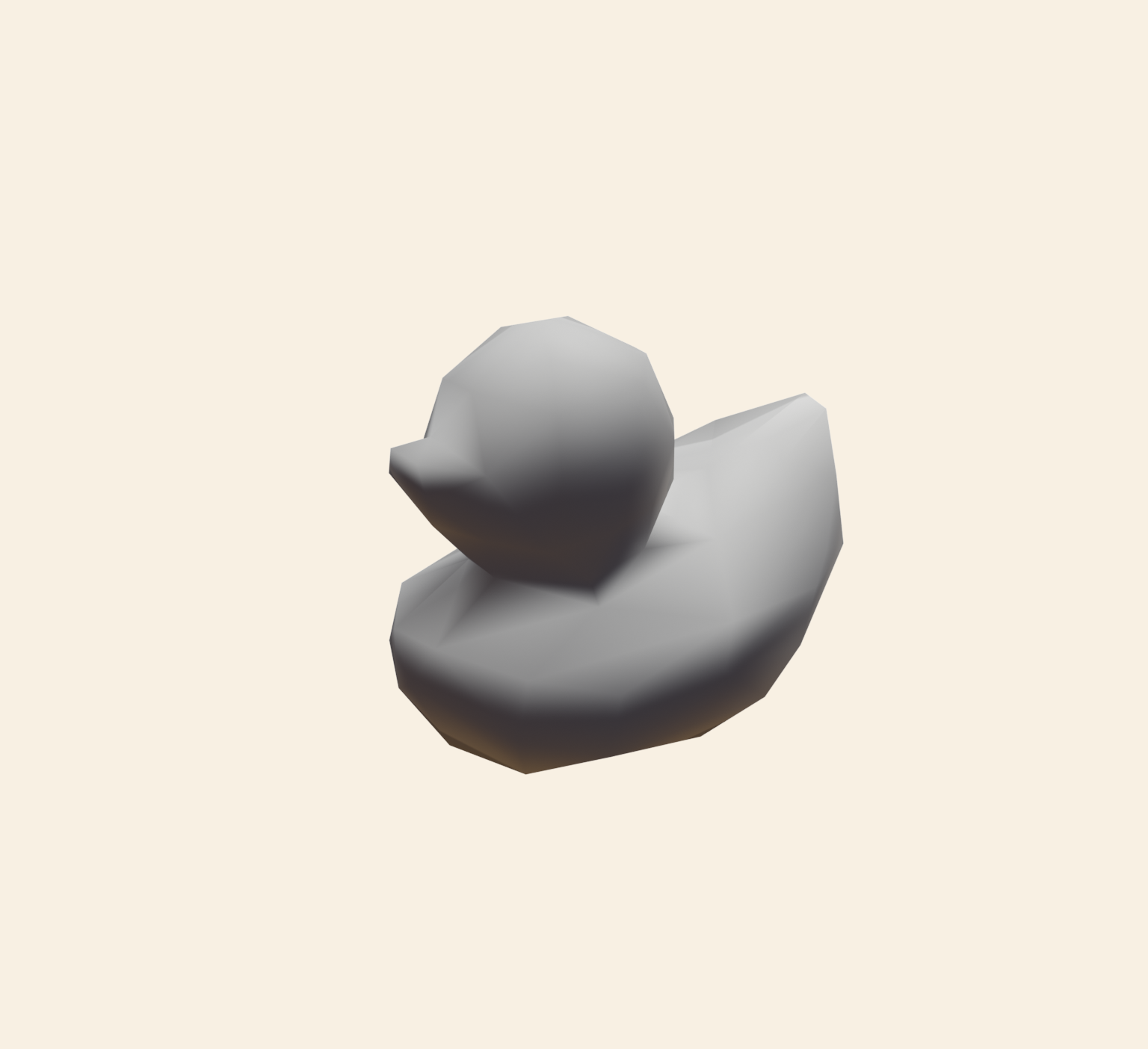} & 
    \includegraphics[width=0.14\columnwidth]{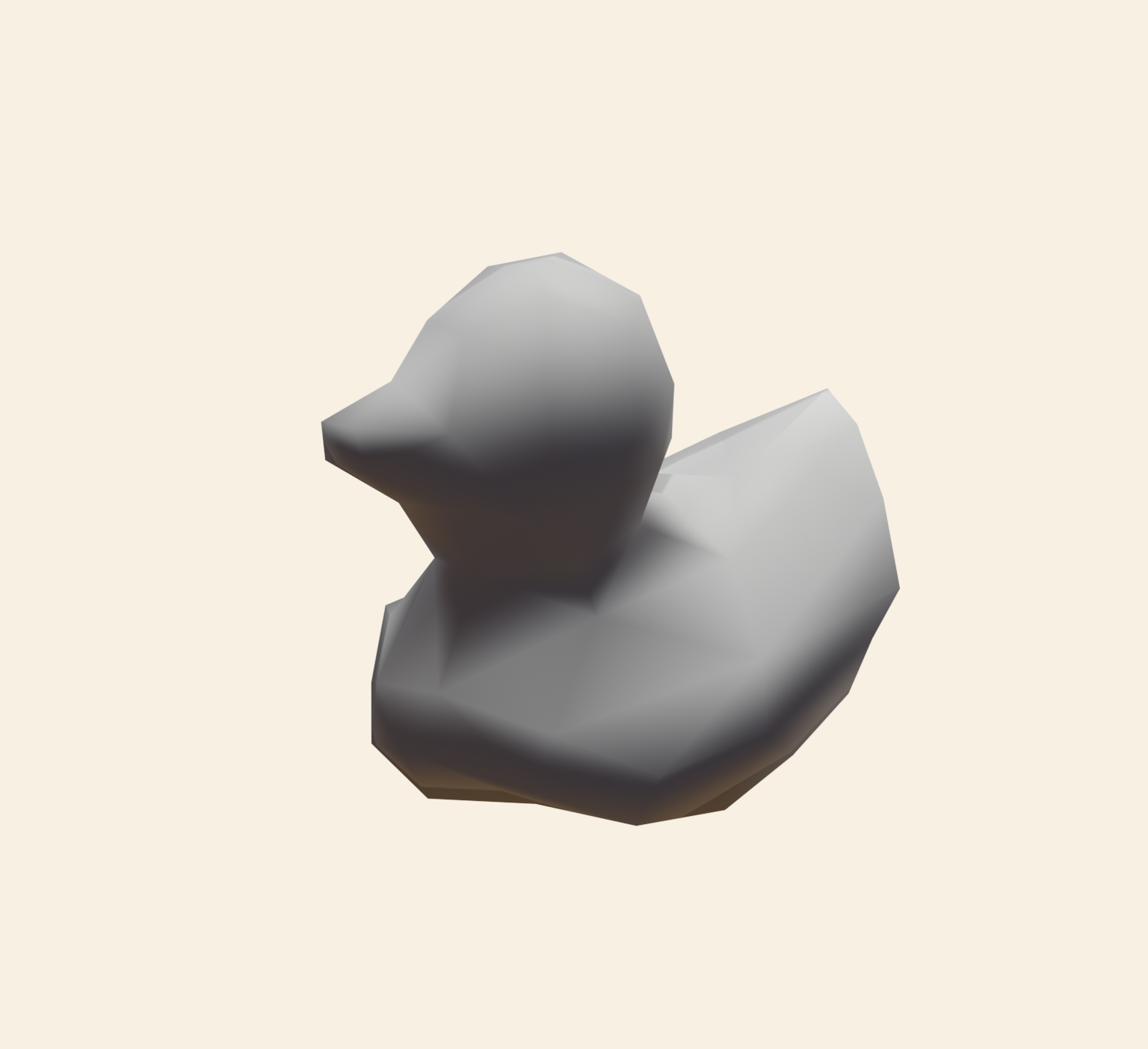} & 
    \includegraphics[width=0.14\columnwidth]{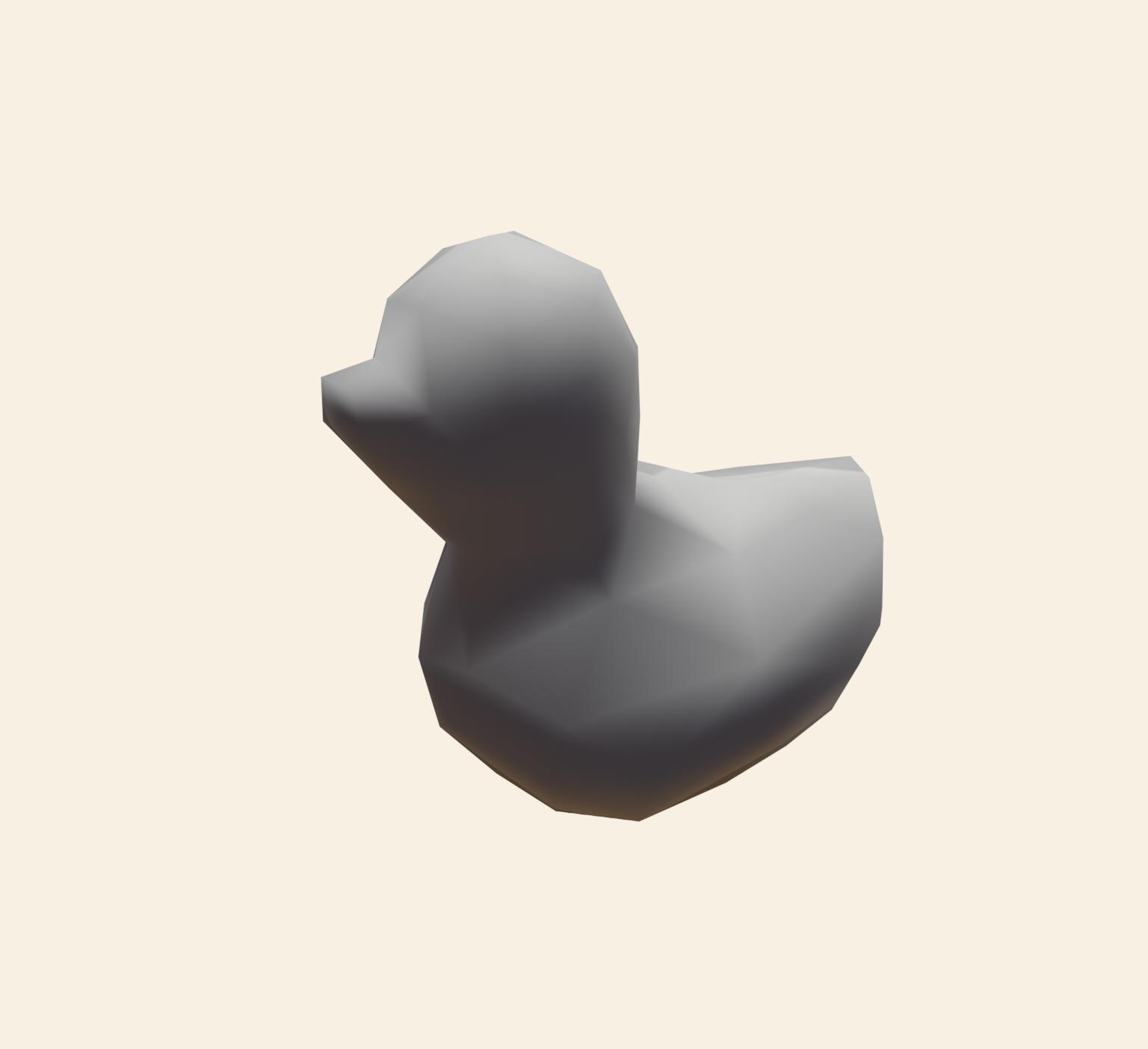} & 
    \includegraphics[width=0.14\columnwidth]{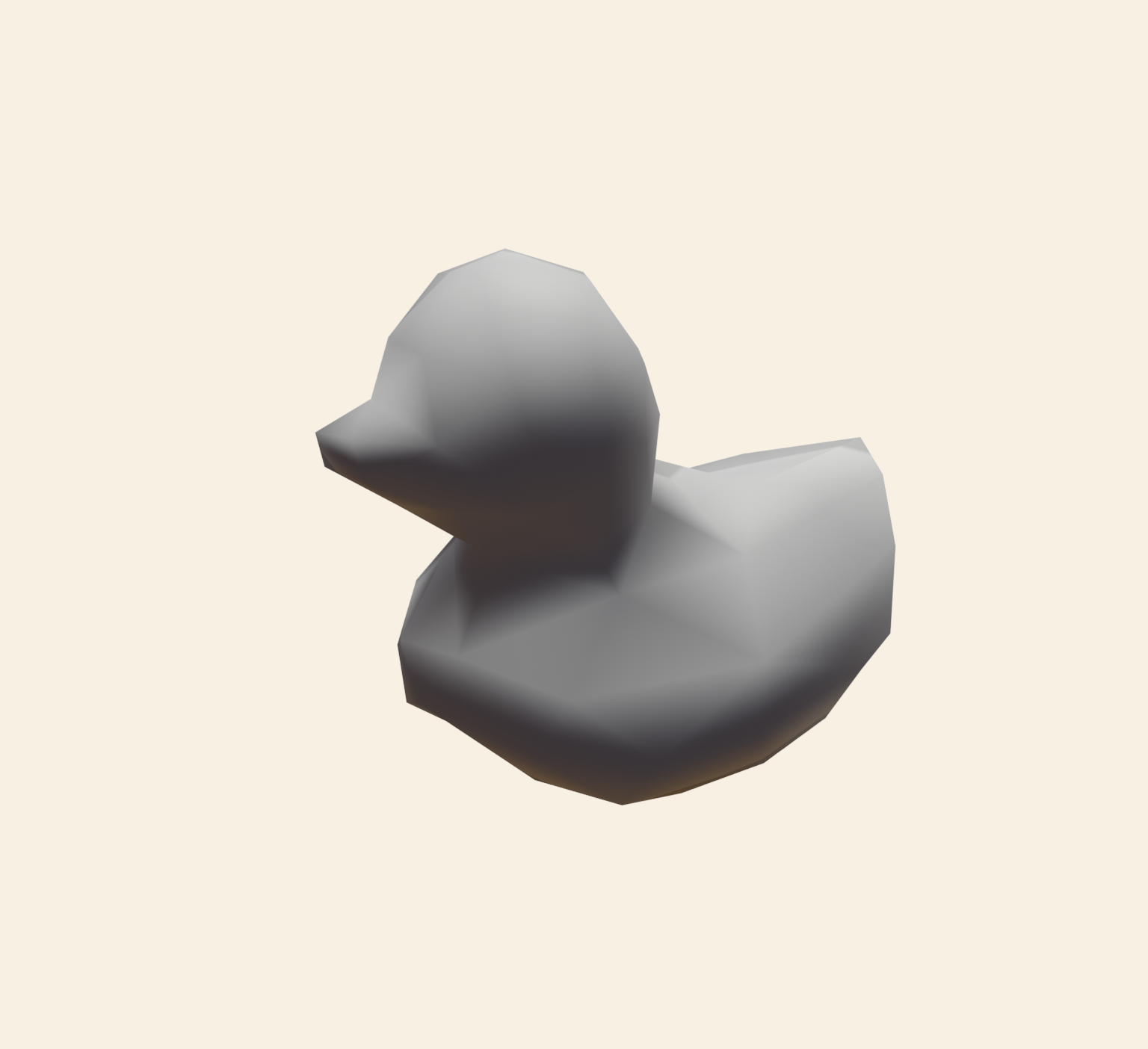} & 
    \includegraphics[width=0.14\columnwidth]{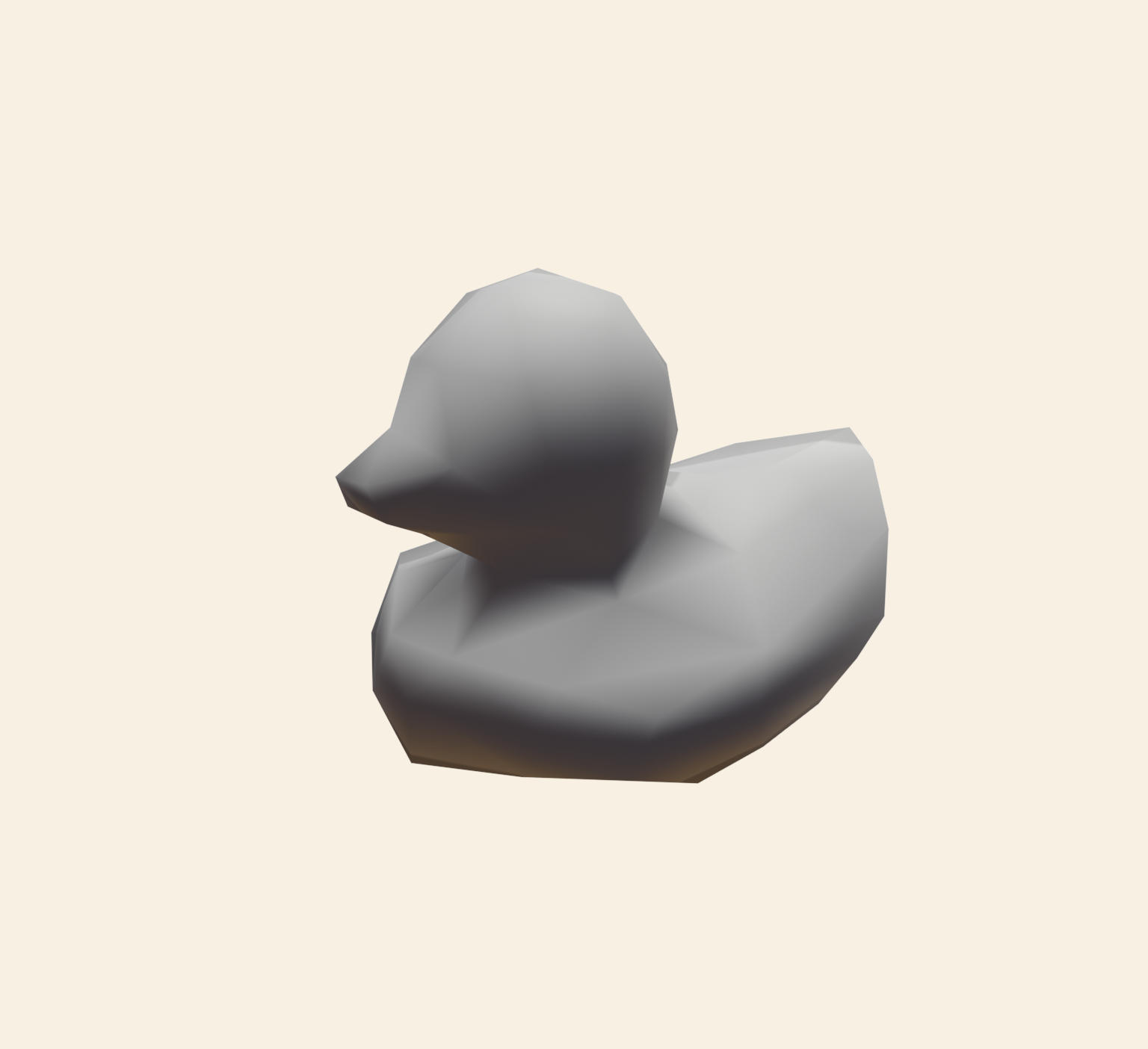} & \includegraphics[width=0.14\columnwidth]{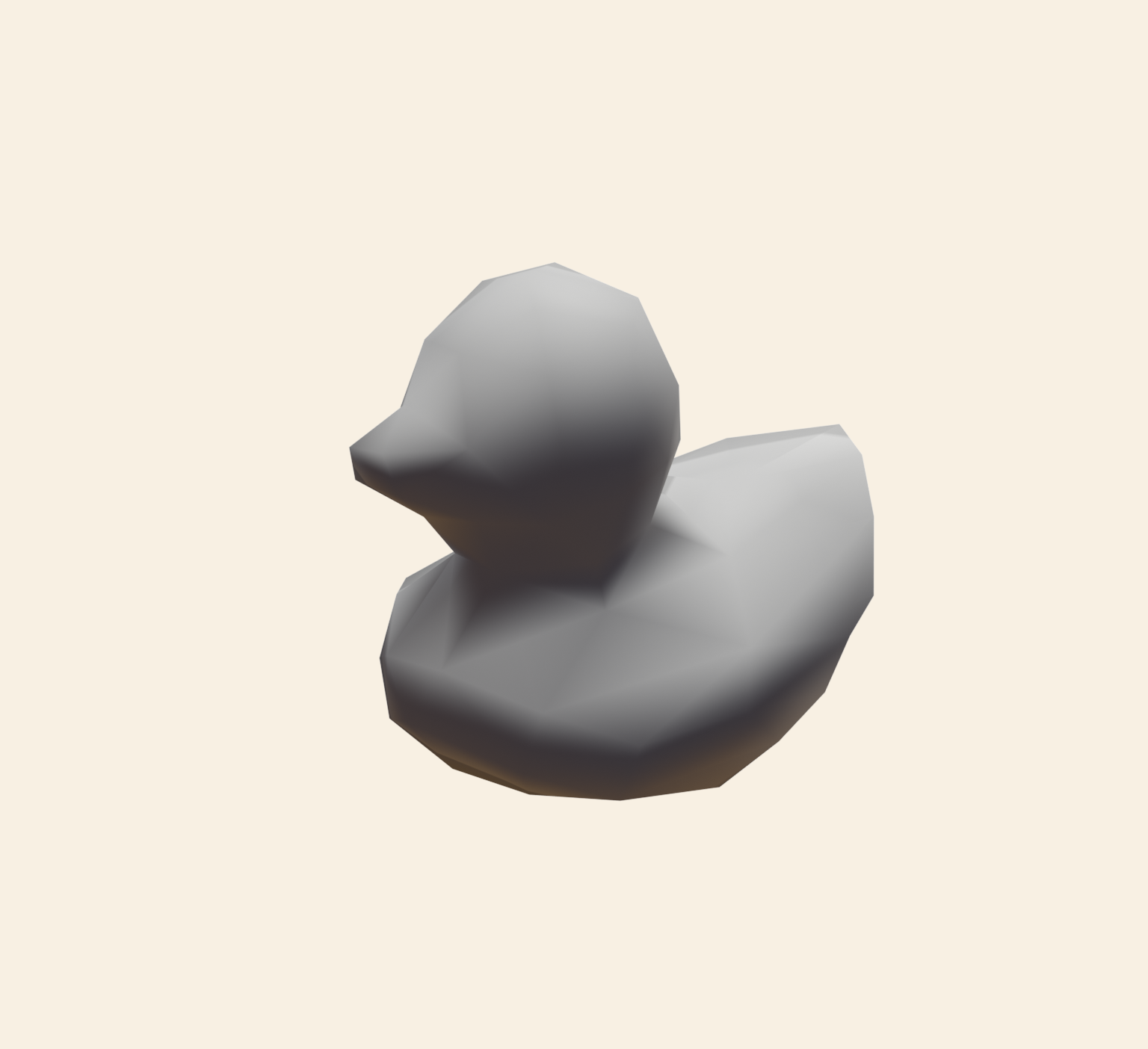}   \\
    \multicolumn{7}{c}{Target Position}\\
        \rotatebox{90}{Target} & 
    \includegraphics[width=0.14\columnwidth]{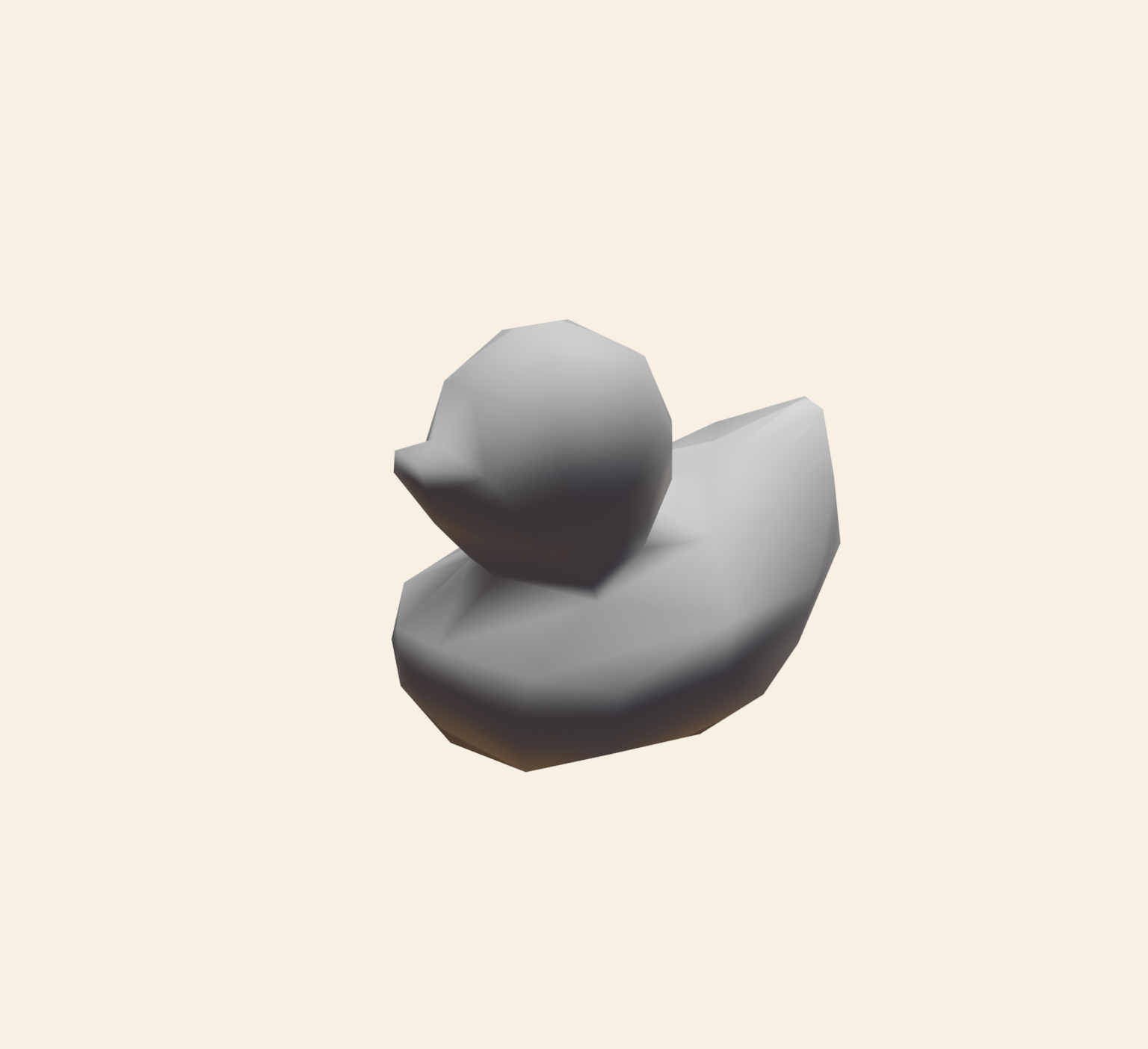} & 
    \includegraphics[width=0.14\columnwidth]{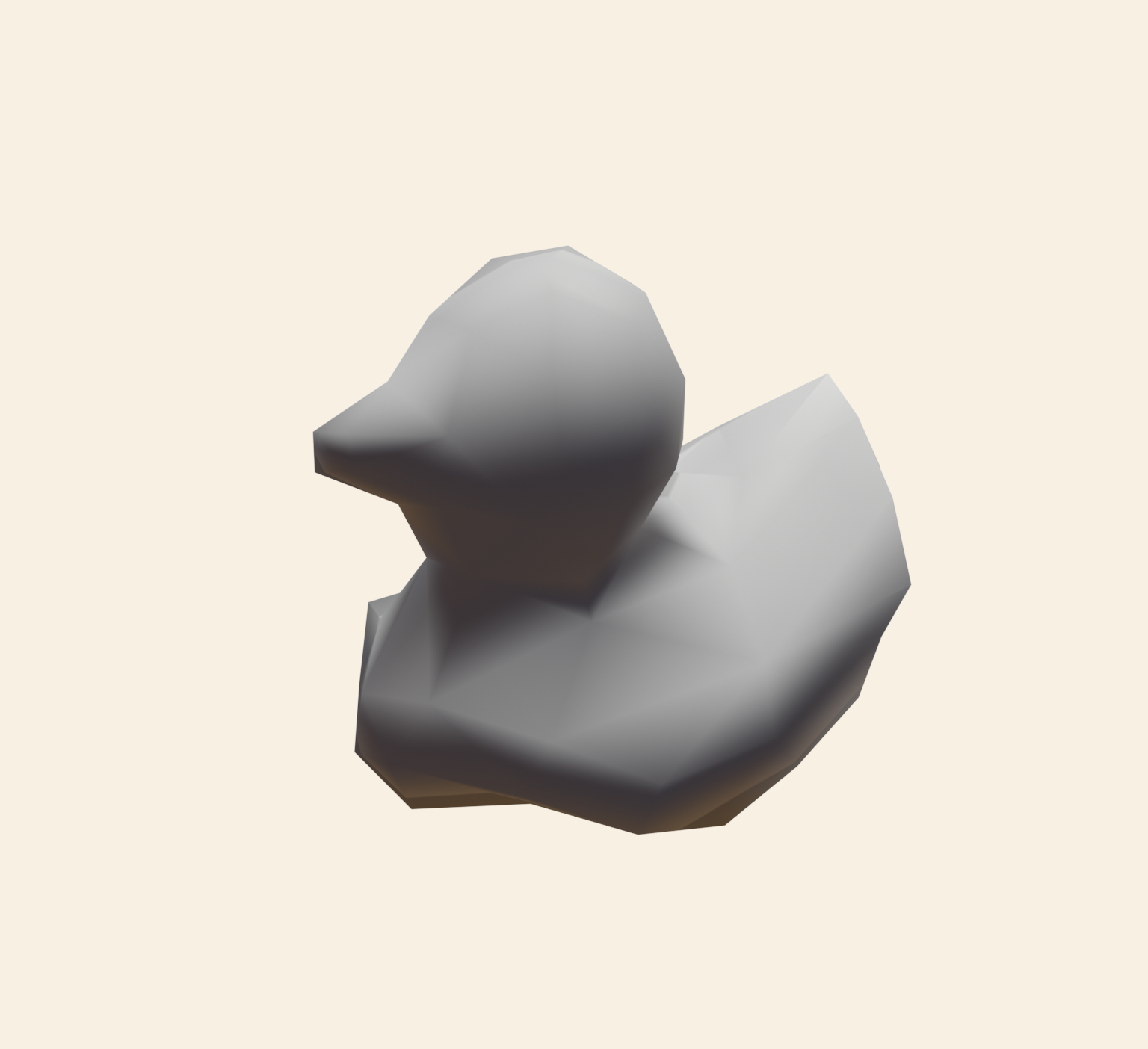} & 
    \includegraphics[width=0.14\columnwidth]{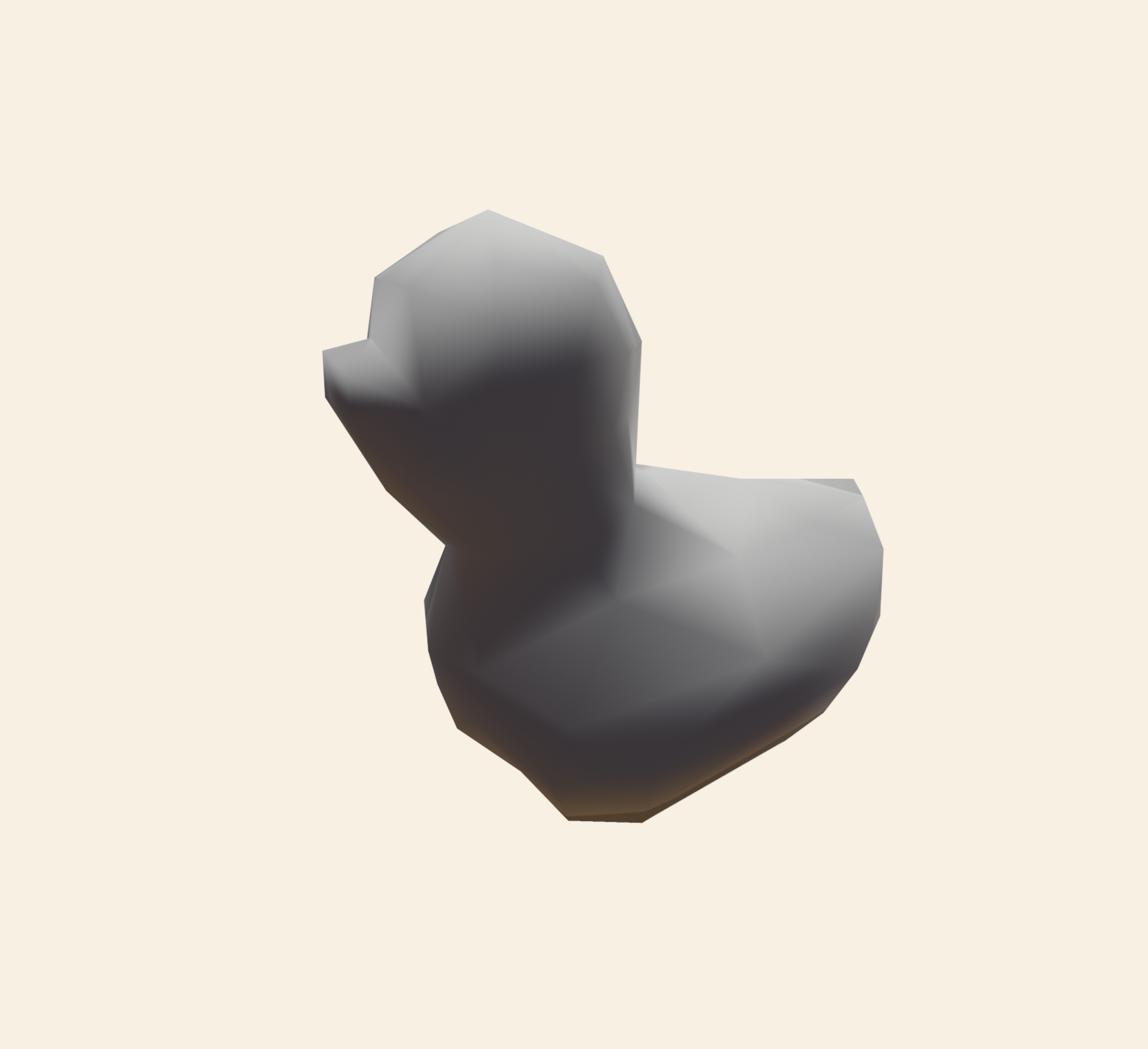} & 
    \includegraphics[width=0.14\columnwidth]{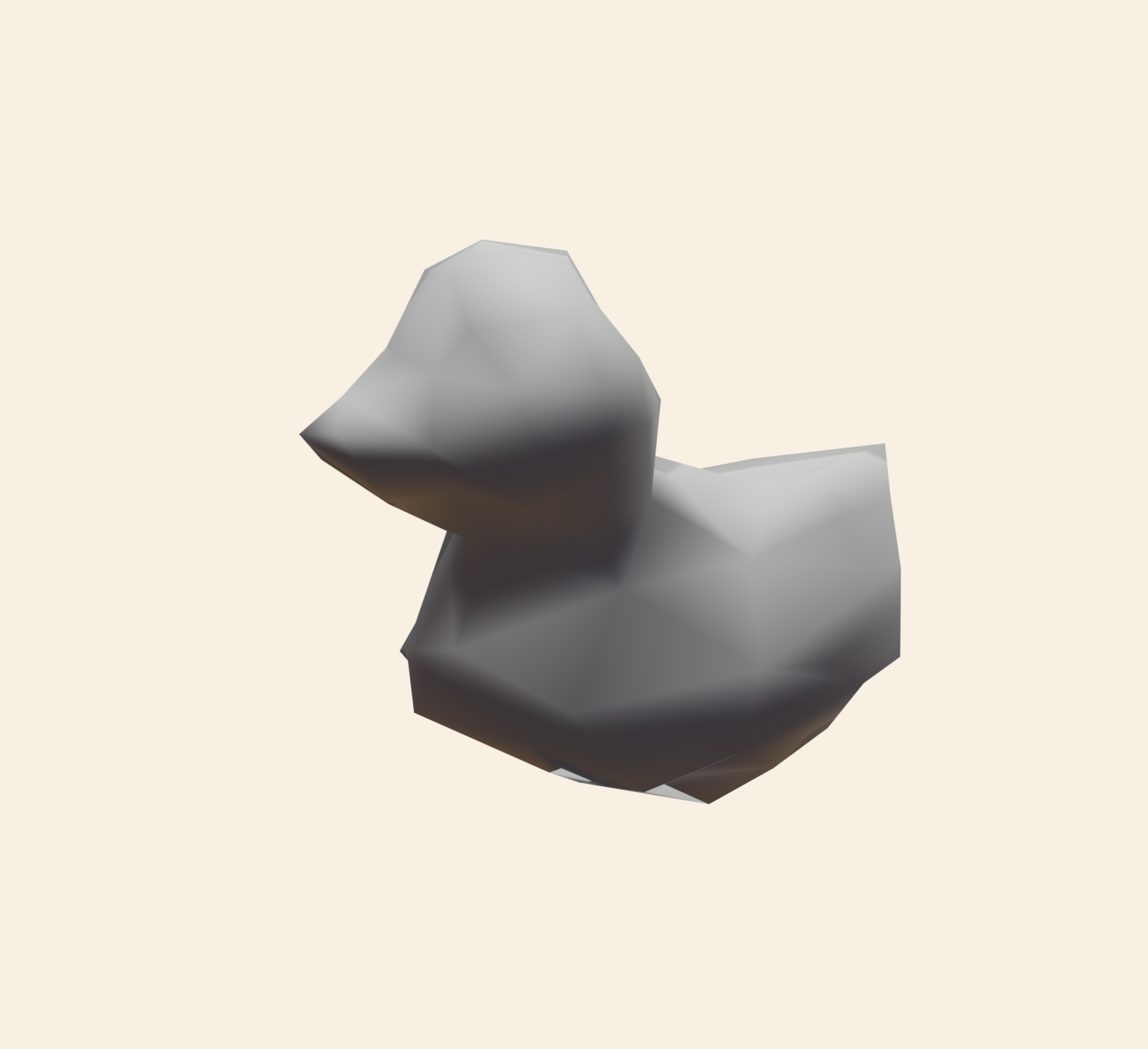} & 
    \includegraphics[width=0.14\columnwidth]{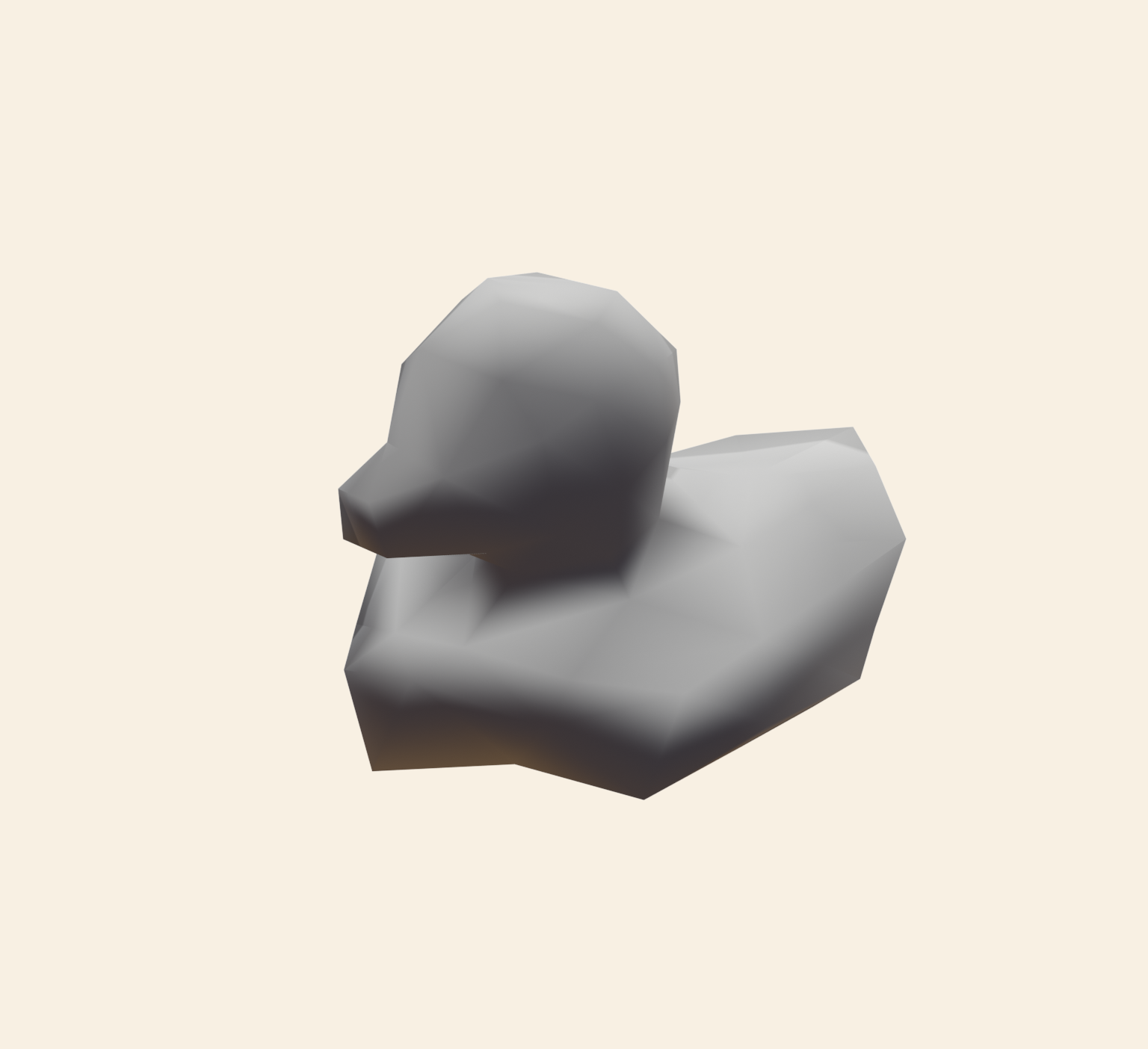} & 
    \includegraphics[width=0.14\columnwidth]{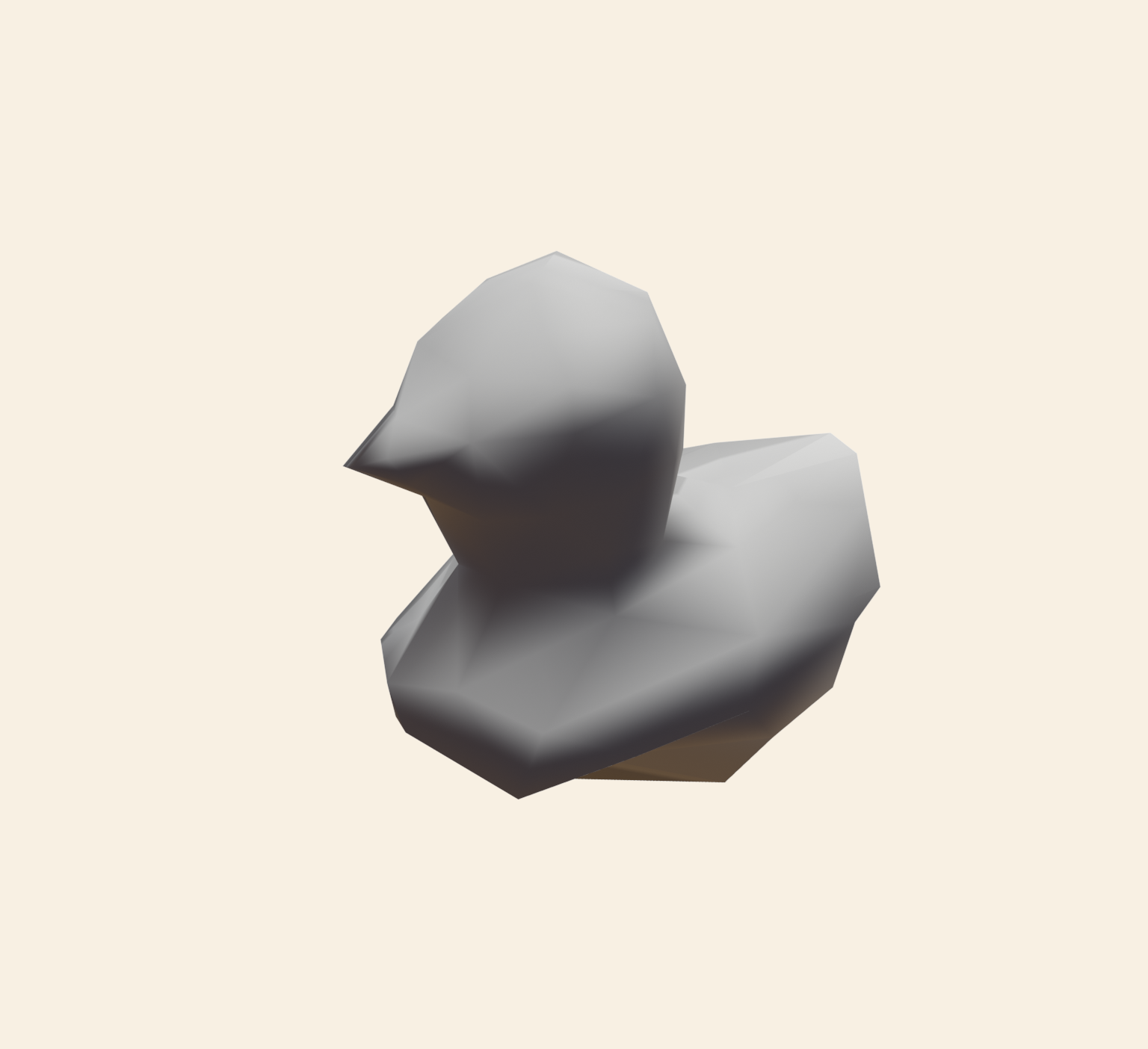}   \\
    &  0 & \qty{1.8}{s} & \qty{3.6}{s} &  \qty{5.4}{s} &  \qty{7.2}{s} &  \qty{9}{s}\\
    \end{tabular}
    \caption{Qualitative illustrations corresponding to our comparison with Verlet for the parameter estimation experiment in Figure~\ref{fig:quantsysidJaxx}. For a large time step, our method estimates a parameter that matches the target while Verlet fails to achieve this.
    \label{fig:qualsysidJaxx}
    }
\end{figure}

Figure~\ref{fig:quantsysidJaxx} plots a comparison of error in our parameter optimization against that obtained using Verlet integration on the model of a duck. The differentiable Verlet integration is implemented in JAXMD\cite{schoenholz2020jax} and we only match the last frame of the simulation. As with the forward model, we observe in Figure~\ref{fig:quantsysidJaxx} that our model outperforms state-of-the-art systems for large time step s. The associated qualitative comparisons are shown in Figure~\ref{fig:qualsysidJaxx}.

The target position is generated by running the forward simulation with Verlet integration. The step is \qty{0.0045}{s} and it is run for \qty{9}{s}. The parameters used were $k_s = \qty{2.5}{N/m}$ and $q = \qty{80}{\uC}$. In the backward problem, we only optimize the charge using $\Loss = ||\mathbf{x}_{sim}-\mathbf{x}_{target}||_2^2/n$, where $\mathbf{x}_{sim}$ and $\mathbf{x}_{target}$ represent the final position of guessed parameters and our target position. We run two experiments with large time step (\qty{0.18}{s}) and small time step (\qty{0.009}{s}). The results show that for the small time step, both methods can optimize to the target position, but for the larger time step, Verlet fails while ours is still able to match the target.

\subsubsection{DDEF runtime}
\begin{figure}[htbp!]
  \centering
\includegraphics[width=0.9\columnwidth]{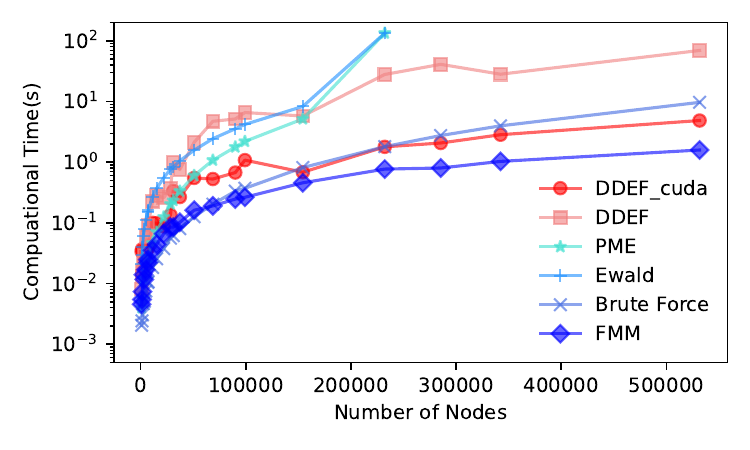}
\caption{Computation speed comparison. FMM method is a CPU paralleled version\cite{FMMgit}. Brute force, PME and Ewald are all from OpenMM\cite{eastman2023openmm}, we use CPU optimised version. DDEF is our CPU paralleled version and DDEF\_cuda is GPU accelerated. 
%We run the mesh from 1000 nodes to 530,000 nodes mesh. FMM is highly optimised which is the fastest among all the method; our GPU accelerated version is the second fastest methods. Interestingly, the OpenMM brute force method is the third fastest method, which is musch faster than the Ewald and PME. Our CPU paralleled methods comes fourth place. PME and Ewald can at most run the mesh with 230,000 meshes(due to memory limit), so it stops early and the trends showed that they are much slower than other methods.
}
\label{fig:DDEF_time}
\end{figure}

We compare the computation speed with other state-of-the-art methods, the result is shown in Figure \ref{fig:DDEF_time}. We use an optimized library for FMM \cite{FMMgit} that is CPU parallel and combines adaptive hierarchical methods. We set the approximation order to 1 for similar accuracy performance. Brute force, PME and Ewald are all from the OpenMM library\cite{eastman2023openmm} which uses highly optimized and CPU parallelized implementation. DDEF is our CPU parallelized version and DDEF\_cuda is the GPU accelerated one (GPU is used to accelerate the far-field gathering). The test uses meshes between $1K$ and $530K$ vertices.  Interestingly, OpenMM brute force is reasonably fast (with memory limits)- much faster than Ewald and PME. Our CPU parallelized method comes fourth place. We capped $m$ at 1000, but can achieve better results with adaptive $m$. 
PME and Ewald failed beyond the $230K$ vertices (memory limit). We tested DDEF up to $3M$ vertices.

\subsection{Mechanisms for artistic control}
\subsubsection{Manipulation of charges}
\begin{figure}[htbp]
    \centering
    \setlength{\tabcolsep}{2pt} % Reduce space between columns
    \renewcommand{\arraystretch}{0.8} % Reduce space between rows
    \begin{tabular}{ccc}
            % Consider removing [6pt] to reduce vertical space
    \includegraphics[width=0.14\textwidth]{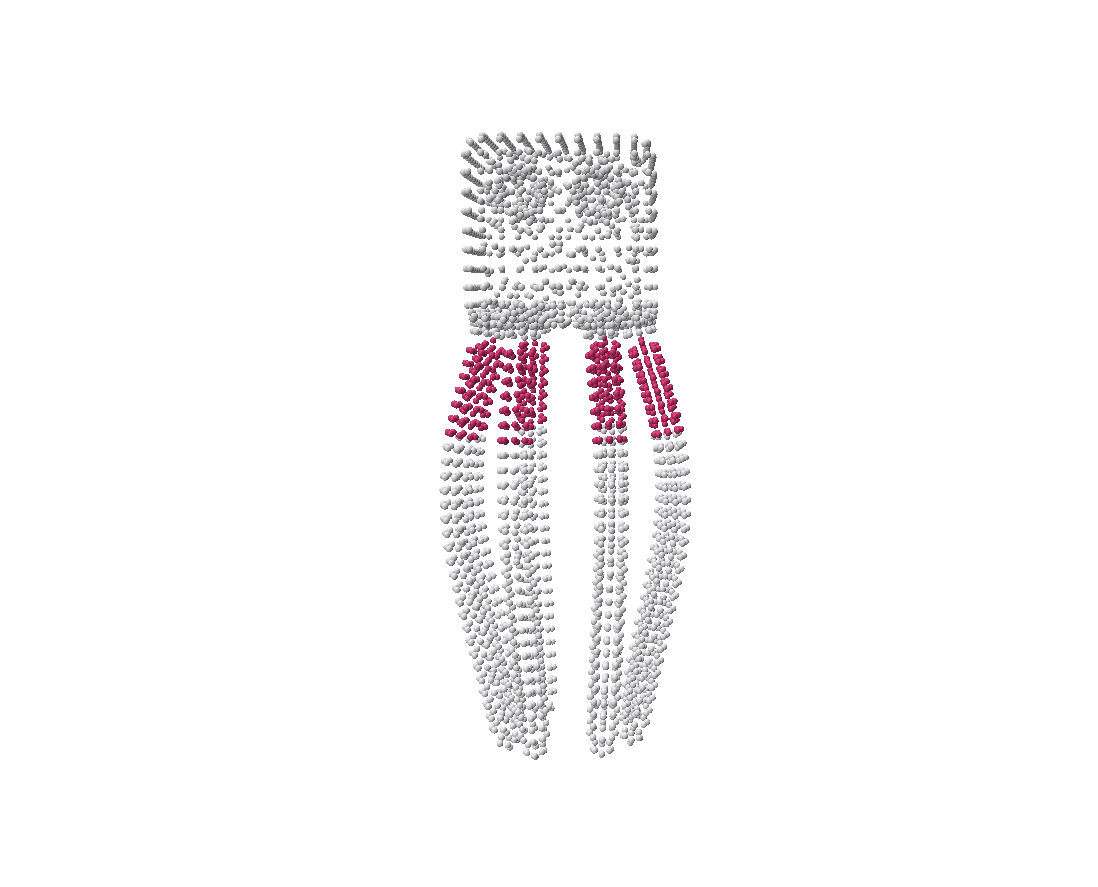} & 
    \includegraphics[width=0.14\textwidth]{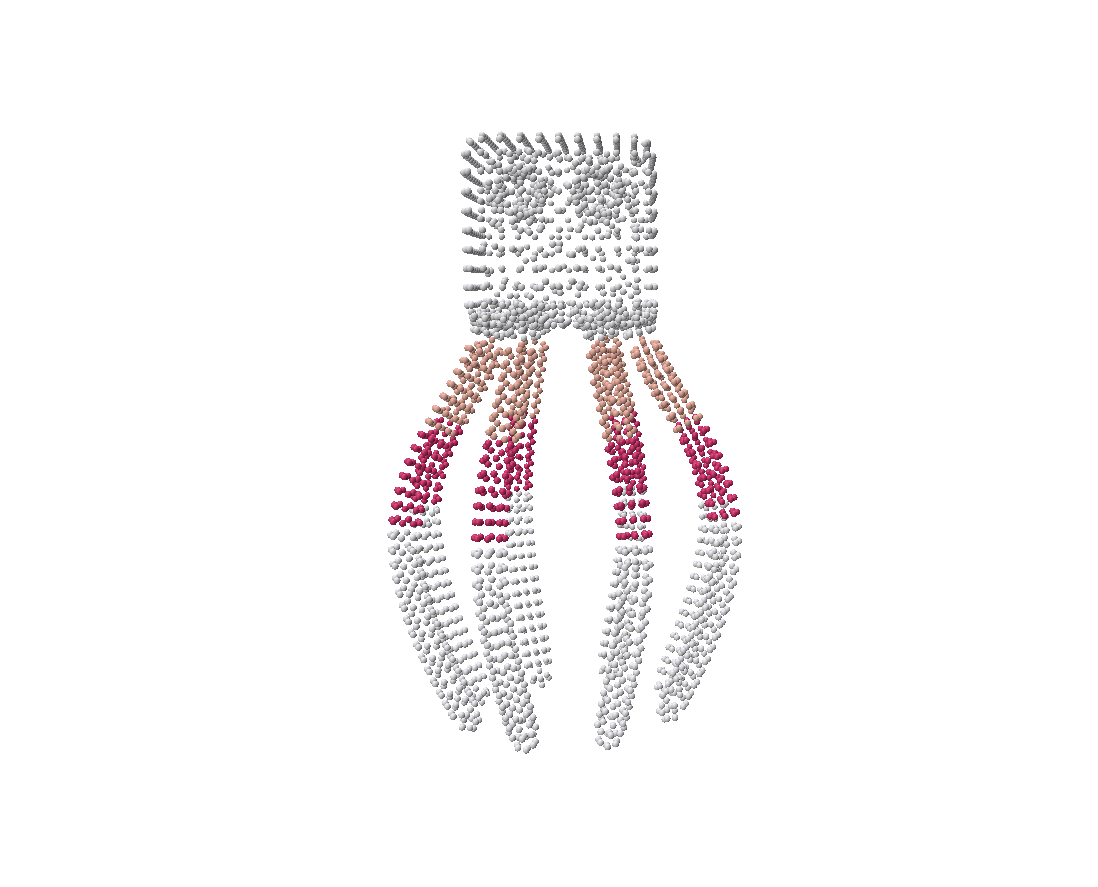} & 
    \includegraphics[width=0.14\textwidth]{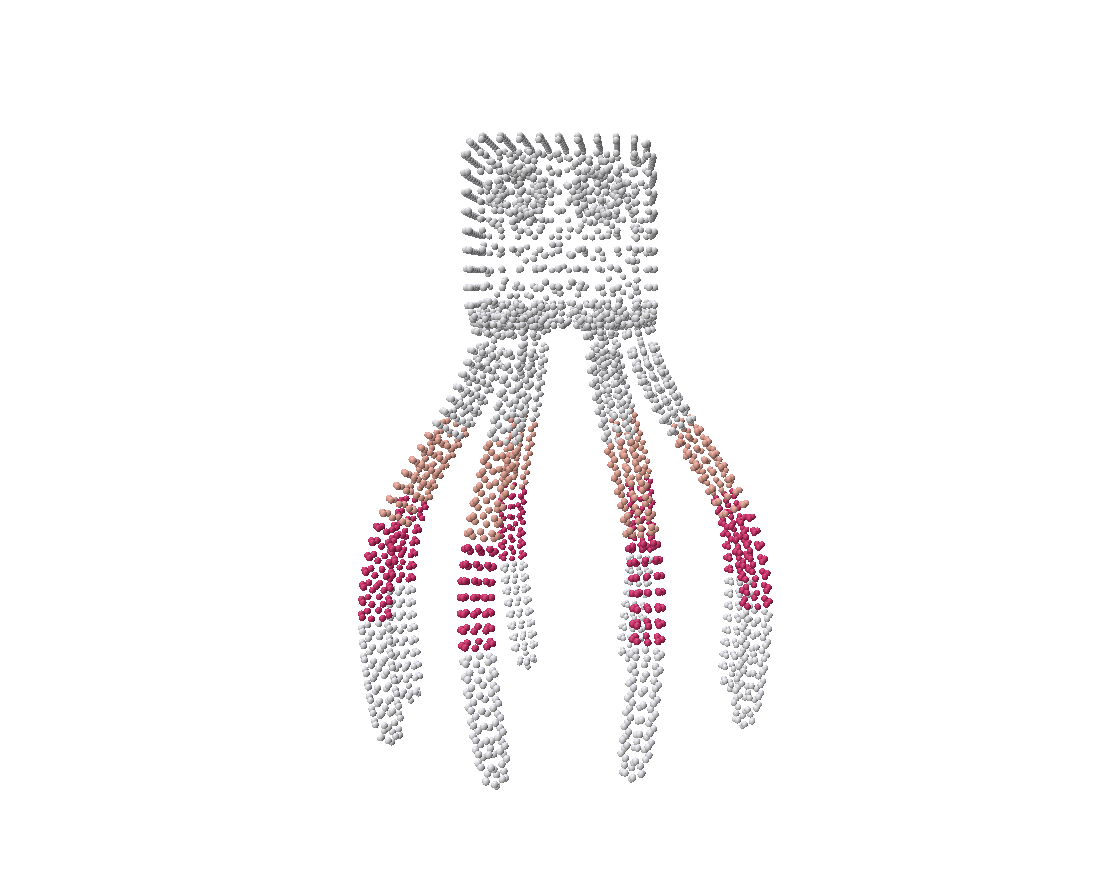} \\ 
    \includegraphics[width=0.14\textwidth]{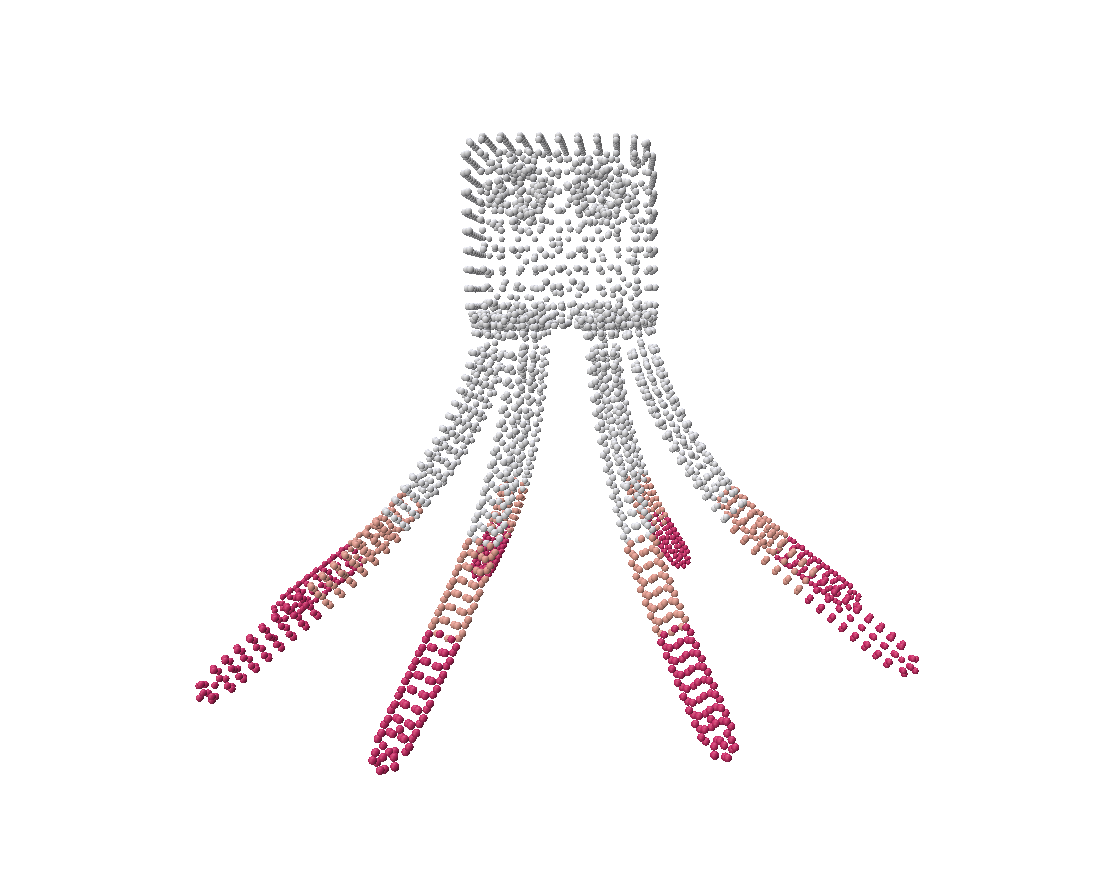} & 
    \includegraphics[width=0.14\textwidth]{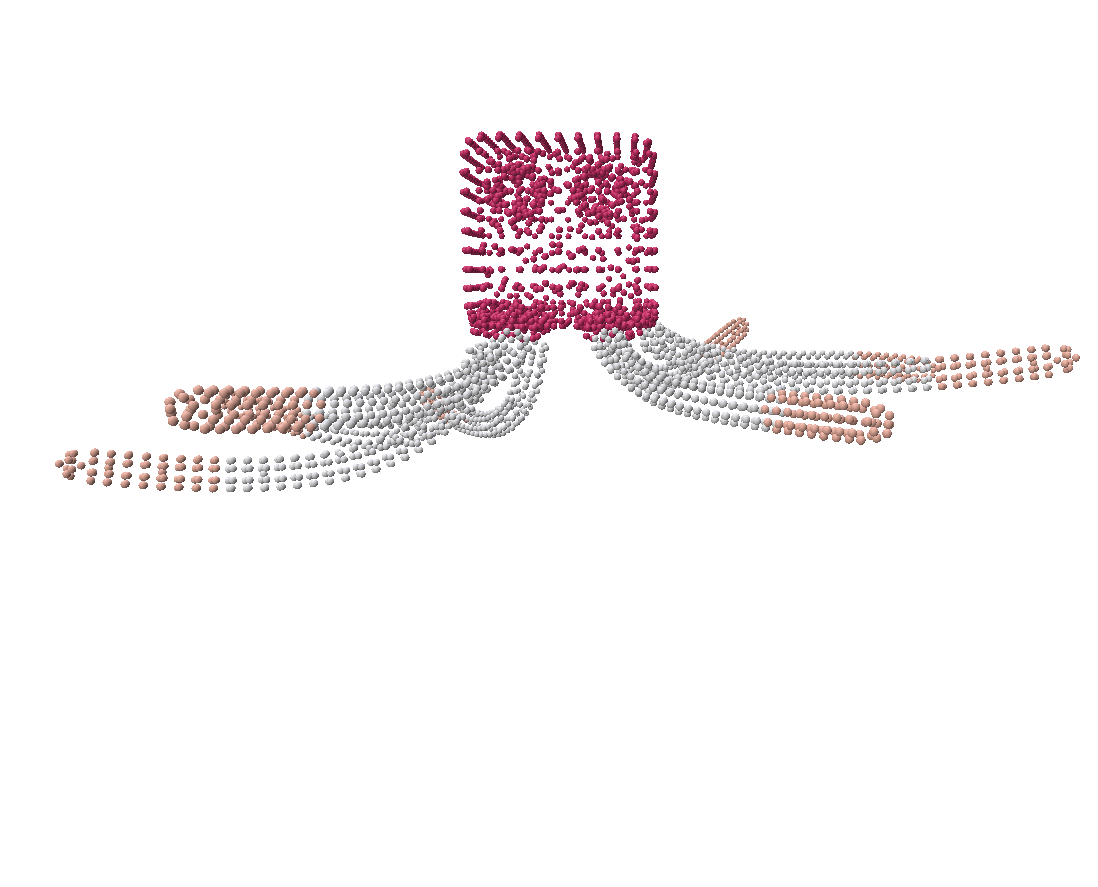} & 
    \includegraphics[width=0.14\textwidth]{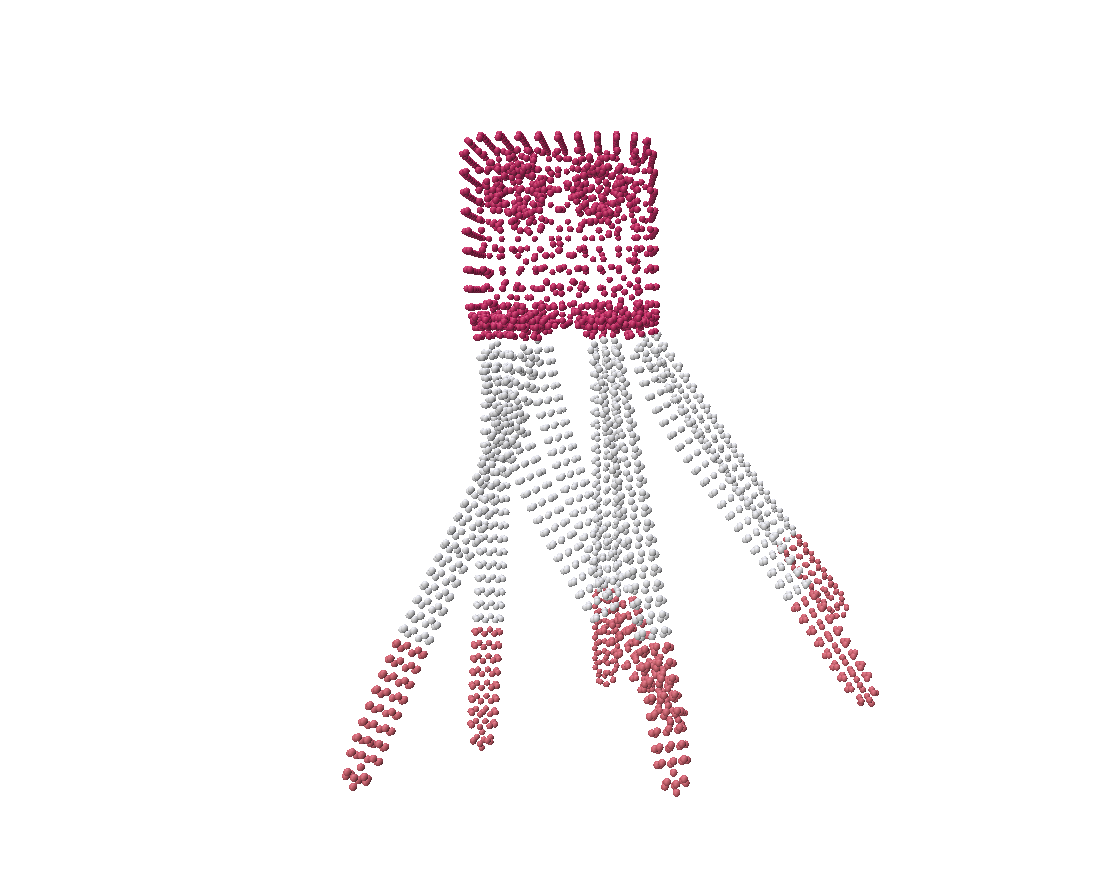} 
    \end{tabular}
    \caption{Manipulating the internal charges of the mesh. Here we progressively increase and decrease the charges in subsequent segments of the octopus' appendages in order to create a 'jellyfish-like' effect. The red color indicates a positive electric charge}
    \label{fig:chargeManipulation}
\end{figure}

The dynamics of the system can be actuated via modification of the charges. 
We developed a simple and intuitive graphical user interface to edit charges and run our simulation in real time, using Polyscope \cite{polyscope}. We performed vertex group assignments during the 3D modelling process in Blender\cite{blender} before loading the mesh as a point cloud in our interface. This process can be considered  `rigging for charged bodies'.  We show the effects of key-framing charge assignments to different vertex groups. Sliders provide artistic control over different parts of the mesh as they attract/repel each other. We demonstrate in Figure \ref{fig:chargeManipulation} the charge manipulations for the dynamics shown in Figure~\ref{fig:teaser}. It is also possible to affect dynamics by the introduction of external charges. These could be just single particles, meshes whose vertices are charged or in the form of primitives such as a sphere with a charge and a Gaussian fall-off of its charge. The effects of these controls are shown in the accompanying video. The video also shows an example of a single external charge on a charged mesh. 

\subsubsection{External charge} 
\begin{figure}[htbp]
    \centering
    \setlength{\tabcolsep}{2pt} % Reduce space between columns
    \renewcommand{\arraystretch}{0.8} % Reduce space between rows
    \begin{tabular}{ccc}
            % Consider removing [6pt] to reduce vertical space
    %\includegraphics[width=0.14\textwidth]{figs/ec/frame_000077.png} & 
    \includegraphics[width=0.14\textwidth]{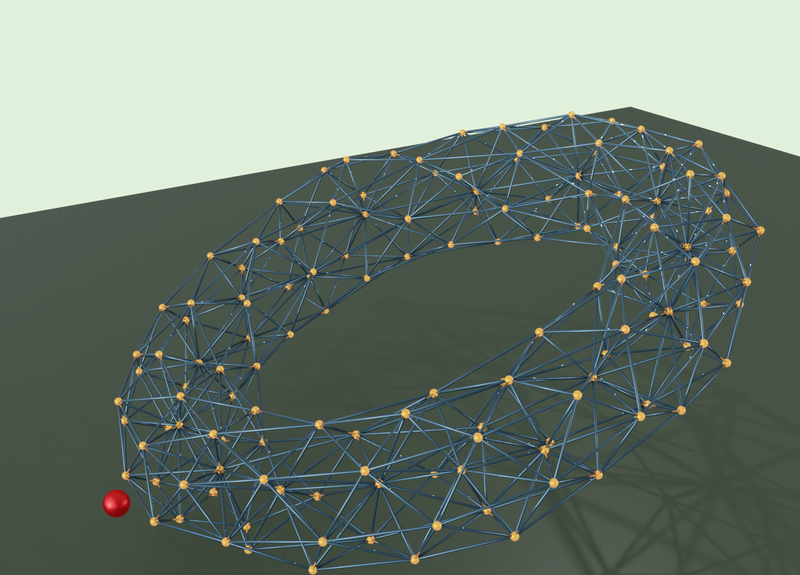} & 
    \includegraphics[width=0.14\textwidth]{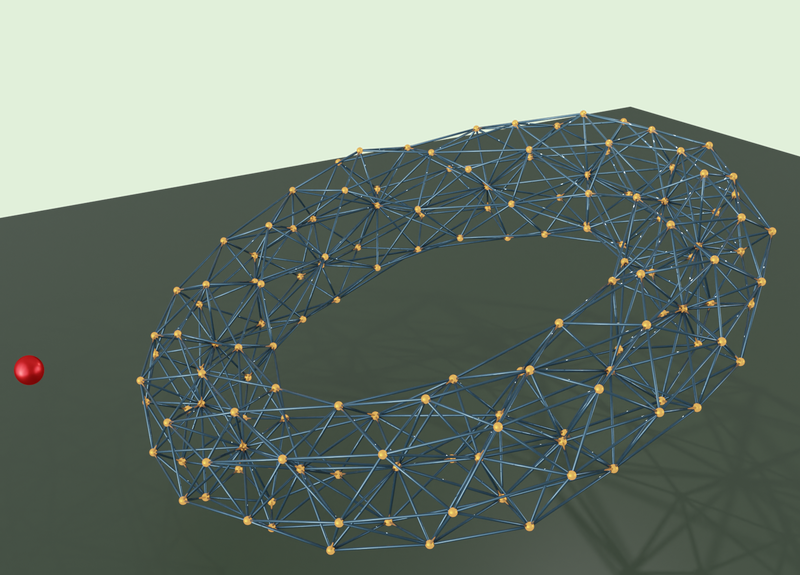} &
    \includegraphics[width=0.14\textwidth]{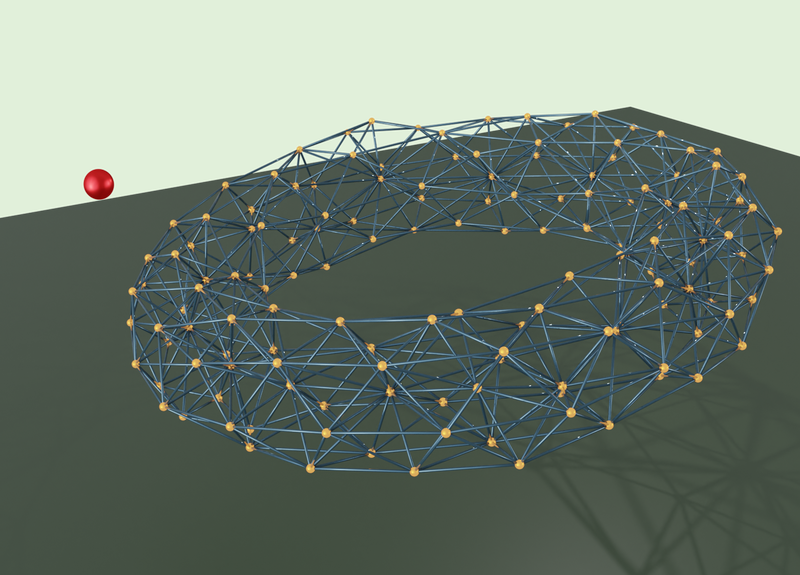} \\
    \includegraphics[width=0.14\textwidth]{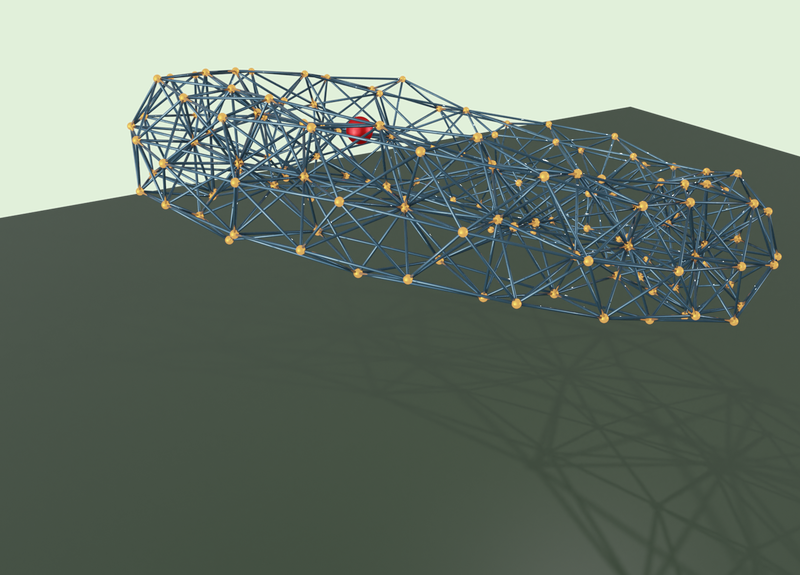} & 
    \includegraphics[width=0.14\textwidth]{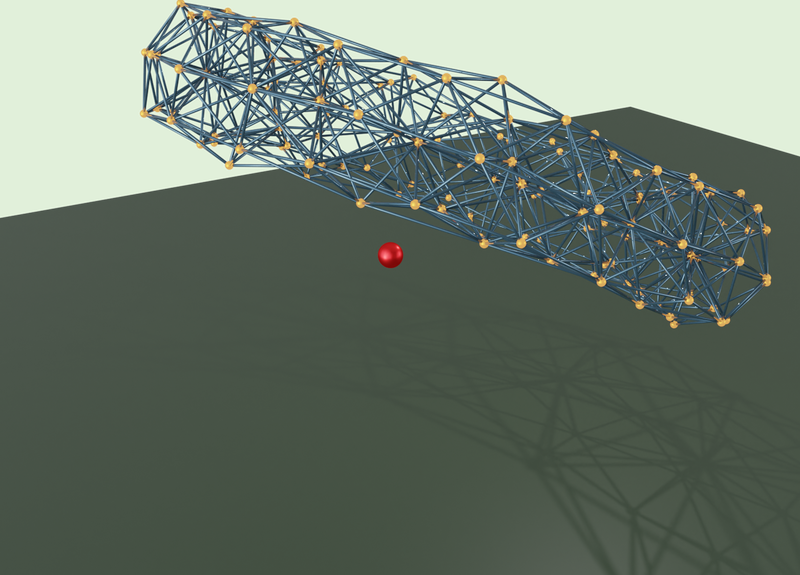}  & 
    \includegraphics[width=0.14\textwidth]{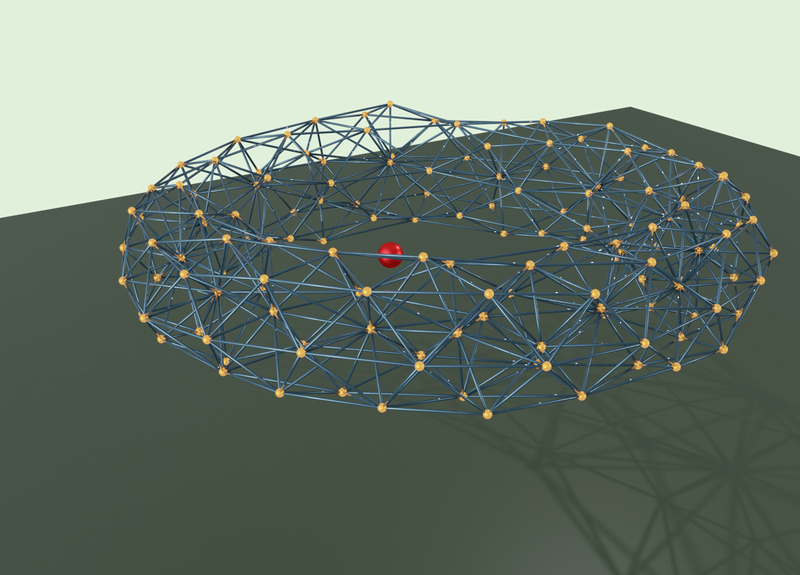} \\
    \end{tabular}
    \caption{Animation using an external charged particle. The red sphere is an external negative charge that is rotated through the positively charged nodes of the torus. The mesh has a charge $80 \mu C$ and a spring constant of $10N/m$. The external charge is $-42\mu C$. We refer to the accompanying video for the entire demonstration.}
    \label{fig:external charge demo}
\end{figure}

\BZ{mark as potential removal if run out of space; this one is mentioned above would appears in the video}Figure~\ref{fig:external charge demo} shows that the charged system can be controlled via fields generated due to external charges. The figure shows a negative external charge (red sphere) interacting with our mesh, whose nodes are positively charged. The mesh has a spring constant of $10N/m$ and a charge of $80\mu C$. We simulate with timestep = 0.06$s$ for 1000 steps. The figure shows from left to right that the mesh moves towards the charge as it rotates. The entire episode is shown in the accompanying video.

\subsubsection{External electric field}

\begin{figure}[htbp]
    \centering
    \setlength{\tabcolsep}{2pt} % Reduce space between columns
    \renewcommand{\arraystretch}{0.8} % Reduce space between rows
    \begin{tabular}{ccc}
            % Consider removing [6pt] to reduce vertical space
    \includegraphics[width=0.14\textwidth]{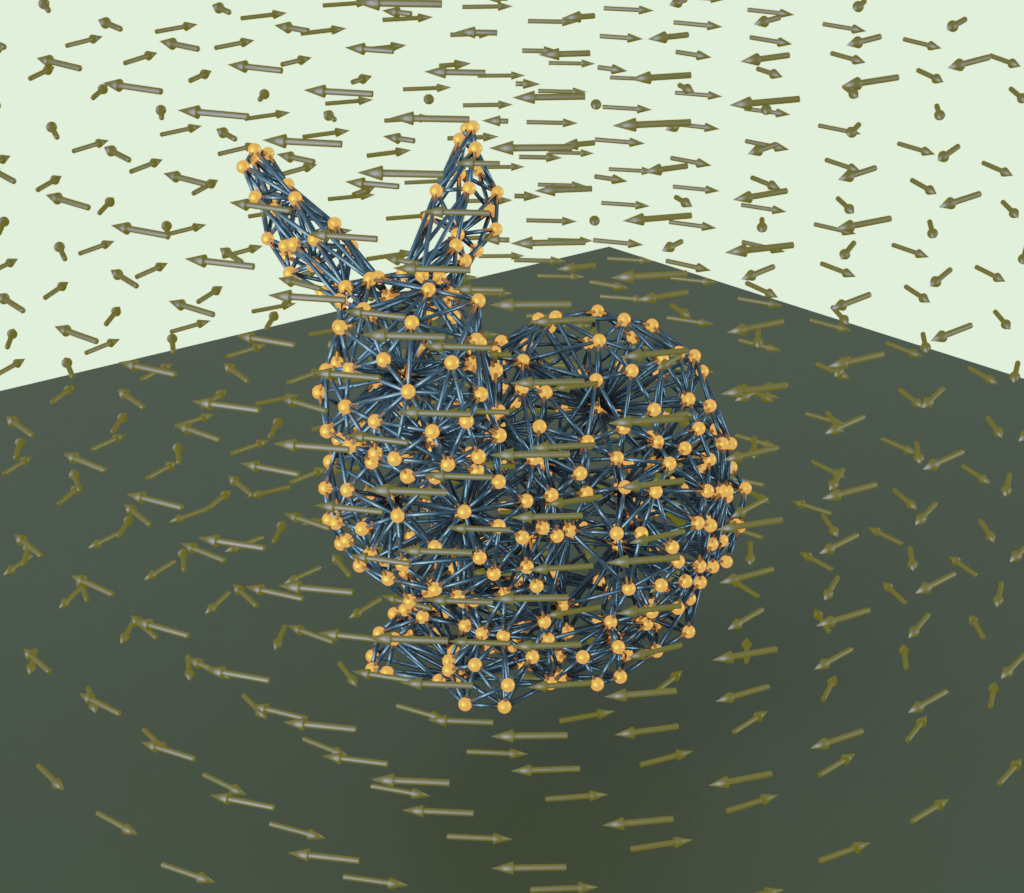} & 
    \includegraphics[width=0.14\textwidth]{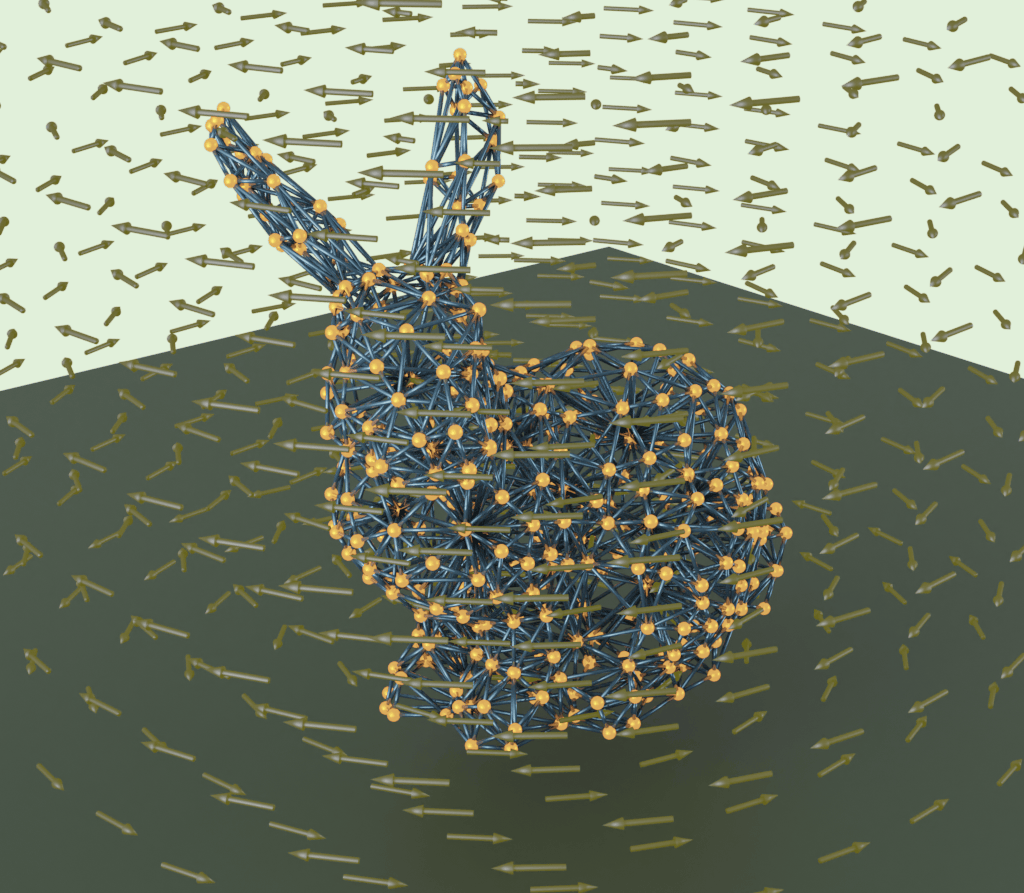} & 
    \includegraphics[width=0.14\textwidth]{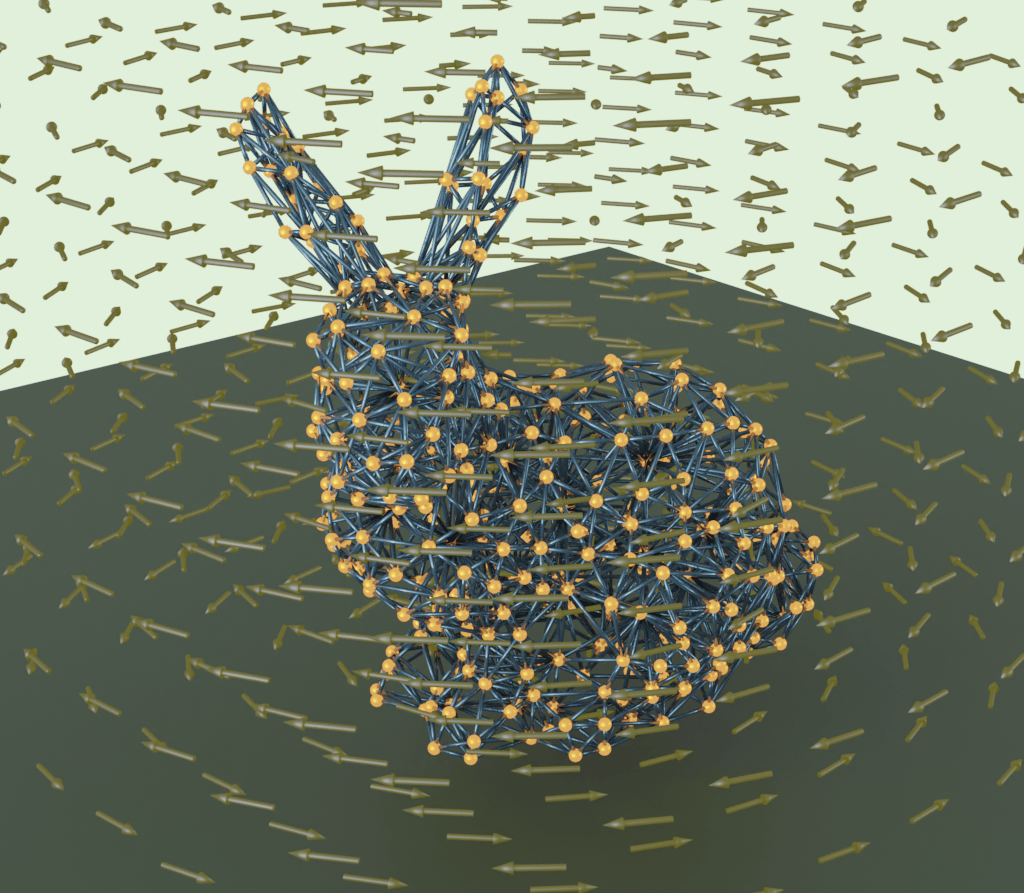} \\ 
    \includegraphics[width=0.14\textwidth]{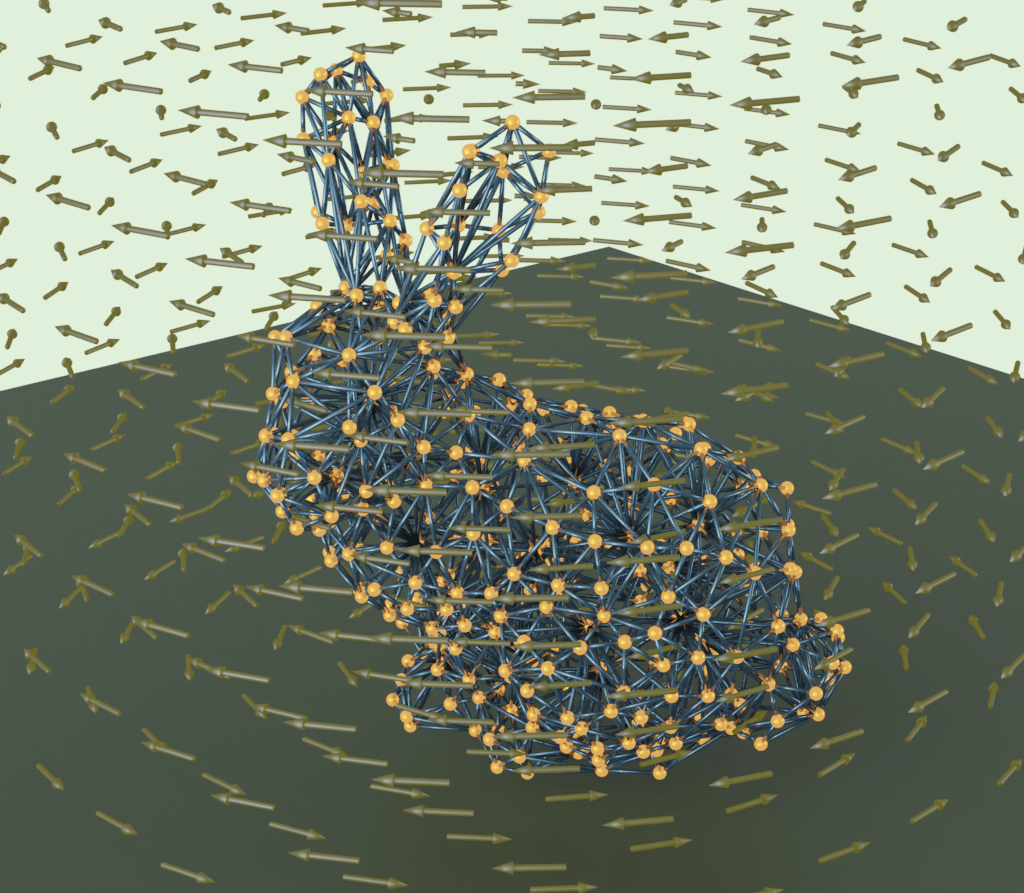} & 
    \includegraphics[width=0.14\textwidth]{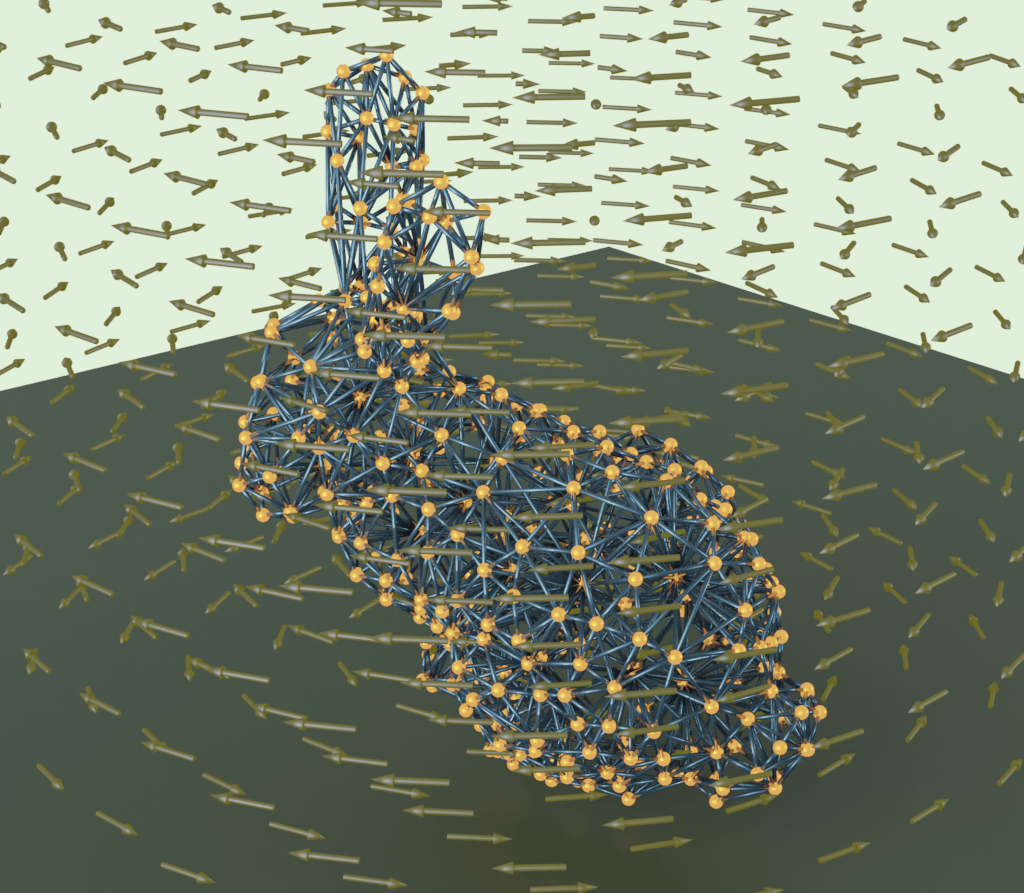} & 
    \includegraphics[width=0.14\textwidth]{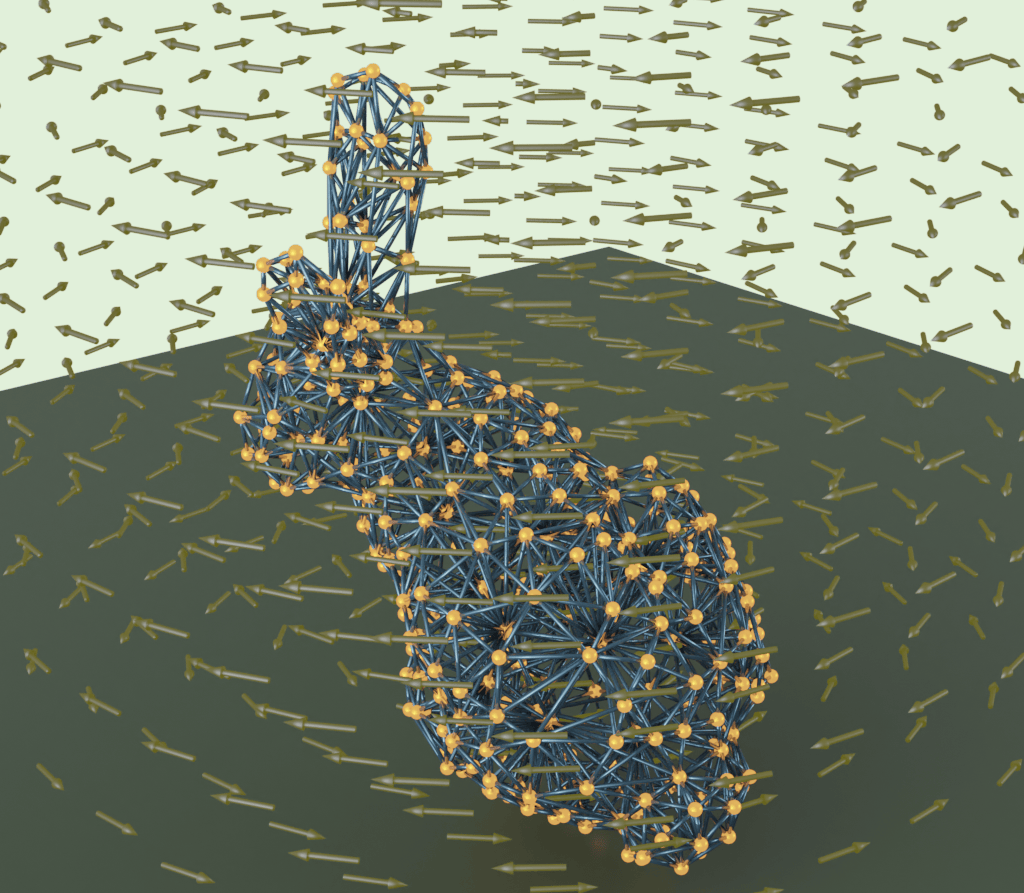} 
    \end{tabular}
    \caption{Using an external electric (vector) field to animate the mesh. Here we chose a vector field defined analytically to cause rotation about the z-axis. The points on the bunny have a charge of $8\mu C$ which leads to a smooth rotation over time. The animation in the accompanying video shows the bunny rotating clockwise.}
    \label{fig:electricField}
\end{figure}

We also demonstrate our system can be controlled with external electric fields, supplied as vector fields. Figure~\ref{fig:electricField} depicts a bunny whose nodes are positively charged embedded in an electric field, the electric field is defined as $\mathbf{e_{\mathrm{ex}}}(x,y,z) = [y/z,\; x/z,\; 0]^T$. In this case, the electric field induces a rotation about the z-axis. The bunny is charged with $8\mu C$, which rotates smoothly. The speed of rotation and direction can be controlled by changing the charge of the bunny or the strength of the external field. Please see the accompanying video for the entire demonstration.

\section{Discussion}

\paragraph{Error in far-field estimation}
The main error in our algorithm (DDEF) for approximating the far-field is due to the interpolation within the tetrahedra. Since we use low-discrepancy sampling to generate the grid points, the tetrahedra are well-shaped and so as the number of grid points $m$ is increased, the grid points are expected to be sufficiently far from charges that contributed far-field to them. Since the magnitude of the electric field decays as the inverse square of distance, the interpolation error cannot grow arbitrarily. Ignoring the far-field, on the other hand, would lead to a significant bias (see Table below). This is not surprising since cut-off methods are avoided for  large systems due to long-tail contributions to the far-field.
%\begin{table}[H]
%    \centering
   
        \begin{tabular}{rccc} 
            \rowcolor[gray]{.93}
            far-field &$m=1$K & $m=2$K & $m=10$K \\
             DDEF (ours) & 0.046 (0.06) & 0.052 (0.03) & 0.059 (0.09)  \\            
              ignored  & 0.078 (0.19) & 0.378 (0.93) & 2.435 (6.68)  
     \end{tabular}
%     \caption{Comparison of mean relative errors (std. dev.) in the Coulomb force calculations when far-field is computed using DDEF (top) vs ignored (bottom), averaged over all the meshes shown in this paper, for different domain discretizations (columns).
%     \label{tab:cutoff}}
%\end{table}

\paragraph{Cuda Implementation}
The far-field computation in DDEF is the computational bottleneck because the far-field charges are the majority in practice. We use CUDA implementation for acceleration while other computations remain on the CPU. The results are up to 10x faster compared to the CPU-parallelized version. It is also possible to evaluate the near-field on the GPU to further optimize the computation speed without extra GPU memory overhead. This is because the far-field computation is separated from the near-field computation. However, in our testing case, even with GPU evaluation of the far-field, the near-field computation still takes a comparatively small amount of time.

\paragraph{N-body Algorithms} 
The existing methods for solving N-body problems, such as Barnes-Hut, FMM, and PMTA, are all efficient on the CPU. However, handling the hierarchical tree structure is challenging for GPU implementation. DDEF is simpler and easier to parallelize, making it suitable for CUDA implementation. FMM can however achieve a slightly better accuracy than DDEF because, instead of using barycentric interpolation, the multipole expansion is more physically accurate and controllable. %Therefore, in future work, we can adopt the multipole expansion in DDEF when interpolating the far-field on charge without breaking the parallelizability of the algorithm.

\paragraph{Projective dynamics}
Although we chose to build on the implicit integrator formulation for springs, it would indeed also be possible to build on a more sophisticated formulation such as projective dynamics (PD) \cite{bouaziz2023projective}. However, since projective dynamics exploits quadratic constraints it is not clear how to incorporate pairwise Coulomb forces into that framework. Another challenge is that a collection of point charges may not have an equilibrium state. We tried a few approaches to address this issue, but found that the proposed approach was the simplest and most robust. We leave the incorporation of Coulomb forces into the projective dynamics framework as future work.

\paragraph{Limitations}
There are two factors affecting error in our combined system: System stiffness and ratio of potentials. When spring constants are low and charges are high, the oscillations caused due to springs are low-frequency while the strong Coulomb interaction (and its inverse dependence on distance) requires high-frequency updating. This is especially important for systems with a large density since the loose springs offer less resistance for point-charges to approach each other closely. The difficulty, in this case, is compounded by the Coulomb potential being much larger than the elastic potential which causes our combined simulation to be dominated by the explicit integrator. Thus, the main limitation of our integrator is that it exhibits large error for systems with low spring constants and high charge densities. A related issue is that simultaneous parameter estimation (for charges and spring constants) for such systems requires careful weighting of the largely different magnitudes of the gradients.

\bibliographystyle{ACM-Reference-Format}
\bibliography{mcs}

\onecolumn
\newpage
% Content that you want to place on a single column page
\twocolumn

\end{document}